\documentclass[prd,aps,twocolumn,preprintnumbers, showpacs, nofootinbib,superscriptaddress,notitlepage]{revtex4-1}
\usepackage{amsmath,epsfig,amssymb}
\usepackage[utf8]{inputenc}
\usepackage{inputenc}
\usepackage{tabularx}
\usepackage[usenames]{color}
\usepackage{ulem} 
\usepackage{bigstrut}
\usepackage{slashed}
\usepackage{multirow}
\usepackage{dsfont}
\usepackage{xcolor}
\usepackage{caption}
\usepackage{graphicx}
\usepackage{float}
\usepackage{subfigure}
\usepackage{subcaption}
\usepackage{svg}
\usepackage{makecell}
\usepackage{lipsum}
\usepackage{import}

\allowdisplaybreaks

\begin{document}

\title{Hybrid renormalization for distribution amplitude of a light baryon in large momentum effective  theory}
\collaboration{\bf{Lattice Parton Collaboration ($\rm {\bf LPC}$)}}

\author{\includegraphics[scale=0.10]{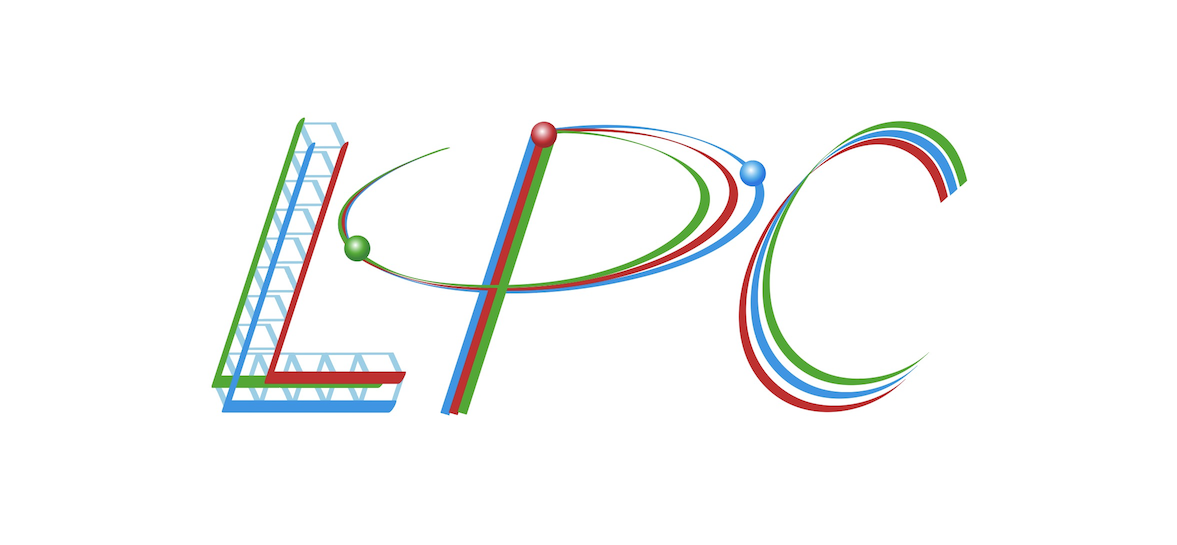}\\Haoyang Bai}
\affiliation{Institute of High Energy Physics, CAS, Beijing 100049, China}
\affiliation{School of Physics, University of Chinese Academy of Sciences, Beijing 100049, China}

\author{Jun Hua}
\email{Corresponding author: junhua@scnu.edu.cn}
\affiliation{Key Laboratory of Atomic and Subatomic Structure and Quantum Control (MOE), 
Guangdong Basic Research Center of Excellence for Structure and Fundamental Interactions of Matter, 
Institute of Quantum Matter, South China Normal University, Guangzhou 510006, China}
\affiliation{Guangdong-Hong Kong Joint Laboratory of Quantum Matter, 
Guangdong Provincial Key Laboratory of Nuclear Science, Southern Nuclear Science Computing Center, 
South China Normal University, Guangzhou 510006, China}

\author{Xiangdong Ji}
\affiliation{Department of Physics, University of Maryland, College 
Park, MD 20742, USA}

\author{Xiangyu Jiang}
\affiliation{CAS Key Laboratory of Theoretical Physics, Institute of Theoretical Physics, Chinese Academy of Sciences, Beijing 100190, China}

\author{Jian Liang}
\affiliation{Key Laboratory of Atomic and Subatomic Structure and Quantum Control (MOE), Guangdong Basic Research Center of Excellence for Structure and Fundamental Interactions of Matter, Institute of Quantum Matter, South China Normal University, Guangzhou 510006, China}
\affiliation{Guangdong-Hong Kong Joint Laboratory of Quantum Matter, Guangdong Provincial Key Laboratory of Nuclear Science, Southern Nuclear Science Computing Center, South China Normal University, Guangzhou 510006, China}

\author{Andreas Sch\"afer}
\affiliation{Institut f\"ur Theoretische Physik, Universit\"at Regensburg, D-93040 Regensburg, Germany}
\affiliation{Department of Physics, National Taiwan University, Taipei, Taiwan 106, China}

\author{Wei Wang}
\email{Corresponding author: wei.wang@sjtu.edu.cn}
\affiliation{State Key Laboratory of Dark Matter Physics, Key Laboratory for Particle Astrophysics and Cosmology (MOE),  Shanghai Key Laboratory for Particle Physics and Cosmology, School of Physics and Astronomy, Shanghai Jiao Tong University, Shanghai 200240, China}
\affiliation{Southern Center for Nuclear-Science Theory (SCNT), Institute of Modern Physics, Chinese Academy of Sciences, Huizhou 516000, Guangdong Province, China}

\author{Yi-Bo Yang}
\affiliation{CAS Key Laboratory of Theoretical Physics, Institute of Theoretical Physics, Chinese Academy of Sciences, Beijing 100190, China}
\affiliation{School of Fundamental Physics and Mathematical Sciences, Hangzhou Institute for Advanced Study, UCAS, Hangzhou 310024, China}
\affiliation{International Centre for Theoretical Physics Asia-Pacific, Beijing/Hangzhou, China}
\affiliation{School of Physical Sciences, University of Chinese Academy of Sciences,
Beijing 100049, China}

\author{Jian-Hui Zhang}
\affiliation{School of Science and Engineering, The Chinese University of Hong Kong, Shenzhen 518172, China}

\author{JiaLu Zhang}
\affiliation{State Key Laboratory of Dark Matter Physics, Key Laboratory for Particle Astrophysics and Cosmology (MOE),  Shanghai Key Laboratory for Particle Physics and Cosmology, Tsung-Dao Lee Institute and School of Physics and Astronomy, Shanghai Jiao Tong University, Shanghai 200240, China}

\author{Mu-Hua Zhang}
\affiliation{State Key Laboratory of Dark Matter Physics, Key Laboratory for Particle Astrophysics and Cosmology (MOE),  Shanghai Key Laboratory for Particle Physics and Cosmology, Tsung-Dao Lee Institute and School of Physics and Astronomy, Shanghai Jiao Tong University, Shanghai 200240, China}

\author{Qi-An Zhang}
\affiliation{School of Physics, Beihang University, Beijing 102206, China}

\begin{abstract}
Lightcone distribution amplitudes for a light baryon can be extracted through the simulation of the quasi-distribution amplitudes (quasi-DAs) on the lattice. 
We implement the hybrid renormalization for the  quasi DAs of light baryons. Lattice simulations are performed using $N_f = 2+1$ stout-smeared clover fermions and a tree-level Symanzik-improved gauge action, with three lattice spacings of ${0.105, 0.077, 0.052}$ fm. By analyzing zero-momentum matrix elements for different lattice spacings, we extract the linear divergence associated with the Wilson-line self-energy. Matching to perturbative matrix elements in the $\overline{\text{MS}}$ scheme yields the residual self-renormalization factors.
Using these factors, we renormalize the quasi-DAs within the hybrid scheme, which combines self-renormalization at large separations and the ratio scheme at short distances. The renormalized results demonstrate effective cancellation of linear divergences and yield smooth, continuum-like coordinate-space distributions suitable for subsequent Fourier transformation and perturbative matching. These results establish the viability of both self and hybrid renormalization frameworks for light baryon quasi-DAs, providing a robust foundation for LaMET-based determinations of light-cone distribution amplitudes in future.
\end{abstract}

\maketitle

\section{Introduction}
Light baryons-particularly protons and neutrons-constitute the vast majority of visible matter in the universe. Understanding their internal structure is one key objective in modern nuclear and particle physics, with profound implications for both theory and experiment. An important theoretical quantity for describing the internal dynamics of baryons at high energy are the baryon light-cone distribution amplitudes (LCDAs), which encode the longitudinal momentum structure of valence quarks in the baryon.

Moreover, the LCDAs of light baryons serve as crucial nonperturbative inputs not only for predicting and interpreting CP-violating observables in weak decays of heavy bottom baryons, but also for the calculation of baryonic form factors and the exploration of possible signals of new physics. For instance,  recently, the LHCb Collaboration reported the first experimental observation of CP violation in a baryonic decay channel, $\Lambda_b^0 \rightarrow p K^{-} \pi^{+} \pi^{-}$~\cite{LHCb:2025ray}, as well as previous evidence for CP violation in the decay $\Lambda_b^0 \rightarrow \Lambda K^{+} K^{-}$~\cite{LHCb:2024yzj}. This milestone result, along with anticipated high-precision measurements in future experiments, places increasing demands on the accuracy of theoretical inputs. 
 {Several related theoretical analyses and predictions can be found in Refs.~\cite{Wang:2015ndk,Han:2022srw,Huang:2022lfr,Han:2024kgz,Han:2025tvc,Lu:2025gjt,Wang:2011uv,Chen:2024fhj}.  
In particular, in Ref.~\cite{Chen:2024fhj} an next-to-leading order analysis of the nucleon form factors using the first moments of baryon LCDAs from Ref.~\cite{Bali:2024oxg} is about an order of magnitude smaller than the experimental value.
Moreover, the analyses in Refs.~\cite{Han:2024kgz,Han:2025tvc} show that parameter uncertainties in model parametrizations for LCDAs already lead to more than $30\%$ uncertainties to physical observables like decay branching fractions and CP violations, while systematic uncertainties are yet unexplored. 
These observations strongly motivate developments of baryon LCDA studies  from first-principles of QCD. }


In contrast to the extensive studies of meson LCDAs-including a few  moments~\cite{Bali:2017ude, RQCD:2019osh} and $x$-dependent distributions~\cite{Zhang:2017bzy,Chen:2017gck,Zhang:2020gaj,Gao:2022vyh,Holligan:2023rex,Hua:2020gnw, LatticeParton:2022zqc,Baker:2024zcd,Cloet:2024vbv} from lattice QCD, as well as phenomenological  approaches such as QCD sum rules~
\cite{Chernyak:1981zz,CLEO:1997fho, BaBar:2009rrj, Belle:2012wwz}
and Dyson-Schwinger equations ~\cite{Chang:2013pq,Cui:2020tdf,Roberts:2021nhw}, the investigation of baryon LCDAs remains significantly less developed. The most commonly used estimates still originate from the Chernyak-Ogloblin-Zhitnitsky (COZ)  model \cite{Chernyak:1987nu}, proposed several decades ago within the QCD sum rule framework. Recently, several lowest moments of octet baryons have been computed on the lattice using the operator product expansion (OPE) \cite{Bali:2015ykx,RQCD:2019hps,Bali:2024oxg}, but these results remain insufficient for the precision required in phenomenological studies of baryonic decays.

 {In the past decade, several inspiring approaches have been proposed to calculate light-cone distributions from lattice QCD, including the large-momentum effective theory (LaMET)~\cite{Ji:2013dva,Ji:2014gla}, the pseudo-distribution approach~\cite{Orginos:2017kos,Radyushkin:2017cyf}, and the lattice cross-section method~\cite{Ma:2014jla}. 
Among these, LaMET has proven particularly effective in enabling the extraction of fully $x$-dependent distribution amplitudes from Euclidean lattice data, and significant progress has been achieved in this framework~\cite{Zhang:2017bzy,Zhang:2020gaj,Holligan:2023rex,Hua:2020gnw,LatticeParton:2022zqc,Baker:2024zcd,Cloet:2024vbv,Xiong:2013bka,Alexandrou:2016eyt,Chen:2017mie,Zhang:2017zfe,Xu:2018mpf,Liu:2018hxv,Wang:2019msf,Zhang:2019qiq,Chen:2020ody,Ji:2020brr,Ji:2020ect,LatticeParton:2020uhz,Lin:2020rxa,Bhattacharya:2021moj,Gao:2021hxl,LatticePartonLPC:2021gpi,Li:2021wvl,Deng:2022gzi,Gao:2022iex,Gao:2022uhg,LatticeParton:2022xsd,LatticePartonCollaborationLPC:2022myp,LatticePartonLPC:2022eev,Zhang:2022xuw,Zhu:2022bja,Deng:2023csv,Ji:2023pba,LatticeParton:2023xdl,Zhao:2023ptv,Liu:2023onm,Avkhadiev:2024mgd,Good:2024iur,Han:2024cht,Han:2024min,Holligan:2024umc,Holligan:2024wpv,Ji:2024hit,Wang:2024wwa,LatticeParton:2024zko,Zhang:2024omt,Bollweg:2025iol,Wang:2025uap,Ji:2025mvk,Chen:2025cxr}. 
Meanwhile, the pseudo-distribution~\cite{Radyushkin:2017lvu,Zhang:2018ggy,Karpie:2018zaz,Joo:2019jct,Joo:2019bzr,HadStruc:2021qdf,Bhat:2022zrw,Kovner:2024pwl,Bhattacharya:2024qpp,HadStruc:2024rix,NieMiera:2025inn} and current-current correlation methods~\cite{Bali:2017gfr,Sufian:2019bol,Bali:2018spj,Sufian:2020vzb,Zimmermann:2024zde} have also produced many interesting and complementary results. }
Recently, the LaMET  has been extended to the leading twist LCDAs of baryons~\cite{LatticeParton:2024vck}, where some preliminary results are given. In this analysis, the quasi DAs are renormalized in a ratio scheme for both short-distance and long-distance spatial separations.  To obtain a robust results for baryon LCDAs,  the renormalization of quasi-DAs for baryons still requires further study, particularly the development of a more systematic scheme to address linear divergences.
Linear divergences pose a significant challenge to the renormalization of nonlocal operators on the lattice, making it difficult to directly extract light-cone distributions from quasi-distributions. In recent years, it has been noticed that  the hybrid scheme and self-renormalization methods~\cite{Ji:2020brr, LatticePartonLPC:2021gpi} enable the proper renormalization of nonlocal operators in lattice QCD. These developments have, in turn, facilitated precision LaMET calculations of hadronic distribution functions. 

The hybrid renormalization scheme, combined with the self-renormalization method, determines the linear divergence by computing nonlocal operators at multiple lattice spacings and extracting the divergence through a parametrized fit of the zero momentum matrix elements. In the perturbative region, these results are then matched to the corresponding perturbative expressions in the $\overline{\text{MS}}$ scheme, yielding renormalization factors that allow conversion to the light-cone scheme. In regions beyond the reach of lattice calculation, asymptotic extrapolation is introduced to circumvent the inverse problem associated with the limited Fourier transform. However, applying this hybrid scheme procedure to baryon LCDAs is significantly more challenging than in the meson cases, as the perturbative expressions for baryon LCDAs typically feature more pronounced physical peaks. This necessitates high-precision numerical simulations at short distances and makes the matching procedure substantially more intricate. In this work, we present a numerical implementation of the hybrid renormalization scheme for baryon LCDAs, laying a solid foundation for future precision calculations.

The rest of this paper is organized as follows. In Sec.~\ref{sec:framework}, we give the definition of the quasi-distribution amplitudes (quasi-DAs) of light baryons and discuss their symmetry properties in coordinate space. Sec.~\ref{sec:frame_hy} reviews the self-renormalization and the hybrid scheme for baryon quasi-DAs. The lattice simulation setup is described in Sec.~\ref{sec:setup}. Numerical results for the self-renormalization procedure, as well as the renormalized quasi-DAs for both the $\Lambda$ and the proton, are presented in Sec.~\ref{sec:numerical}. The final section provides a summary and discusses prospects for future work.
 
\section{Quasi Distribution Amplitudes for light baryon} \label{sec:framework}

\subsection{Definitions of LCDAs and quasi DAs for a  light baryon}

\begin{figure}
\centering
\includegraphics[scale=0.11]{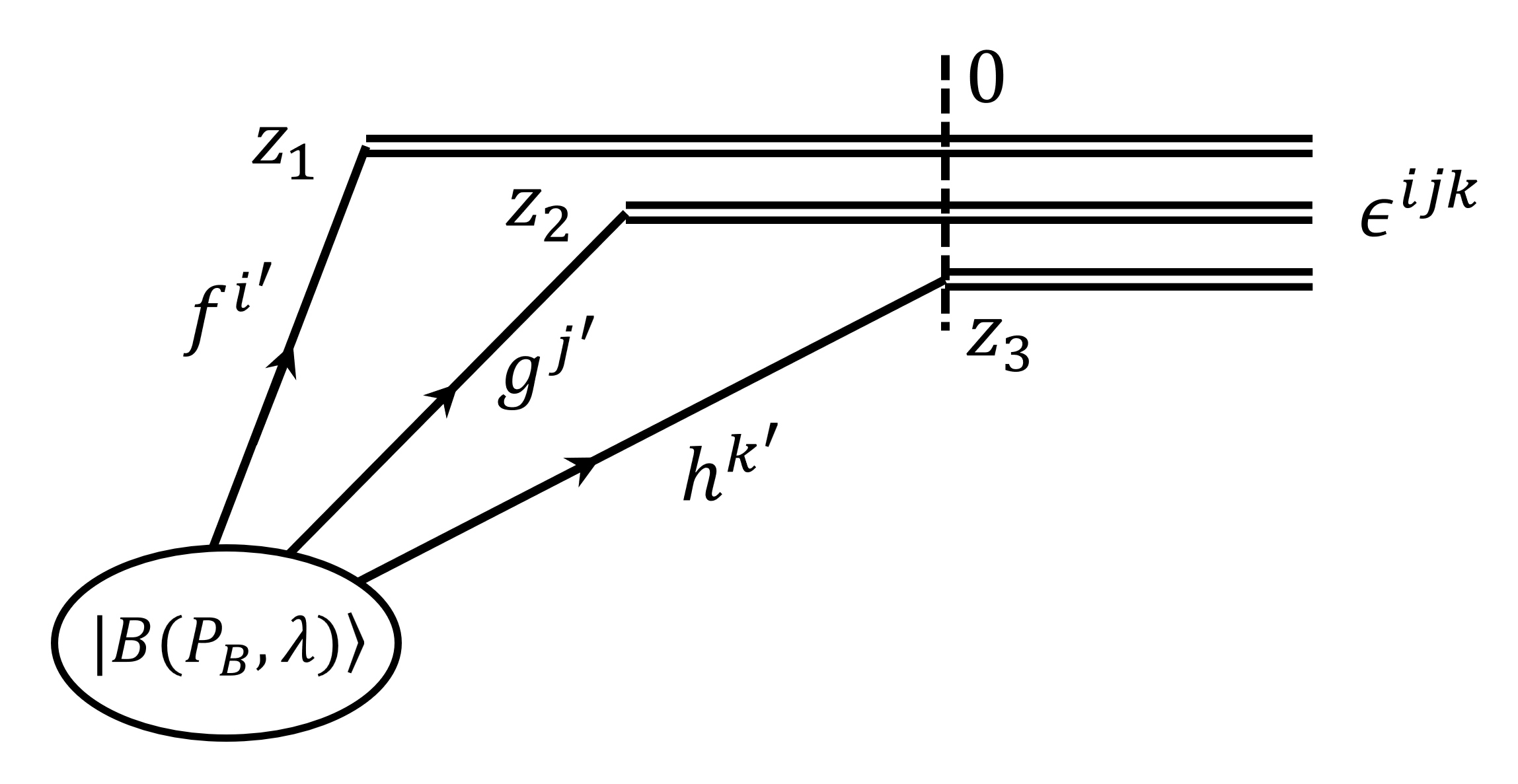}
\caption{The structure of the light baryon LCDA. Three quark fields are connected with Wilson lines to a reference position. $z_3$ can be set to zero for simplification.}
\label{fig:baryonplot}
\end{figure}

The baryon light-cone distribution amplitudes are defined as the hadron-to-vacuum matrix elements of nonlocal three-quark operators at light-like separations \cite{Braun:1999te,Han:2024ucv}: 
\begin{equation}
\begin{split}
{H(z_1,z_2,z_3)}_{\alpha\beta\gamma} 
& = \epsilon^{ijk} \langle 0 | 
f_{\alpha}^{i'}(z_1 n) W^{i'i}(z_1 n, z_0 n) \\
& \times g_{\beta}^{j'}(z_2 n) W^{j'j}(z_2 n, z_0 n) \\
& \times h_{\gamma}^{k'}(z_3 n) W^{k'k}(z_3 n, z_0 n) | B(P_B,\lambda) \rangle,
\label{eq:baryon_matrix}
\end{split}
\end{equation}
where $| B(P_B,\lambda) \rangle$ represents a baryon state with momentum $P_B^\mu=P_B^{+} \bar n^\mu = (P_B^z,0,0,P_B^z)$ and helicity $\lambda$. 
These $\alpha$, $\beta$, $\gamma$ are Dirac indices. $i^{(\prime)}$, $j^{(\prime)}$, $k^{(\prime)}$ refer to color indices, and $f$, $g$, $h$ are the quark fields in each baryon state. 
The light-like Wilson lines \( W_{ij} \) connect different nonlocal quark fields to a general reference position $z_0$ to preserve gauge invariance. Two light-cone unit vectors are defined as $n^\mu=(1,0,0,-1)/\sqrt{2}$ and $\bar n^\mu=(1,0,0,1)/\sqrt{2}$. The above structure can be depicted in Fig.~\ref{fig:baryonplot} for light baryon LCDA.

The matrix element of the LCDAs for an octet baryon in Eq.~(\ref{eq:baryon_matrix}) can be decomposed into three functions at leading twist (twist-3):
\begin{equation}
\begin{split}
\langle 0 | f_{\alpha}(z_1 n) & g_{\beta}(z_2 n) h_{\gamma}(z_3 n) | B(P_B) \rangle \\
&= \frac{1}{4} f_V \Big[ (\slashed{P}_B C)_{\alpha \beta} (\gamma_5 u_B)_{\gamma} V^B (z_i n \cdot P_B) \\
&\quad \quad + (\slashed{P}_B \gamma_5 C)_{\alpha \beta} (u_B)_{\gamma} A^B (z_i n \cdot P_B) \Big] \\
&+ \frac{1}{4} f_T (i \sigma_{\mu \nu} P_B^{\nu} C)_{\alpha \beta} (\gamma^{\mu} \gamma_5 u_B)_{\gamma} T^B (z_i n \cdot P_B),
\end{split}
\end{equation}
where \( C \equiv i \gamma^2 \gamma^0 \) signifies the charge conjugation matrix and \( u_B \) stands for the baryon spinor.

\begin{figure}
\centering
\includegraphics[scale=0.09]{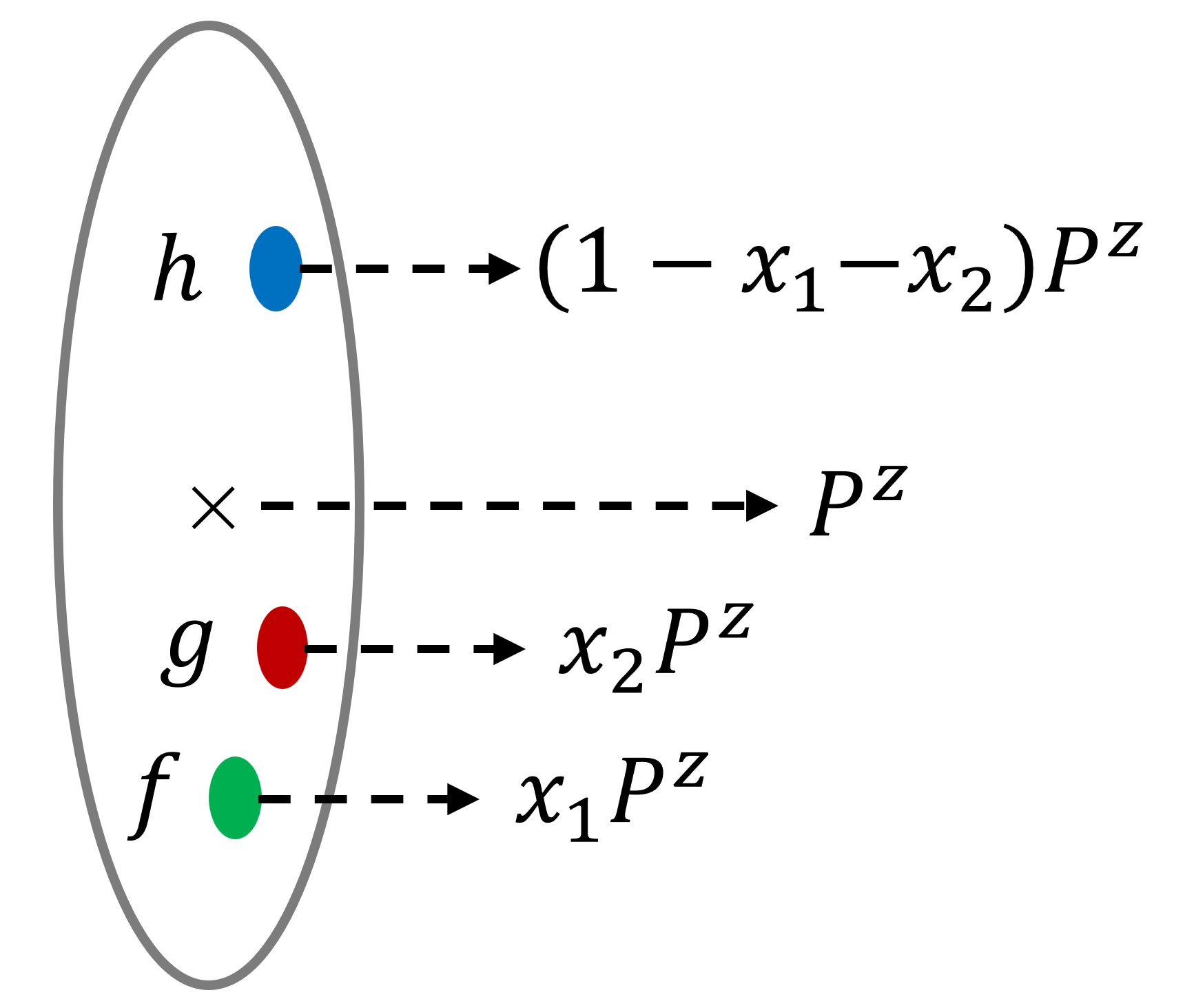}
\caption{The structure of a fast moving light baryon with valance quarks.}
\label{fig:distribution}
\end{figure}

Therefore, the different leading twist components can be projected out from light-cone nonlocal matrix elements as:
\begin{equation}
\begin{split}
&{{\Phi}_V^B}(z_i n \cdot P_B,\mu) (-f_V) P_B^{+} \gamma_5 u_B \\
&= \langle 0 | f^{T}(z_1 n)(C \slashed{n}) g(z_2 n) h(z_3 n) | B(P_B) \rangle,\\[1.5ex]
&{{\Phi}_A^B}(z_i n \cdot P_B,\mu) f_V P_B^{+} u_B \\
&= \langle 0 | f^{T}(z_1 n)(C \gamma_5 \slashed{n}) g(z_2 n) h(z_3 n) | B(P_B) \rangle,\\[1.5ex]
&{{\Phi}_T^B}(z_i n \cdot P_B,\mu) (2 f_T) P_B^{+} \gamma_5 u_B \\
&= \langle 0 | f^{T}(z_1 n)(i C \sigma_{\mu \nu} n^{\nu}) g(z_2 n) \gamma^{\mu} h(z_3 n) | B(P_B) \rangle.
\label{eq:LCDA_terms}
\end{split}
\end{equation}
where $\mu$ represents the renormalization scale.
To simplify the notation, we can set $z_3$ to zero as shown in Fig.~\ref{fig:baryonplot}.
With this convention, the LCDA generally defined in momentum space can be linked to the nonlocal matrix elements in light-cone coordinates by Fourier transformation:
\begin{equation}
\begin{split}
\phi_{V/A/T}(x_1,x_2,\mu) &= \int \frac{P_B^+ dz_1 }{2\pi} \int \frac{P_B^+ dz_2 }{2\pi}  \\
\times & \ e^{i(x_1z_1+x_2z_2)P_B^+} \Phi_{V/A/T}^B(z_1 n,z_2 n,\mu),
\end{split}
\end{equation}
where \( x_1 \) and \( x_2 \) denote the longitudinal momentum fractions of the $f$ and $g$ quarks,  while the remaining quark $h$ carries the fraction \( x_3 = 1 - x_1 - x_2 \), as shown in Fig.~\ref{fig:distribution}. In this work, we focus on the leading twist A-term $\phi^B_A$ of the $\Lambda$ and the V-term $\phi^B_V$ of the proton, as these components remain nonvanishing in the local limit, as will be discussed in the following subsection.

Since lattice QCD cannot directly access time-like operators, an approach proposed by LaMET is to compute quasi-DAs defined in Euclidean space at finite but large hadron momentum. These quasi-DAs can then be matched to the corresponding LCDAs in the large momentum limit via an effective field theory framework. In Euclidean space, the leading twist quasi-DAs of baryons are defined as follows:
\begin{equation}
\begin{split}
& {\tilde{\Phi}_V}^{B}(z_1, z_2, z_3, P_B^{z},\mu) (-f_V) P_B^{z} \gamma_5 u_B \\
&= \left\langle 0 \left| f^{T}(z_1 n_z)(C \gamma^\nu) g(z_2 n_z) h(z_3 n_z) \right| B(P_B) \right\rangle _R, \\[1.5ex]
&{\tilde{\Phi}_A}^{B}(z_1, z_2, z_3, P_B^{z},\mu) (f_A) P_B^{z} u_B \\
&= \left\langle 0 \left| f^{T}(z_1 n_z)(C \gamma_5 \gamma^\nu) g(z_2 n_z) h(z_3 n_z) \right| B(P_B) \right\rangle _R, \\[1.5ex]
&{\tilde{\Phi}_T}^{B}(z_1, z_2, z_3, P_B^{z},\mu) (2 f_T) P_B^{z} \gamma_5 u_B \\
&= \left\langle 0 \left| f^{T}(z_1 n_z) \frac{1}{2} C [\gamma^\nu, \gamma^{\mu}] g(z_2 n_z) \gamma_{\mu} h(z_3 n_z) \right| B(P_B) \right\rangle _R,
\label{eq:quasiDA_terms}
\end{split}
\end{equation}
where $n_z$ is a unit space-like vector along the \( z \) direction, and \( P_B^{z} \) is the hadron momentum along the same direction. The subscript \( R \) refers to an appropriate non-perturbative renormalization scheme on the lattice. 

The definition of the quasi-DA is very similar to the LCDA in Eq.~(\ref{eq:LCDA_terms}), while the quark fields are separated by a Euclidean space distance $z_i$. {The operator’s Dirac structure can be chosen as $\gamma^\nu = \gamma^t$ or $\gamma^\nu = \gamma^z$, both approaching $\gamma^+$ in the large momentum limit.  A possible origin of lattice artefacts for these operators is operator mixing effects. Unfortunately, no dedicated  analysis of such effects exists. However, insights from quark bilinear operator studies in Refs.~\cite{Chen:2017mie,Zhang:2022xuw} indicate that $\gamma^t$-based operators exhibit less mixing than $\gamma^z$-based ones. For this reason,  we choose $\gamma^\nu = \gamma^t$. }
Similarly, the momentum space quasi-DAs can be obtained from coordinate space matrix elements through a Fourier transformation:
\begin{equation}
\begin{split}
\tilde{\phi}_{V/A/T}(x_1,x_2,\mu) &= \int \frac{P^z dz_1}{2\pi} \int \frac{P^z dz_2}{2\pi}  \\
\times & \ e^{-i(x_1z_1+x_2z_2)P^z} \tilde{\Phi}_{V/A/T}^B(z_1,z_2,\mu),
\end{split}
\end{equation}

In lattice simulations, the ground state of nonlocal matrix elements of quasi-DAs in Eq.~\ref{eq:quasiDA_terms} can be extracted by reduction of the corresponding two point correlation functions, defined as: 
\begin{equation}
\begin{split}
C_2(z_1,z_2;t,\vec{P})&=\int d^3xe^{-i\vec{P}\vec{x}}\langle0|\hat{O}_{Sink}(\vec{x},t;z_1,z_2)_{\gamma}\\
&\times{\bar{\hat O}}_{source}(0,0;0,0)_{\gamma'}T^{\gamma\gamma'}|0\rangle,
\label{eq:2pt_definition}
\end{split}
\end{equation}
where $T^{\alpha\beta}$ denotes the projection operator, which is optional for difference cases.
The sink operator $\hat{O}_{Sink}(\vec{x},t;z_1,z_2)_\gamma$ depends on the specific leading twist term \(V/A/T\), as illustrated by the construction of the A-term operator:
\begin{equation}
\begin{split}
&\hat{O}^{A}_{Sink}(\vec{x},t;z_1,z_2)_\gamma \\
&= \epsilon^{ijk}W^{ii'}(z_0,\vec{x}+z_1n_z)f_{\alpha}^{i'}(\vec{x}+z_1n_z,t) \\
&\quad \times {(C \gamma^t)}_{\alpha\beta}W^{jj'}(z_0,\vec{x}+z_2n_z)g_\beta^{j'}(\vec{x}+z_2n_z,t) \\
&\quad \times W^{kk'}(z_0,\vec{x})h^{k'}_{\gamma}(\vec{x},t).
\end{split}
\end{equation}
Here, $W^{ij}(z_0,\vec{x})$ represents the space-like Wilson line along the $z$ direction. By setting the reference point $z_0$ at the h-quark position, it simplifies to:
\begin{equation}
\begin{split}
&\hat{O}^{A}_{Sink}(\vec{x},t;z_1,z_2)_\gamma \\
&= \epsilon^{ijk}W^{ii'}(\vec{x},\vec{x}+z_1n_z)f_{\alpha}^{i'}(\vec{x}+z_1n_z,t) \\
&\quad \times {(C \gamma^t)}_{\alpha\beta}W^{jj'}(\vec{x},\vec{x}+z_2n_z)g_\beta^{j'}(\vec{x}+z_2n_z,t) \\
&\quad \times h^k_{\gamma}(\vec{x},t).
\label{eq:sink_operator}
\end{split}
\end{equation}

\subsection{Symmetries in baryon quasi-DAs}

\begin{table*}[t]
\centering
\begin{tabular}{|c|c|c|c|c|c|c|c|c|}
\hline
\textbf{Octet} & \( n \) & \( p \) & \( \Sigma^- \) & \( \Sigma^0 \) & \( \Sigma^+ \) & \( \Xi^- \) & \( \Xi^0 \) & \( \Lambda \) \\ \hline
\( f \; g \; h \) & d d u & u u d & d d s & \(\frac{(u d s + d u s)}{\sqrt{2}} \) & u u s & s s d & s s u & u d s \\ \hline
\end{tabular}
\caption{The corresponding valence quark \( f, g, h \) for each light octet baryon.}
\label{table:valence_quarks}
\end{table*}

The quasi-DAs of the light octet baryons, defined as nonlocal matrix elements in Eq.~(\ref{eq:quasiDA_terms}), exhibit specific symmetry properties governed by their flavor structures. These symmetries can be utilized to streamline lattice computations. By treating the quasi-DAs ${\tilde{\Phi}}(z_1, z_2)$ as functions of the quark separation coordinates \( z_1 \) and \( z_2 \), we can distinguish different regions in the \( z_1-z_2 \) plane as shown in Fig.~\ref{fig:sym_region}, to highlight the symmetry properties of the baryon quasi-DAs :

\begin{figure}
\centering
\includegraphics[scale=0.17]{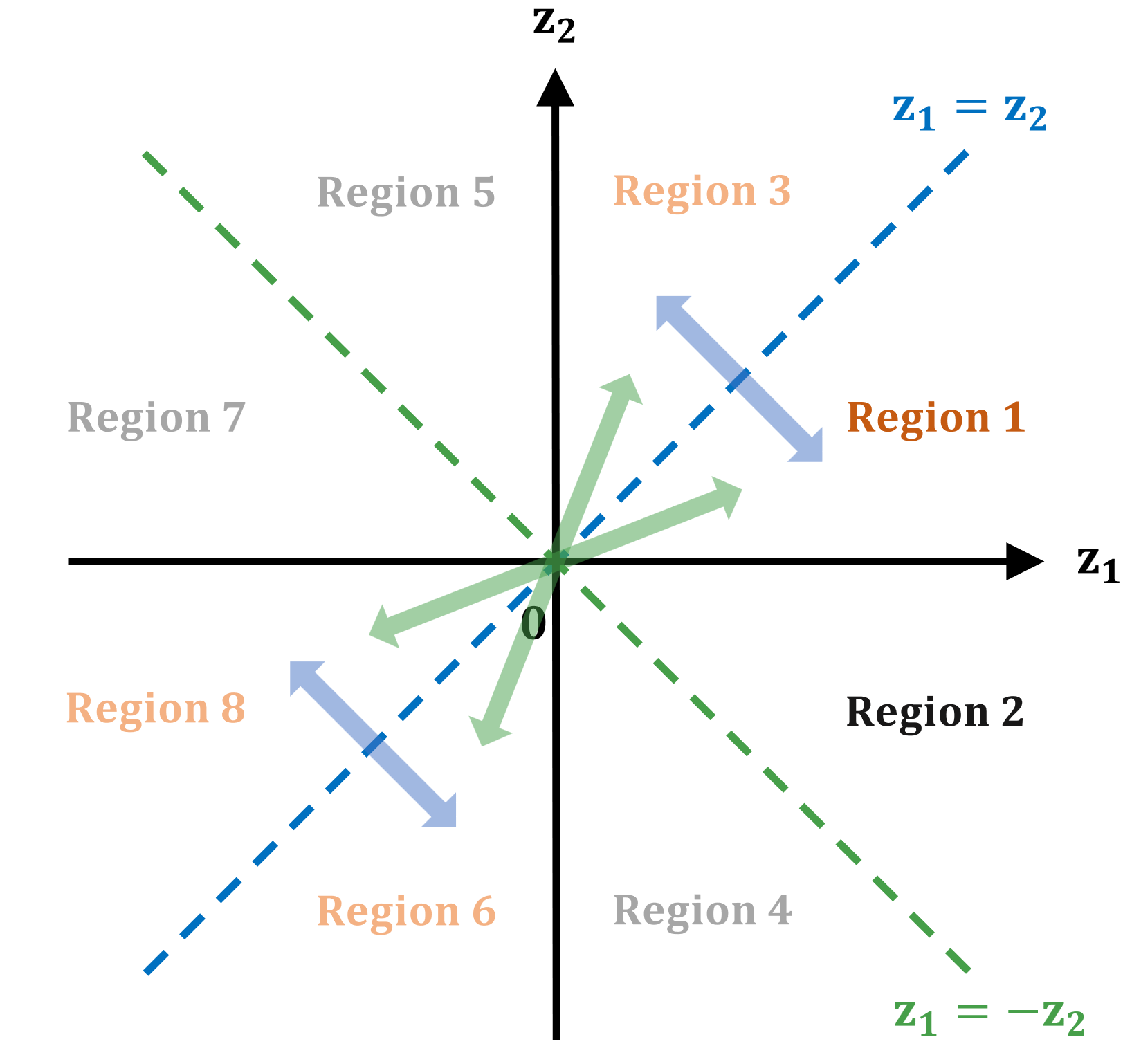}
\caption{The diagonals and regional divisions on the $z_1-z_2$ plane for quasi-DAs.}
\label{fig:sym_region}
\end{figure}

\begin{itemize}
\item The identical quark fields in the matrix elements induce symmetry under the exchange $z_1 \longleftrightarrow z_2$ for all octet baryons except $\Lambda$ and $\Sigma^0$, where $f$ and $g$ are quarks of the same flavor. As a result, the three structures \( V, A, T \) obey specific symmetries:
\begin{align}
    &\begin{aligned}
        \tilde{\Phi}_V^{B}(z_1, z_2) &= \tilde{\Phi}_V^{B}(z_2, z_1), \\
        \tilde{\Phi}_A^{B}(z_1, z_2) &= -\tilde{\Phi}_A^{B}(z_2, z_1), \\
        \tilde{\Phi}_T^{B}(z_1, z_2) &= \tilde{\Phi}_T^{B}(z_2, z_1)
    \end{aligned} \label{sym1}
\end{align}

\item After incorporating isospin symmetry where $f$ and $g$ are light quarks, additional constraints arise for the 
$\Lambda$ and $\Sigma^0$:
\begin{align}
    &\begin{aligned}
        \tilde{\Phi}_V^{\Lambda}(z_1, z_2) &= -\tilde{\Phi}_V^{\Lambda}(z_2, z_1), \\
        \tilde{\Phi}_A^{\Lambda}(z_1, z_2) &= \tilde{\Phi}_A^{\Lambda}(z_2, z_1), \\
        \tilde{\Phi}_T^{\Lambda}(z_1, z_2) &= -\tilde{\Phi}_T^{\Lambda}(z_2, z_1).
    \end{aligned} \label{sym2}
\end{align}
\begin{align}
    &\begin{aligned}
        \tilde{\Phi}_V^{\Sigma^0}(z_1, z_2) &= \tilde{\Phi}_V^{\Sigma^0}(z_2, z_1), \\
        \tilde{\Phi}_A^{\Sigma^0}(z_1, z_2) &= -\tilde{\Phi}_A^{\Sigma^0}(z_2, z_1), \\
        \tilde{\Phi}_T^{\Sigma^0}(z_1, z_2) &= \tilde{\Phi}_T^{\Sigma^0}(z_2, z_1).
    \end{aligned} \label{sym3}
\end{align}

\item Another symmetry originates from the physical requirement that the LCDA in momentum space is pure real. Since the coordinate-space matrix elements are related to the momentum-space quasi-DAs via a Fourier transform:
\begin{align}
    &\begin{aligned}
    \tilde{\Phi}\left(z_1, z_2, \mu\right)= & \int_0^1 d x_1 \int_0^1 d x_2 e^{i\left(x_1 z_1 P^z+x_2 z_2 P^z\right)} \\
    & \times \tilde{\phi}\left(x_1, x_2, \mu\right).
    \end{aligned} \label{eq:quasi_mom}
\end{align}
This imposes a complex-conjugation constraint:
\begin{align}
\tilde{\Phi}(z_1,z_2)=\tilde{\Phi}^*(-z_1,-z_2).
\label{eq:sys_2}
\end{align}
\end{itemize}

Therefore, for the A-term of the $\Lambda$ and V/T-terms of the proton, the identical symmetry properties in coordinate space can be illustrated with regions in Fig.~\ref{fig:sym_region} by the following equalities:
\begin{align}
&\tilde{\Phi}_{1}(z_1,z_2) = \tilde{\Phi}_{3}(z_2,z_1) = \tilde{\Phi}^*_{6}(-z_2, -z_1) = \tilde{\Phi}^*_{8}(-z_1,-z_2), \nonumber \\
&\tilde{\Phi}_{2}(z_1,z_2) = \tilde{\Phi}^*_{4}(-z_2,-z_1) = \tilde{\Phi}_{5}(z_2,z_1) = \tilde{\Phi}^*_{7}(-z_1,-z_2).
\label{eq:sym_prop}
\end{align}
For those V/T-terms of the $\Lambda$ and A-term of the proton, the symmetry properties imply:
\begin{align}
&\tilde{\Phi}_{1}(z_1,z_2) =- \tilde{\Phi}_{3}(z_2,z_1) = -\tilde{\Phi}^*_{6}(-z_2, -z_1) = \tilde{\Phi}^*_{8}(-z_1,-z_2), \nonumber \\
&\tilde{\Phi}_{2}(z_1,z_2) = -\tilde{\Phi}^*_{4}(-z_2,-z_1) = -\tilde{\Phi}_{5}(z_2,z_1) = \tilde{\Phi}^*_{7}(-z_1,-z_2).
\label{eq:sym_prop_2}
\end{align}

{Based on the symmetries discussed above, it is straightforward to see that only regions 1 and 2 in the $z_1$-$z_2$ plane are independent. Since our lattice simulation does not distinguish between u and d quarks, the quasi-DAs are strictly symmetric under the exchange $z_1 \longleftrightarrow z_2$ as illustrated in Eqs. (10)-(12). These relations are satisfied on every individual configuration, thus allowing us to restrict calculations to the lower-right triangular region.   In contrast to the flavor symmetries in Eqs.(10–12),  the symmetry shown in Eq.(14) is a spacetime symmetry and  holds after the configuration average.  This symmetry stems from the fact that the matrix element is purely real in momentum space. On the lattice, the spacetime discretization and quantum fluctuations will disrupt this symmetry on each individual configuration.  In our analysis, we will use the data both from regions 4 and 6 and from regions 1 and 2, and average over them.  }

In addition, Eqs.(15) and (16) can impose further constraints on the quasi-DAs. Specifically, Eq.~\ref{eq:sym_prop} implies:
\begin{equation}
\operatorname{Im}\left[\tilde{\Phi}\left(z_1, z_2\right)\right]_{(z_1=-z_2)} \equiv 0,
\end{equation}
while Eq.~\ref{eq:sym_prop_2} leads to
\begin{align}
&\begin{aligned}
\left[\tilde{\Phi}\left(z_1, z_2\right)\right]_{(z_1=z_2)} &\equiv 0, \\
\operatorname{Re}\left[\tilde{\Phi}\left(z_1, z_2\right)\right]_{(z_1=-z_2)} &\equiv 0.
\end{aligned} 
\end{align}
The corresponding consequence is that the V/T terms of the $\Lambda$ and the A term of the proton vanish in the local limit.

\section{Framework of Hybrid Renormalization}\label{sec:frame_hy}

\subsection{Review of hybrid renormalization}
In this work, we implement the hybrid renormalization method~\cite{Ji:2020brr}, a well-defined scheme for subtracting UV divergences without introducing extra non-perturbative effects at large distances. At short distances, it divides the large momentum matrix element by the zero momentum matrix element (which defines the  the ratio scheme), which eliminates the UV divergences, introduces some of the perturbatively controllable $z$-dependences, and preserves the normalization of a distribution. At large distances, the renormalization factor extracted from self-renormalization~\cite{LatticePartonLPC:2021gpi} is introduced to cancel UV divergences, without introducing uncontrolled IR effects. Over the years this scheme has found a wide range of applications~\cite{Hua:2020gnw,Gao:2021dbh,LatticeParton:2022zqc,Chou:2022drv,Hua:2022wop,Gao:2022iex,LatticeParton:2022xsd,Su:2022fiu,Ji:2022ezo,Gao:2022ytj,Gao:2022uhg,Ji:2022thb,Zhang:2023tnc,Holligan:2023rex,Zhang:2023bxs,Gao:2023lny,Chen:2024rgi,Ji:2024hit,Baker:2024zcd,Cloet:2024vbv,Holligan:2025ydm,Good:2025daz}. 

For the quasi-distribution amplitudes of a light baryon, there are difficulties in applying the hybrid renormalization, both on the theoretical and numerical side. The theoretical issue arises from the mixing regions involving both short and large distances, such as ($|z_1| \ll 1/\Lambda_{\rm QCD}$ and $|z_2| \sim 1/\Lambda_{\rm QCD}$),  ($|z_2| \ll 1/\Lambda_{\rm QCD}$ and $|z_1| \sim 1/\Lambda_{\rm QCD}$), or ($|z_1-z_2| \ll 1/\Lambda_{\rm QCD}$, $|z_1| \sim 1/\Lambda_{\rm QCD}$ and $|z_2| \sim 1/\Lambda_{\rm QCD}$). In those regions, neither the ratio scheme nor self-renormalization can be simply used. However, as shown in one-loop perturbation theory~\cite{Deng:2023csv,Han:2023xbl,Han:2023hgy}, the UV logarithms $\ln(z_1^2)$, $\ln(z_2^2)$, and $\ln((z_1-z_2)^2)$ are factorized out, which means that short and long distances can be treated independently in those mixing regions. Following the logic mentioned above, the hybrid renormalization designed for quasi distribution amplitudes of a light baryon has been proposed in~\cite{Han:2023xbl}, together with the hybrid counterterm at one loop.

{The numerical issue lies in fitting the scheme conversion factor between the lattice calculations and perturbative calculations in the $\overline{\rm{MS}}$ scheme, which is performed in the short-distance region $0 < |z_1|, |z_2|, |z_1-z_2| \ll 1/\Lambda_{\rm QCD}$, for continuum perturbation theory to work.  } This window hardly exists if the lattice spacing $a$ is not small enough. Therefore, the ratio scheme is temporarily used in all regions in a preliminary study of baryon DA~\cite{LatticeParton:2024vck}. In this work, with more precise data at smaller lattice spacings $a$, we implement the hybrid renormalization, which avoids introducing extra non-perturbative effects compared to the ratio scheme.   

\subsection{Self renormalization}
Let us denote the bare lattice matrix element as $M\left(z_1, z_2, 0, P^z, a\right)$, which is related to Eq.~(\ref{eq:quasiDA_terms}) setting $z_3=0$ without subscript $R$. The normalized matrix element is denoted as $\hat{M}\left(z_1, z_2, 0, P^z, a\right) = M\left(z_1, z_2, 0, P^z, a\right)/M\left(0, 0, 0, P^z, a\right)$.

The starting point of the hybrid renormalization scheme is self-renormalization, which extracts the renormalization factor $Z_{R}(z_1,z_2,a,\mu)$ by converting the lattice data to the $\overline{\rm MS}$ scheme. The renormalization factor is an asymptotic series expansion at small lattice spacing with both logarithmic and power dependences, inspired by perturbation theory and parametrized as follows~\cite{Han:2023xbl},
\begin{align}\label{eq:ZRm0}
&Z_{R}(z_1,z_2,a,\mu) = \exp\Big[\left(\frac{k}{a \ln[a \Lambda_{\overline{\rm QCD}}]} 
- m_{0}\right) \tilde{z} \nonumber\\
&+\frac{\gamma_0}{b_0} \ln \bigg[\frac{\ln [1 /(a \Lambda_{\overline{\rm QCD}})]}{\ln [\mu / \Lambda_{\rm \overline{MS}}]}\bigg]+\ln \left[1+\frac{d}{\ln (a \Lambda_{\overline{\rm QCD}})}\right] \nonumber\\
&+ f(z_1,z_2)a^2 \Big]\ ,
\end{align}
where $\displaystyle\left(\frac{k}{a \ln[a \Lambda_{\overline{\rm QCD}}]} 
- m_{0}\right) \tilde{z}$ is the linear divergence~\cite{Chen:2016fxx,Ji:2017oey,Ishikawa:2017faj,Green:2017xeu, Ji:2020brr} and the mass renormalization parameter~\cite{Ji:1995tm,Beneke:1998ui,Bauer:2011ws,Bali:2013pla,Zhang:2023bxs}, $\displaystyle\frac{\gamma_0}{b_0} \ln \bigg[\frac{\ln [1 /(a \Lambda_{\overline{\rm QCD}})]}{\ln [\mu / \Lambda_{\rm \overline{MS}}]}\bigg]+\ln \left[1+\frac{d}{\ln (a \Lambda_{\overline{\rm QCD}})}\right]$ contains the log divergence. 
{$f(z_1,z_2)a^2$ contains discretization effects. In general the explicit forms of discretization effects depend on the details of the lattice action and the relevant operator. 
The lattice configuration has been generated with stout smeared clover fermion action and Symanzik gauge actions. Both the fermion and gauge actions are tadpole improved self-consistently and the anticipated discretization effects are ${\cal O}(a^2)$, see \cite{CLQCD:2023sdb}.  For the operators,  the choice of an ${\cal O}(a^2)$ order parameterization is inspired by an analysis within lattice perturbation theory for the non-local operator with a Wilson line~\cite{Chen:2016fxx}.  } $\tilde{z}$ is the effective length for the linear divergence, which is defined as follows
\begin{eqnarray}
\tilde{z} = \left\{
        \begin{array}{ll}
            |z_1-z_2|, & \quad z_1 z_2 < 0 \\
            {\rm max}\left(|z_1|,|z_2|\right), & \quad z_1 z_2 \geq 0.
        \end{array}
    \right.
\end{eqnarray}
The effective length $\tilde{z}$ can be justified by the effective length of Wilson links~\cite{Han:2023xbl,LatticeParton:2024vck}.

The parameters $b_0=\frac{11 C_A - 2 n_f}{6\pi}$, $\displaystyle\gamma_0=\frac{C_F}{2\pi}\left(5- \frac{7}{4}\delta_{z_1,0}-\frac{7}{4}\delta_{z_2,0}-\frac{3}{2}\delta_{z_1-z_2,0}\right)$ are determined from perturbation theory as shown in Ref.~\cite{Han:2023xbl}.
{We use $\Lambda_{\overline{\mathrm{MS}}}=0.338$ GeV for $n_f=3$, as determined by FLAG2024 \cite{FlavourLatticeAveragingGroupFLAG:2024oxs}, in the running of the renormalized quantities to the ${\overline{\mathrm{MS}}}$ scheme.}
The parameters $k$, $\Lambda_{\overline{\rm QCD}}$ and $f(z_1,z_2)$ are obtained by fitting the $a$-dependence of the bare lattice data $\hat{M}\left(z_1, z_2, 0, P^z=0, a\right)$.
According to the study in Ref.~\cite{LatticePartonLPC:2021gpi}, $k$ and $\Lambda_{\overline{\rm QCD}}$ are strongly correlated. Therefore, a reasonable and convenient approach is to fix $\Lambda_{\overline{\rm QCD}}$ at a physically sensible value and fit $k$ accordingly. In this work, we adopt $\Lambda_{\overline{\rm QCD}}=0.2$ GeV.
The parameters $m_0$ and $d$ are calculated by matching the renormalized lattice data to perturbation theory at short distances,
\begin{align}
\frac{\hat{M}\left(z_1, z_2, 0, 0, a\right)}{Z_{R}(z_1,z_2,a,\mu)}=\hat{M}_{p}\left(z_1, z_2, 0, 0, \mu\right), \label{Eq:SelfMS}
\end{align}
where the perturbative zero-momentum matrix element $\hat{M}_{p}$ at one-loop is
\begin{align}
&\hat{M}_{p}\left(z_1, z_2, 0, 0, \mu\right) = 1 + \frac{\alpha_s C_F}{2 \pi} \left[ \frac{7}{8} \ln\left(\frac{z_1^2 \mu^2 e^{2 \gamma_E}}{4}\right) \right.\nonumber\\
&\left.+ \frac{7}{8} \ln\left(\frac{z_2^2 \mu^2 e^{2 \gamma_E}}{4}\right) + \frac{3}{4} \ln\left(\frac{(z_1-z_2)^2 \mu^2 e^{2 \gamma_E}}{4}\right) + 4 \right],
\label{eq:MSp0}
\end{align}
which can be found in Ref.~\cite{Han:2023xbl}. 

After extracting the renormalization factor $Z_{R}(z_1,z_2,a,\mu)$, the renormalized lattice matrix elements in $\overline{\rm MS}$ scheme can be defined as:
\begin{align}\label{eq:latinMSbar}
\hat{M}_{\overline{\rm MS}}\left(z_1, z_2, 0, P^z, \mu\right) = \frac{\hat{M}\left(z_1, z_2, 0, P^z, a\right)}{Z_{R}(z_1,z_2,a,\mu)},
\end{align}
which will be utilized to calculate the hybrid renormalization scheme matrix elements in the next subsection. 

\subsection{Implementation of hybrid renormalization for baryon quasi-DAs}
\label{subsec:Hybrid renormalization}
After obtaining the renormalization parameters from the self-renormalization procedure, one can apply them to renormalize the bare nonlocal matrix elements of quasi-DAs with nonzero momentum. However, $Z_R\left(z_1, z_2, a, \mu\right)$ in self-renormalization will introduce peaks at $z_{1,2} = 0$ and $z_2 = z_1$, reflecting the short-distance logarithmic behavior of the perturbative quasi-DA. Such singularities not only impact the numerical quasi-DAs but also manifest in the matching kernel for converting the self renormalized scheme to the light-cone scheme, introducing intractability to both theoretical calculations and numerical processing~\cite{LatticePartonCollaborationLPC:2021xdx}. To address these singularities in scheme conversions, we implement a hybrid renormalization scheme~\cite{Ji:2020brr}.

The hybrid renormalization scheme can be viewed as an improvement of self-renormalization, by combining the ratio scheme at short distances with the self-renormalization at large distances. Three typical regions exist in the hybrid renormalization method for baryons: the short-distance, large-distance, and mixing regions. As shown in~\cite{Han:2023xbl}, the overall matrix elements in the hybrid renormalization scheme for baryon quasi-DAs follow:
\begin{widetext}
\begin{align}\label{eq:Mhybrid}
&\hat{M}_{H}(z_1,z_2,0,P^z) = \frac{\hat{M}_{\overline{\rm MS}}\left(z_1, z_2, 0, P^z,\mu\right)}{\hat{M}_{\overline{\rm MS}}\left(z_1, z_2, 0, 0,\mu\right)} \left(\theta(2z_s-|z_1|)\theta(z_s-|z_2|)+\theta(z_s-|z_1|)\theta(|z_2|-z_s)\theta(2z_s-|z_2|)\right) \nonumber\\
&+\frac{\hat{M}_{\overline{\rm MS}}\left(z_1, z_2, 0, P^z, \mu\right)}{\hat{M}_{\overline{\rm MS}}\left(z_1, {\rm sign}(z_2)2z_s, 0, 0, \mu\right)}\theta(z_s-|z_1|)\theta(|z_2|-2z_s)+\frac{\hat{M}_{\overline{\rm MS}}\left(z_1, z_2, 0, P^z, \mu\right)}{\hat{M}_{\overline{\rm MS}}\left({\rm sign}(z_1)2z_s, z_2, 0, 0, \mu\right)}\theta(|z_1|-2z_s)\theta(z_s-|z_2|) \nonumber\\
&+\frac{\hat{M}_{\overline{\rm MS}}\left(z_1, z_2, 0, P^z, \mu\right)}{\hat{M}_{\overline{\rm MS}}\left(z_s+(z_1-z_2)\theta(z_1-z_2), z_s+(z_2-z_1)\theta(z_2-z_1), 0, 0, \mu\right)}\theta(|z_1|-z_s)\theta(|z_2|-z_s)\theta(z_s-|z_1-z_2|) \nonumber\\
&+ \frac{\hat{M}_{\overline{\rm MS}}\left(z_1, z_2, 0, P^z, \mu\right)}{\hat{M}_{\overline{\rm MS}}\left(z_s+(z_1+z_2)\theta(z_1+z_2), -z_s+(z_2+z_1)\theta(-z_2-z_1), 0, 0, \mu\right)}\theta(|z_1|-z_s)\theta(|z_2|-z_s)\theta(z_s-|z_1+z_2|) \nonumber\\
&+\frac{\hat{M}_{\overline{\rm MS}}\left(z_1, z_2, 0, P^z, \mu\right)}{\hat{M}_{\overline{\rm MS}}\left({\rm sign}(z_1)z_s, {\rm sign}(z_2)2z_s, 0, 0, \mu\right)}\theta(|z_1|-z_s)\theta(|z_2|-z_s)\theta(|z_1-z_2|-z_s)\theta(|z_1+z_2|-z_s) \ .
\end{align}
\end{widetext}
Here, \(\hat{M}_{\overline{\rm MS}}\left(z_1, z_2, 0, P^z,\mu\right)\) and \(\hat{M}_{\overline{\rm MS}}\left(z_1, z_2, 0, 0,\mu\right)\) represent the large-momentum and zero-momentum matrix elements, respectively, renormalized to the \(\overline{\rm MS}\) scheme using the self-renormalization factor $Z_{R}$. \(z_s\) is the hybrid cut-off chosen to satisfy \(a \ll 2z_s \ll 1/\Lambda_{\rm QCD}\). The \(\theta\) functions are unit step functions that define region divisions and maintain continuity conditions.  For example, \(\theta(z_s - |z_1|)\) corresponds to the short distance interval \(z_1 \in [-z_s, z_s]\). Terms in the denominator such as \({\rm sign}(z_1)z_s\) indicate the values at points \(z_1 = \pm z_s\), with the specific sign chosen to match the quadrant in which \(z_1\) lies.

\begin{figure}[htb]
    \centering
    \includegraphics[width=0.49\textwidth]{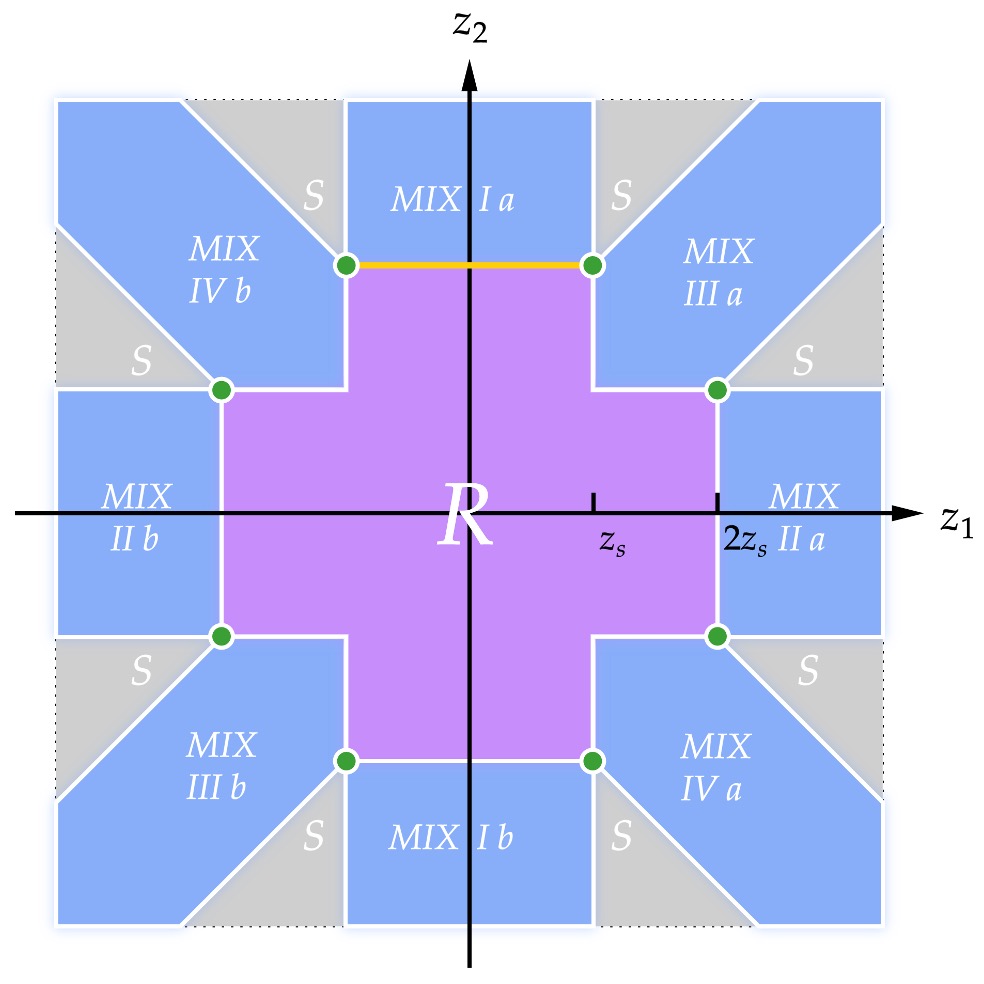}
    \caption{A schematic diagram of renormalization regions. Distinct colors denote different region types: short-distance region (purple), long-distance region (gray), and mixing regions (blue). Separate renormalization schemes are applied to each region.}
    \label{pic:Renorm}
\end{figure}  

To carefully implement the hybrid renormalization in the three different regions above, we first partition the $z_1$-$z_2$ coordinate plane according to the perturbative and non-perturbative regimes of $|z_1|$, $|z_2|$, and $|z_1 - z_2|$, while also considering $|z_1 + z_2|$ for completeness. The regions are categorized as short-distance (perturbative), large-distance (non-perturbative), and mixing regions, as illustrated in Fig.~\ref{pic:Renorm}. Then, we can apply different renormalization procedures in each region according to Eq.~\ref{eq:Mhybrid} with Fig.~\ref{pic:Renorm}.

\begin{itemize}
\item \textbf{Short-distance region} (purple area in Fig.~\ref{pic:Renorm}):
For this region, we apply the ratio scheme:
\begin{equation}
\frac{\hat{M}_{\overline{\rm MS}}\left(z_1, z_2, 0, P^z,\mu\right)}{\hat{M}_{\overline{\rm MS}}\left(z_1, z_2, 0, 0,\mu\right)} \cdot S_{\text{short}}(z_1,z_2)
\end{equation}
where $S_{\text{short}}$ is the step function defining the short-distance region:
\begin{align}
S_{\text{short}} &= \theta(2z_s-|z_1|)\theta(z_s-|z_2|) \\ \nonumber
&+ \theta(z_s-|z_1|)\theta(|z_2|-z_s)\theta(2z_s-|z_2|)
\end{align}
Here, we choose $z_s = 0.20$ fm which satisfies $a \ll 2z_s \ll 1/\Lambda_{\rm QCD}$. This choice cancels singular logarithmic terms in the perturbative region while preserving normalization.

\item \textbf{Long-distance region} (gray areas in Fig.~\ref{pic:Renorm}):  
We use the self renormalized large momentum matrix element at this region. 
\begin{equation}
\frac{\hat{M}_{\overline{\rm MS}}\left(z_1, z_2, 0, P^z, \mu\right)}{\hat{M}_{\overline{\rm MS}}\left({\rm sign}(z_1)z_s, {\rm sign}(z_2)2z_s, 0, 0, \mu\right)} \cdot S_{\text{long}}(z_1,z_2)
\label{eq:long_region}
\end{equation}  
where $S_{\text{long}}$ is the step functions to separate the short or medium distance and long distance regions, and combining with the denominator to address the continuity condition:
\begin{align}
S_{\text{long}}&= \theta(|z_1|-z_s)\theta(|z_2|-z_s)\\ \nonumber
& \times \theta(|z_1-z_2|-z_s)\theta(|z_1+z_2|-z_s)
\end{align}
The value in the denominator is the self renormalized zero-momentum matrix elements at inner boundary points of the gray region (green points in Fig.~\ref{pic:Renorm}), which ensures continuity with the ratio scheme. Here we can only consider the results on four points $({\rm sign}(z_1)z_s, {\rm sign}(z_2)2z_s)$ in Eq.~(\ref{eq:long_region}). The other four points can be derived based on the isospin symmetry shown in Eq.~(\ref{eq:sym_prop}).

\item \textbf{Mixing regions} (blue areas in Fig.~\ref{pic:Renorm}):  
In these regions, one of the nonlocal separation directions lies within the perturbative regions while the others do not. For example, in the blue vertical region \textit{MIX I a} ($z_1$ is perturbative, while $z_2$ and $|z_1-z_2|$ are non-perturbative). We will introduce the ratio scheme for $z_1$ and the self-renormalization for $z_2$ within this region, as shown in the following equation:  
\begin{equation}
\frac{\hat{M}_{\overline{\rm MS}}\left(z_1, z_2, 0, P^z, \mu\right)}{\hat{M}_{\overline{\rm MS}}\left(z_1, {\rm sign}(z_2)2z_s, 0, 0, \mu\right)} \cdot \theta(z_s-|z_1|)\theta(|z_2|-2z_s)
\end{equation}  
The numerator part represents the self renormalized large-momentum matrix elements in this region, while the denominator part represents the self renormalized zero-momentum matrix elements on the boundary line between the self renormalization region and the perturbative region (yellow line in Fig.~\ref{pic:Renorm}).
Therefore, the ratio at identical short-distance $z_1$ dependence cancels the $\ln(z_1^2)$ singularities. The $z_2$ dependence in the numerator is handled by self-renormalization, wile the denominator truncates $z_2$-dependence at $\mathrm{sign}(z_2)2z_s$, avoiding extra non-perturbative effects. Similar treatments apply to other blue regions:  
\begin{widetext}
\begin{align}
&\frac{\hat{M}_{\overline{\rm MS}}\left(z_1, z_2, 0, P^z, \mu\right)}{\hat{M}_{\overline{\rm MS}}\left({\rm sign}(z_1)2z_s, z_2, 0, 0, \mu\right)}\theta(|z_1|-2z_s)\theta(z_s-|z_2|) \nonumber\\
&+\frac{\hat{M}_{\overline{\rm MS}}\left(z_1, z_2, 0, P^z, \mu\right)}{\hat{M}_{\overline{\rm MS}}\left(z_s+(z_1-z_2)\theta(z_1-z_2), z_s+(z_2-z_1)\theta(z_2-z_1), 0, 0, \mu\right)}\theta(|z_1|-z_s)\theta(|z_2|-z_s)\theta(z_s-|z_1-z_2|) \nonumber\\
&+ \frac{\hat{M}_{\overline{\rm MS}}\left(z_1, z_2, 0, P^z, \mu\right)}{\hat{M}_{\overline{\rm MS}}\left(z_s+(z_1+z_2)\theta(z_1+z_2), -z_s+(z_2+z_1)\theta(-z_2-z_1), 0, 0, \mu\right)}\theta(|z_1|-z_s)\theta(|z_2|-z_s)\theta(z_s-|z_1+z_2|)
\end{align}
\end{widetext}
\end{itemize}

\section{Simulation Setup}\label{sec:setup}

\subsection{Lattice setup}
The numerical simulations in this work are based on the $N_f=2+1$ flavor gauge ensembles generated by the CLQCD Collaboration~\cite{CLQCD:2023sdb}. These ensembles have been extensively validated in previous studies of hadron spectroscopy~\cite{Liu:2022gxf,Xing:2022ijm,Yan:2024yuq}, heavy meson physics~\cite{Zhang:2021oja,Liu:2023pwr}, and nucleon structure~\cite{Chen:2024rgi}. In this work, we use three ensembles with lattice spacings of $a=\{0.105,0.077,0.052\}$ fm, which allow for self-renormalization and systematic control for discretization effects. More details are summarized in Table~\ref{tab:ensembles}.
\begin{table}[h]
\centering
\caption{Key parameters for the simulations on the three lattice ensembles. Statistics $N_{\text{cfg}} \times N_{\text{src}}$ denotes the number of measurements on each ensemble. }
\label{tab:ensembles}
\begin{tabular}{lccc}
\hline
\hline
Parameter & C24P29 & F32P30 & H48P32 \\
\hline
Volume ($n_s^3 \times n_t$)  & $24^3 \times 72$ & $32^3 \times 96$ & $48^3 \times 144$ \\
$a$ (fm) & 0.10530(18) & 0.07746(18) & 0.05187(26) \\
$m_{\pi}$ (MeV) & 292.7(1.2) & 303.2(1.3) & 317.2(0.9) \\
$m_K$ (MeV) & 509.4(1.1) & 524.6(1.8) & 536.1(3.0) \\
Statistics ($N_{\rm cfg} \times N_{\rm src}$) & $864 \times 4$ & $777 \times 4$ & $555 \times 6$ \\
\hline
\end{tabular}
\end{table}

To improve data quality at large momenta, we employ point-source propagators with momentum smearing~\cite{Bali:2016lva}. To check the renormalization for both small and large momentum, we compute the quasi-DAs at three different momenta: 
\begin{align*}
\text{C24P29:}\ & P^z = \{0, 0.49, 1.96\}\ \text{GeV}, \\
\text{F32P30:}\ & P^z = \{0, 0.50, 2.00\}\ \text{GeV}, \\
\text{H48P32:}\ & P^z = \{0, 0.50, 1.99\}\ \text{GeV}.
\end{align*}
The zero-momentum case is used to extract renormalization factors, while the nonzero-momentum cases serve to examine the behavior of the renormalized quasi-DAs.

In addition, to enhance the signal-to-noise ratio at large z-separations, we apply single-step hypercubic (HYP) smearing~\cite{Hasenfratz:2001hp,DeGrand:2002vu} to the spatial Wilson lines. It is important to note that HYP smearing modifies the linear divergence of the system, which is handled through self-renormalization. In Appendix~A, we demonstrate that quasi-DAs with different iterations of HYP smearing, despite having distinct linear divergences, yield consistent results after renormalization.

\subsection{Interpolators and projection operators}
\label{subsec:Interpolators and projection operators}
To extract the ground state of the hadron-to-vacuum matrix elements defined in Eq.~(\ref{eq:2pt_definition}), we construct a separate set of composite quark operators at the source side. The structure of the sink side is already fixed by the leading twist structure Eq.~(\ref{eq:quasiDA_terms}) of the baryon LCDAs. To optimize the signal-to-noise ratio (SNR) and better extract the ground state while suppressing the excited-state contamination, we adopt two distinct strategies tailored to different momenta.

\begin{figure}[htbp]
\centering
\includegraphics[width=0.47\textwidth]{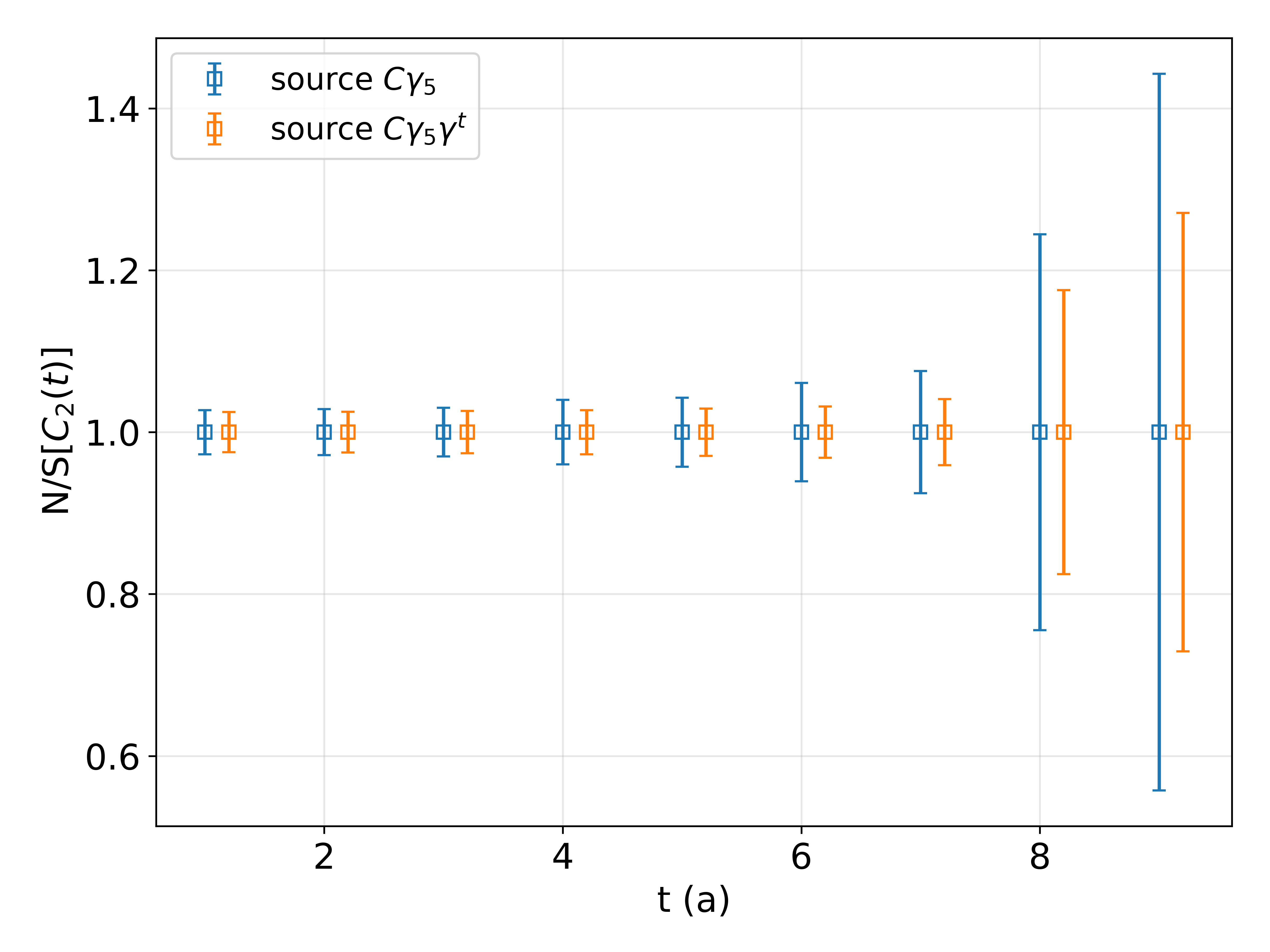}
\caption{Comparison of noise-to-signal ratios for the $\Lambda$ baryon two-point functions at $P^z=2.5$ GeV and F32P30 ($a=0.077$ fm) using traditional (Eq.~\ref{eq:source_zero_p}) and modified (Eq.~\ref{eq:source_boost_p_mod}) source interpolators.}
\label{fig:source_compare}
\end{figure}

For small boost momenta, we typically choose the following operator combinations as interpolators for protons and $\Lambda$ baryons:
\begin{align} \label{eq:source_zero_p}
&\bar{O}^P = (u^TC\gamma_5d)u, \\ \nonumber
&\bar{O}^\Lambda = \frac{ 2(u^TC\gamma_5d)s + (u^TC\gamma_5s)d + (s^TC\gamma_5d)u}{\sqrt{6}} .
\end{align} 
Inspired by Ref.~\cite{Zhang:2025hyo}, for the highly boosted momenta ($P^z \geq 2$ GeV), we adopt the modified interpolating operators that are constructed to better overlap with the leading Fock state in the boosted frame:
\begin{align} \label{eq:source_boost_p_mod}
&\bar{O}^P_{\text{mod}} = (u^TC\gamma_5\gamma^t d)u,\\ \nonumber
&\bar{O}^\Lambda_{\text{mod}} = \frac{ 2(u^TC\gamma_5\gamma^t d)s + (u^TC\gamma_5\gamma^t s)d + (s^TC\gamma_5\gamma^t d)u}{\sqrt{6}}. 
\end{align}
In Fig.~\ref{fig:source_compare}, we show an SNR comparison for large-momentum 2pt correlation functions with the typically source interpolator and the modified operator.

For the two-point baryon correlation functions, the projection operator $T^{\gamma\gamma'}$ shown in Eq.~(\ref{eq:2pt_definition}) is also required for matrix components. Conventionally, the projection operator $T = (I + \gamma^t)/2$  has been widely used in studies of light octet baryons at zero momentum, as it eliminates contributions from negative-parity excited states while projecting onto positive-parity states, including the ground state.

However, with nonzero momentum, the parity can no longer be easily identified through operator constructions, as contributions from states of different parity become mixed. Unlike the zero-momentum case, it is no longer possible to use a fixed projection operator to isolate only the positive-parity states; we can only ensure that the ground state corresponds to the desired positive-parity baryon. Therefore, at large momenta, the choice of projection operator is no longer unique. Instead, we can employ different projection operators and combine them to better extract the ground-state of leading twist.

In summary, our lattice simulation of two-point correlation function employs the following baryon interpolators and projection operators, as listed in Table~\ref{tab:operators_table}.

\begin{table}[htbp]  
\centering  
\caption{Baryon interpolators and projection operators for different boosted momenta.}  
\label{tab:operators_table}  
\begin{tabular}{llll}  
\hline
\hline
Momentum & Baryon & Interpolator & Projection \\  
\hline
\( P^z = 0, 1 \) & Proton & \( (u^T C \gamma_5 d) u \) & \( (I + \gamma^t)/2 \) \\  
& \(\Lambda\) & \( \frac{(2(u^T C \gamma_5 d) s + \cdots) }{\sqrt{6}}\) &  \\  
\hline
\( P^z \geq 4 \) & Proton & \( (u^T C \gamma_5 \gamma^t d) u \) & \( \gamma^t +  \gamma^z \) \\  
& \(\Lambda\) & \( \frac{(2(u^T C \gamma_5 \gamma^t d) s + \cdots) }{\sqrt{6}}\) & \\  
\hline
\end{tabular}  
\end{table}

\subsection{Extraction of ground state matrix elements}
\label{sec:Extraction of matrix elements}
To better suppress the excited-state contamination and reliably extract the ground-state contribution, we perform a two-state fit to the two-point correlation functions in this work. The parameterization follows:
\begin{align}
& C^{\rm norm}_2(z_1,z_2;t,P^z) = \frac{C_2(z_1,z_2;t,P^z)}{C_2(0,0;t,P^z)} \\ \nonumber
& = \tilde\Phi(z_1,z_2,P^z)\left(1 + \Delta A(z_1,z_2,P^z)e^{-\Delta E \cdot t}\right),
\label{eq:two_state_fit}
\end{align}
where \( \tilde\Phi(z_1,z_2,P^z) \) represents the ground-state matrix element, and the \( \Delta A e^{-\Delta E \cdot t} \) term accounts for excited-state contamination. However, it is well known that this fit can be sensitive to the fit range in time $t$. 

\begin{figure}[htbp]
\centering
\includegraphics[width=0.47\textwidth]{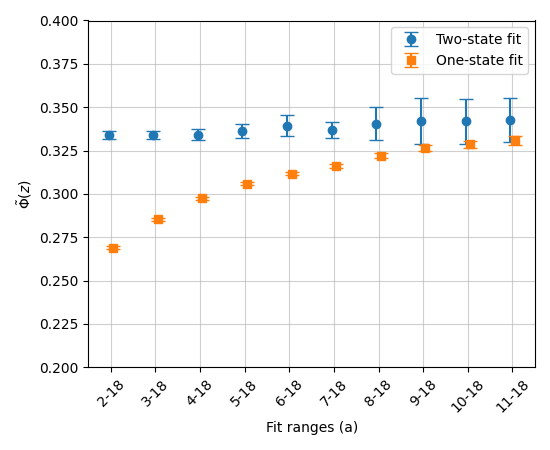}
\caption{Comparison of one-state (constant) and two-state fits for the ground-state quasi-DAs on the normalized two-point function at \( \{z_1,z_2,P^z\} = \{6a,2a,0.5\ \text{GeV}\} \) of F32P30 ($a=0.077$ fm). }
\label{fig:two_state_fit}
\end{figure}

\begin{figure}[htbp]
\centering
\includegraphics[width=0.45\textwidth]{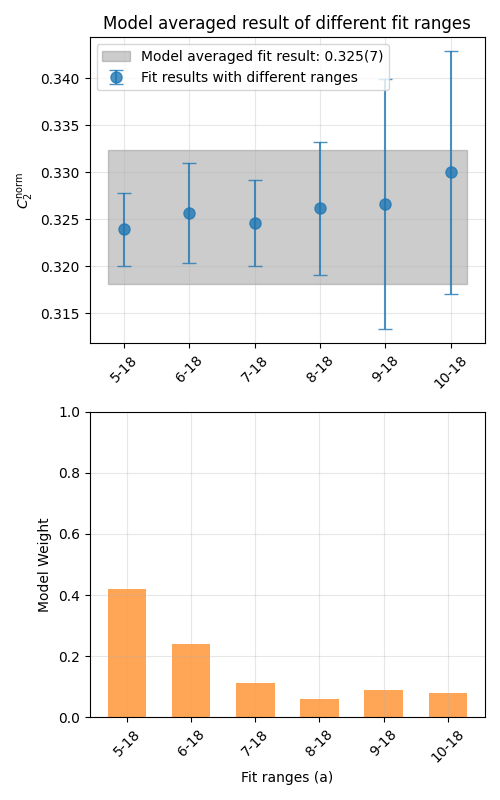}
\caption{Model averaging of $\Lambda$ ground-state matrix elements at \( \{z_1,z_2,P^z\} = \{6a,0a,0.5\ \text{GeV}\} \) for F32P30 ($a=0.077$ fm): (Upper) Fit results for different fit ranges (colored points) and weighted average result with total uncertainty including systematics (gray band); (Lower) Corresponding weights \( w_i \) (bars) of each fit range.}
\label{fig:model_average}
\end{figure}

\begin{figure}[htbp]
\centering
\includegraphics[width=0.47\textwidth]{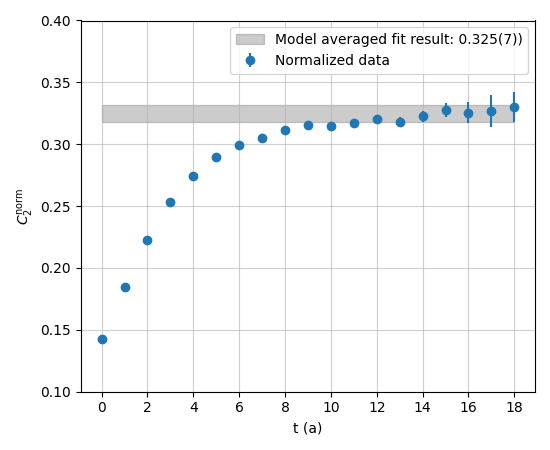}
\caption{Result of the model-averaged two-state fit on normalized matrix element of $\Lambda$ at \( \{z_1,z_2,P^z\} = \{6a,0a,0.5\ \text{GeV}\} \) for F32P30 ($a=0.077$ fm). The colored points are the normalized data while the gray band shows the model averaged fitting result.}
\label{fig:fit_quality}
\end{figure}

To systematically address the fit-range dependence, we implement the model averaging procedure introduced in Ref.~\cite{Jay:2020jkz}, instead of a two-state fit with fixed range. This approach can combine two-state fit results over different fit ranges \( [t_{\min}, t_{\max}] \), with weights for each fit range determined by both the fit quality (\( \chi^2/{\rm dof} \)) and the initial point $t_{min}$, as shown in Eq.~(\ref{eq:weights}). The weighted average result and its uncertainty are given by:
\begin{align}
\bar{\Psi} = \sum_i w_i \Psi_i, \quad \sigma^2_{\bar{\Psi}} = \sum_i w_i \Psi_i^2 - \bar{\Psi}^2 + w_i \sigma_i^2, 
\end{align}
\begin{align}
w_i \propto \exp\left(-\frac{1}{2}\chi^2_i- t_{\min}^{(i)}\right),
\label{eq:weights}
\end{align}
where $\Psi_i$ and $\sigma_i^2$ are the result and uncertainty of each fit,  \( t_{\min}^{(i)} \) represents the initial time point of each fit, $w_i$ are the weights to average all cases. This method accounts for potential systematic uncertainties arising from the choice of fit ranges while avoiding artificially inflated errors. An example comparing the model-averaged fit with two-state fits over different t ranges is shown in Fig.~\ref{fig:model_average}. The upper panel displays the matrix elements extracted from the ground state using various fit strategies, and the lower panel shows the corresponding weights assigned to different values $t_{min}$ in the model averaging procedure.

This combined approach (two-state fitting with model averaging) yields stable extractions of ground-state matrix elements and reliable uncertainty estimates, enabling a reliable determination of the quasi-DA matrix elements necessary for the subsequent renormalization and matching procedures. An example of the fitting results using this method is shown in Fig.~\ref{fig:fit_quality}. Therefore, we will use the quasi-DA matrix elements determined by this method in the following analysis.

\subsection{Dispersion relation}
\label{sec:dispersion}
Since our study involves two-point functions at large momenta, we examine the dispersion relation to assess both discretization effects and the reliability of the calculation. The effective energies \(E(P^z)\) are extracted from the local two-point correlation functions (Eq.~\ref{eq:2pt_definition}) at local points \(z_1=z_2=0\) by applying the model-averaged two-state fit for both the \(\Lambda\) and the proton, with the highest boosted momenta up to \(P^z = 7 \times 2\pi/(n_s a) \simeq 3.5\) GeV on all three ensembles.  

\begin{figure}[htbp]
\centering
\subfigure[\ $\Lambda$]{
    \centering
    \includegraphics[scale=0.36]{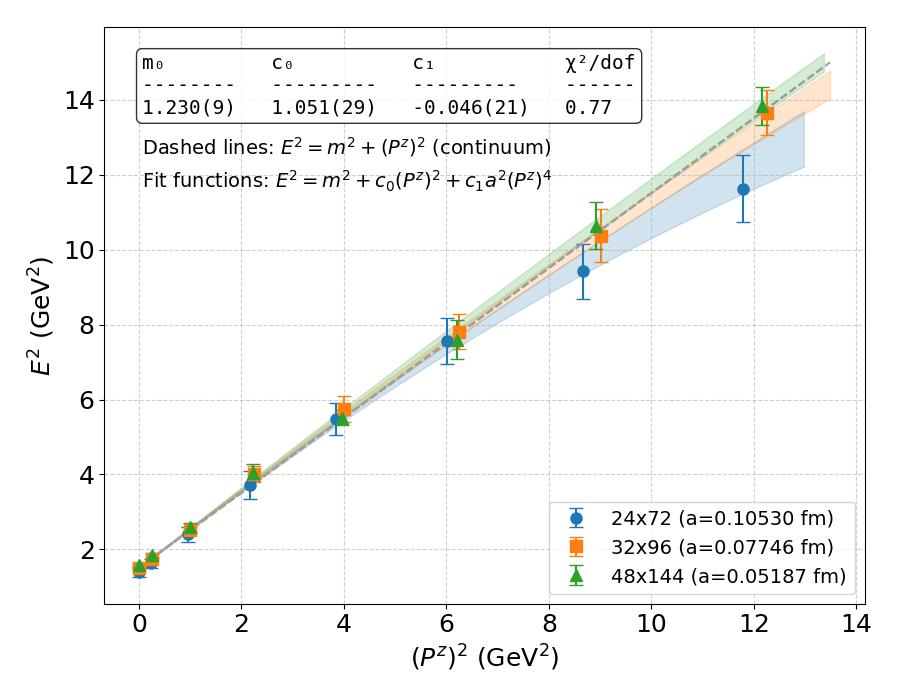}
    }
\vspace{0.0cm} 

\subfigure[\ Proton]{
    \centering
    \includegraphics[scale=0.36]{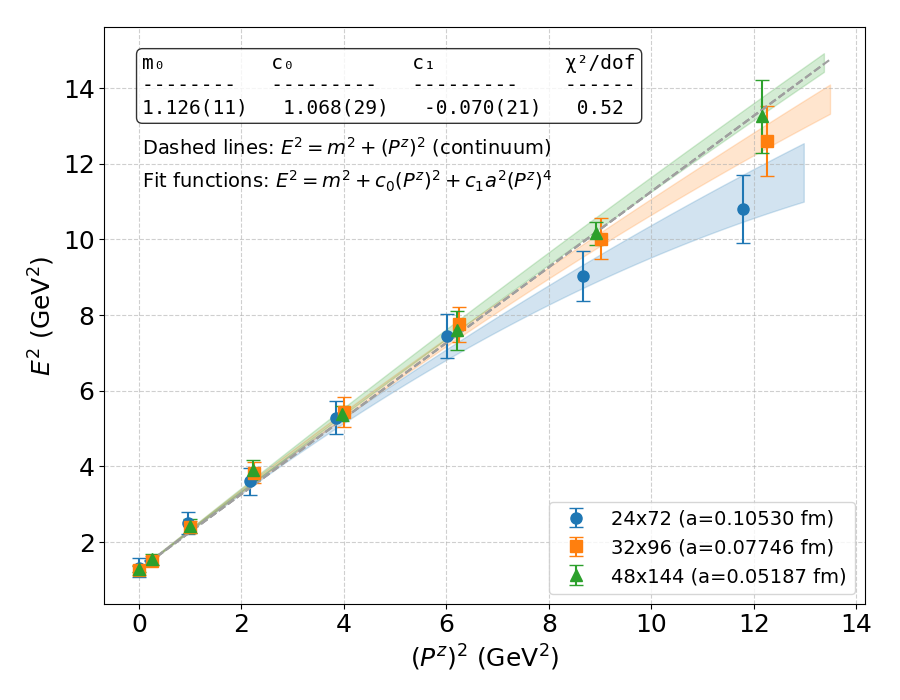}
    }
\caption{Dispersion relations for the $\Lambda$ baryon and the proton on C24P29 ($a=0.105$ fm), F32P30 ($a=0.077$ fm), and H48P32 ($a=0.052$ fm). The dashed lines show the continuum relativistic prediction $E^2 = m_{\Lambda}^2 + (P^z)^2$. }
\label{fig:dispersion_combined}
\end{figure}

The dispersion relations are quantified through the parameterization:  
\begin{align}  
E^2 = m_{B}^2 + c_0 (P^z)^2 + c_1 a^2 (P^z)^4,  
\label{eq:dispers_fit}  
\end{align}  
where \(c_0\) and \(c_1\) parameterize deviations from the continuum relativistic expectation \(E^2 = m_{B}^2 + (P^z)^2\). Figure~\ref{fig:dispersion_combined} displays the results for the three ensembles.  

We find that the values of \(c_0\) and \(c_1\) are consistent with the continuum expectation within 3$\sigma$. The slight deviation of \(c_0\) from 1 is primarily due to complicated excited-state contamination of the $\Lambda$ and the proton, while the deviation of \(c_1\) from 0 reflects discretization effects. Moreover, the results exhibit a trend towards the continuum limit as the lattice spacing decreases, providing theoretical support for the subsequent continuum extrapolation.

\section{Numerical Application for Hybrid Renormalization}\label{sec:numerical}
\begin{figure*}[htbp]
\centering
\subfigure[\ $\Lambda$, $z_1 =0$]{
    \centering
    \includegraphics[scale=0.38]{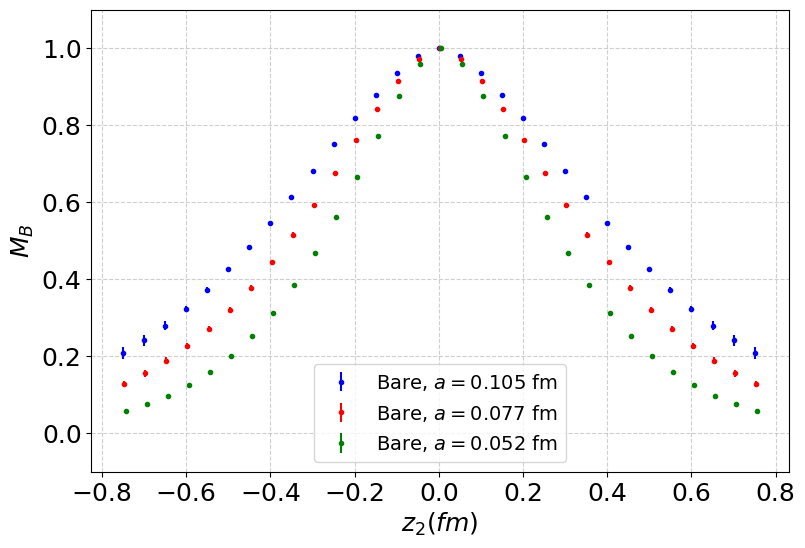}
    }
\vspace{0.0cm} 
\subfigure[\ $\Lambda$, $z_1 = z_2$]{
    \centering
    \includegraphics[scale=0.38]{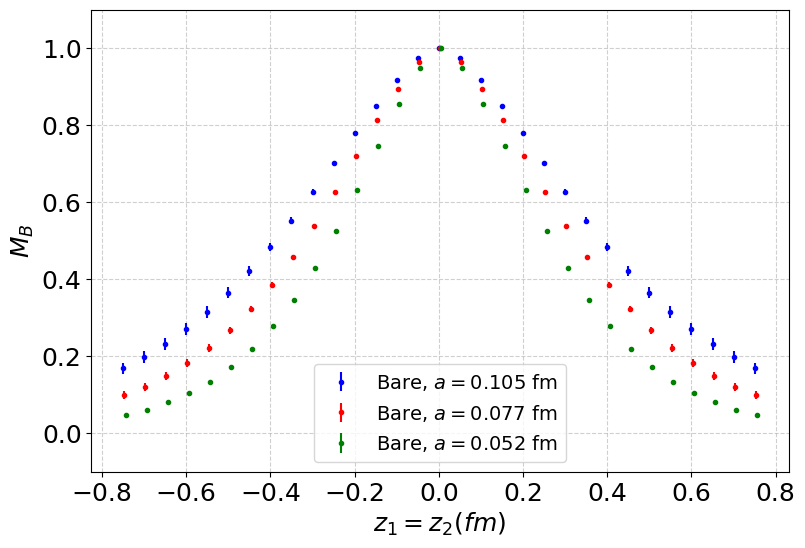}
    }
\vspace{0.0cm} 

\subfigure[\ Proton, $z_1 =0$]{
    \centering
    \includegraphics[scale=0.38]{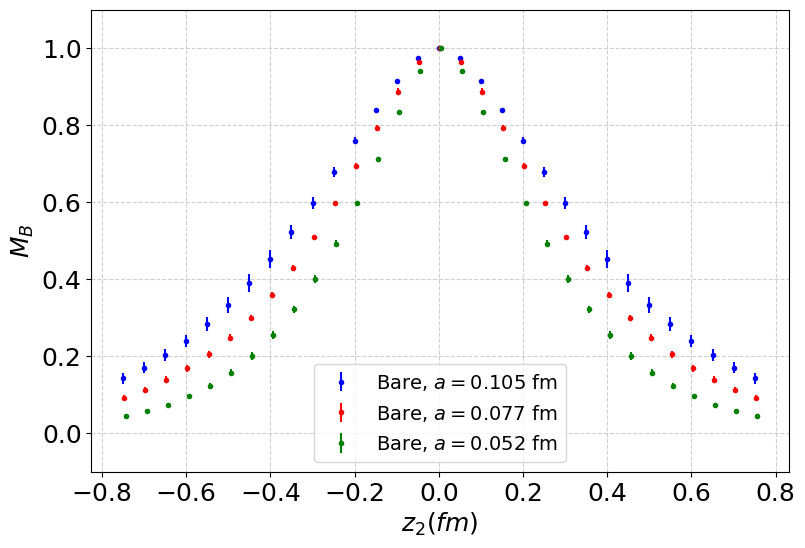}
    }
\vspace{0.0cm} 
\subfigure[\ Proton, $z_1 = z_2$]{
    \centering
    \includegraphics[scale=0.38]{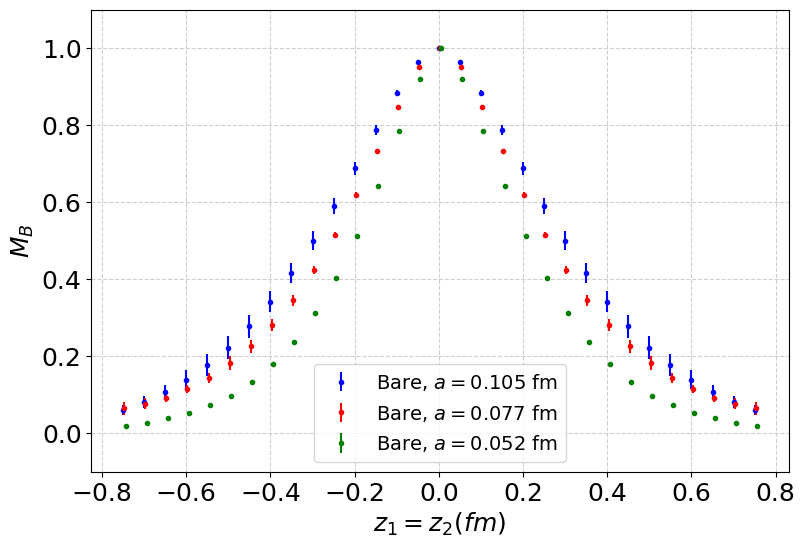}
    }
\caption{Normalized bare zero-momentum quasi-DA for the proton and the $\Lambda$ at three different lattice spacings.}
\label{fig:Normed_Bare}
\end{figure*}

As discussed in the framework of hybrid renormalization, lattice calculations involving nonlocal operators suffer from linear divergences. These divergences manifest themselves as significant deviations in the computed values of the same physical quantity at different lattice spacings. In the cases of baryon quasi-DAs, which involve nonlocal separations in two directions, we illustrate the impact of linear divergence more clearly by fixing either 
$z_1$ or $z_2$. As shown in Fig.~\ref{fig:Normed_Bare}, the left panel presents the bare quasi-DA of the $\Lambda$ baryon at fixed $z_1=0$, while the right panel shows the case with $z_1=z_2$. One can observe significant differences among the results obtained at three different lattice spacings, far exceeding the expectations of discretization effects. Therefore, a proper renormalization of linear divergences is essential to ensure reliable matching to the $\overline{\mathrm{MS}}$ scheme and to obtain a meaningful continuum limit.
{To facilitate comparison and enable consistent analysis of lattice spacing dependence within the self-renormalization scheme, we interpolate the bare matrix element results from each ensemble onto a common grid with a spacing of 0.05 fm. }

\subsection{Linear and logarithmic divergence extraction}
The matrix element of a baryon quasi-DA can be parameterize as 
\begin{equation}
\begin{aligned}
& M\left(z_1, z_2, a, \mu\right)=\exp \left[\left(\frac{k}{a \ln \left[a \Lambda_{\overline{\rm QCD}}\right]}\right) \tilde{z} + g(z_1,z_2)\right. \\
& +\frac{\gamma_0}{b_0} \ln \left[\frac{\ln \left[1 /\left(a \Lambda_{\overline{\rm QCD}}\right)\right]}{\ln \left[\mu / \Lambda_{\overline{\mathrm{MS}}}\right]}\right]+\ln \left[1+\frac{d}{\ln \left(a \Lambda_{\overline{\rm QCD}}\right)}\right] \\
& \left.+f\left(z_1, z_2\right) a^2\right], \label{Eq:SelfMz}
\end{aligned}
\end{equation}
which is similar to Eq.~\ref{eq:ZRm0}, but with 
\begin{equation}
\begin{aligned}
g(z_1,z_2) = \mathrm{In}\left[\hat{M}_{\overline{\mathrm{MS}}}\left(z_1, z_2, 0,0, \mu\right) + m_0 \tilde{z} \right].\label{Eq:Selfgz}
\end{aligned}
\end{equation}
Here $\hat{M}_{\overline{\mathrm{MS}}}\left(z_1, z_2, 0,0, \mu\right)$ is the numerical zero-momentum matrix elements in $\overline{\mathrm{MS}}$ scheme, which is consistent with the perturbative $\hat{M}_p\left(z_1, z_2, 0,0, \mu\right)$ at short distance. The $\mathrm{\overline{MS}}$ renormalization scale $\mu$ is chosen as $\mu=2$ GeV in our following analysis.

\begin{figure*}[htbp]
\centering
\subfigure[\ $\Lambda$, $z_1 = -0.20$ fm]{
    \centering
    \includegraphics[scale=0.36]{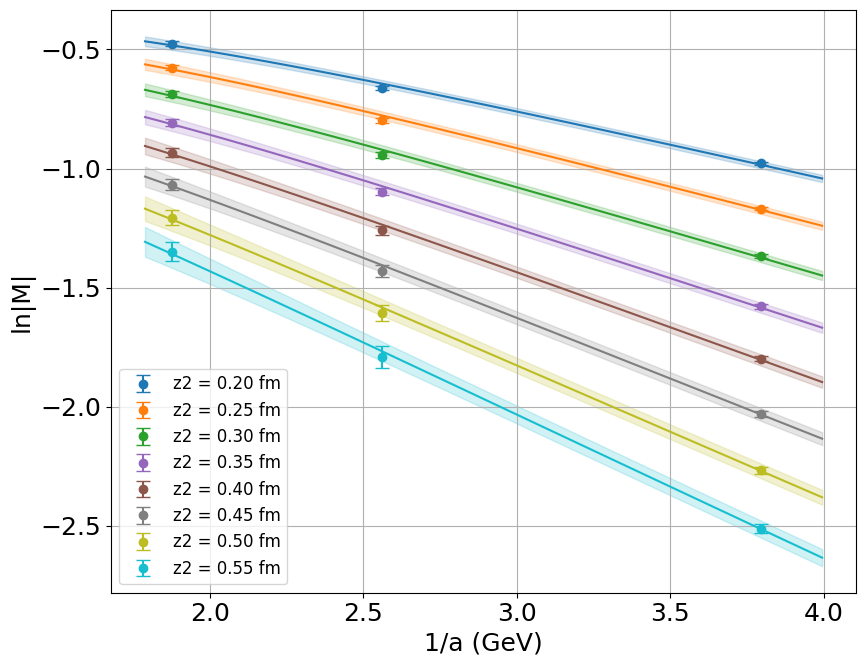}
    }
\vspace{0.0cm} 
\subfigure[\ Proton, $z_1 = -0.20$ fm]{
    \centering
    \includegraphics[scale=0.36]{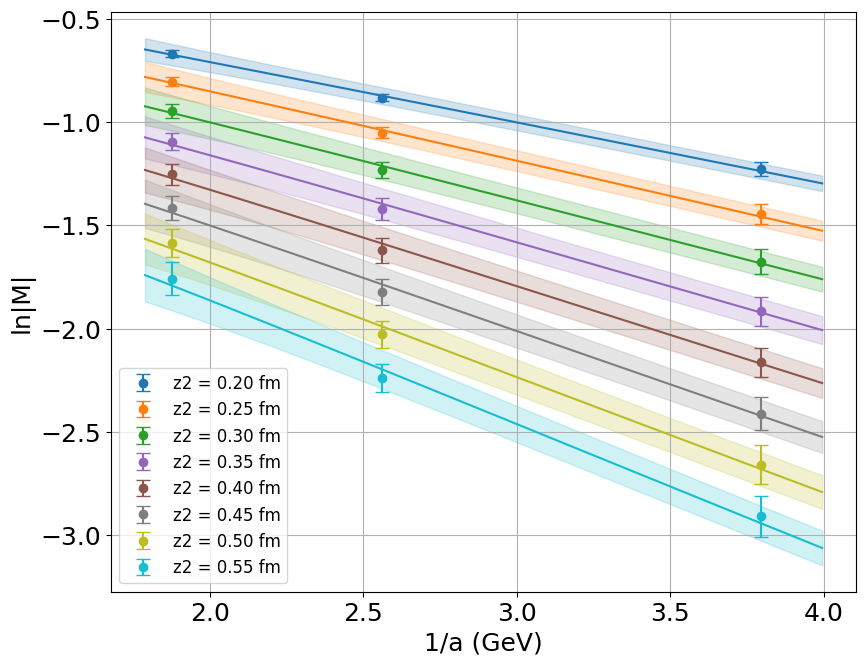}
    }
\caption{Extraction of the linear divergence with lattice spacing dependence on zero-momentum matrix elements of the proton and the $\Lambda$ at $z_1 = -0.20$ fm.}
\label{fig:linear_divergence_z1z2_0}
\end{figure*}

From the analytical expression for the matrix element Eq.~\ref{Eq:SelfMz}, it is obvious that for a given nonlocal matrix element, when the separation $z$ is relatively large while the lattice spacing is small, the dominant contribution in the matrix element arises from the linear divergence. To better illustrate both the impact of this linear divergence on the numerical lattice data, we can examine the dependence of the matrix elements on $1/a$ using a logarithmic scale. As shown in Fig.~\ref{fig:linear_divergence_z1z2_0}, both the $\Lambda$ and the proton exhibit a clear linear dependence on 1/a. Since the baryon LCDA involves two nonlocal separations,  $z_1$ and $z_2$, we present several representative cases with fixed $z_1 = -0.20$ fm, respectively, to illustrate the linear behavior, more cases are shown in Appendix \ref{Linear_Dis}.

Therefore, following the analytical parameterization form in Eq.~\ref{Eq:SelfMz}, we perform a global fit to the numerical matrix elements of the quasi-DAs. Compared to the meson case, which involves only a single nonlocal separation direction, the baryon quasi-DA features two nonlocal separation directions, imposing stronger constraints on determining the linear divergence parameter. This also provides more flexibility in selecting the fitting region, allowing us to avoid potential lattice artifacts, such as discretization effects. Specifically, we use the region satisfying $|z_{1,2}|, |z_1-z_2|>0.15$ fm to fit the linear divergence.
The extracted linear divergence parameters for the $\Lambda$ and the proton are $k^{\Lambda}=0.779(03)$ with $\chi^2/\mathrm{d.o.f.}=1.0$, and $k^P=0.781(05)$ with $\chi^2/\mathrm{d.o.f.}=0.9$, respectively. These consistent values indicate that the linear divergence is largely independent of the external state. 
More consistency checks of linear divergence can be seen in the Appendix \ref{Linear_Dis}. Therefore, we take the extracted linear divergence values $k^{\Lambda,P}$ and perform self-renormalization for the $\Lambda$ and the proton separately.

\begin{figure}[htbp]
\centering
\subfigure[\ $\Lambda$]{
    \centering
    \includegraphics[scale=0.36]{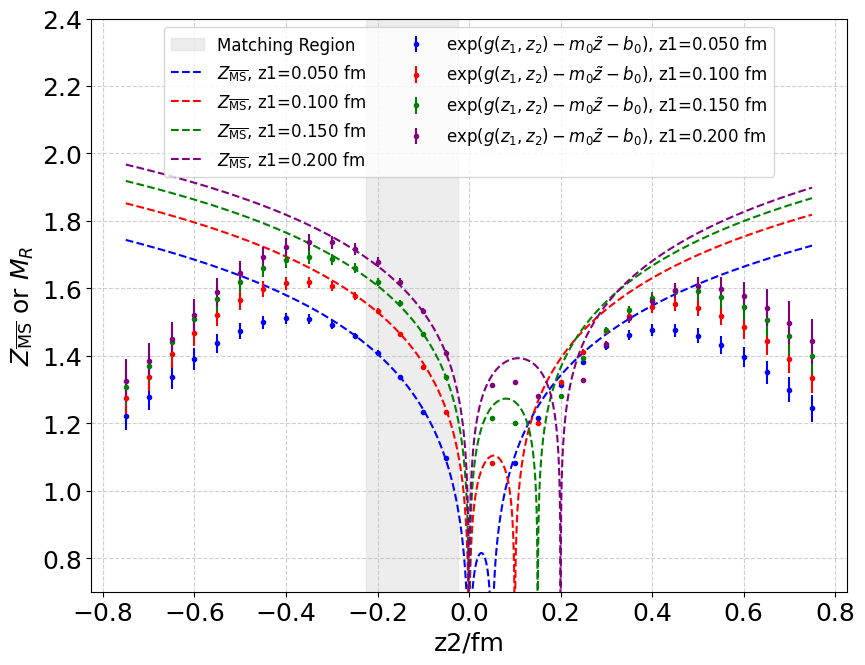}
    }
\vspace{0.0cm} 

\subfigure[\ Proton]{
    \centering
    \includegraphics[scale=0.36]{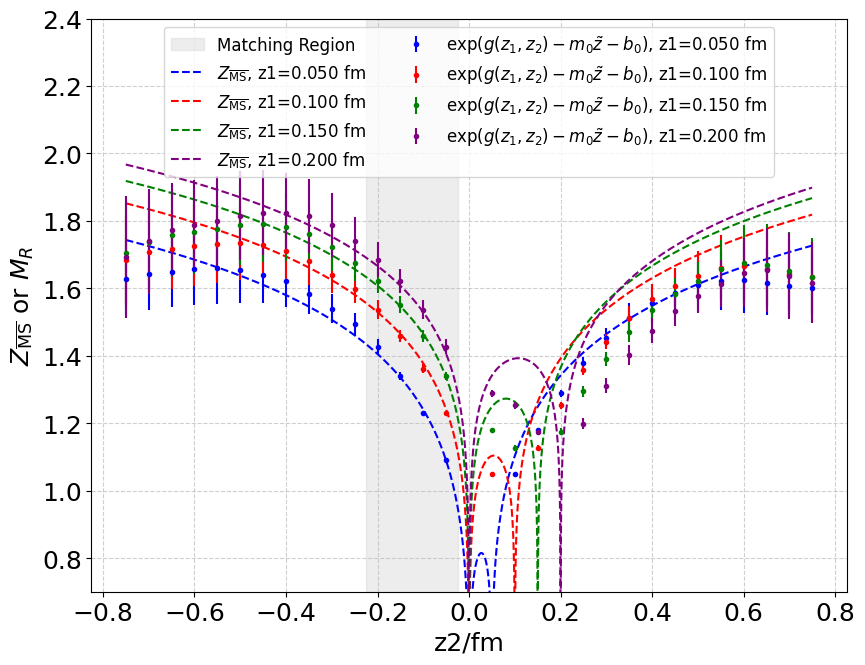}
    }
\caption{The fitting result of the lattice matrix elements to the $\overline{\mathrm{MS}}$ scheme perturbative one-loop result in the perturbative region for the proton and the $\Lambda$.}
\label{fig:self_fitting}
\end{figure}

To further extract the renormalization parameters required for self-renormalization, one needs to match the lattice-calculated quasi-DA to the perturbative expression in the $\overline{\text{MS}}$ scheme within the perturbative region, as discussed in Eq.~\ref{eq:MSp0}. {The matching is performed in the short-distance region to allow the application of perturbation theory.} It is important to note that the perturbative expressions exhibit peaks at $z_1=z_2$, while the numerical lattice data are smooth and continuous for all regions. It makes the matching between discrete lattice data and the continuum expressions particularly challenging, not only for the discrete points along $z_1=z_2$ but also for those nearby. To suppress the impact of perturbative peaks on the matching, we fit the data from quadrants II and IV. The extracted renormalization parameters are general and can be applied to test cases in quadrants I and III, including $z_1 = z_2$. As shown in Fig.~\ref{fig:self_fitting}, for the cases with $z_1 = 0.05, 0.10, 0.15, 0.20$ fm, the lattice data for both the $\Lambda$ and the proton agree well with the perturbative zero-momentum matrix elements when $z_2 < 0$. However, for $z_2 > 0$, only the data points away from the $z_1 = z_2$ region show good agreement with the perturbative results, for example, $z_1 = 0.05$ with $z_2 = 0.15, 0.20$ fm, or $z_1 = 0.20$ with $z_2 = 0.05$ fm. The renormalization parameters extracted and applied in Fig.~\ref{fig:self_fitting} are $m^{\Lambda}_0=0.859(09)$ with $d^{\Lambda}=0.476(04)$ for the $\Lambda$ and $m^{P}_0=0.073(22)$ with $d^{P}=0.381(09)$ for the proton. 

\subsection{Hybrid renormalization results for baryon quasi-DA}

\begin{figure*}[htbp]
\centering
\subfigure[\ Bare result of $\Lambda$ at $P=0.5$ GeV]{
    \centering
    \includegraphics[scale=0.38]{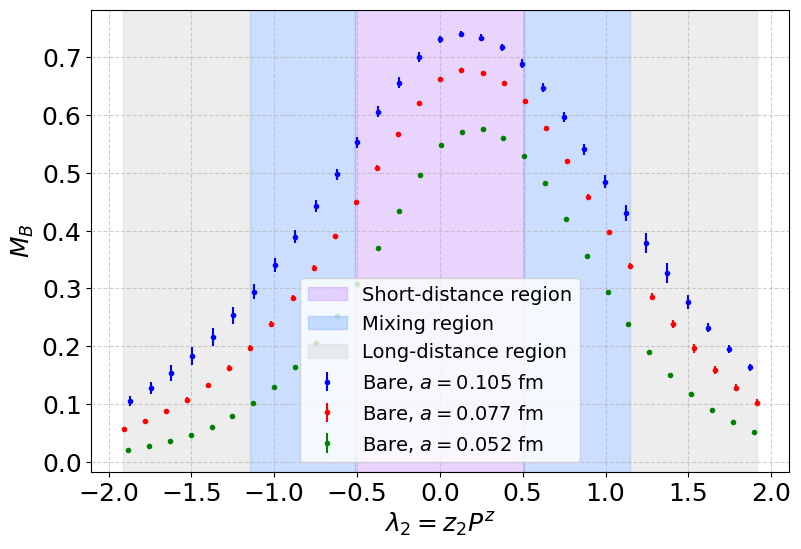}
    }
\vspace{0.0cm} 
\subfigure[\ Ratio scheme result of $\Lambda$ at $P=0.5$ GeV]{
    \centering
    \includegraphics[scale=0.38]{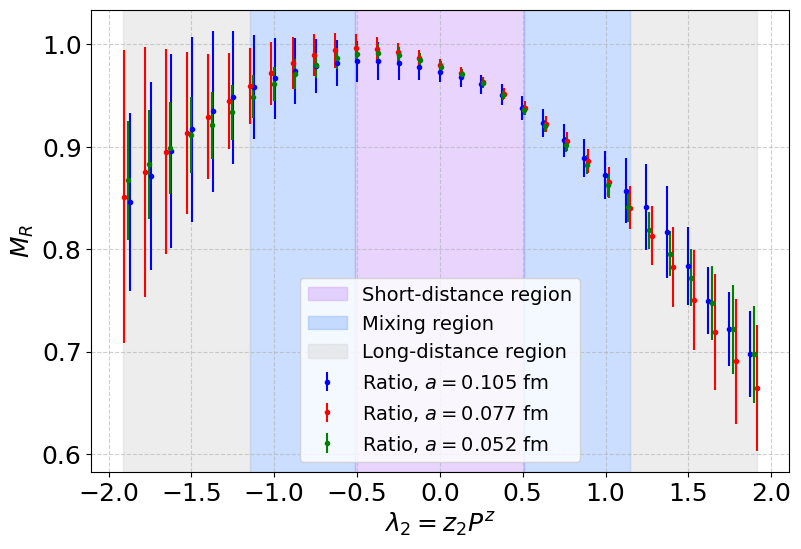}
    }
\vspace{0.0cm} 
\subfigure[\ Self renormalized result of $\Lambda$ at $P=0.5$ GeV]{
    \centering
    \includegraphics[scale=0.38]{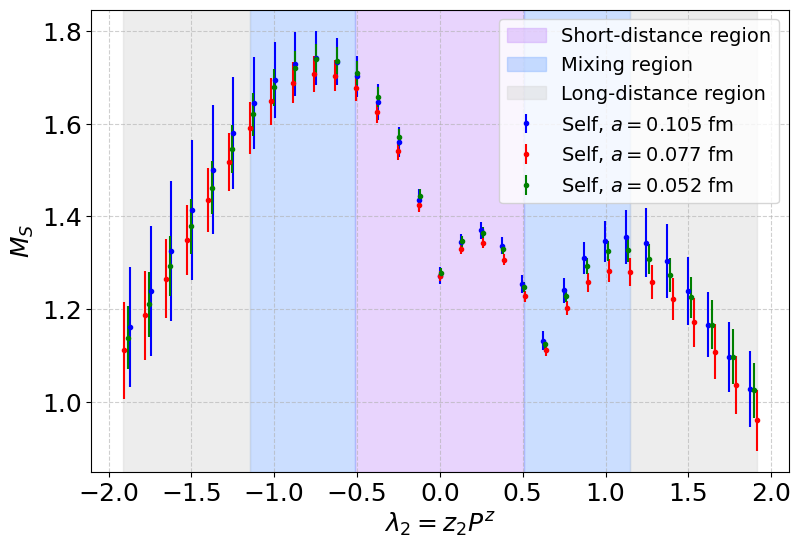}
    }
\vspace{0.0cm} 
\subfigure[\ Hybrid scheme result of $\Lambda$ at $P=0.5$ GeV]{
    \centering
    \includegraphics[scale=0.38]{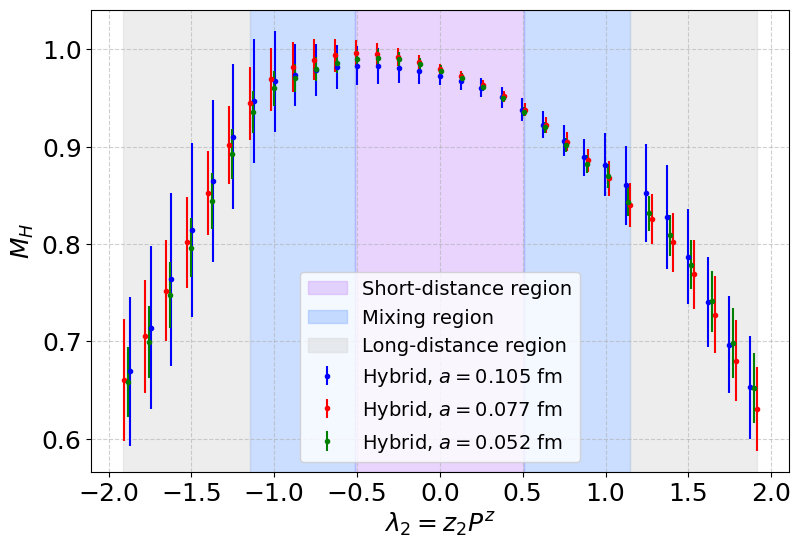}
    }
\caption{Bare, ratio scheme, self renormalized \& hybrid renormalization scheme results of the $\Lambda$ quasi-DA matrix elements at $P^z=0.5$ GeV, fix $z_1=0.250$ fm ($\lambda_1 = 0.631$).}
\label{fig:schemes_lambda_pz1}
\end{figure*}

\begin{figure*}[htbp]
\centering
\subfigure[\ Bare result of $\Lambda$ at $P=2.0$ GeV]{
    \centering
    \includegraphics[scale=0.38]{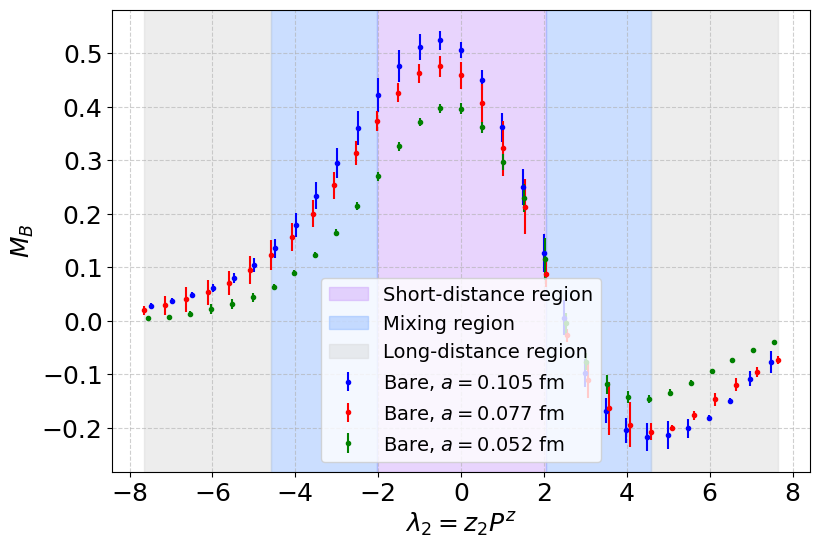}
    }
\vspace{0.0cm} 
\subfigure[\ Ratio scheme result of $\Lambda$ at $P=2.0$ GeV]{
    \centering
    \includegraphics[scale=0.38]{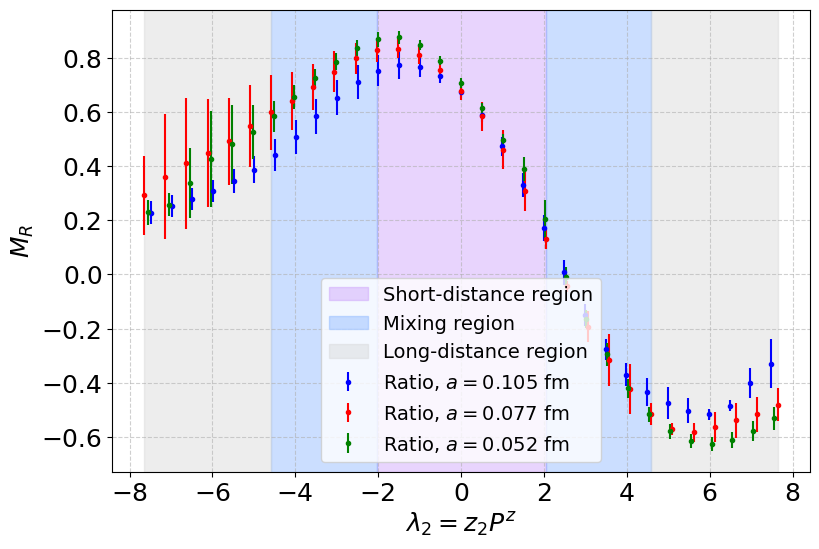}
    }
\vspace{0.0cm} 
\subfigure[\ Self renormalized result of $\Lambda$ at $P=2.0$ GeV]{
    \centering
    \includegraphics[scale=0.38]{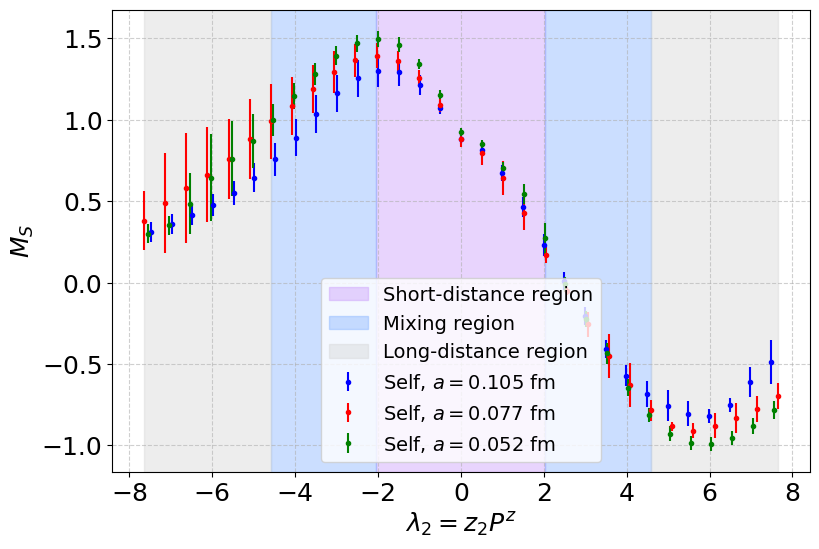}
    }
\vspace{0.0cm} 
\subfigure[\ Hybrid scheme result of $\Lambda$ at $P=2.0$ GeV]{
    \centering
    \includegraphics[scale=0.38]{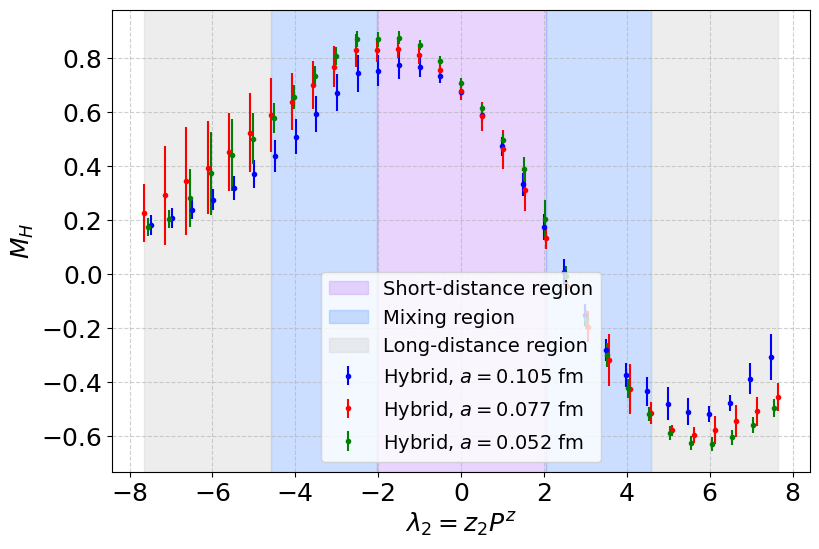}
    }
\caption{Bare, ratio scheme, self renormalized \& hybrid renormalization scheme results of the $\Lambda$ quasi-DA matrix elements at $P^z=2.0$ GeV, fix $z_1=0.250$ fm ($\lambda_1 = 2.522$).}
\label{fig:schemes_lambda_pz4}
\end{figure*}

With the above fitting process, we can extract the self-renormalization factor \(Z_R(z_1, z_2, a, \mu)\) from the zero-momentum matrix elements and then use it to renormalize large-momentum matrix elements, yielding the self renormalized quasi-DA. In Fig.~\ref{fig:schemes_lambda_pz1} and Fig.~\ref{fig:schemes_lambda_pz4}, we show the comparison of bare, ratio renormalized and self renormalized matrix elements in subfigures (a, b, c) for the $\Lambda$ with both $P^z=0.5$ GeV and $P^z=2.0$ GeV. 

As evident from these figures, the ratio scheme results exhibit continuity and smoothness across all regions, because taking the ratio to the zero-momentum matrix elements eliminates both short-distance logarithms and UV divergences. However, this division inevitably introduces non-perturbative IR effects at large distances. The self renormalized results demonstrate effective removal of UV divergences at large distances without introducing additional uncontrollable non-perturbative effects. Yet at short distances, the self renormalized results exhibit two peaks at $z_2=0$ and $z_2=z_1$, reflecting the short-distance behavior of the perturbative quasi-DA as discussed before. This issue naturally leads us to employ the hybrid renormalization framework discussed in Sec.~\ref{subsec:Hybrid renormalization}.

The hybrid renormalization is implemented as follows: After extracting the self-renormalization factor \(Z_R(z_1, z_2, a, \mu)\) from the zero-momentum matrix elements, this factor is applied to renormalize both large-momentum and zero-momentum matrix elements, yielding \(\hat{M}_{\overline{\rm MS}}(z_1, z_2, 0, P^z, \mu)\) and \(\hat{M}_{\overline{\rm MS}}(z_1, z_2, 0, 0, \mu)\). Then perform region-specific ratios as previously described in Sec.~\ref{sec:frame_hy} to achieve continuous hybrid renormalization scheme results.

The hybrid renormalized results for the \(\Lambda\) quasi-DA matrix elements at \(P^z = 0.5\) GeV and \(P^z = 2.0\) GeV are shown in subfigures (d) of Figs.~\ref{fig:schemes_lambda_pz1} and \ref{fig:schemes_lambda_pz4}.  Compared to the self renormalized results (subfigures c), the hybrid renormalization scheme effectively eliminates the singularities in the short-distance region. 
In the long-distance region, it naturally transitions to self-renormalization, which provides optimal UV divergence removal. Crucially, the hybrid renormalized results exhibit smooth continuity across the transition regions while retaining proper normalization, which facilitates subsequent limited Fourier transforms~\cite{Chen:2025cxr, Xiong:2025obq} and effective matching procedures. The corresponding results for the proton quasi-DA are presented in Appendix~\ref{sec:appendix_quasi-DA}.

\begin{figure}[htbp]
\centering
\subfigure[\ $\Lambda$, Re, $P=2.0$ GeV]{
    \centering
    \includegraphics[scale=0.56]{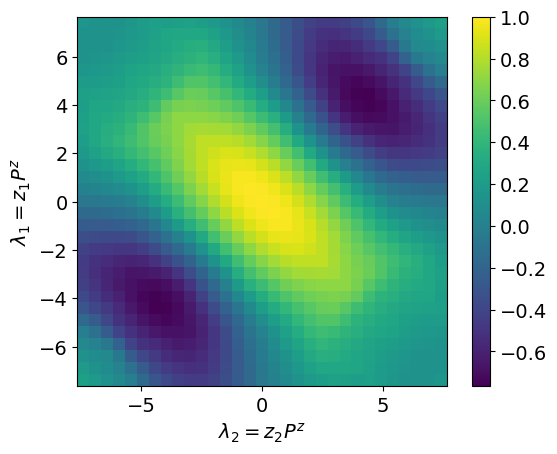}
    }
\vspace{0.0cm} 
\subfigure[\ $\Lambda$, Im, $P=2.0$ GeV]{
    \centering
    \includegraphics[scale=0.56]{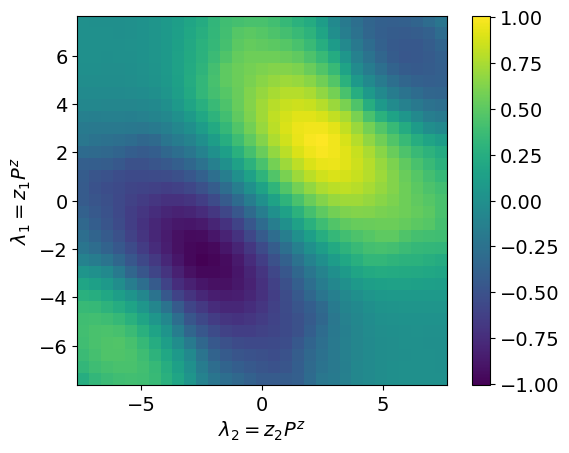}
    }
\caption{2D heatmaps of the hybrid renormalized quasi-DA matrix elements (central values) of the $\Lambda$ at $P=2.0$ GeV.}
\label{fig:quasi_da_heatmap_lambda_pz4}
\end{figure}

\begin{figure}[htbp]
\centering
\subfigure[\ Proton, Re, $P=2.0$ GeV]{
    \centering
    \includegraphics[scale=0.56]{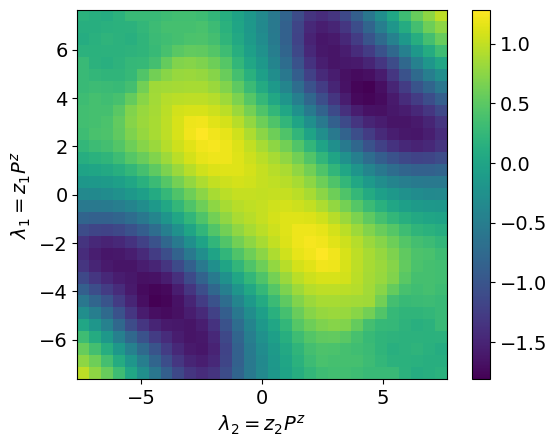}
    }
\vspace{0.0cm} 
\subfigure[\ Proton, Im, $P=2.0$ GeV]{
    \centering
    \includegraphics[scale=0.56]{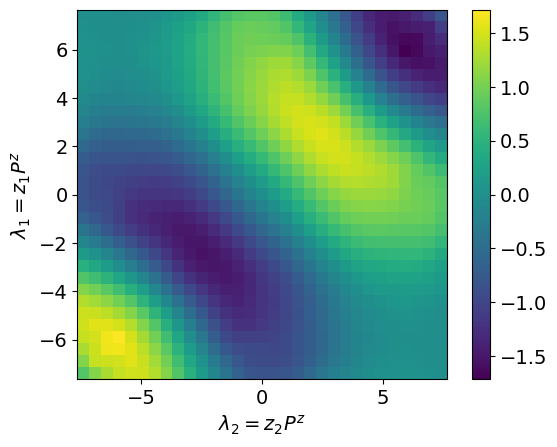}
    }
\caption{2D heatmaps of the hybrid renormalized quasi-DA matrix elements (central values) of the proton at $P=2.0$ GeV.}
\label{fig:quasi_da_heatmap_proton_pz4}
\end{figure}

Figures ~\ref{fig:quasi_da_heatmap_lambda_pz4} and ~\ref{fig:quasi_da_heatmap_proton_pz4} display 2D heat maps of hybrid renormalized quasi-DA matrix elements (central values) for the \(\Lambda\) (A-term) and the proton (V-term) across all four quadrants at \(P^z = 2.0\) GeV. These visualizations further demonstrate the smoothness and continuity of the hybrid renormalized results over the entire coordinate space. Moreover, the quasi-DAs at $2$ GeV exhibit a clear and pronounced oscillatory behavior across all quadrants, consistent with the expected coordinate space structure of baryon quasi-DAs.

\section{Summary}\label{sec:summary}
In this work, we numerically implement the hybrid renormalization scheme based on self-renormalization for the leading twist A-term of the 
$\Lambda$ quasi-DAs and the V/T-terms of the proton quasi-DAs. The linear divergence is extracted by analyzing zero-momentum matrix elements at multiple lattice spacings, $a=\{0.105, 0.077, 0.052\}$ fm. The residual self-renormalization factors are then determined by matching to perturbative matrix elements in the $\overline{\mathrm{MS}}$ scheme. To suppress instabilities caused by the peaked structure in the self-renormalization scheme, we adopt the hybrid renormalization scheme, which combines the ratio scheme at short distances with the self renormalization at large distances. Finally, we present the renormalized quasi-DAs at both a small momentum of $0.5$ GeV and a large momentum of $2.0$ GeV.

The renormalized results indicate that the linear divergence associated with the Wilson line self-energy exhibits negligible dependence on the external state, as expected. Moreover, renormalized quasi-DAs show good convergence with respect to lattice spacing, consistent with the expected behavior of residual discretization effects at large $P^z$. This confirms that the hybrid renormalization scheme results in a reliable continuum limit. Additionally, by properly renormalizing all divergences, the hybrid renormalization scheme yields smooth and continuous quasi-DAs, significantly reducing the complexity of the subsequent Fourier transform and effective matching procedures. 

Therefore, this work provides an important and reliable foundation for the precise and systematically correct computation of baryon LCDAs within the LaMET framework. Building upon the robust renormalization procedure established here, the next goal is to obtain reliable baryon LCDAs in the continuum limit.

\section*{Acknowledgement} 
We thank Fusheng Yu, Jiajie Han and Yuelong Shen for useful discussions.  We thank the CLQCD collaborations for providing us the gauge configurations with dynamical fermions~\cite{CLQCD:2023sdb}, which are generated on the HPC Cluster of ITP-CAS, the Southern Nuclear Science Computing Center(SNSC), the Siyuan-1 cluster supported by the Center for High Performance Computing at Shanghai Jiao Tong University and the Dongjiang Yuan Intelligent Computing Center. 
This work is supported in part by Guangdong Major Project of Basic and Applied Basic Research No. 2020B0301030008, Natural Science Foundation of China under grant No. 12205106, 12125503, 12305103, 12375080, 12375069, 12293060, 12293062, 12435002, 12447101, 12175073, 12275277, 12435004 and 12222503.  YBY is supported in part National Key R\&D Program of China No.2024YFE0109800, and by Strategic Priority Research Program of Chinese Academy of Sciences, Grant No. YSBR-101. JHZ is supported in part by Ministry of Science and Technology of China under Grant No. 2024YFA1611004, and by CUHK-Shenzhen under grant No. UDF01002851. QAZ is supported in part by the Fundamental Research Funds for the Central Universities. The computations in this paper were run on the Siyuan-1 cluster supported by the Center for High Performance Computing at Shanghai Jiao Tong University, and Advanced Computing East China Sub-center. The LQCD simulations were performed using the PyQUDA software suite~\cite{Jiang:2024lto} and QUDA~\cite{Clark:2009wm,Babich:2011np,Clark:2016rdz} through HIP programming model~\cite{Bi:2020wpt}.

\clearpage

\appendix
\begin{widetext}
\section{Symmetry in baryon distributions}

In lattice QCD calculations, it is often useful to exploit symmetry relations derived from theoretical considerations to check the correctness and consistency of numerical results. Since the proton is an isospin-$\frac{1}{2}$ state, the following identity holds:
\begin{align}
\left\langle 0\left|\left(T^2 - \frac{1}{2}\left(\frac{1}{2} + 1\right)\right) \left(\varepsilon^{ijk} u^i_\alpha(1) u^j_\beta(2) d^k_\gamma(3)\right)\right| P \right\rangle = 0,
\end{align}
where
\begin{equation}
T^2 = \frac{1}{2}\left(T_+ T_- + T_- T_+\right) + T_3^2
\end{equation}
is the quadratic Casimir operator of the isospin $\mathfrak{su}(2)$ algebra in the adjoint representation. This condition leads to the constraint~\cite{Braun:2000kw}:
\begin{equation}
\begin{aligned}
\label{isospin}
& \left\langle 0\left|\varepsilon^{ijk} u^i_\alpha(1) u^j_\beta(2) d^k_\gamma(3)\right| P \right\rangle 
+ \left\langle 0\left|\varepsilon^{ijk} u^i_\alpha(1) u^j_\gamma(3) d^k_\beta(2)\right| P \right\rangle \\
& \quad + \left\langle 0\left|\varepsilon^{ijk} u^i_\gamma(3) u^j_\beta(2) d^k_\alpha(1)\right| P \right\rangle = 0.
\end{aligned}
\end{equation}

To bring all terms into the same spinor index order $(\alpha,\beta,\gamma)$, one can employ the following Fierz identities~\cite{Braun:2000kw},
\begin{equation}
\label{Fierz}
\begin{aligned}
(v_1)_{\gamma\beta,\alpha} &= \frac{1}{2}(v_1 - a_1 - t_1)_{\alpha\beta,\gamma}, \\
(a_1)_{\gamma\beta,\alpha} &= \frac{1}{2}(-v_1 + a_1 - t_1)_{\alpha\beta,\gamma}, \\
(t_1)_{\gamma\beta,\alpha} &= - (v_1 + a_1)_{\alpha\beta,\gamma},
\end{aligned}
\end{equation}
where the Lorentz structures are defined as
\begin{align*}
(v_1)_{\alpha\beta,\gamma} = (\slashed{P} C)_{\alpha\beta}(\gamma_5 N)_\gamma,\, 
(a_1)_{\alpha\beta,\gamma} = (\slashed{P} \gamma_5 C)_{\alpha\beta}(N)_\gamma,\, 
(t_1)_{\alpha\beta,\gamma} = (P^\nu i\sigma_{\mu\nu} C)_{\alpha\beta}(\gamma^\mu\gamma_5 N)_\gamma.
\end{align*}
Applying this decomposition to Eq.~(\ref{isospin}), one arrives at the following identity:
\begin{equation}
\begin{aligned}
0 = & \Big[2 T(x_1,x_2,x_3) - V(x_1,x_3,x_2) + A(x_1,x_3,x_2) \\
& \quad - V(x_3,x_2,x_1) - A(x_3,x_2,x_1)\Big] (t_1)_{\alpha\beta,\gamma} \\
+ & \Big[2 V(x_1,x_2,x_3) + 2 V(x_1,x_3,x_2) + 2 A(x_1,x_3,x_2) - 2 T(x_1,x_3,x_2) \\
& \quad + V(x_3,x_2,x_1) - A(x_3,x_2,x_1) - 2 T(x_3,x_2,x_1)\Big] (v_1)_{\alpha\beta,\gamma} \\
+ & \Big[2 A(x_1,x_2,x_3) + 2 V(x_1,x_3,x_2) + 2 A(x_1,x_3,x_2) + 2 T(x_1,x_3,x_2) \\
& \quad - V(x_3,x_2,x_1) + A(x_3,x_2,x_1) - 2 T(x_3,x_2,x_1)\Big] (a_1)_{\alpha\beta,\gamma}.
\end{aligned}
\end{equation}

By making use of the known symmetry properties of the functions $V$, $A$, and $T$, this equation reduces to a single non-trivial constraint. For baryons not of the $\Lambda$ type, this leads to the well-known relation~\cite{Chernyak:1987nu}
\begin{equation}
2T(x_1, x_2, x_3) = (V - A)(x_1, x_3, x_2) + (V - A)(x_2, x_3, x_1).
\end{equation}

For the $\Lambda$ baryon, the corresponding relation becomes
\begin{equation}
2T(x_1, x_2, x_3) = (V - A)(x_1, x_3, x_2) - (V - A)(x_2, x_3, x_1).
\end{equation}

Besides, in some lattice implementations, the order of valence quarks in the interpolating operator may be changed to simplify the contraction structure. For example, the flavor ordering $udu$ may be used for the proton instead of the canonical $uud$. In such cases, the Fierz identities in Eq.~\eqref{Fierz} can be used to relate the LCDA defined in the new flavor ordering to those defined in the standard basis.

\section{Details for HYP Smearing}
\subsection{Improvement of the signal}
Hyper-cubic (HYP) smearing~\cite{Hasenfratz:2001hp,DeGrand:2002vu} is a gauge-field processing technique that applies weighted averaging of adjacent gauge links to obtain smoother gauge configurations. In our calculation of baryon quasi-DAs with nonlocal shifts, replacing the gauge links between quark fields with HYP-smeared links can significantly improve the signal-to-noise ratio of nonlocal matrix element.

Figure~\ref{hyp_p0} (a) displays the zero-momentum quasi-DA for the $\Lambda$ (A-term) along the $z_1=z_2$ direction with HYP smearing iterations $n_{\rm HYP}={0,1,2}$, and the subfigure (b) shows the corresponding noise-to-signal ratios. With single-step HYP smearing, the statistical errors at large distances decrease considerably, while a second smearing step provides marginal additional improvement. Therefore, we choose single-step HYP smearing in our lattice simulation.

Further more, figure~\ref{hyp_effmass} shows the effective mass of the $\Lambda$ baryon matrix element at a specific nonlocal point ($z_1=6a, z_2=0$) using both original and HYP-smeared gauge links, for momenta $P^z=0$ and $P^z=1.49$ GeV. The use of HYP-smeared links substantially reduces the statistical uncertainties, while the effective mass behavior remains consistent across different momenta. This indicates that smearing the sink-side gauge links does not alter the excited-state contamination.

\begin{figure}[htbp]
\centering
\subfigure[\ Bare quasi-DA, $\Lambda$, $P=0$]{
    \centering
    \includegraphics[width=0.45\textwidth]{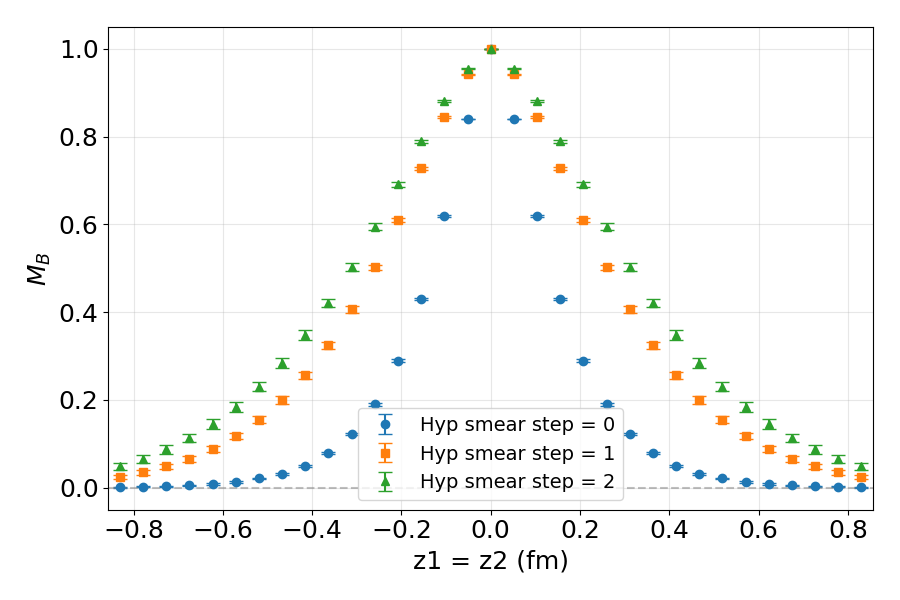}
    }
\vspace{0.0cm} 
\subfigure[\ Noise-to-signal ratio, $\Lambda$, $P=0$]{
    \centering
    \includegraphics[width=0.45\textwidth]{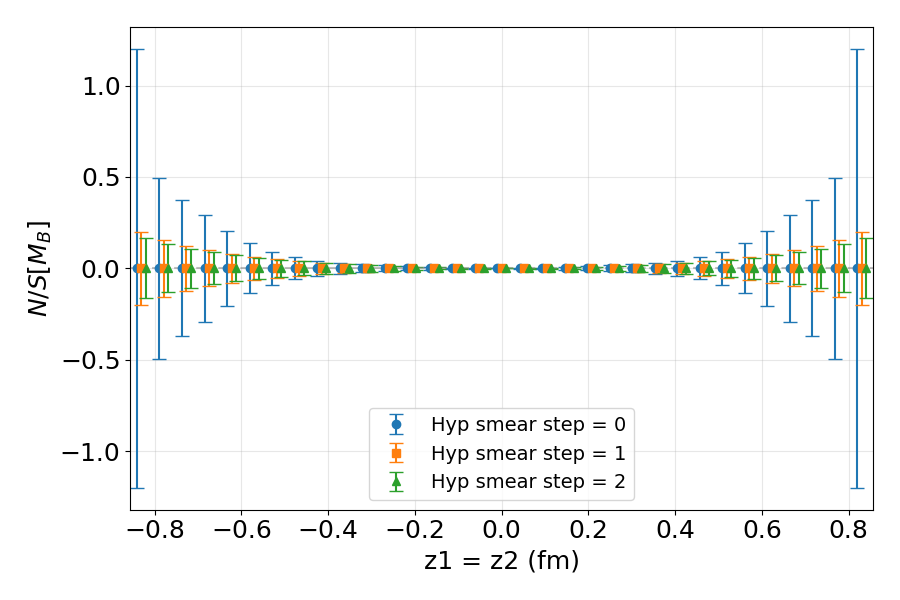}
    }
\caption{Zero-momentum quasi-DA and its corresponding noise-to-signal ratio for the $\Lambda$ (A-term) along $z_1=z_2$, for H48P32 ($a=0.052$ fm),  with HYP smearing iterations $n_{\rm HYP}=0$ (blue), $1$ (orange), and $2$ (green).}
\label{hyp_p0}
\end{figure}

\begin{figure}[htb]
\centering
\includegraphics[width=0.45\textwidth]{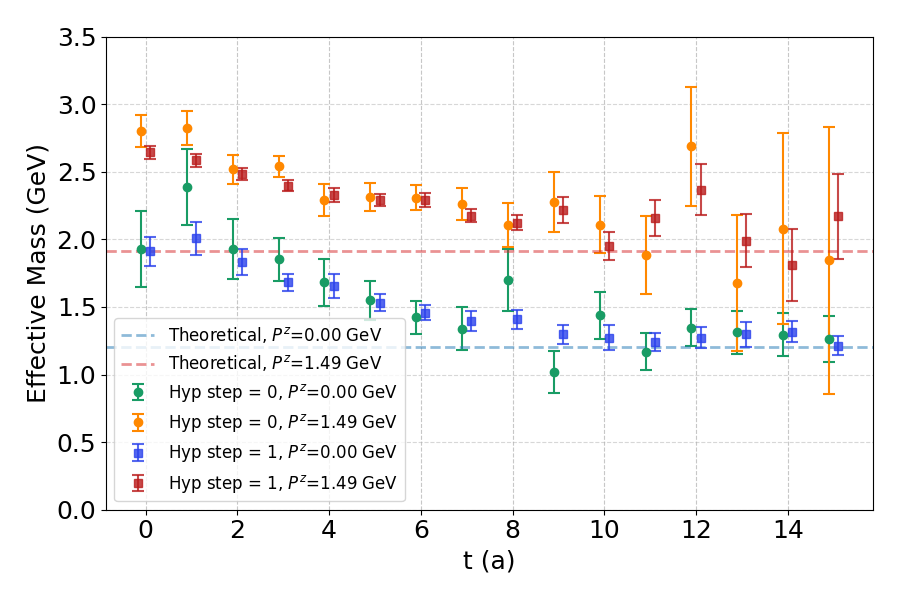}
\caption{Effective mass plots for the $\Lambda$ at a nonlocal point ($z_1=0.36$ fm,$z_2=0$) with original and HYP-smeared gauge links, for H48P32 ($a=0.052$ fm), with $P^z=0$ and $P^z=1.49$ GeV.}
\label{hyp_effmass}
\end{figure}

\subsection{Impact on linear divergence}
HYP smearing smooths the short-distance behavior of gauge links, including the modification of the linear divergence. However, this modification does not affect the renormalized physical results in the ratio or self-renormalization schemes. The parametric form of self-renormalization remains valid for smeared results, as established in Ref.~\cite{LatticePartonLPC:2021gpi}.

To ensure that HYP smearing does not affect the physical results, we also performed a simple check using the ratio scheme renormalization. Figure~\ref{hyp_p3} (a) displays the bare quasi-DA for the $\Lambda$ (A-term) at $P^z = 1.49$ GeV with $n_{\rm HYP} = 0, 1, 2$, while subfigure (b) shows the corresponding ratio scheme quasi-DA results (renormalized by the zero-momentum matrix elements). Within uncertainties, the renormalized quasi-DAs remain consistent across different smearing levels, confirming that the ultraviolet divergences introduced by HYP smearing are mainly linear divergence and can be fully removed by renormalization.

\begin{figure}[htbp]
\centering
\subfigure[\ Bare quasi-DA, $\Lambda$, $P=1.49$ GeV]{
    \centering
    \includegraphics[width=0.45\textwidth]{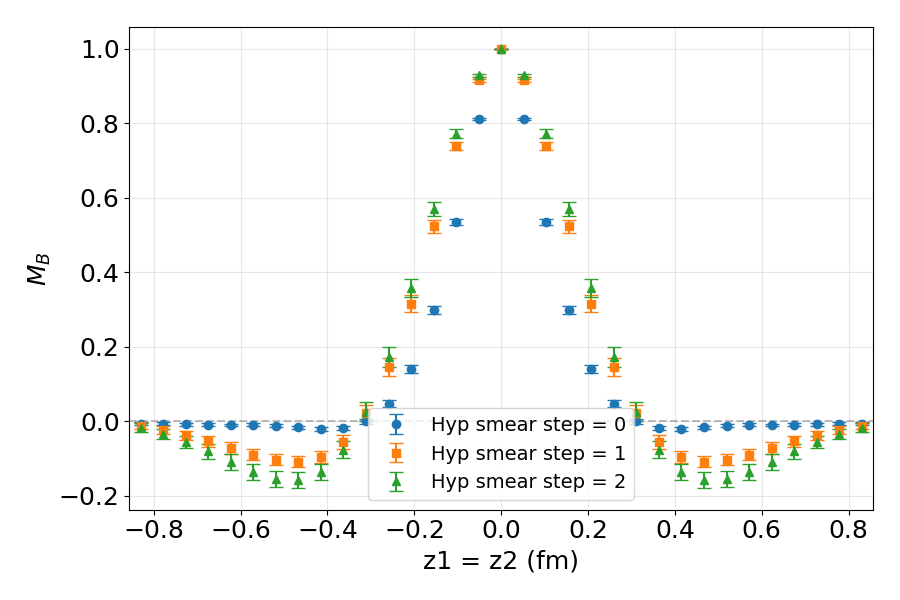}
    }
\vspace{0.0cm} 
\subfigure[\ Ratio scheme quasi-DA, $\Lambda$, $P=1.49$ GeV]{
    \centering
    \includegraphics[width=0.45\textwidth]{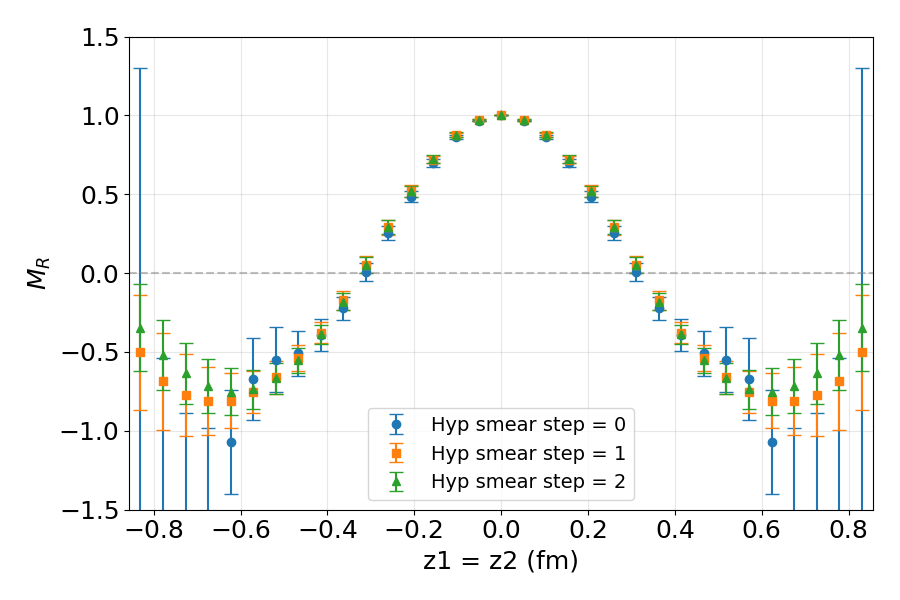}
    }
\caption{Bare quasi-DA and ratio-scheme results for the $\Lambda$ (A-term) at $P^z=1.49$ GeV along $z_1=z_2$, for H48P32 ($a=0.052$ fm), with HYP smearing iterations $n_{\rm HYP}=0$ (blue), $1$ (orange), and $2$ (green).}
\label{hyp_p3}
\end{figure}

\section{Linear divergences in the $\Lambda$ and the proton LCDAs} 

\label{Linear_Dis}

To illustrate the clear linear dependence on lattice spacings for both the $\Lambda$ and the proton, we show more matrix elements plotted against 1/a using a logarithmic scale in Fig~.\ref{fig:linear_divergence_z_dependence}.

\begin{figure*}[htbp]
\centering
\subfigure[\ $\Lambda$, $z_1 =-0.25$ fm]{
    \centering
    \includegraphics[scale=0.35]{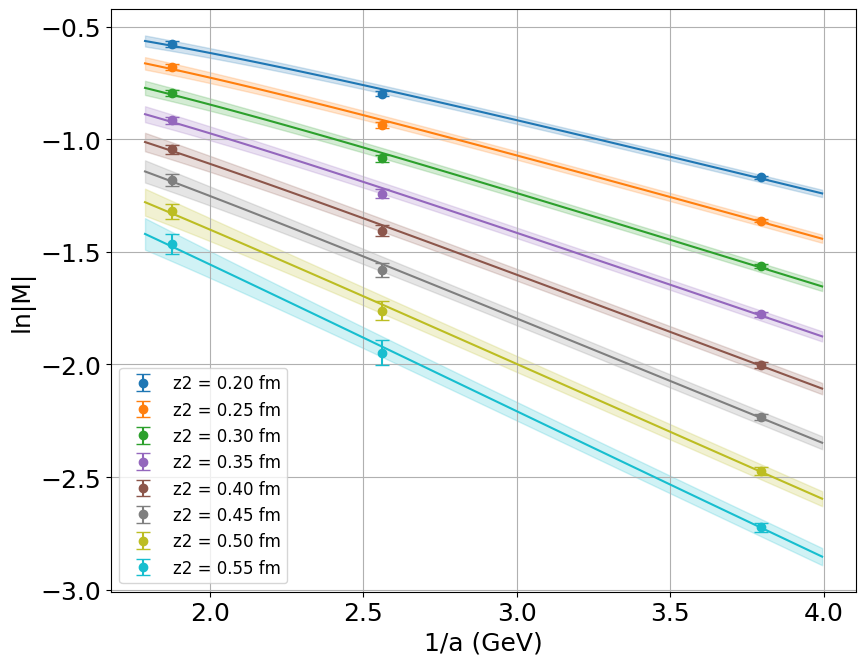}
    }
\vspace{0.0cm} 
\subfigure[\ Proton, $z_1 =-0.25$ fm]{
    \centering
    \includegraphics[scale=0.35]{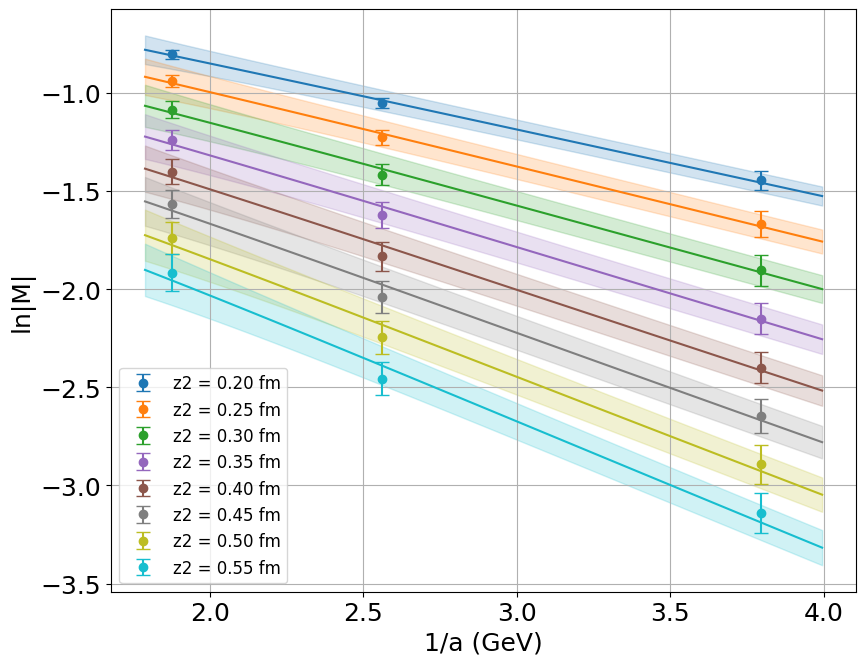}
    }
\vspace{0.0cm} 

\subfigure[\ $\Lambda$, $z_1 = 0.20$ fm]{
    \centering
    \includegraphics[scale=0.35]{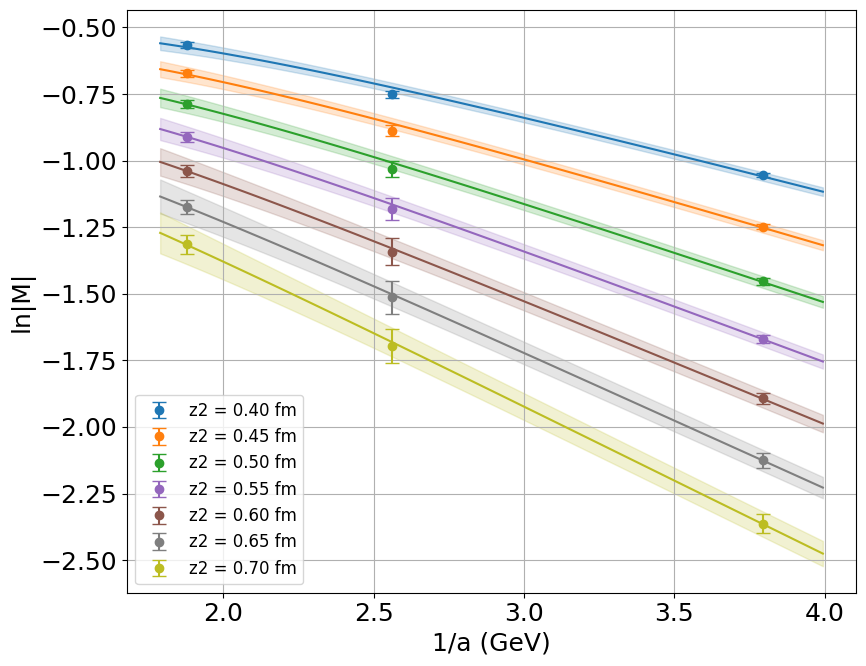}
    }
\vspace{0.0cm} 
\subfigure[\ Proton, $z_1 = 0.20$ fm]{
    \centering
    \includegraphics[scale=0.35]{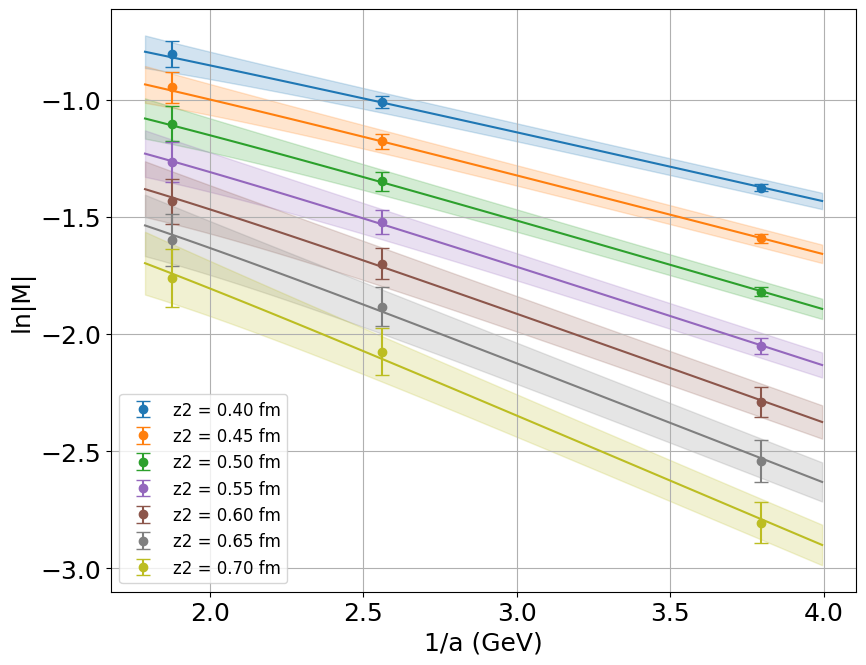}
    }
\caption{Extraction of the linear divergence with lattice spacing dependence for zero-momentum matrix elements of the proton and the $\Lambda$.}
\label{fig:linear_divergence_z_dependence}
\end{figure*}

Theoretically, the linear divergence dominated by the lattice gauge field should be independent of the external states. This can be verified by comparing the ratios of bare quasi-DA matrix elements for the proton (V-term) and the $\Lambda$ (A-term) across three ensembles. As shown in Figs.~\ref{fig:quasi_ratio}, we present the ratio results for both $P^z=0$ and $P^z=0.5$ GeV cases on three ensembles. It can be observed that the ratios are consistent within uncertainties for three different lattice spacings. Consequently, these linear divergences largely cancel out in the ratio, with only minor discrepancies appearing in the short-distance region, suggesting that the linear divergence indeed does not depend on the external states.

\begin{figure}[htbp]
\centering
\subfigure[\ $z_1 =0$, $P^z=0$]{
    \centering
    \includegraphics[scale=0.38]{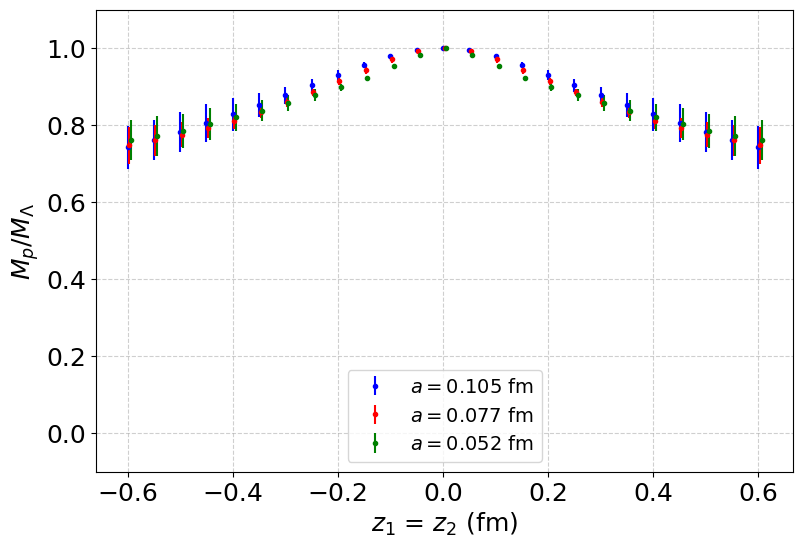}
    }
\vspace{0.0cm} 
\subfigure[\ $z_1 = z_2$, $P^z=0$]{
    \centering
    \includegraphics[scale=0.38]{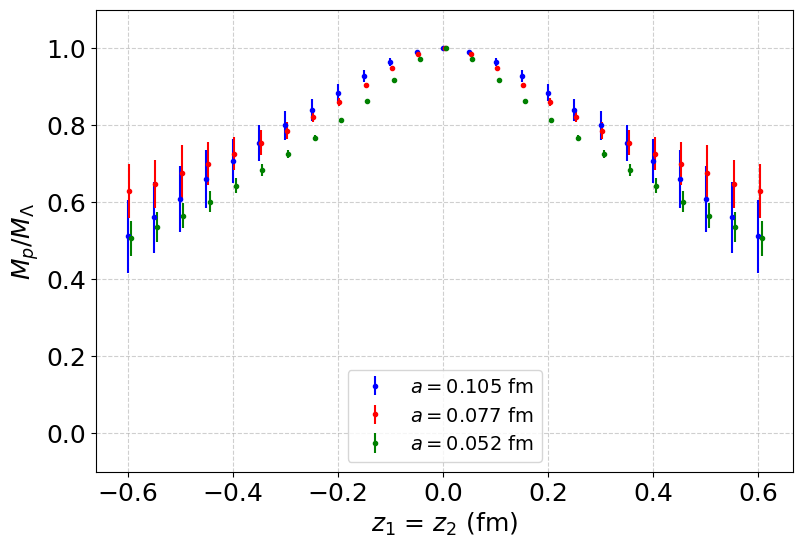}
    }
\vspace{0.0cm} 

\subfigure[\ $z_1 =0$, $P^z=0.5$ GeV]{
    \centering
    \includegraphics[scale=0.38]{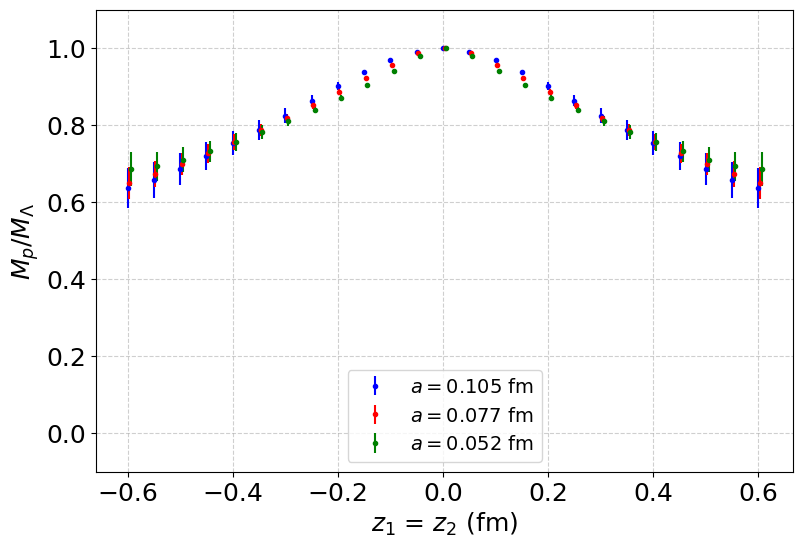}
    }
\vspace{0.0cm} 
\subfigure[\ $z_1 = z_2$, $P^z=0.5$ GeV]{
    \centering
    \includegraphics[scale=0.38]{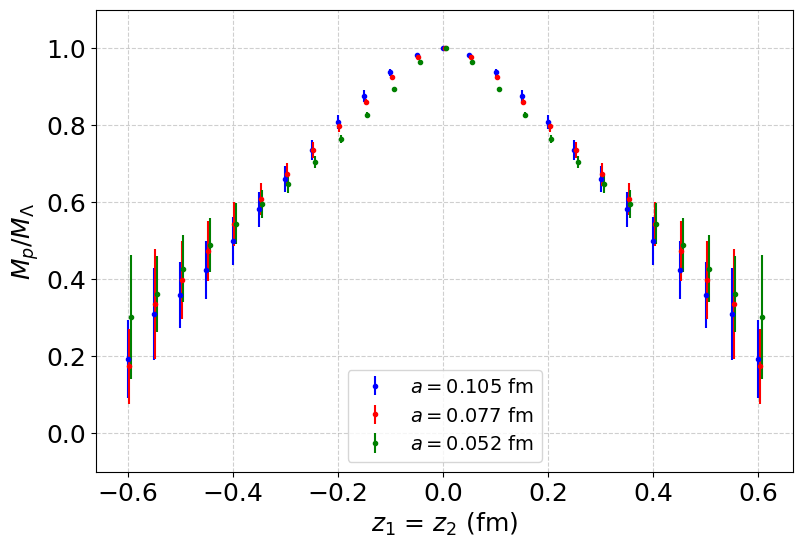}
    }
\vspace{0.0cm} 

\caption{Ratio of the proton (V-term) quasi-DA to the $\Lambda$ (A-term) quasi-DA on different lattice spacings and momenta.}
\label{fig:quasi_ratio}
\end{figure}

\clearpage

\section{More results of baryon quasi-DAs using different schemes}
\label{sec:appendix_quasi-DA}
More cases for the $\Lambda$ (A-term) and the proton (V-term) quasi-DAs in different schemes are shown in this section.

\subsection{More results of the $\Lambda$ (A-term) Quasi-DA at $P^z = 0.5$ GeV in different schemes}

\begin{figure}[htbp]
\centering
\subfigure[\ Bare result of $\Lambda$ at $P=0.5$ GeV]{
    \centering
    \includegraphics[scale=0.185]{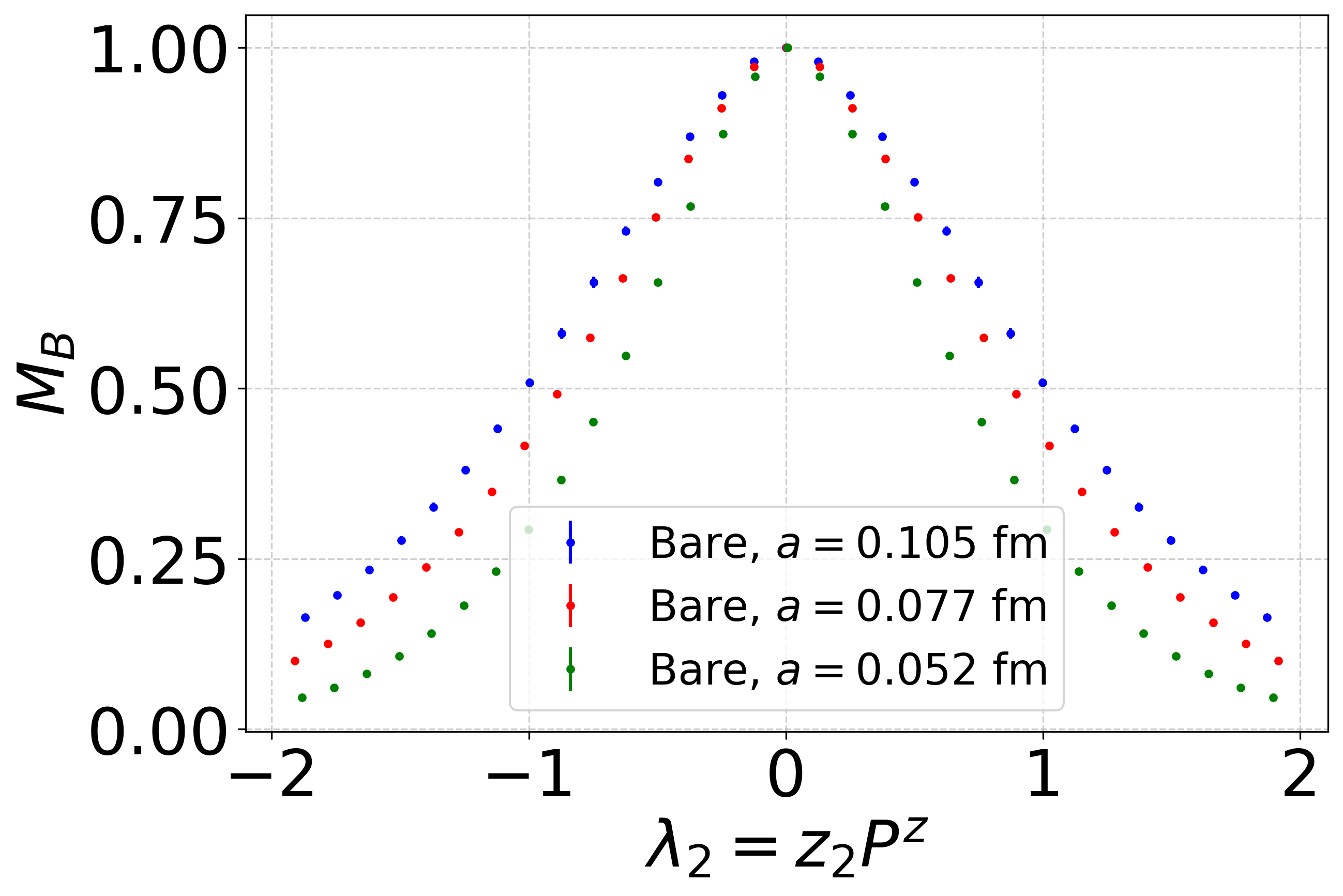}
    }
\vspace{0.0cm} 
\subfigure[\ Ratio scheme result of $\Lambda$ at $P=0.5$ GeV]{
    \centering
    \includegraphics[scale=0.185]{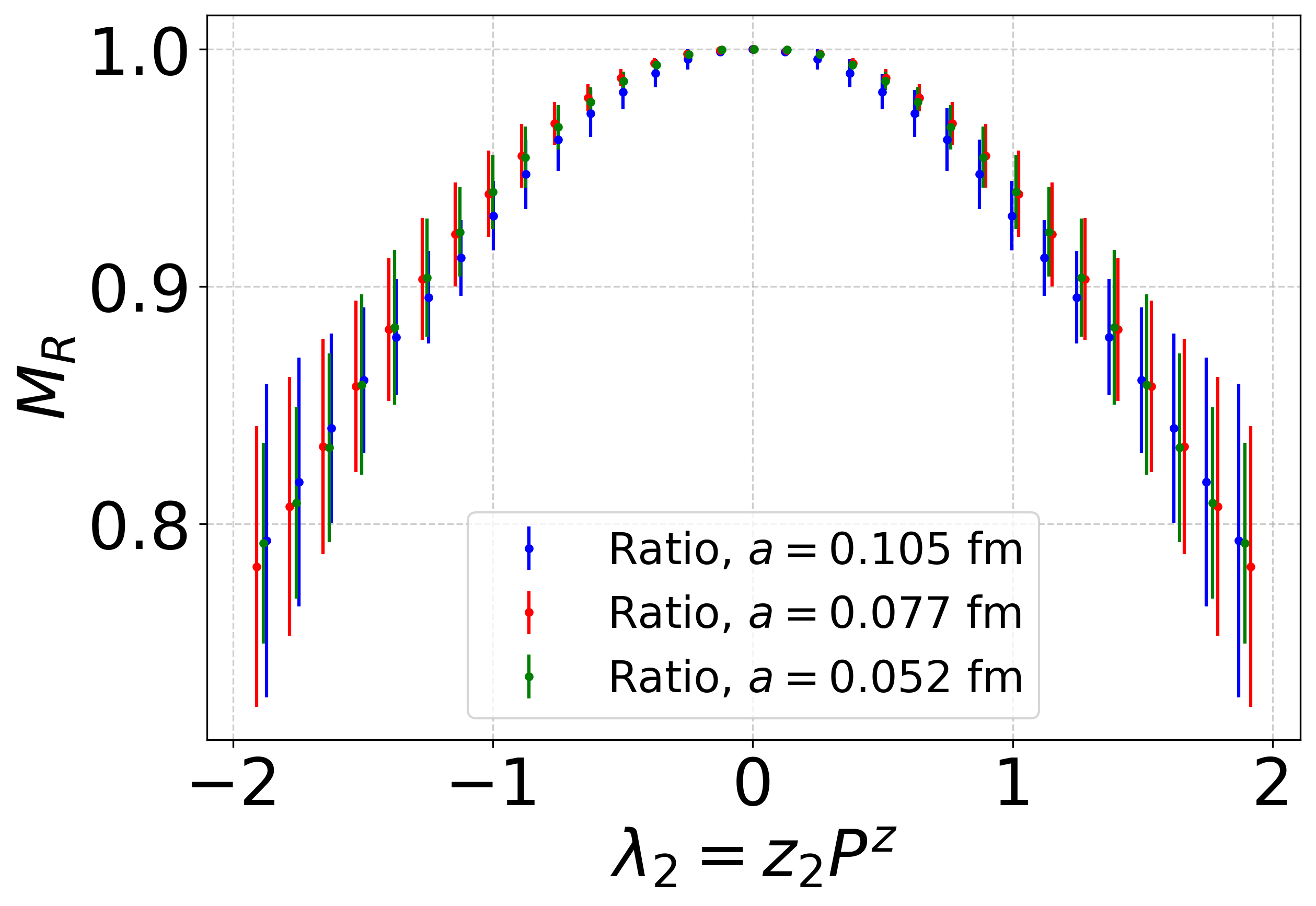}
    }
\vspace{0.0cm} 
\subfigure[\ Self scheme result of $\Lambda$ at $P=0.5$ GeV]{
    \centering
    \includegraphics[scale=0.185]{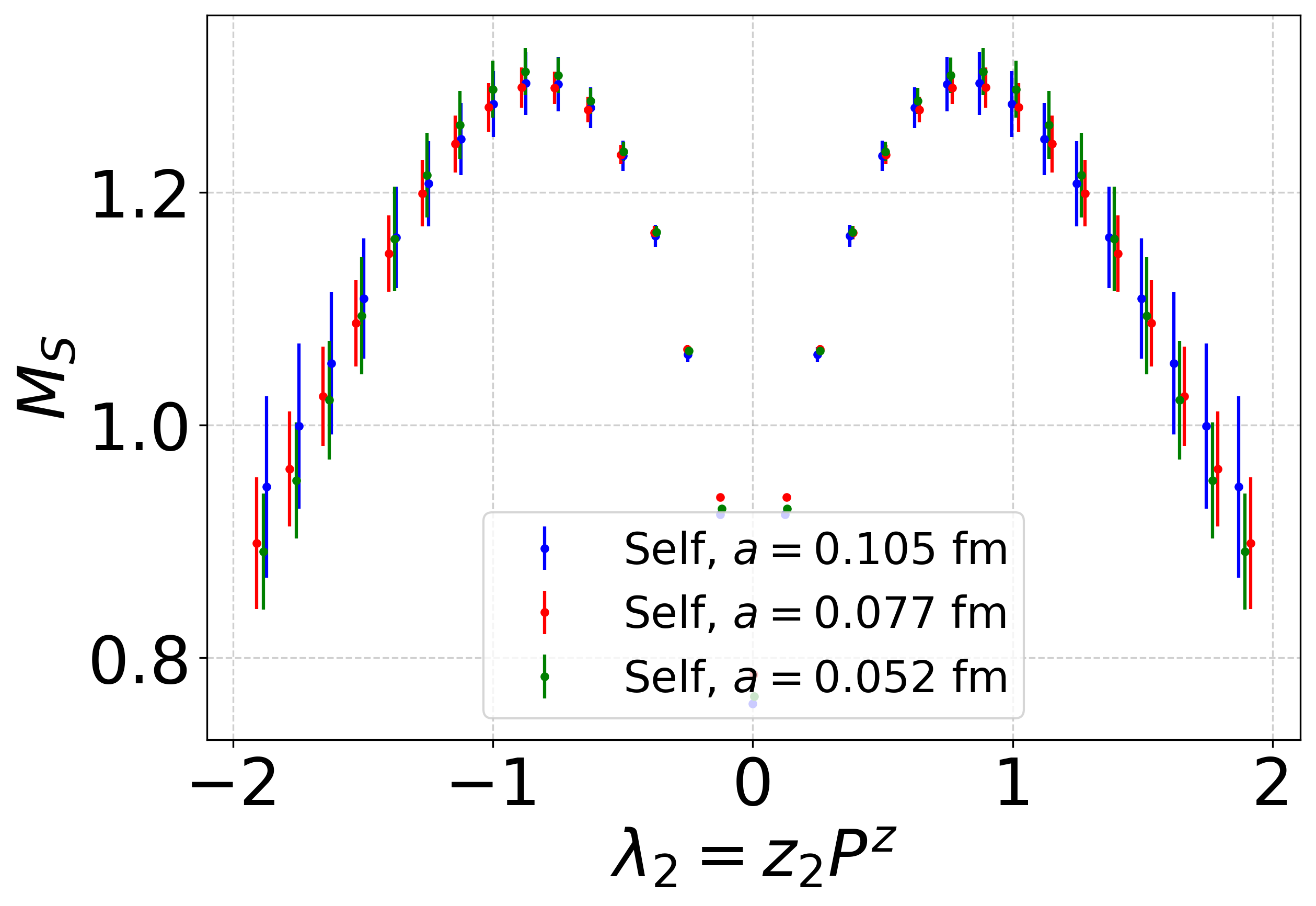}
    }
\vspace{0.0cm} 
\subfigure[\ Hybrid scheme result of $\Lambda$ at $P=0.5$ GeV]{
    \centering
    \includegraphics[scale=0.185]{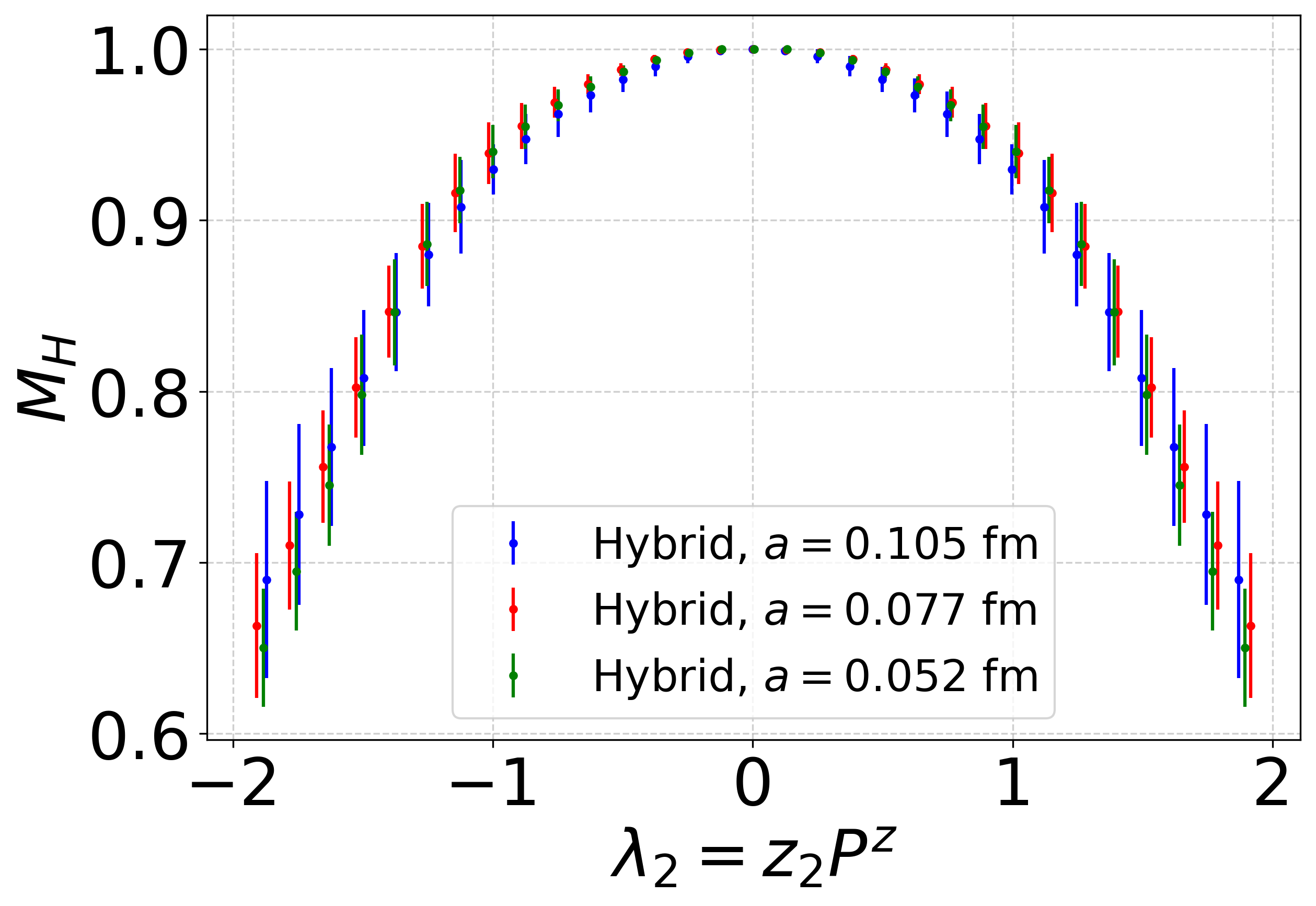}
    }
\caption{Results of the $\Lambda$ quasi-DA matrix elements in different schemes and with $P^z=0.5$ GeV, $z_1=0.000$ fm}
\label{fig:lambda_p1_z0}
\end{figure}

\begin{figure}[htbp]
\centering
\subfigure[\ Bare result of $\Lambda$ at $P=0.5$ GeV]{
    \centering
    \includegraphics[scale=0.185]{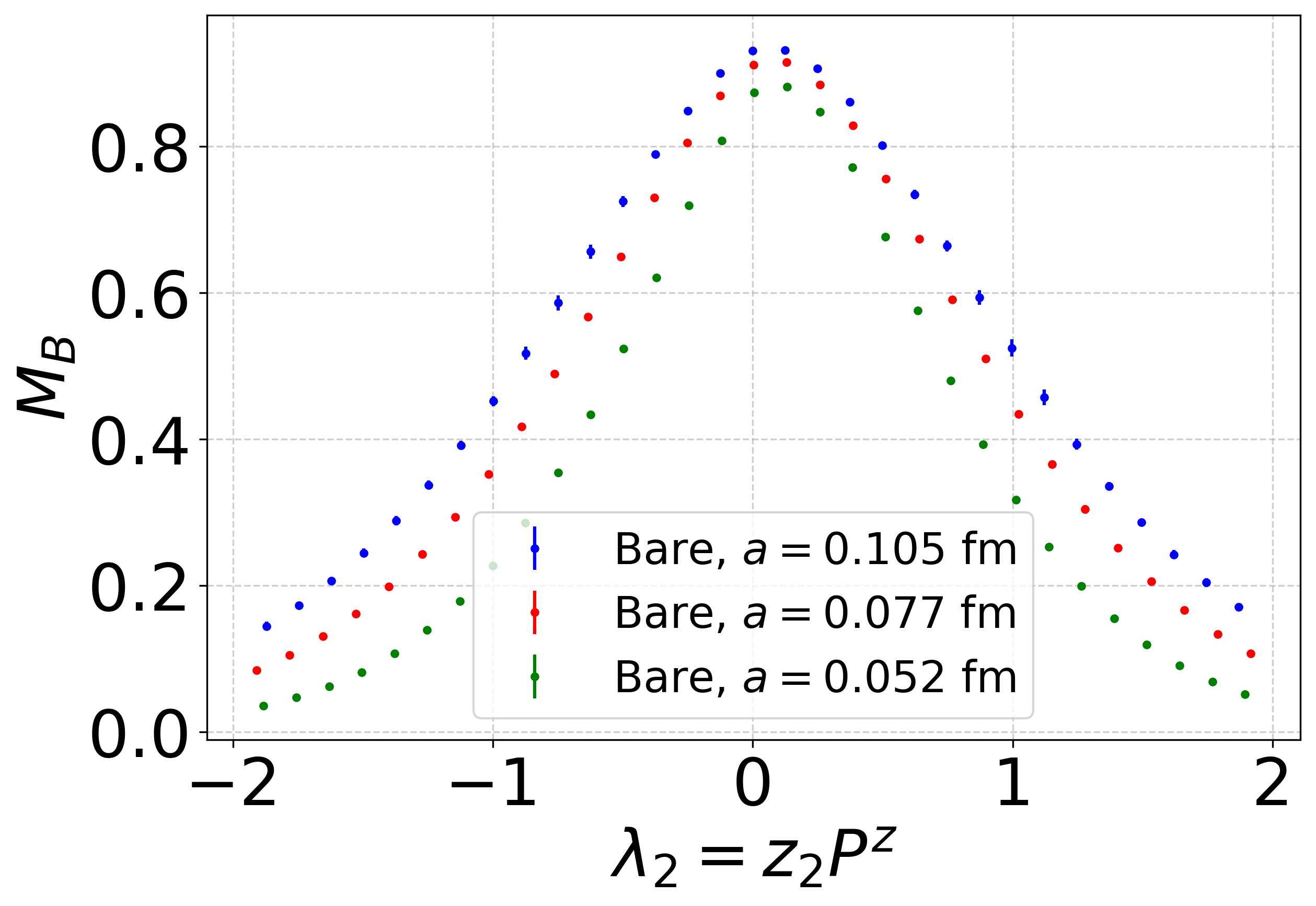}
    }
\vspace{0.0cm} 
\subfigure[\ Ratio scheme result of $\Lambda$ at $P=0.5$ GeV]{
    \centering
    \includegraphics[scale=0.185]{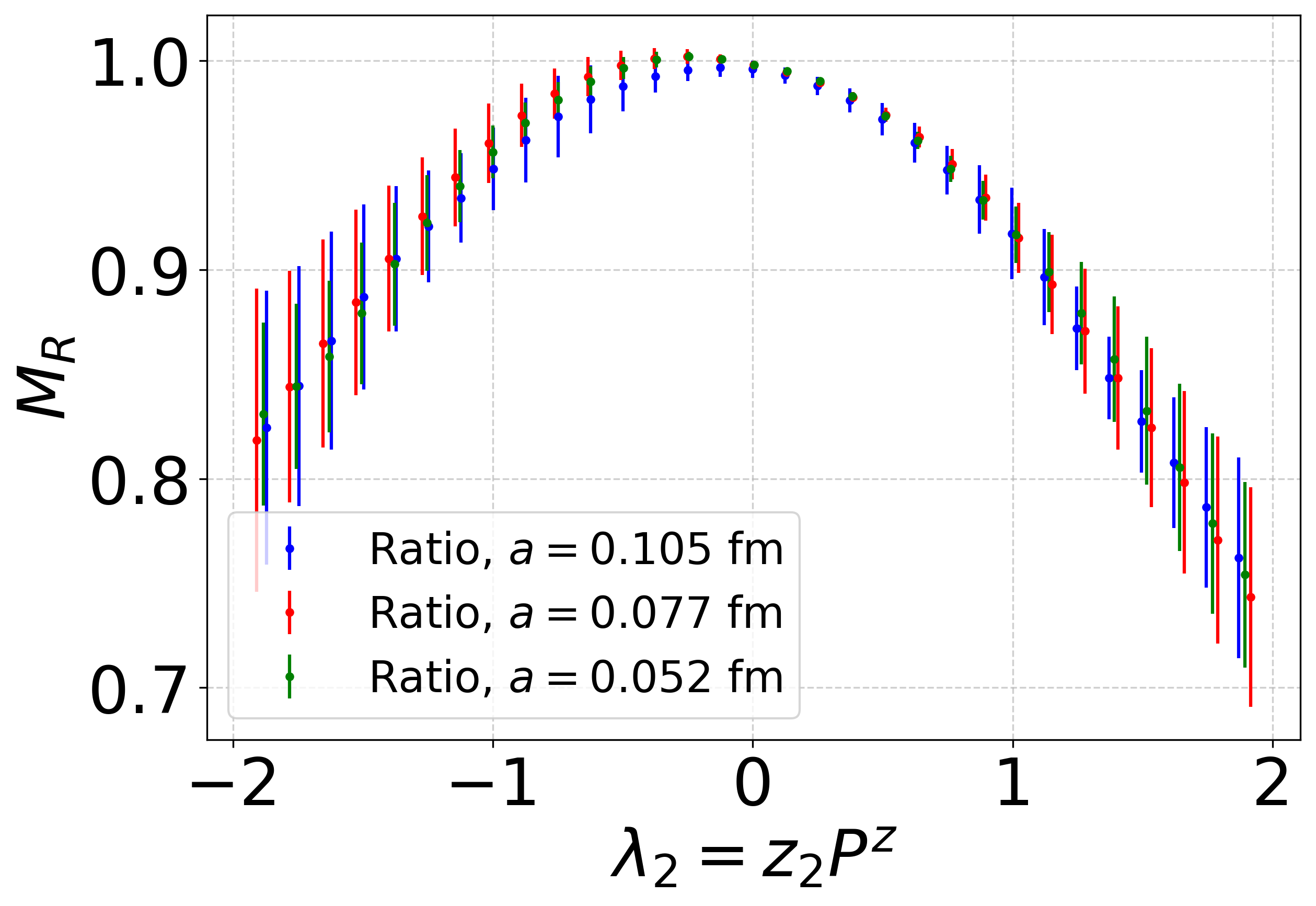}
    }
\vspace{0.0cm} 
\subfigure[\ Self scheme result of $\Lambda$ at $P=0.5$ GeV]{
    \centering
    \includegraphics[scale=0.185]{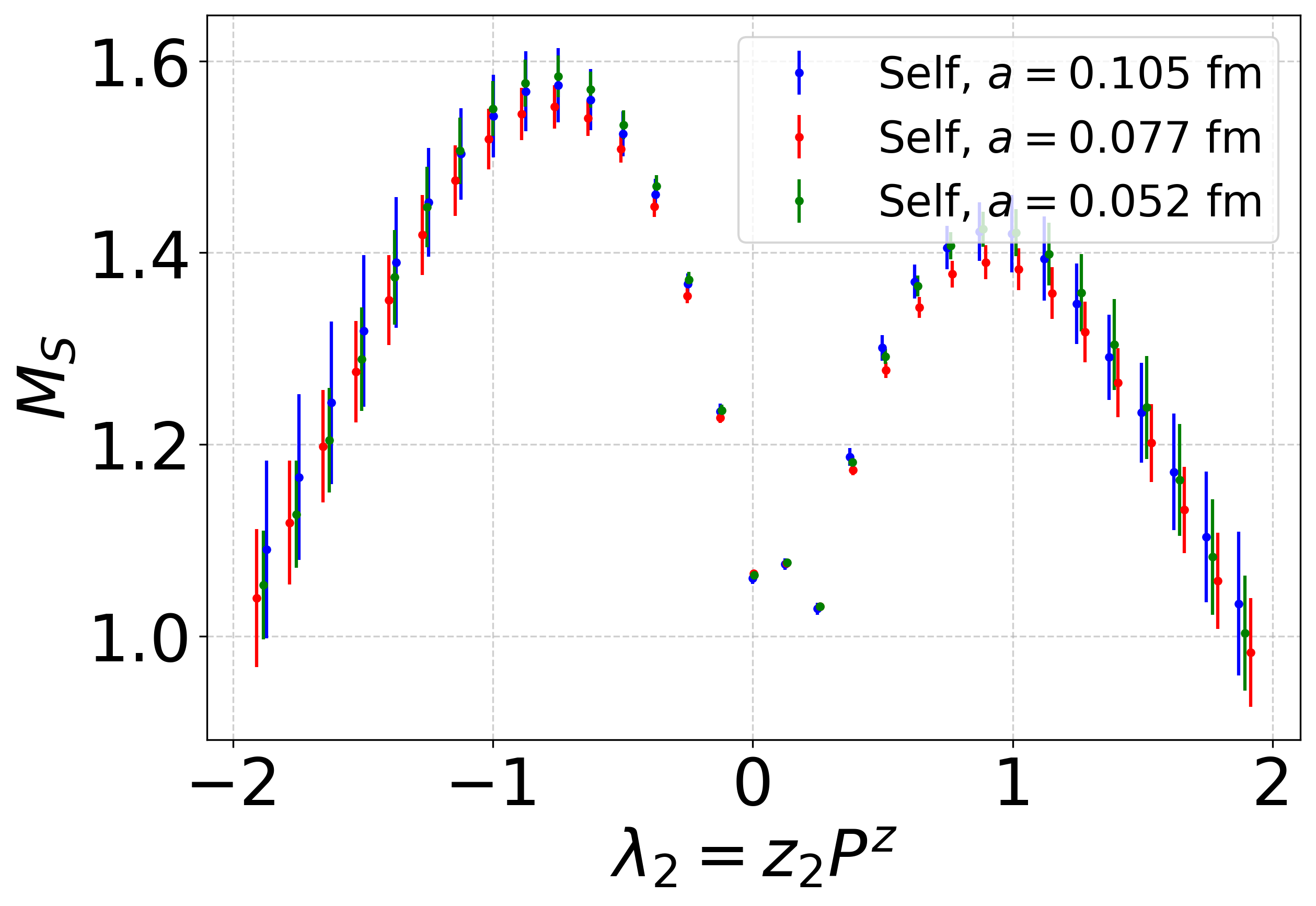}
    }
\vspace{0.0cm} 
\subfigure[\ Hybrid scheme result of $\Lambda$ at $P=0.5$ GeV]{
    \centering
    \includegraphics[scale=0.185]{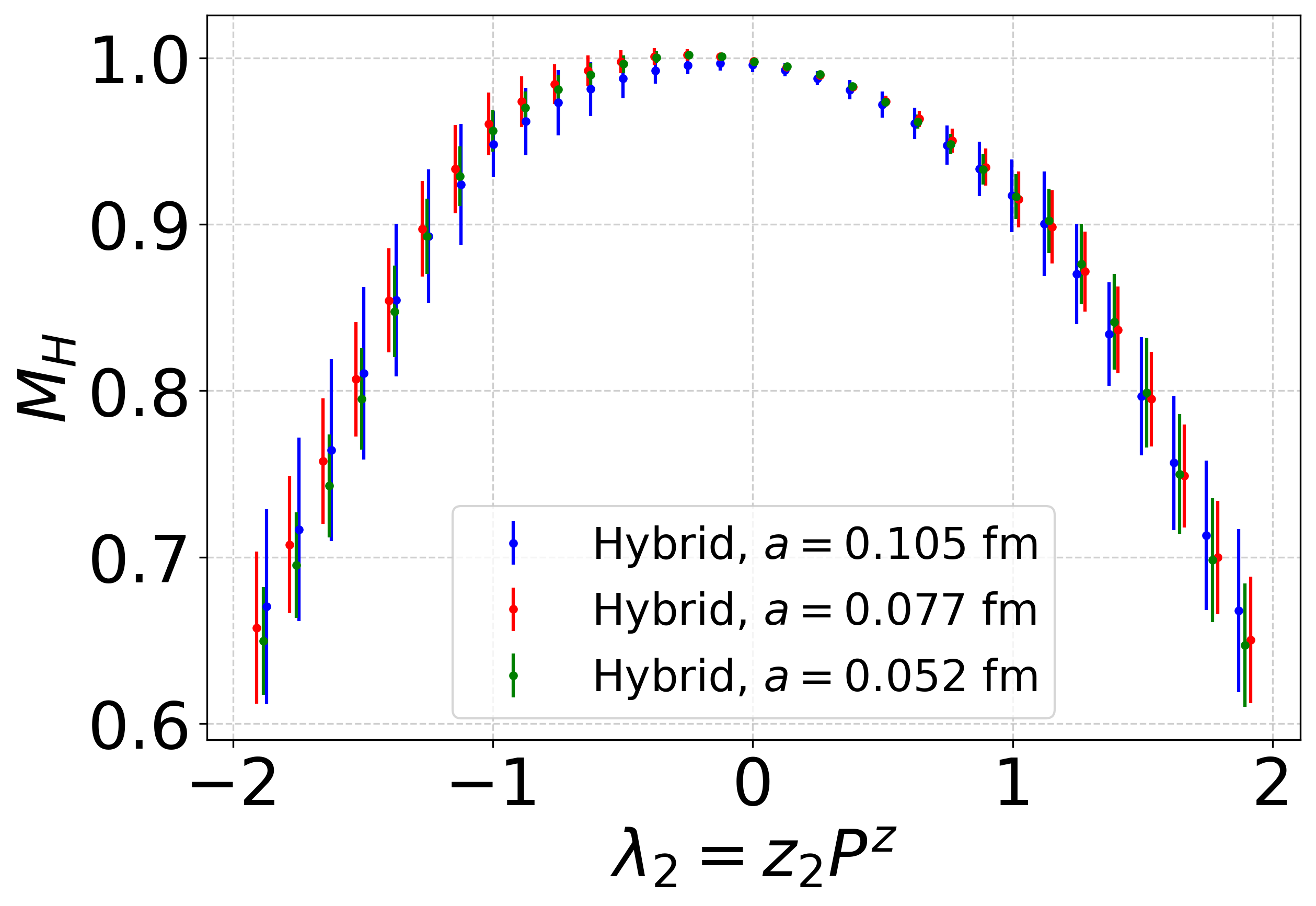}
    }
\caption{Results of the $\Lambda$ quasi-DA matrix elements in different schemes and with $P^z=0.5$ GeV, $z_1=0.100$ fm}
\label{fig:lambda_p1_z2}
\end{figure}

\begin{figure}[htbp]
\centering
\subfigure[\ Bare result of $\Lambda$ at $P=0.5$ GeV]{
    \centering
    \includegraphics[scale=0.185]{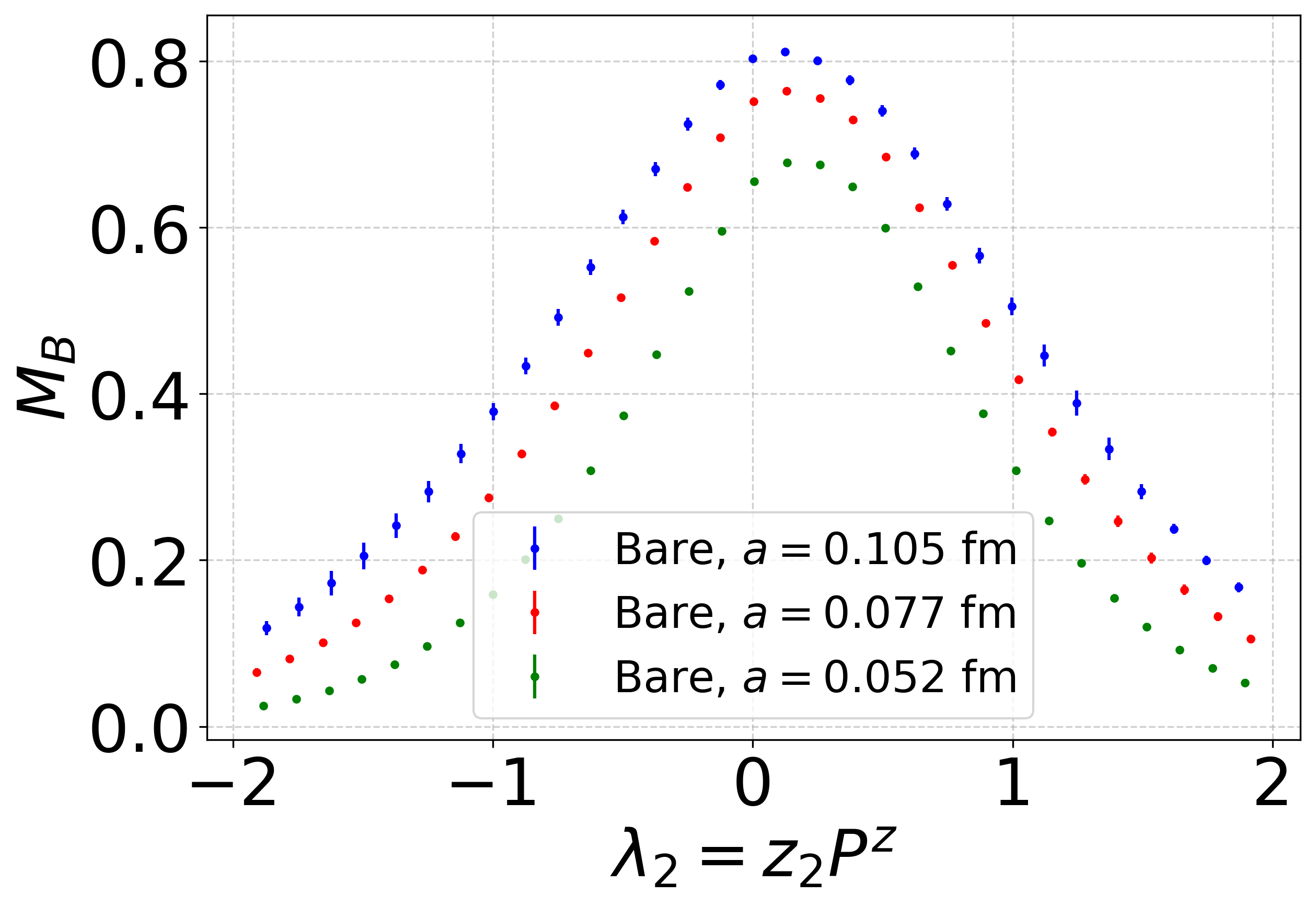}
    }
\vspace{0.0cm} 
\subfigure[\ Ratio scheme result of $\Lambda$ at $P=0.5$ GeV]{
    \centering
    \includegraphics[scale=0.185]{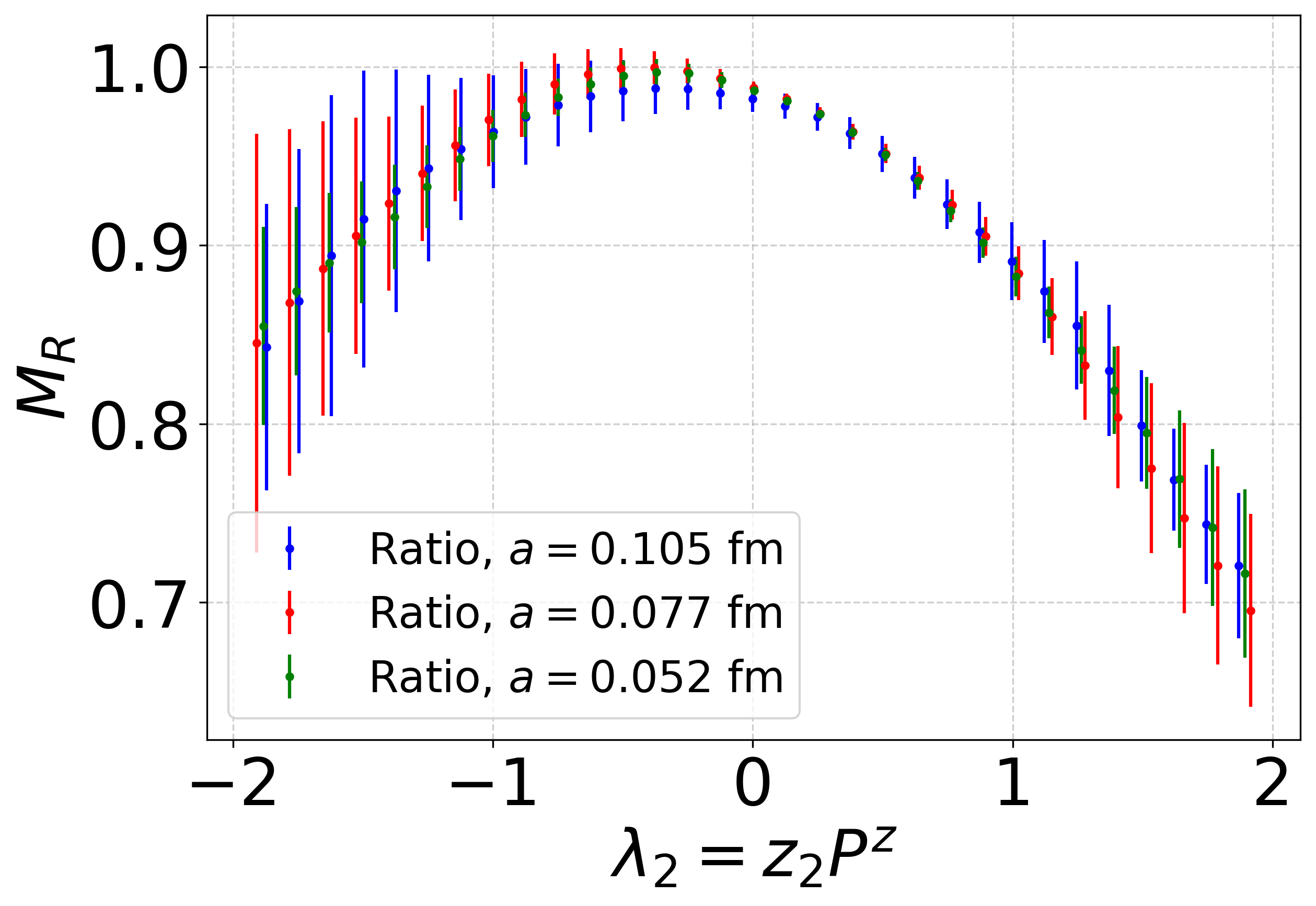}
    }
\vspace{0.0cm} 
\subfigure[\ Self scheme result of $\Lambda$ at $P=0.5$ GeV]{
    \centering
    \includegraphics[scale=0.185]{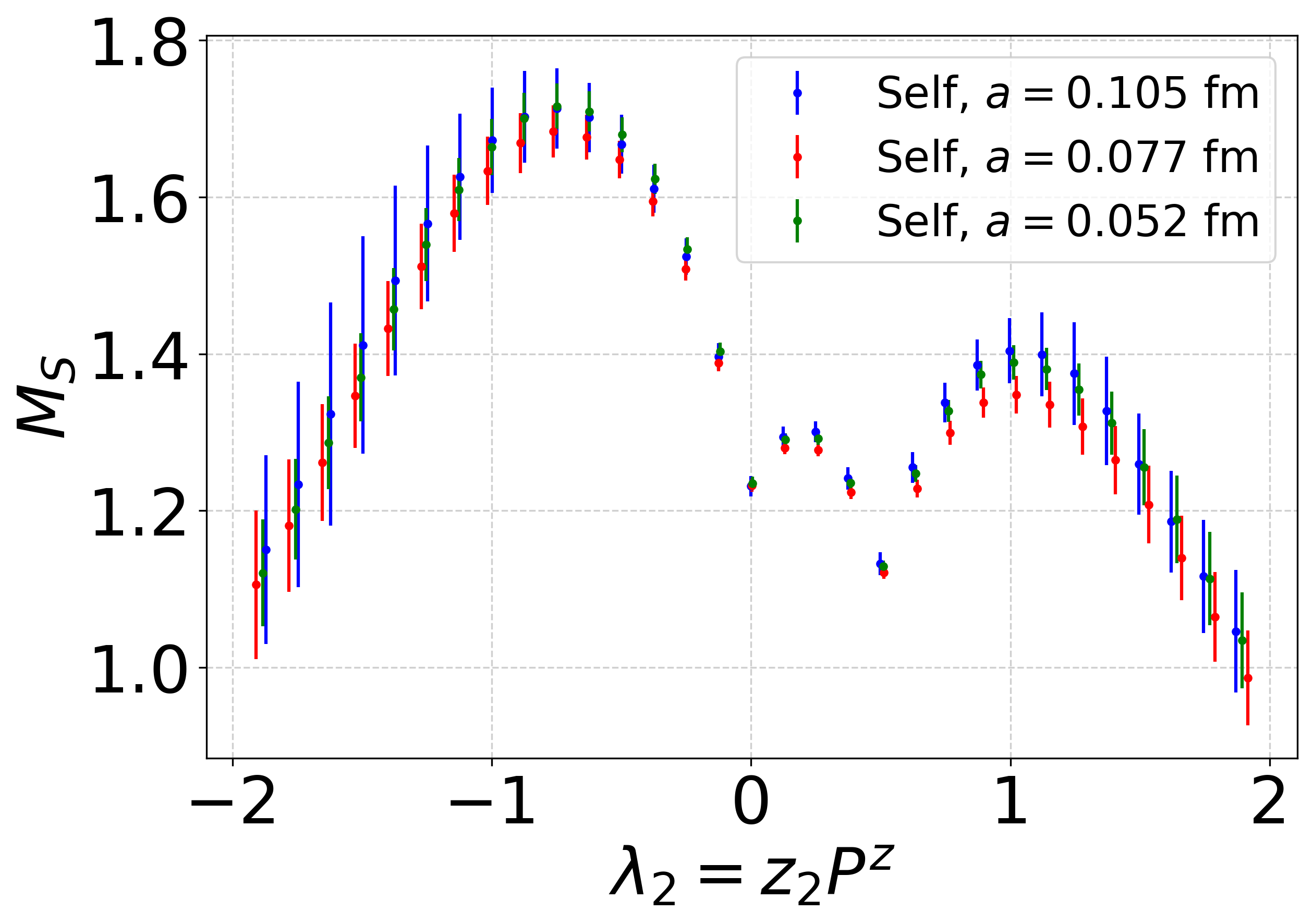}
    }
\vspace{0.0cm} 
\subfigure[\ Hybrid scheme result of $\Lambda$ at $P=0.5$ GeV]{
    \centering
    \includegraphics[scale=0.185]{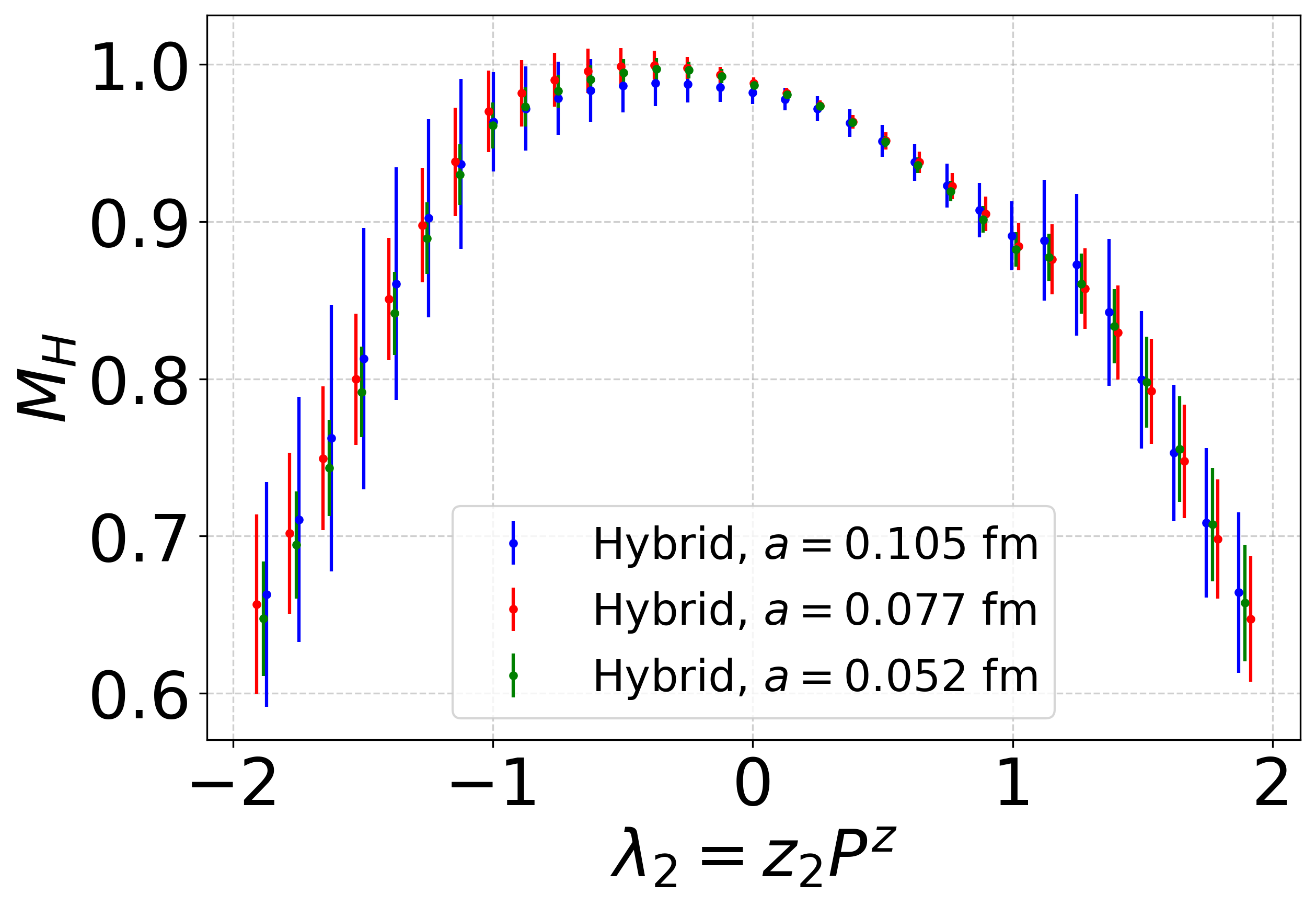}
    }
\caption{Results of the $\Lambda$ quasi-DA matrix elements in different schemes and with $P^z=0.5$ GeV, $z_1=0.200$ fm}
\label{fig:lambda_p1_z4}
\end{figure}

\begin{figure}[htbp]
\centering
\subfigure[\ Bare result of $\Lambda$ at $P=0.5$ GeV]{
    \centering
    \includegraphics[scale=0.185]{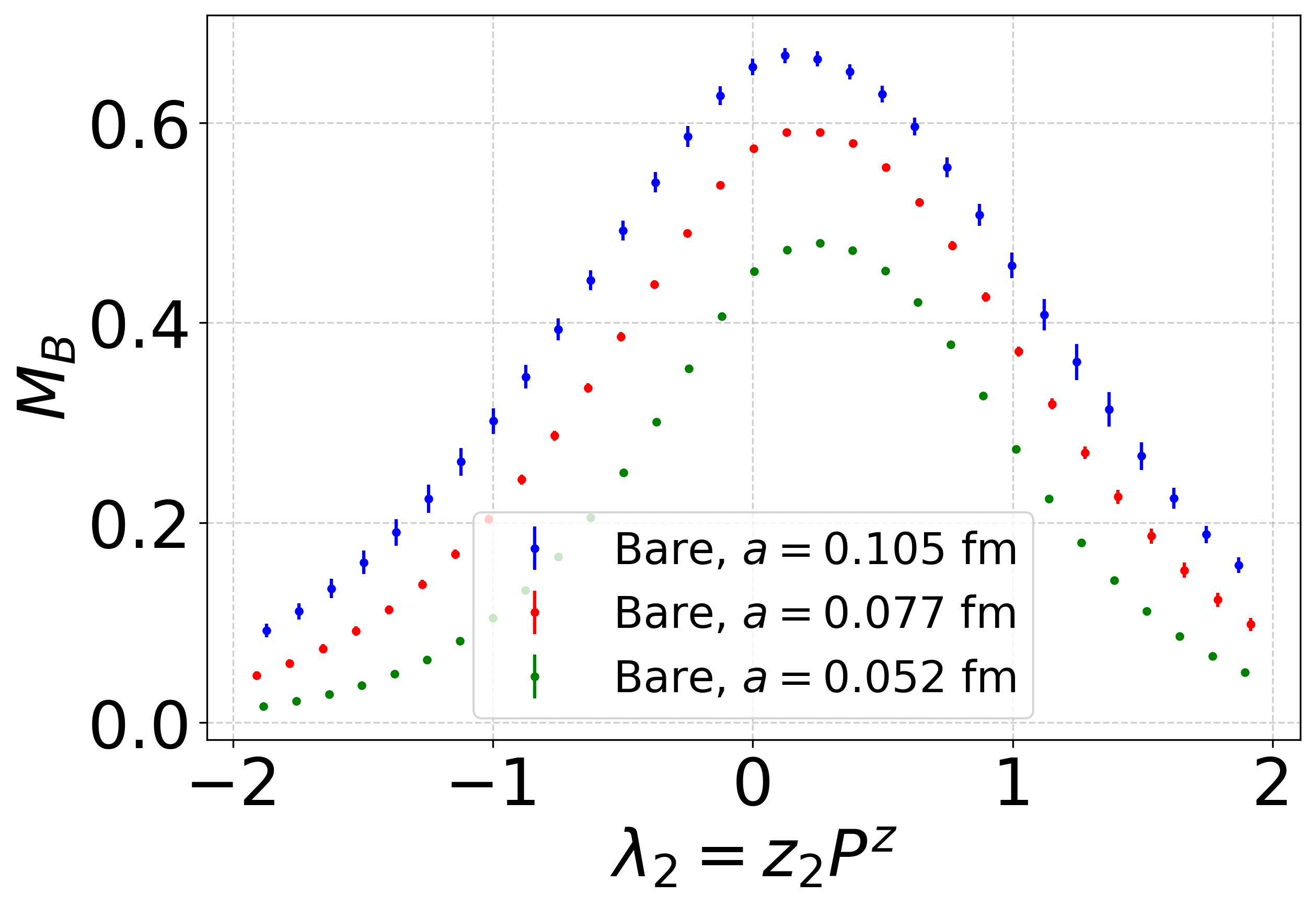}
    }
\vspace{0.0cm} 
\subfigure[\ Ratio scheme result of $\Lambda$ at $P=0.5$ GeV]{
    \centering
    \includegraphics[scale=0.185]{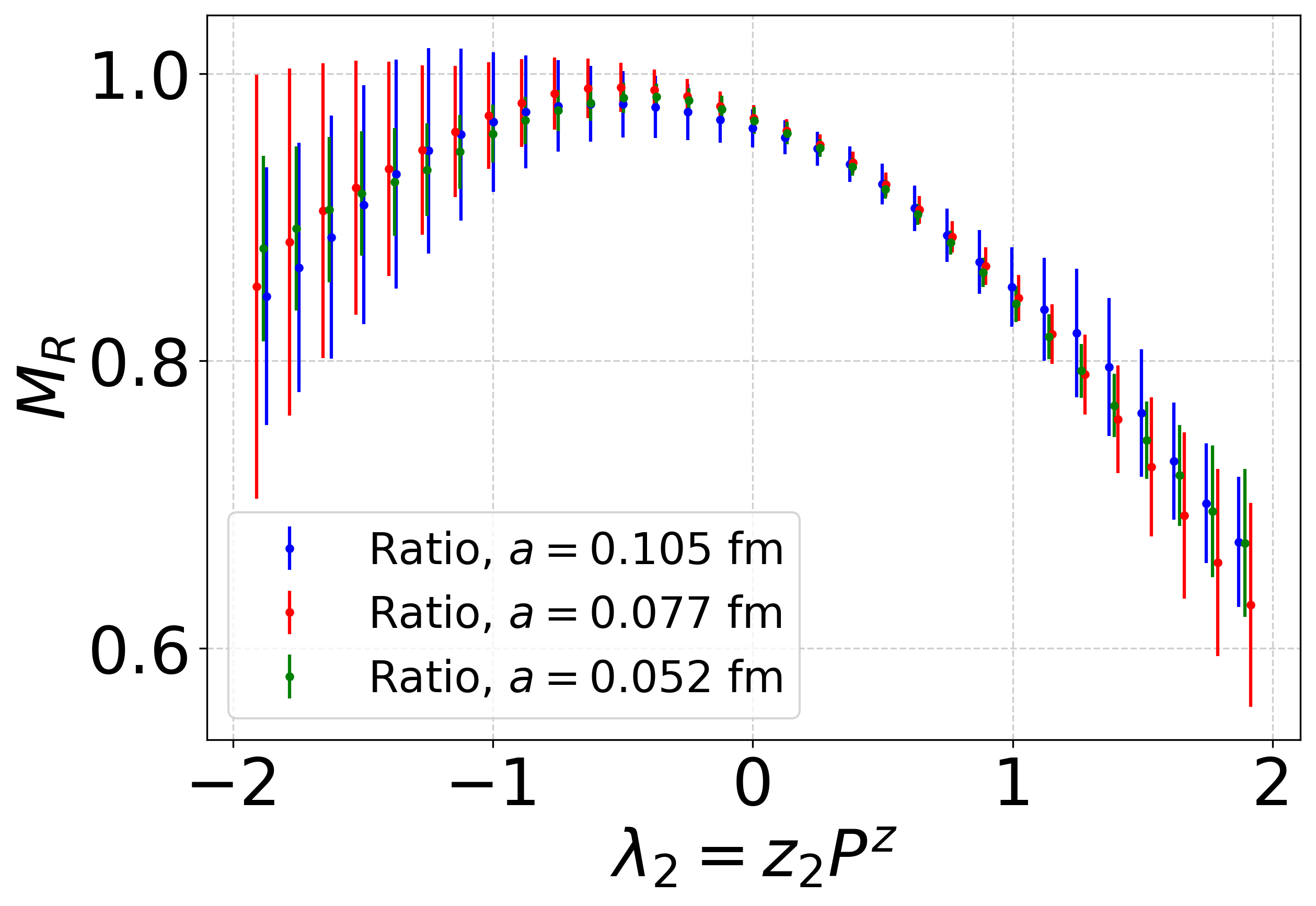}
    }
\vspace{0.0cm} 
\subfigure[\ Self scheme result of $\Lambda$ at $P=0.5$ GeV]{
    \centering
    \includegraphics[scale=0.185]{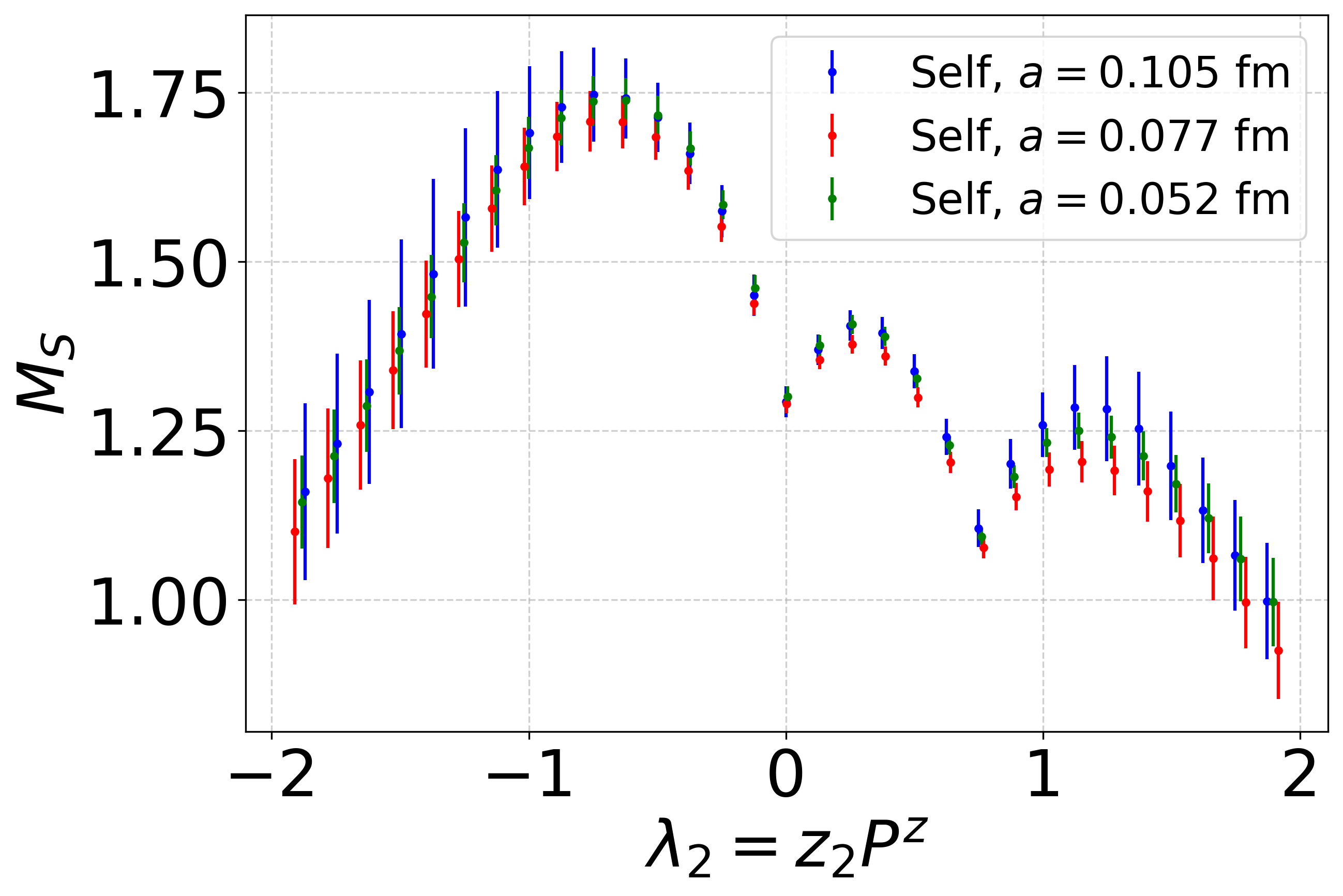}
    }
\vspace{0.0cm} 
\subfigure[\ Hybrid scheme result of $\Lambda$ at $P=0.5$ GeV]{
    \centering
    \includegraphics[scale=0.185]{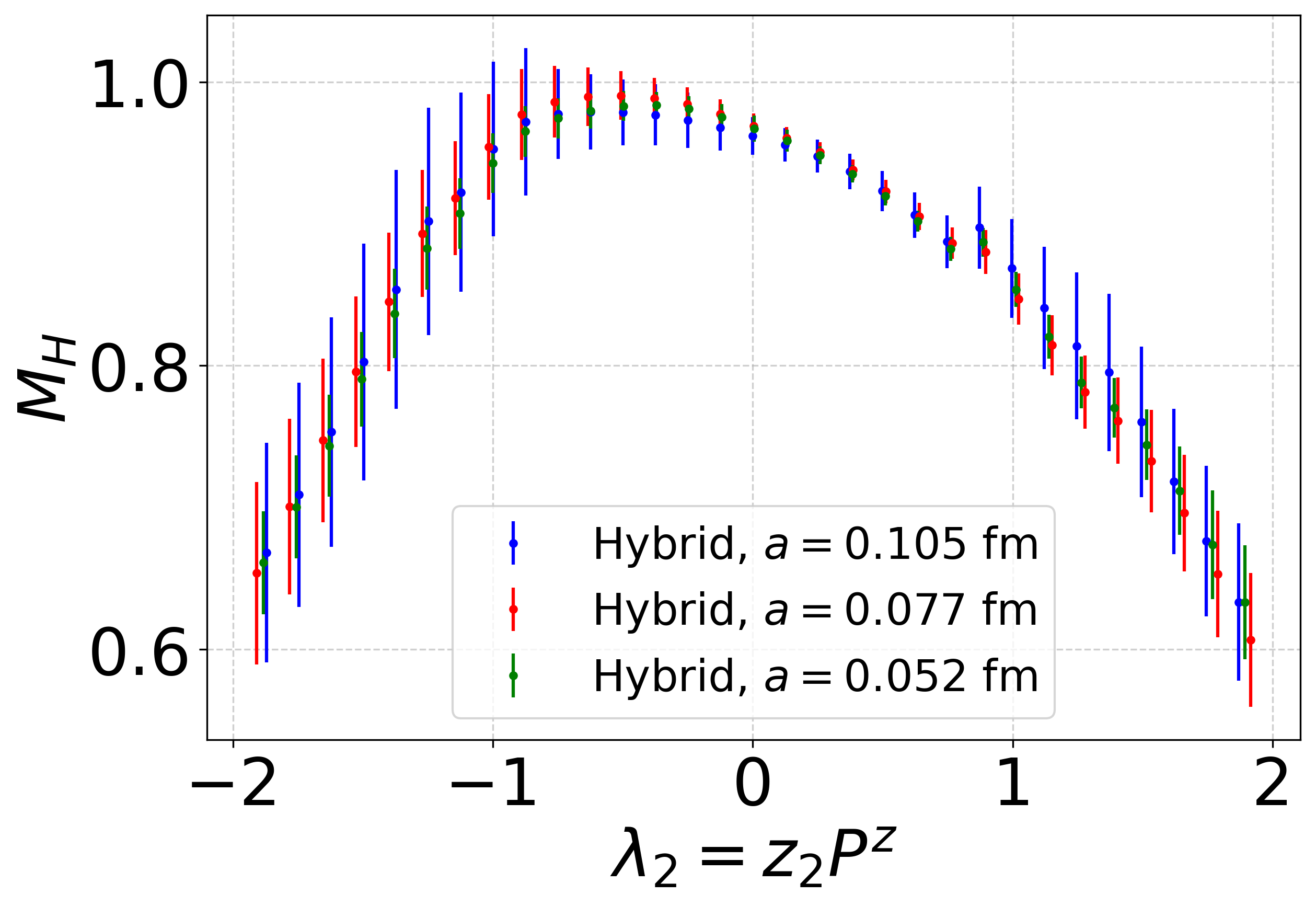}
    }
\caption{Results of the $\Lambda$ quasi-DA matrix elements in different schemes and with $P^z=0.5$ GeV, $z_1=0.300$ fm}
\label{fig:lambda_p1_z6}
\end{figure}

\begin{figure}[htbp]
\centering
\subfigure[\ Bare result of $\Lambda$ at $P=0.5$ GeV]{
    \centering
    \includegraphics[scale=0.185]{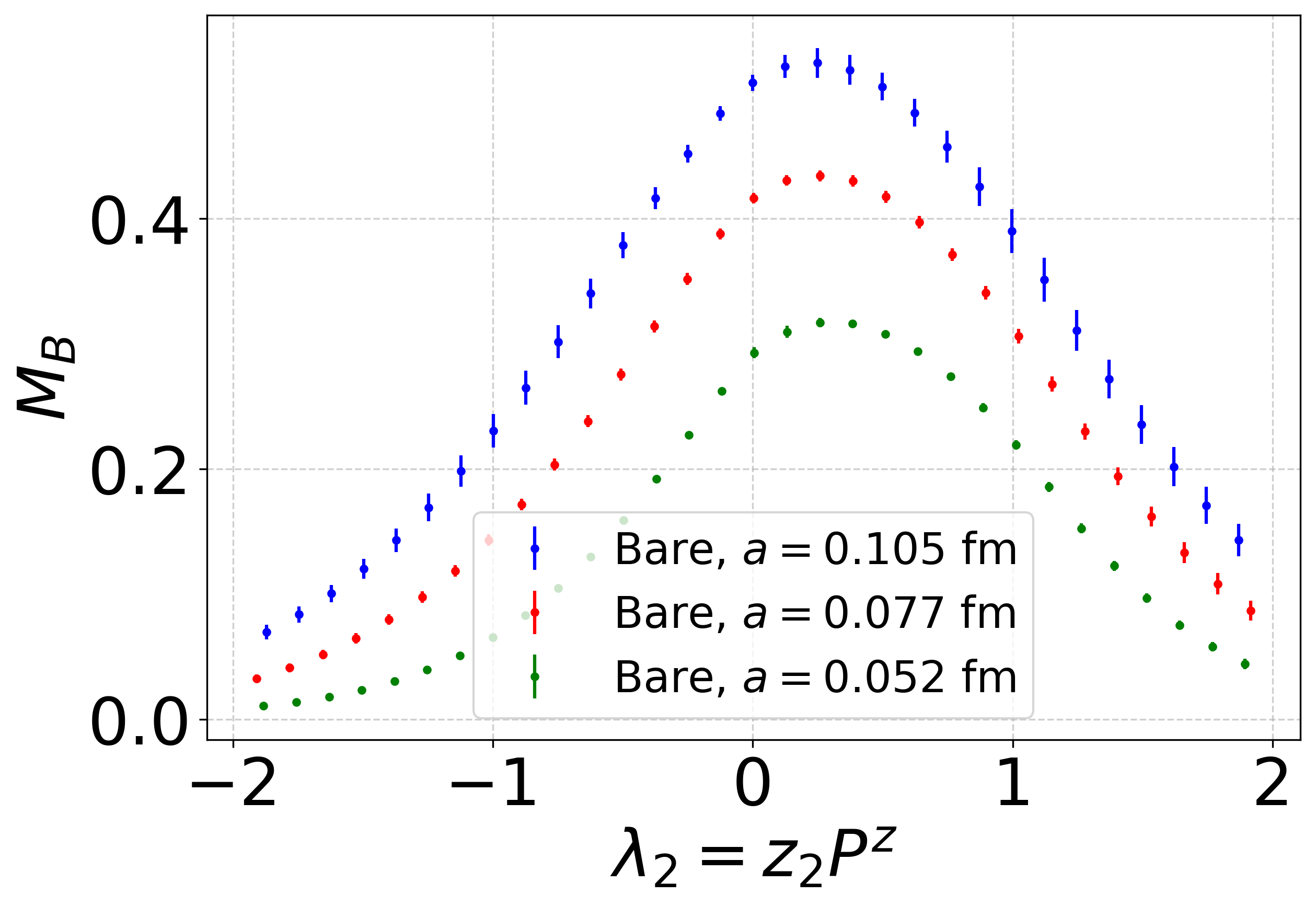}
    }
\vspace{0.0cm} 
\subfigure[\ Ratio scheme result of $\Lambda$ at $P=0.5$ GeV]{
    \centering
    \includegraphics[scale=0.185]{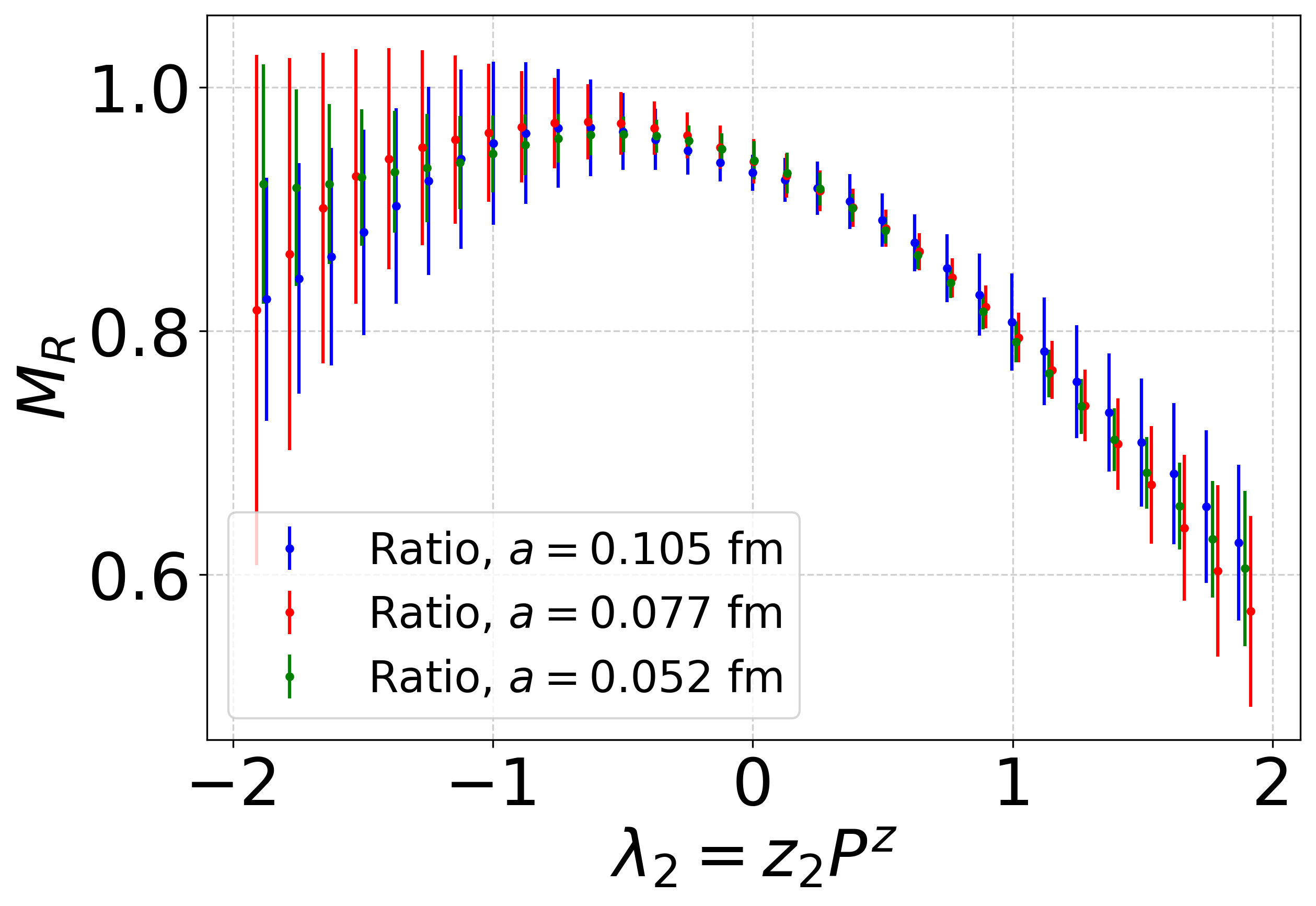}
    }
\vspace{0.0cm} 
\subfigure[\ Self scheme result of $\Lambda$ at $P=0.5$ GeV]{
    \centering
    \includegraphics[scale=0.185]{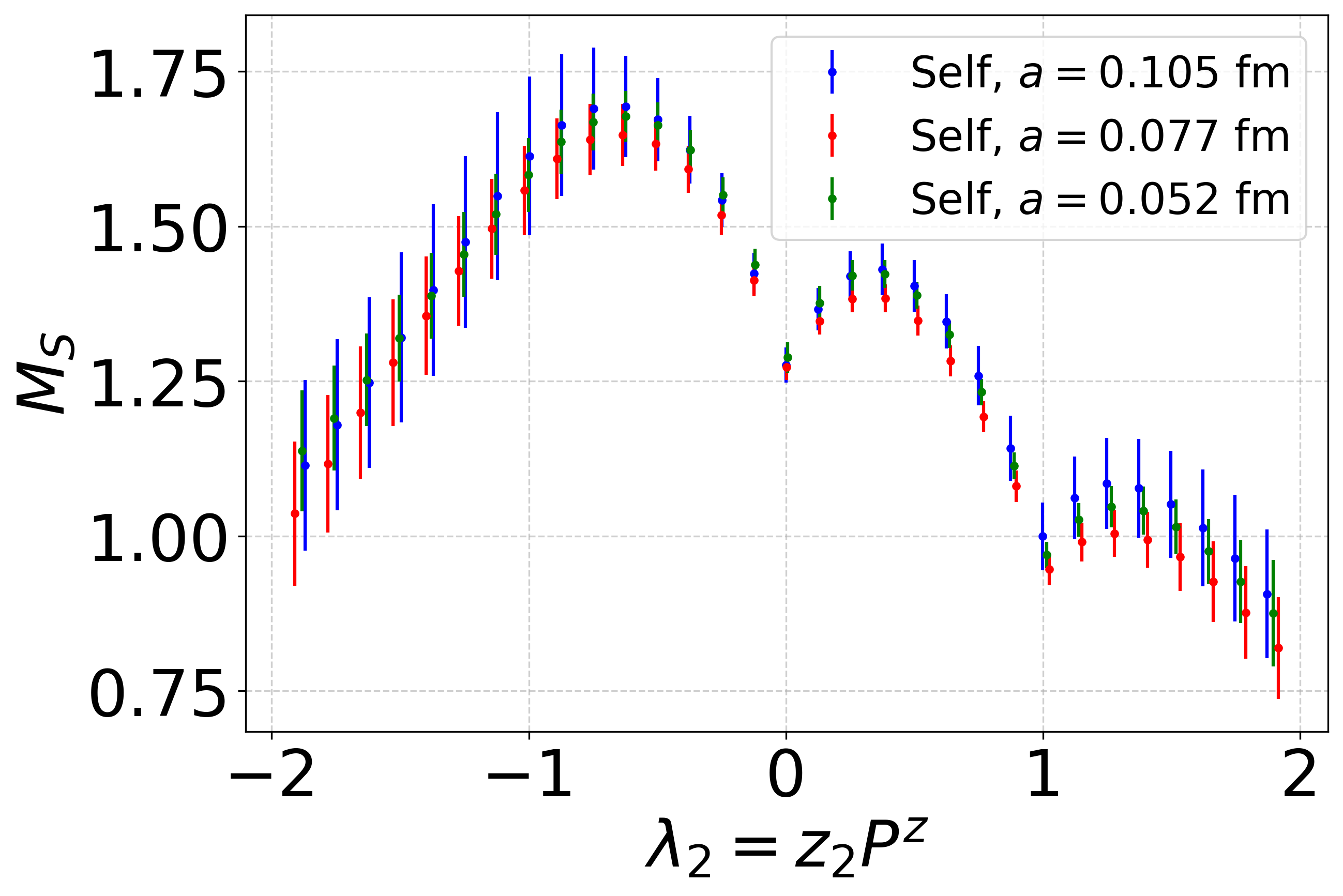}
    }
\vspace{0.0cm} 
\subfigure[\ Hybrid scheme result of $\Lambda$ at $P=0.5$ GeV]{
    \centering
    \includegraphics[scale=0.185]{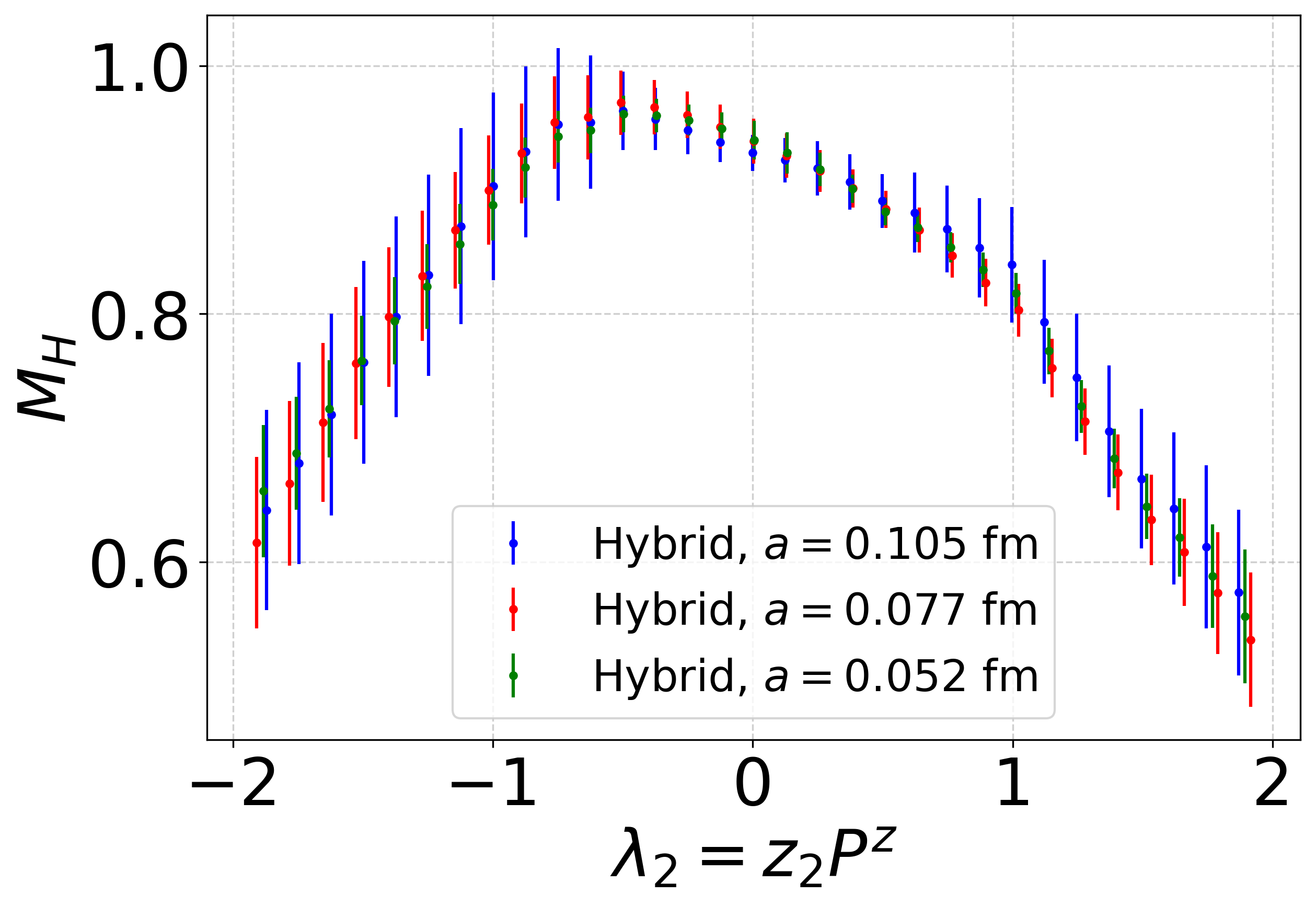}
    }
\caption{Results of the $\Lambda$ quasi-DA matrix elements in different schemes and with $P^z=0.5$ GeV, $z_1=0.400$ fm}
\label{fig:lambda_p1_z8}
\end{figure}

\begin{figure}[htbp]
\centering
\subfigure[\ Bare result of $\Lambda$ at $P=0.5$ GeV]{
    \centering
    \includegraphics[scale=0.185]{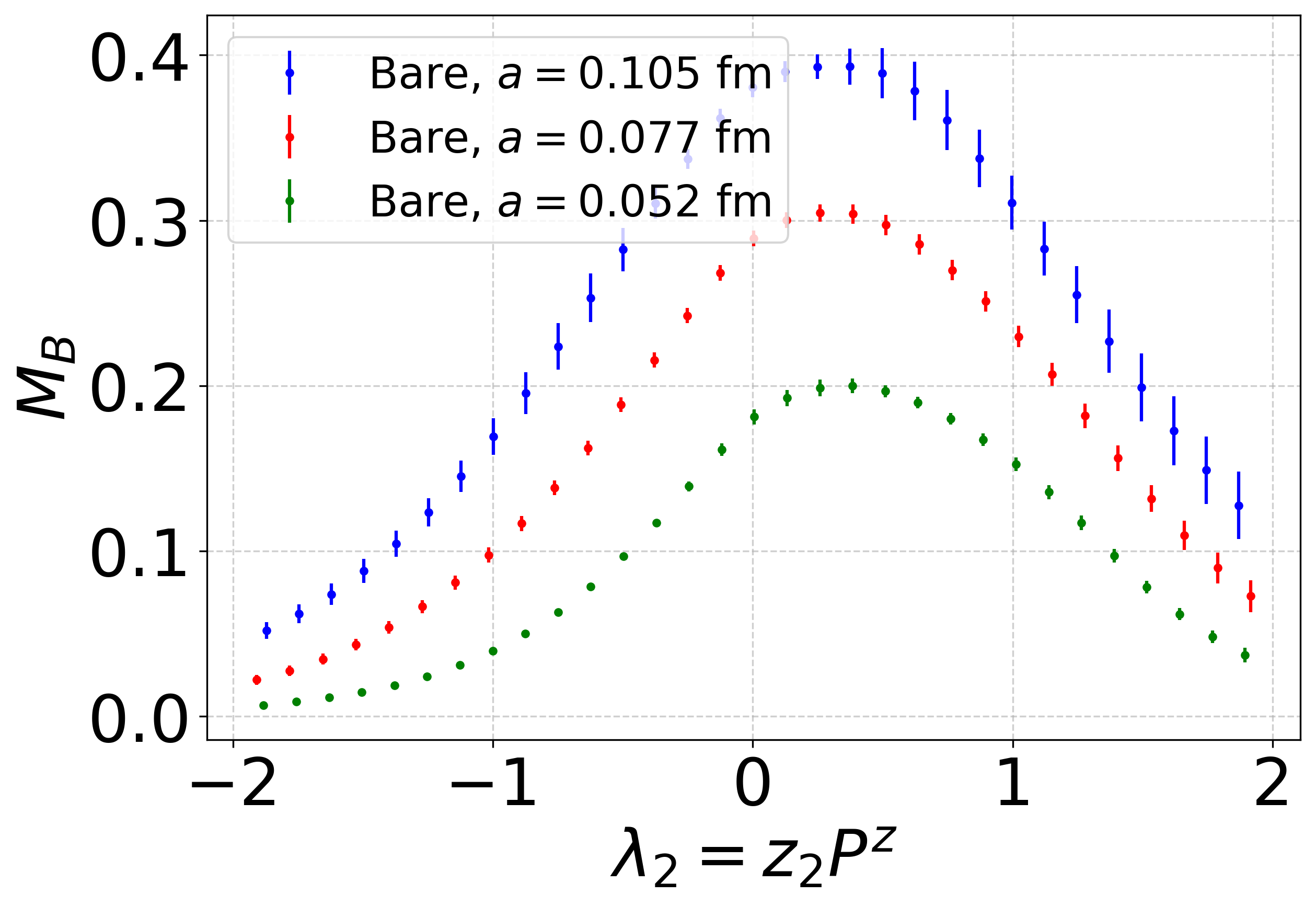}
    }
\vspace{0.0cm} 
\subfigure[\ Ratio scheme result of $\Lambda$ at $P=0.5$ GeV]{
    \centering
    \includegraphics[scale=0.185]{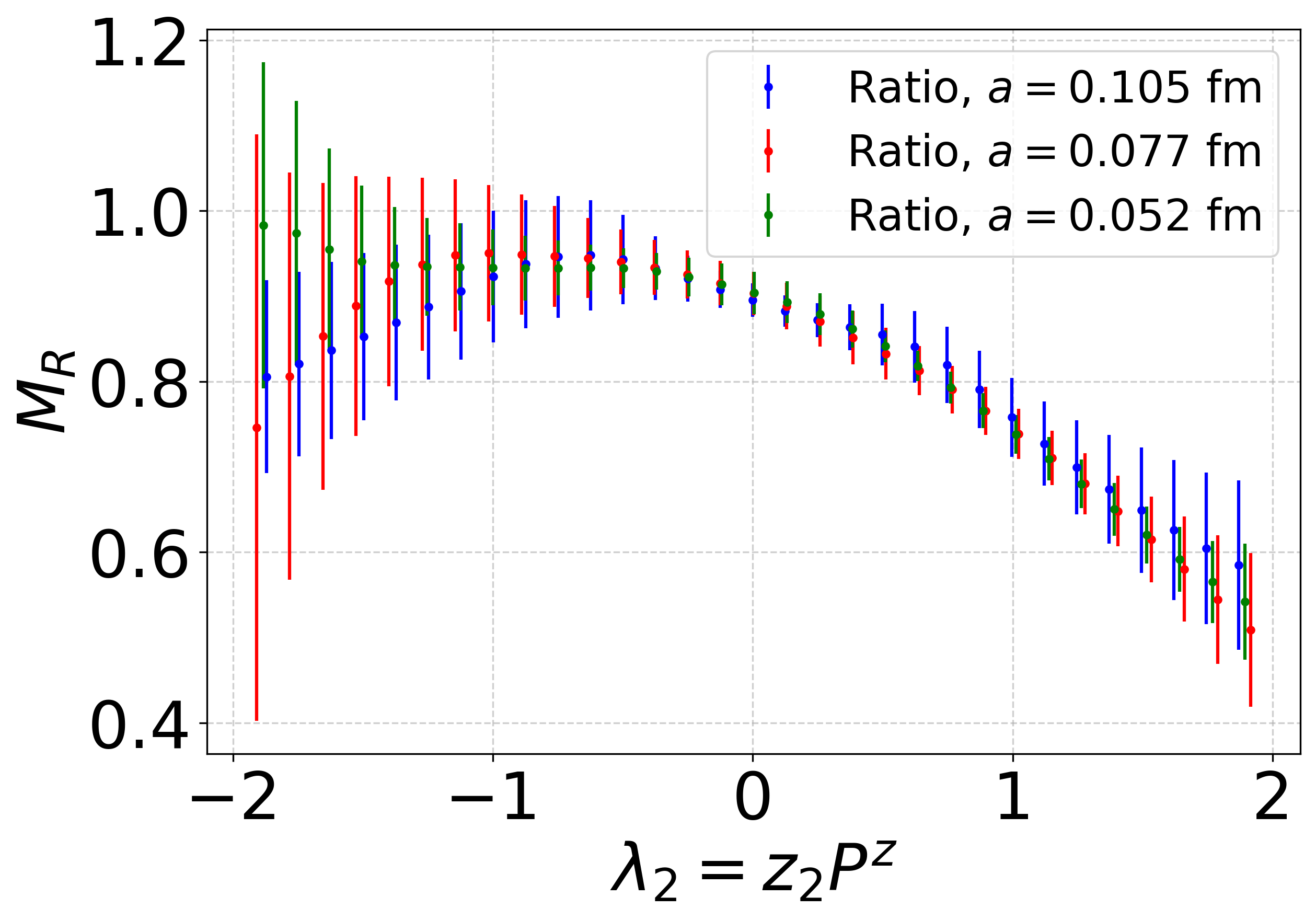}
    }
\vspace{0.0cm} 
\subfigure[\ Self scheme result of $\Lambda$ at $P=0.5$ GeV]{
    \centering
    \includegraphics[scale=0.185]{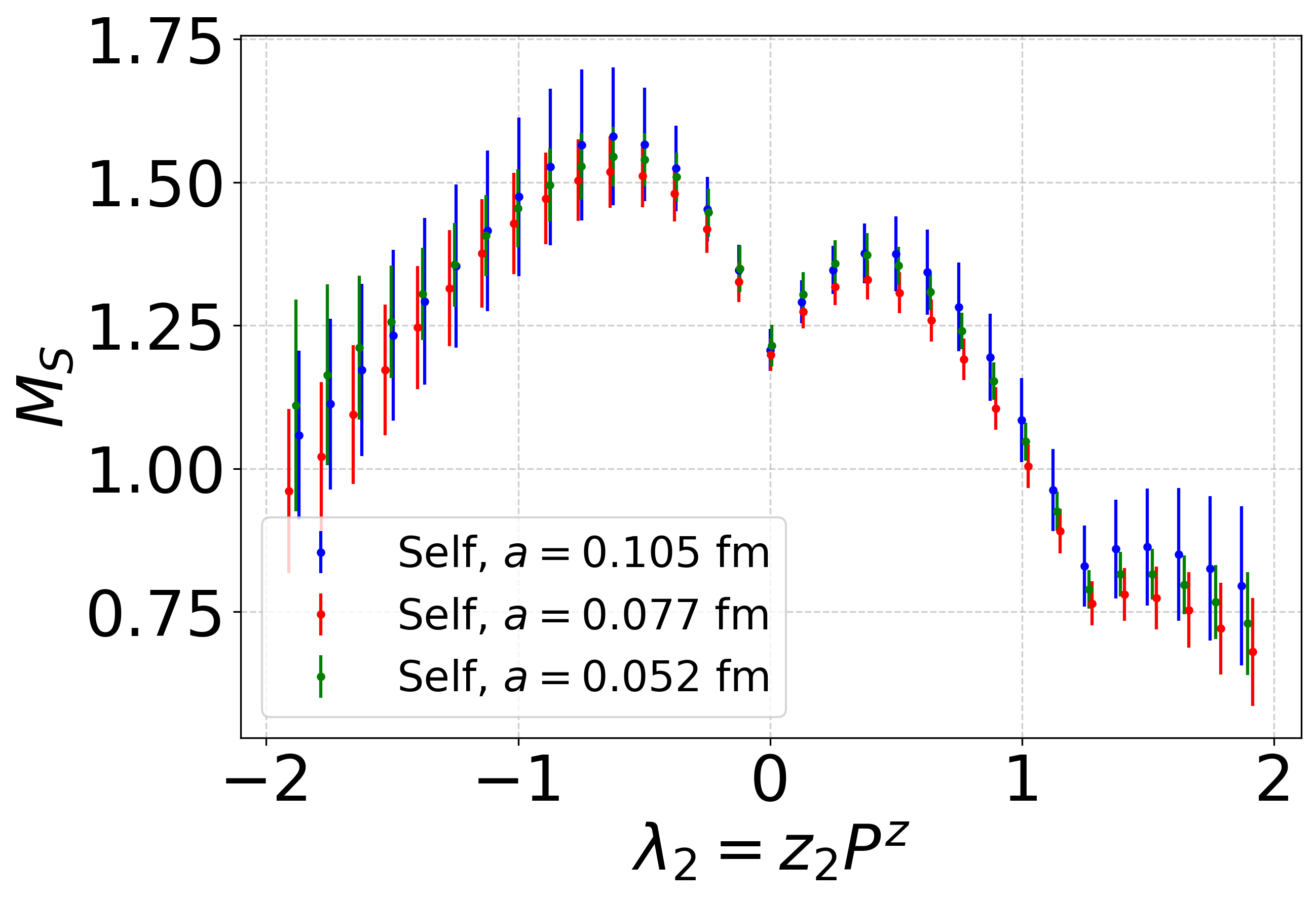}
    }
\vspace{0.0cm} 
\subfigure[\ Hybrid scheme result of $\Lambda$ at $P=0.5$ GeV]{
    \centering
    \includegraphics[scale=0.185]{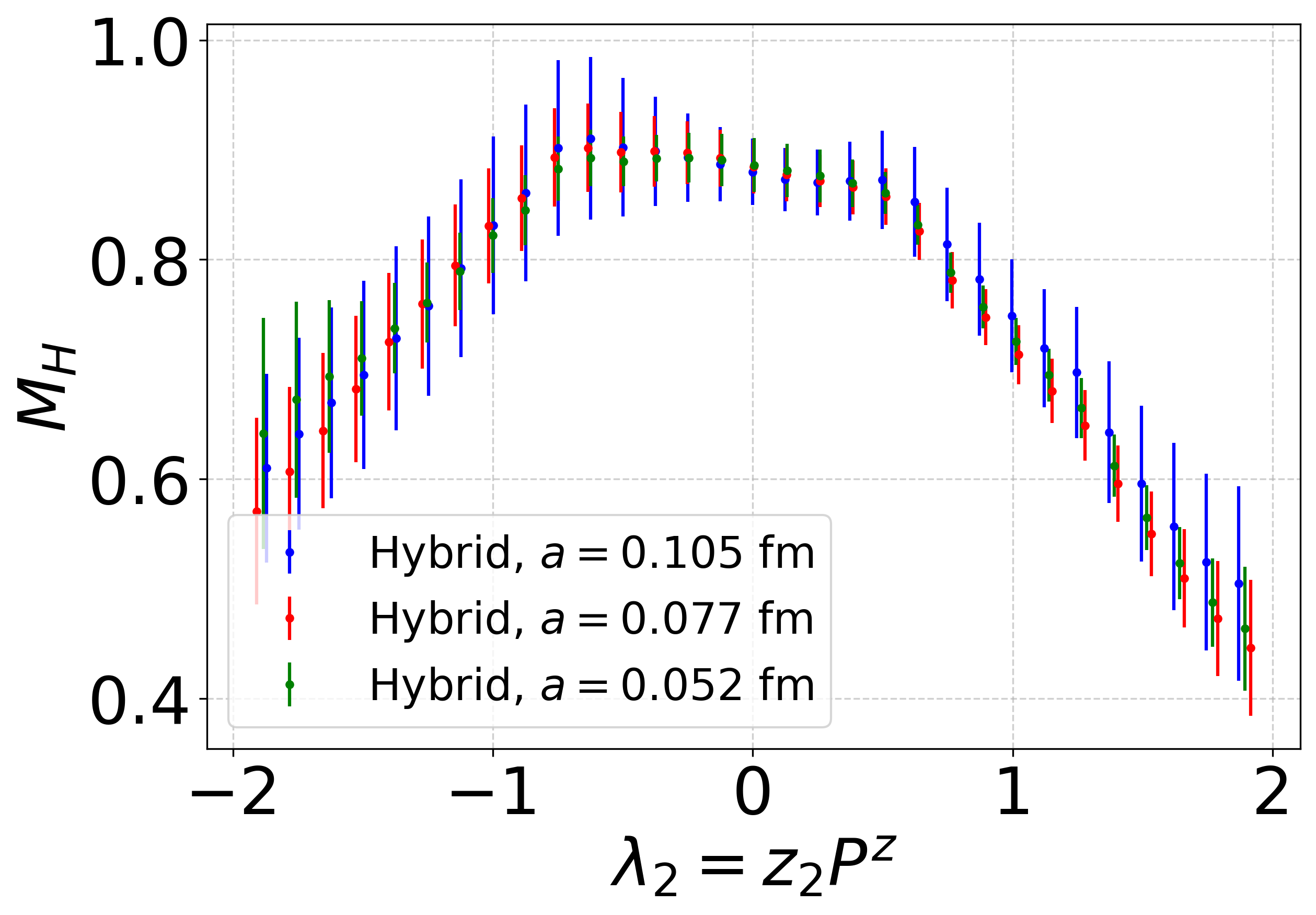}
    }
\caption{Results of the $\Lambda$ quasi-DA matrix elements in different schemes and with $P^z=0.5$ GeV, $z_1=0.500$ fm}
\label{fig:lambda_p1_z10}
\end{figure}

\clearpage

\subsection{More results of the $\Lambda$ (A-term) Quasi-DA at $P^z = 2.0$ GeV in different schemes}
\begin{figure}[htbp]
\centering
\subfigure[\ Bare result of $\Lambda$ at $P=2.0$ GeV]{
    \centering
    \includegraphics[scale=0.185]{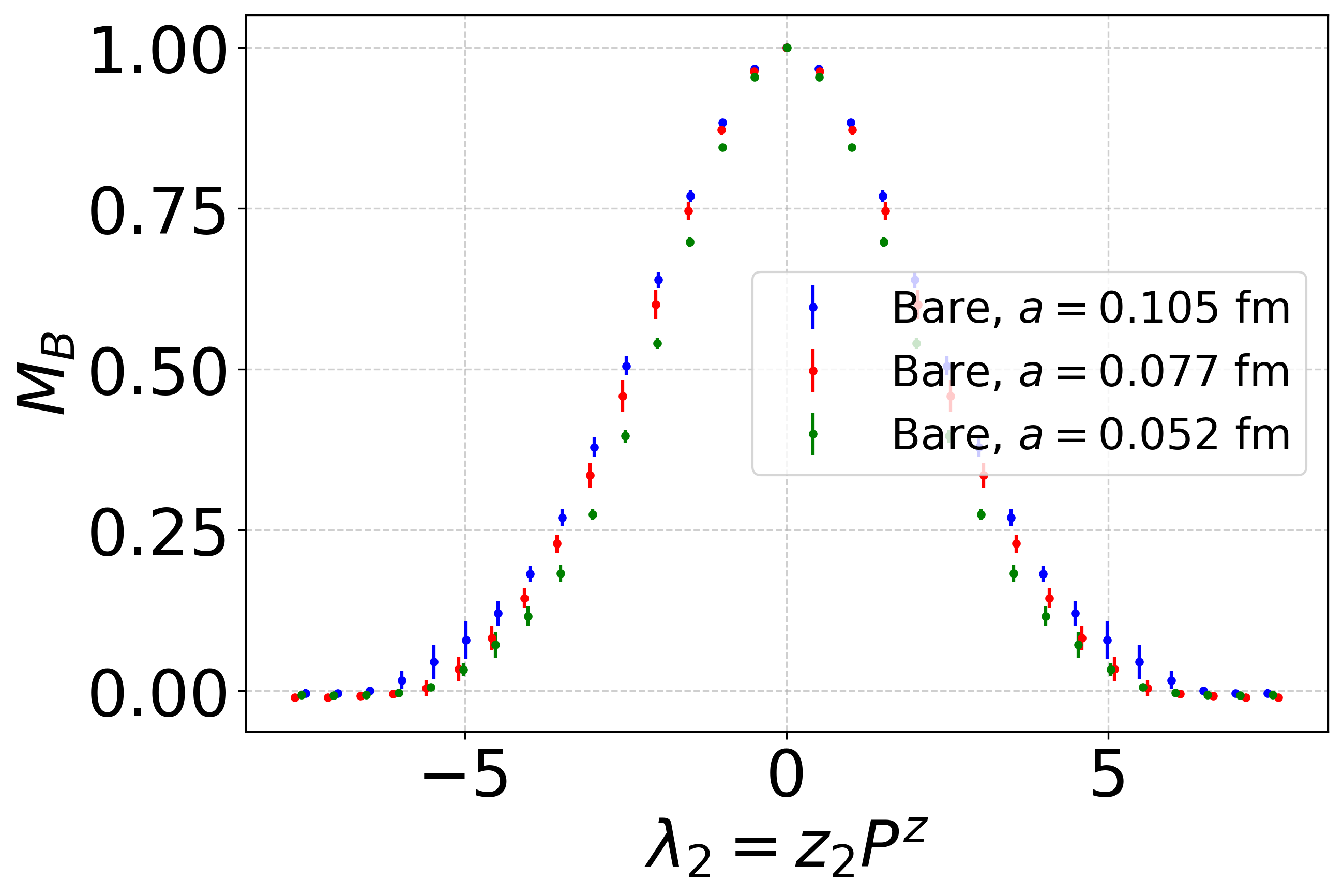}
    }
\vspace{0.0cm} 
\subfigure[\ Ratio scheme result of $\Lambda$ at $P=2.0$ GeV]{
    \centering
    \includegraphics[scale=0.185]{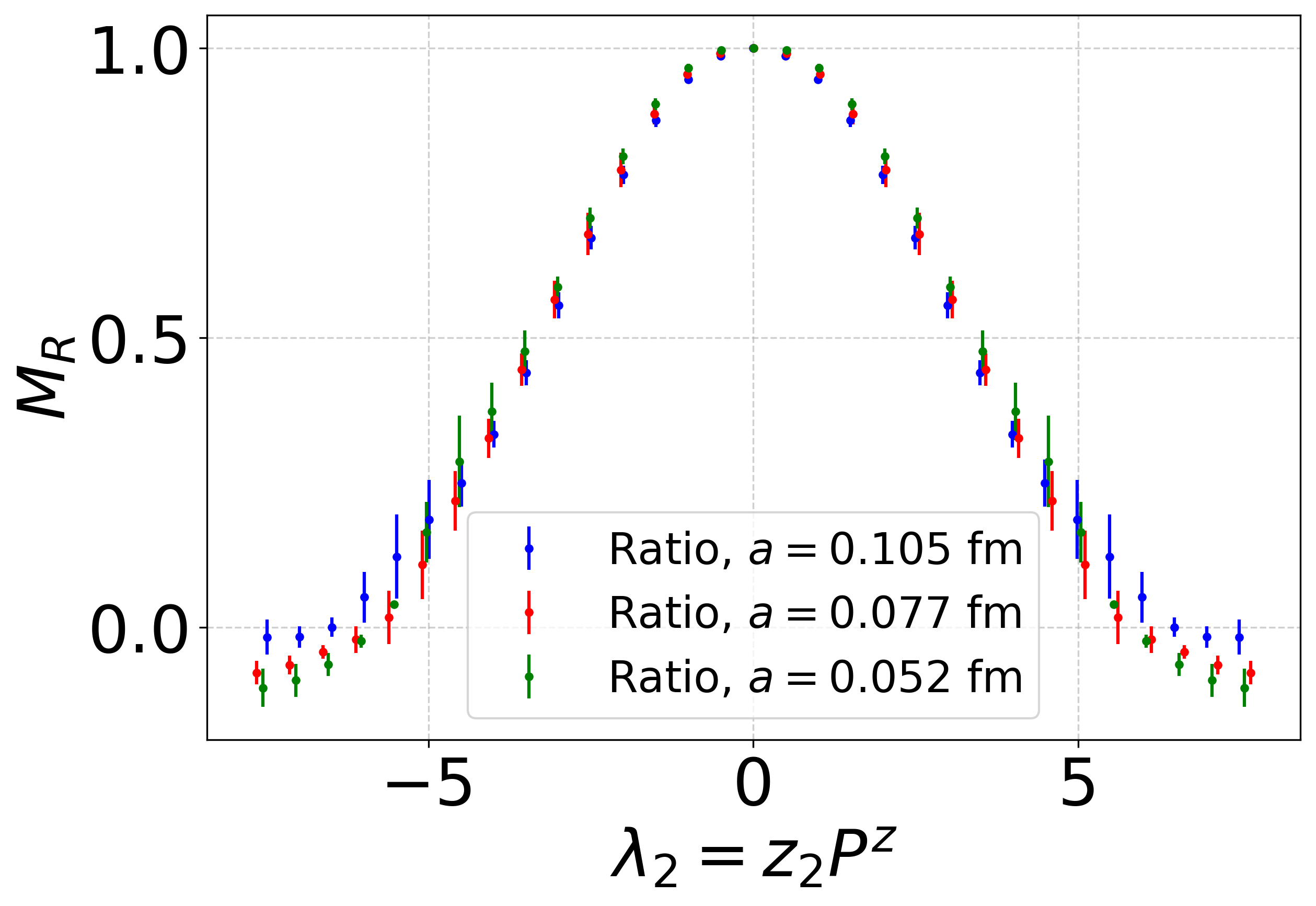}
    }
\vspace{0.0cm} 
\subfigure[\ Self scheme result of $\Lambda$ at $P=2.0$ GeV]{
    \centering
    \includegraphics[scale=0.185]{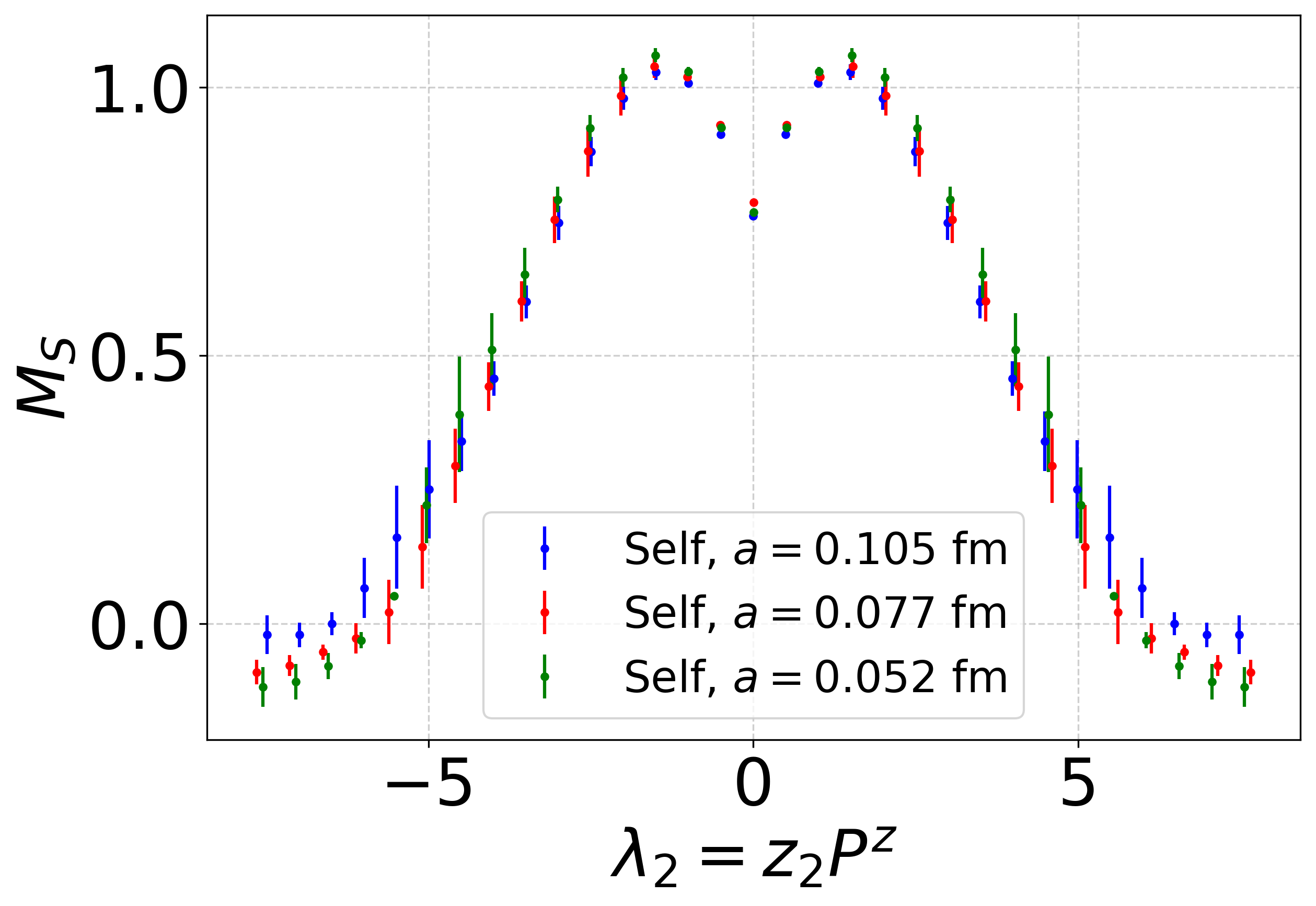}
    }
\vspace{0.0cm} 
\subfigure[\ Hybrid scheme result of $\Lambda$ at $P=2.0$ GeV]{
    \centering
    \includegraphics[scale=0.185]{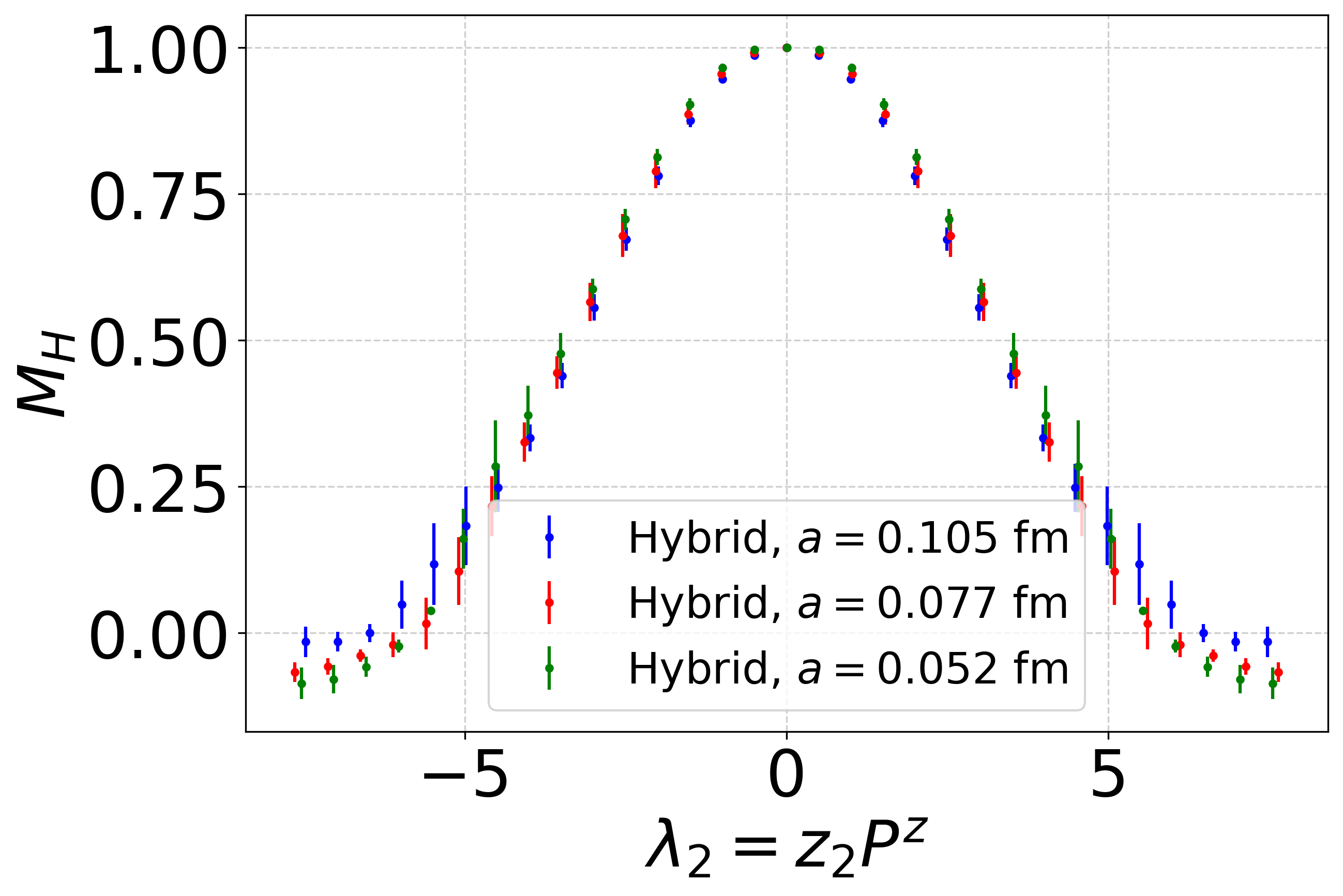}
    }
\caption{Results of the $\Lambda$ quasi-DA matrix elements in different schemes and with $P^z=2.0$ GeV, $z_1=0.000$ fm}
\label{fig:lambda_p4_z0}
\end{figure}

\begin{figure}[htbp]
\centering
\subfigure[\ Bare result of $\Lambda$ at $P=2.0$ GeV]{
    \centering
    \includegraphics[scale=0.185]{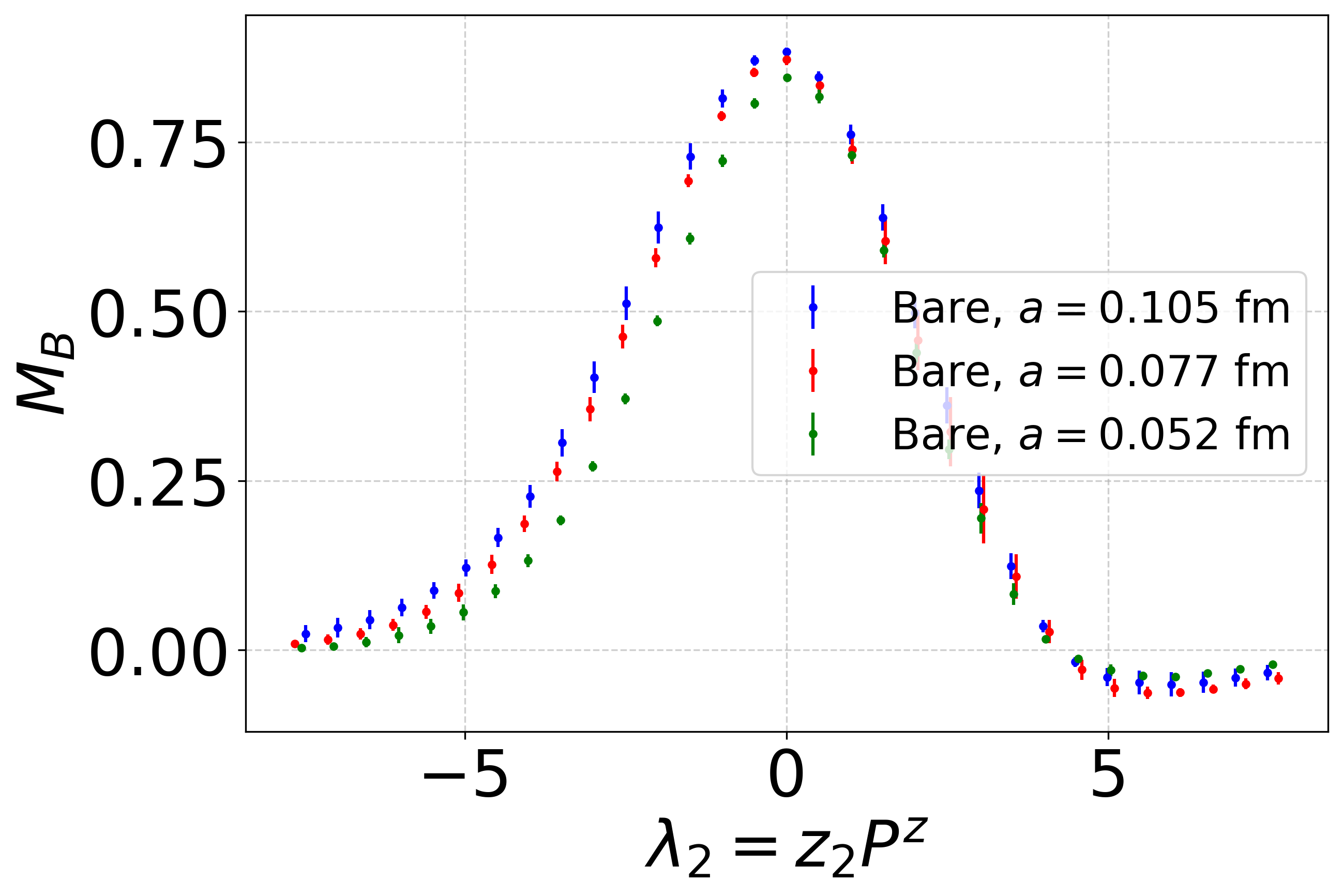}
    }
\vspace{0.0cm} 
\subfigure[\ Ratio scheme result of $\Lambda$ at $P=2.0$ GeV]{
    \centering
    \includegraphics[scale=0.185]{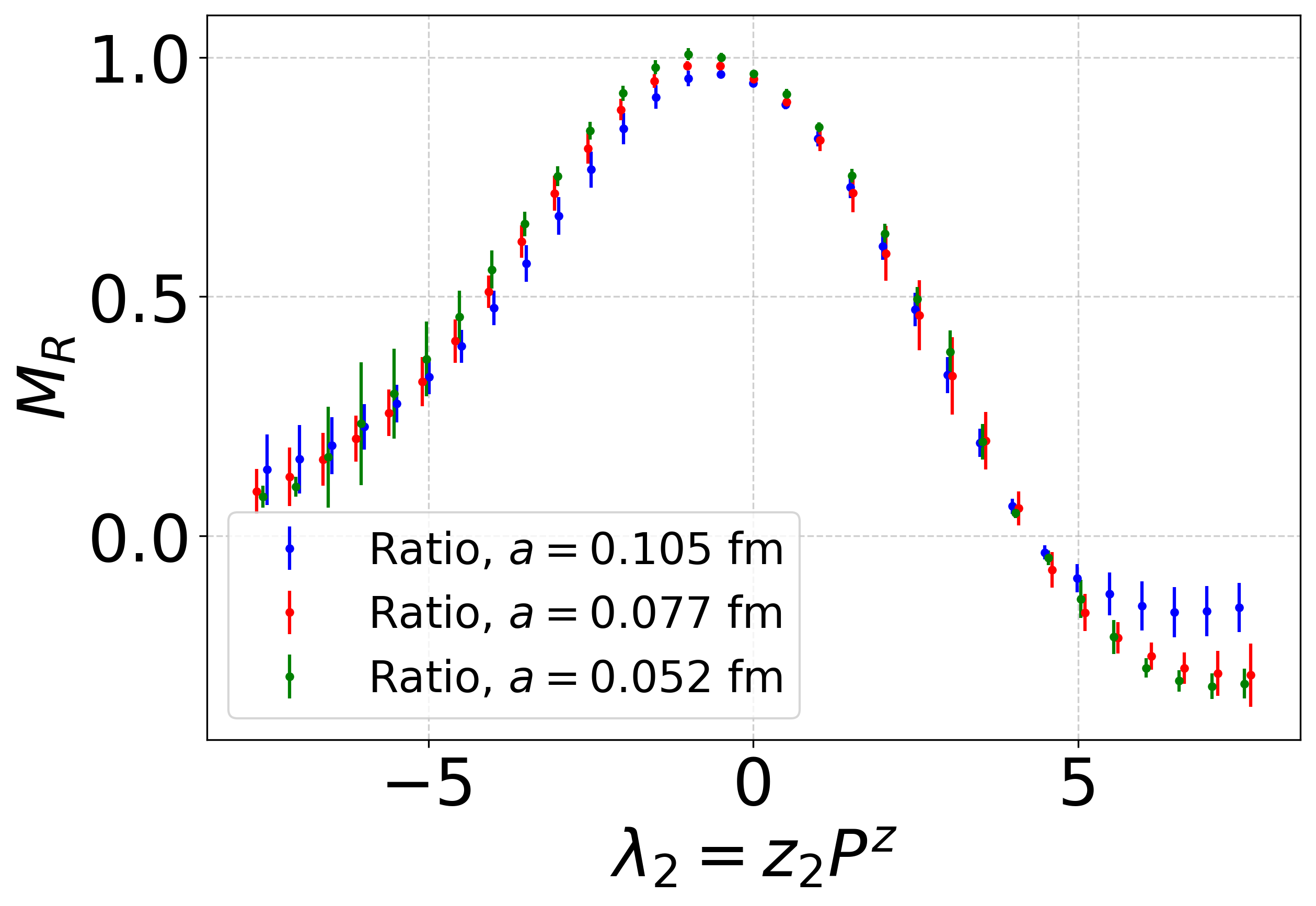}
    }
\vspace{0.0cm} 
\subfigure[\ Self scheme result of $\Lambda$ at $P=2.0$ GeV]{
    \centering
    \includegraphics[scale=0.185]{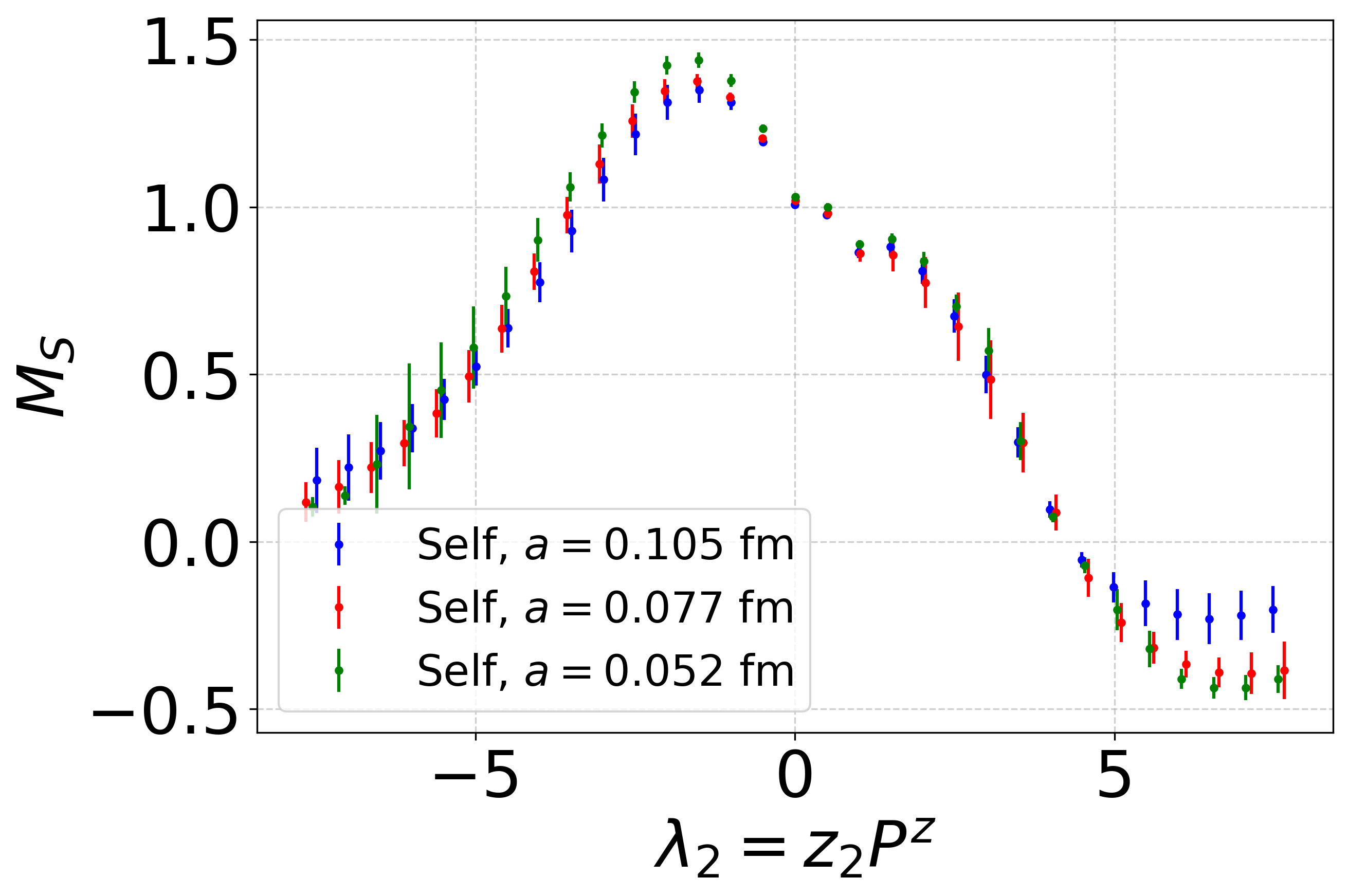}
    }
\vspace{0.0cm} 
\subfigure[\ Hybrid scheme result of $\Lambda$ at $P=2.0$ GeV]{
    \centering
    \includegraphics[scale=0.185]{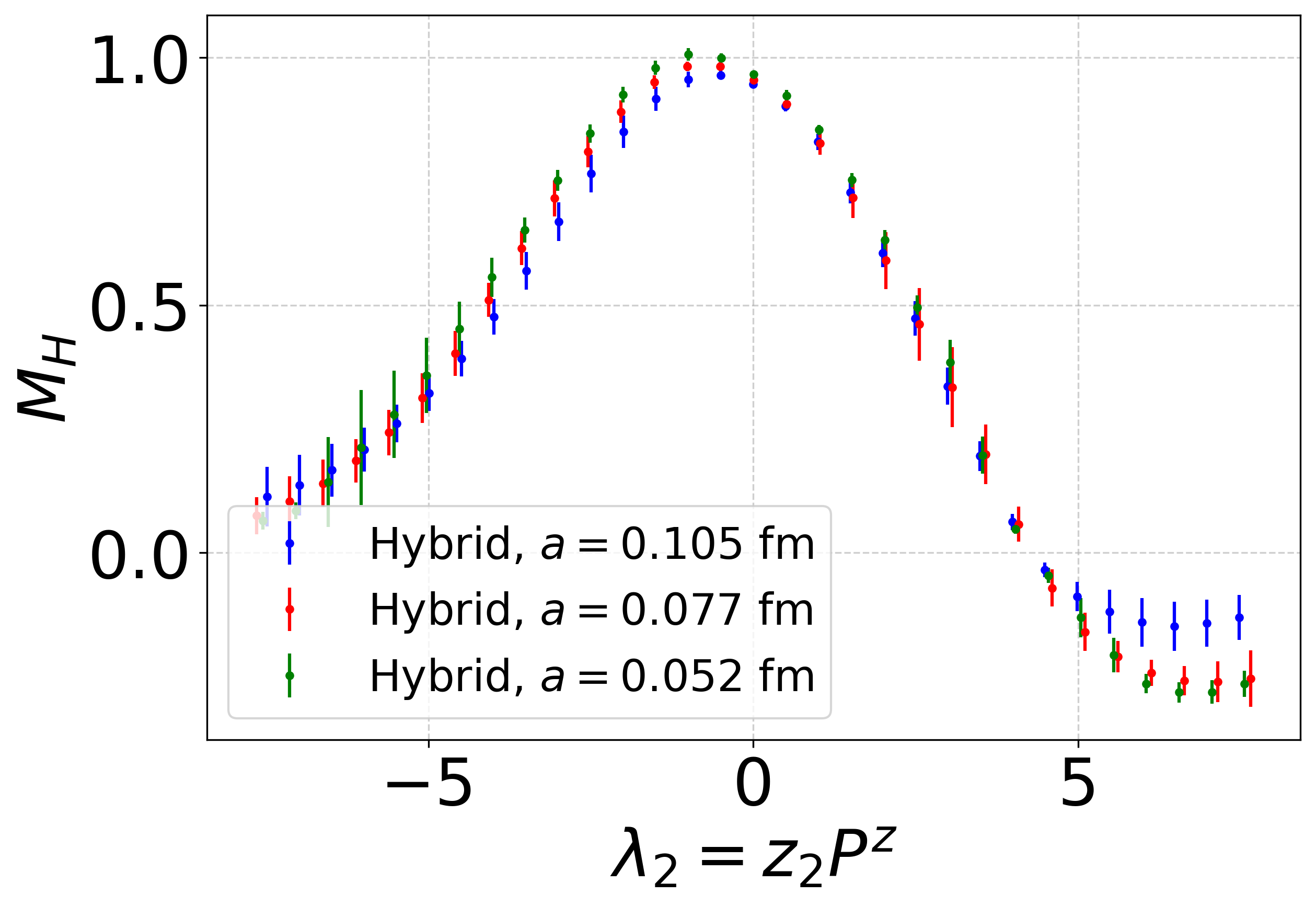}
    }
\caption{Results of the $\Lambda$ quasi-DA matrix elements in different schemes and with $P^z=2.0$ GeV, $z_1=0.100$ fm}
\label{fig:lambda_p4_z2}
\end{figure}

\begin{figure}[htbp]
\centering
\subfigure[\ Bare result of $\Lambda$ at $P=2.0$ GeV]{
    \centering
    \includegraphics[scale=0.185]{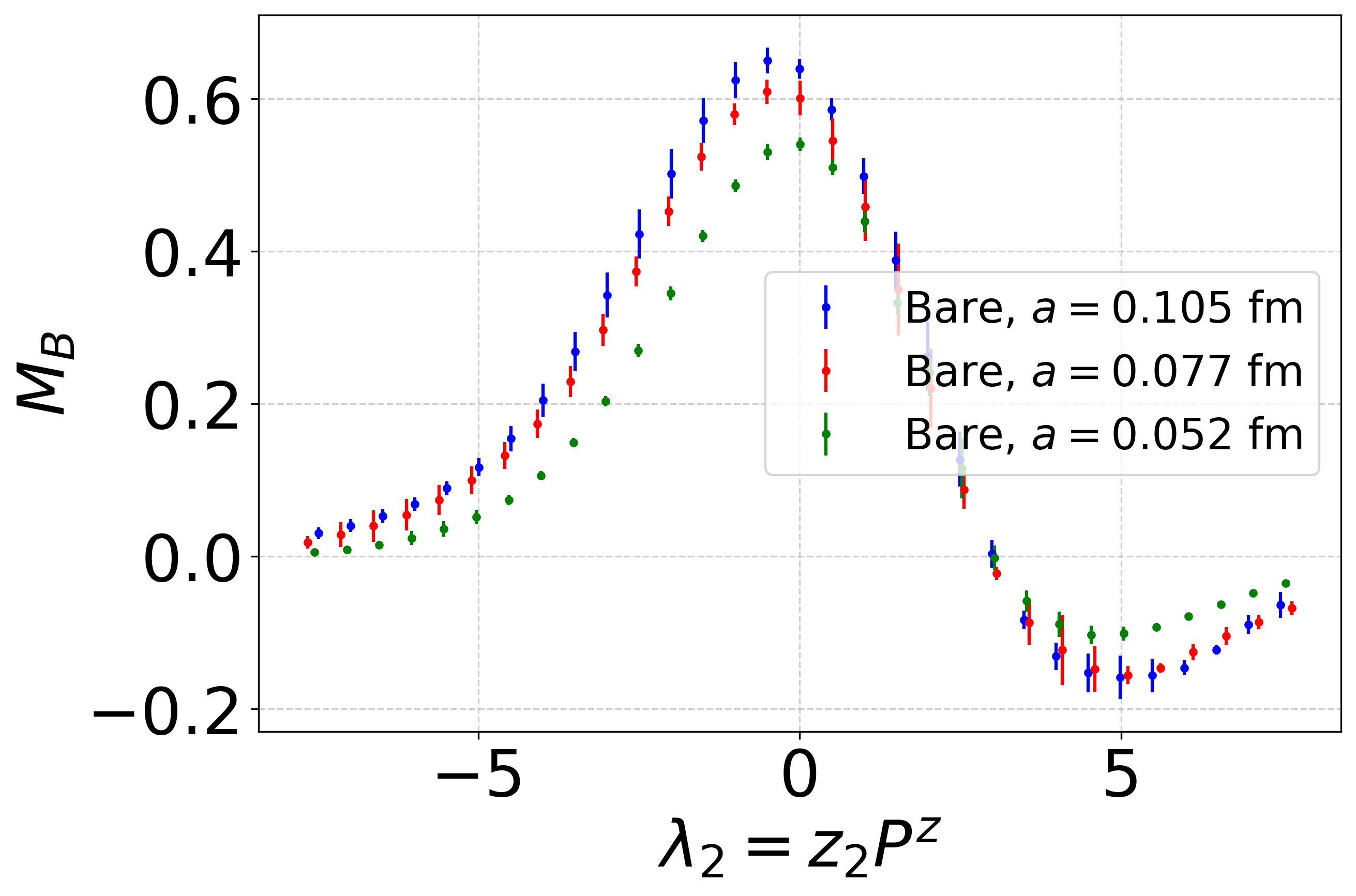}
    }
\vspace{0.0cm} 
\subfigure[\ Ratio scheme result of $\Lambda$ at $P=2.0$ GeV]{
    \centering
    \includegraphics[scale=0.185]{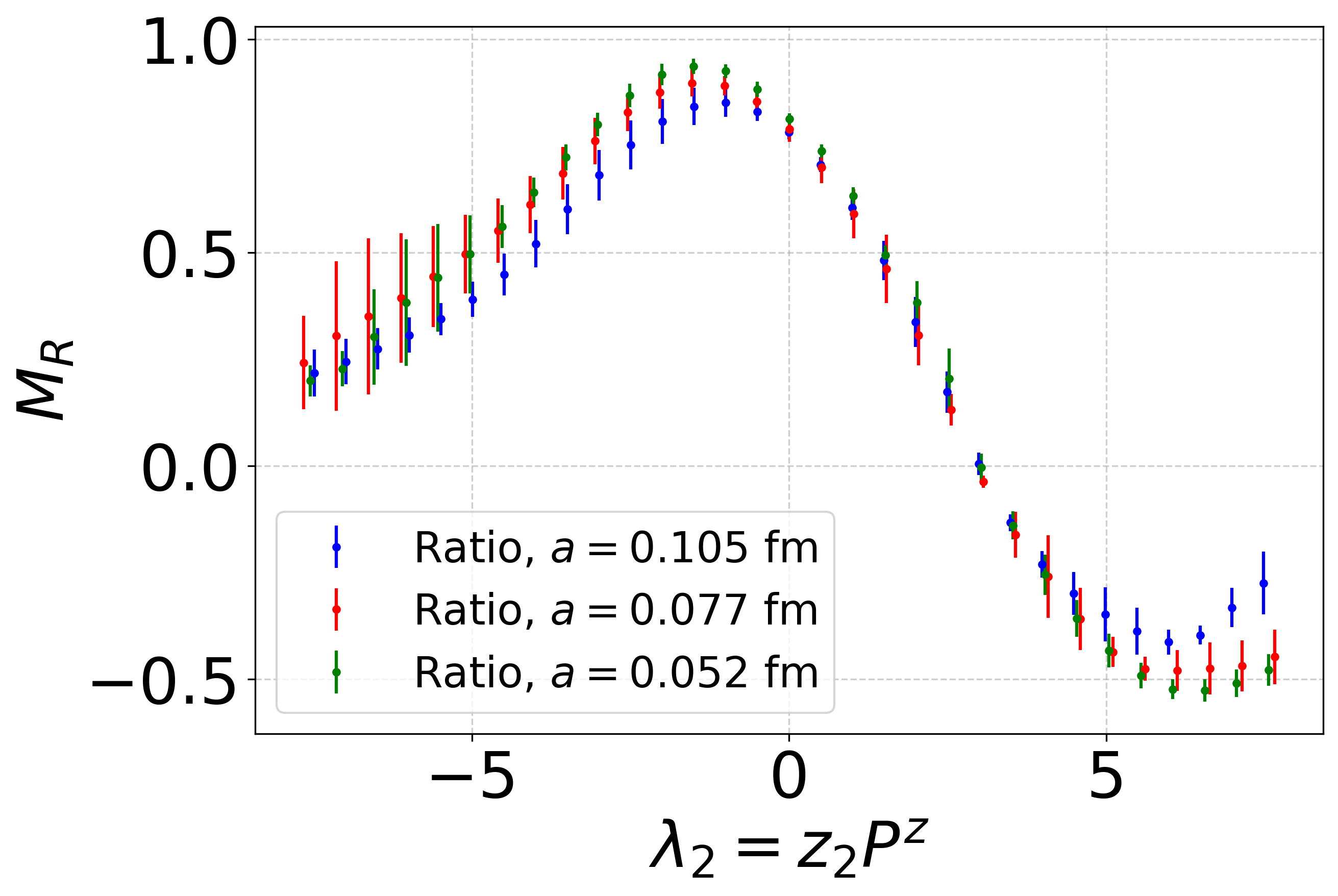}
    }
\vspace{0.0cm} 
\subfigure[\ Self scheme result of $\Lambda$ at $P=2.0$ GeV]{
    \centering
    \includegraphics[scale=0.185]{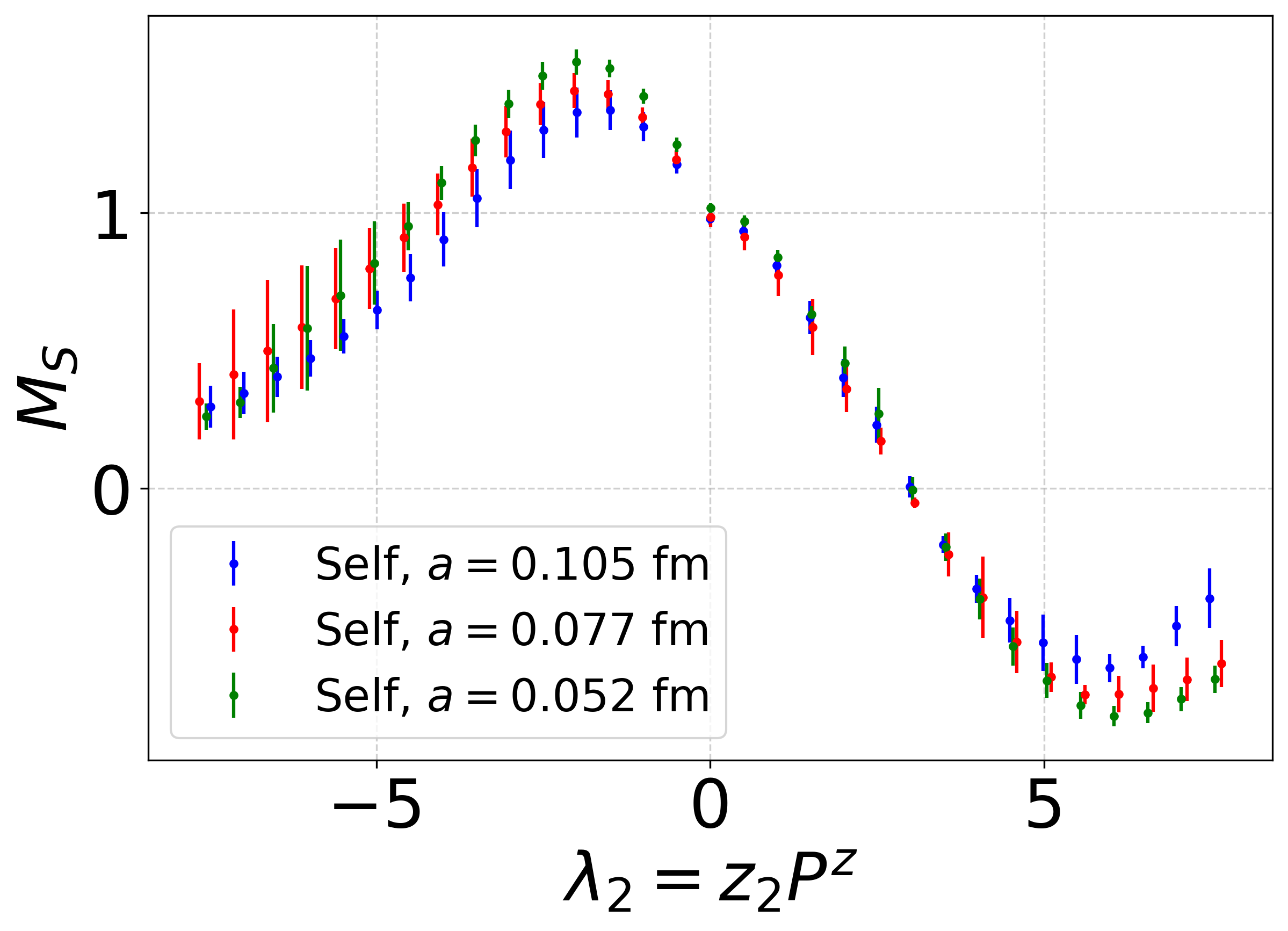}
    }
\vspace{0.0cm} 
\subfigure[\ Hybrid scheme result of $\Lambda$ at $P=2.0$ GeV]{
    \centering
    \includegraphics[scale=0.185]{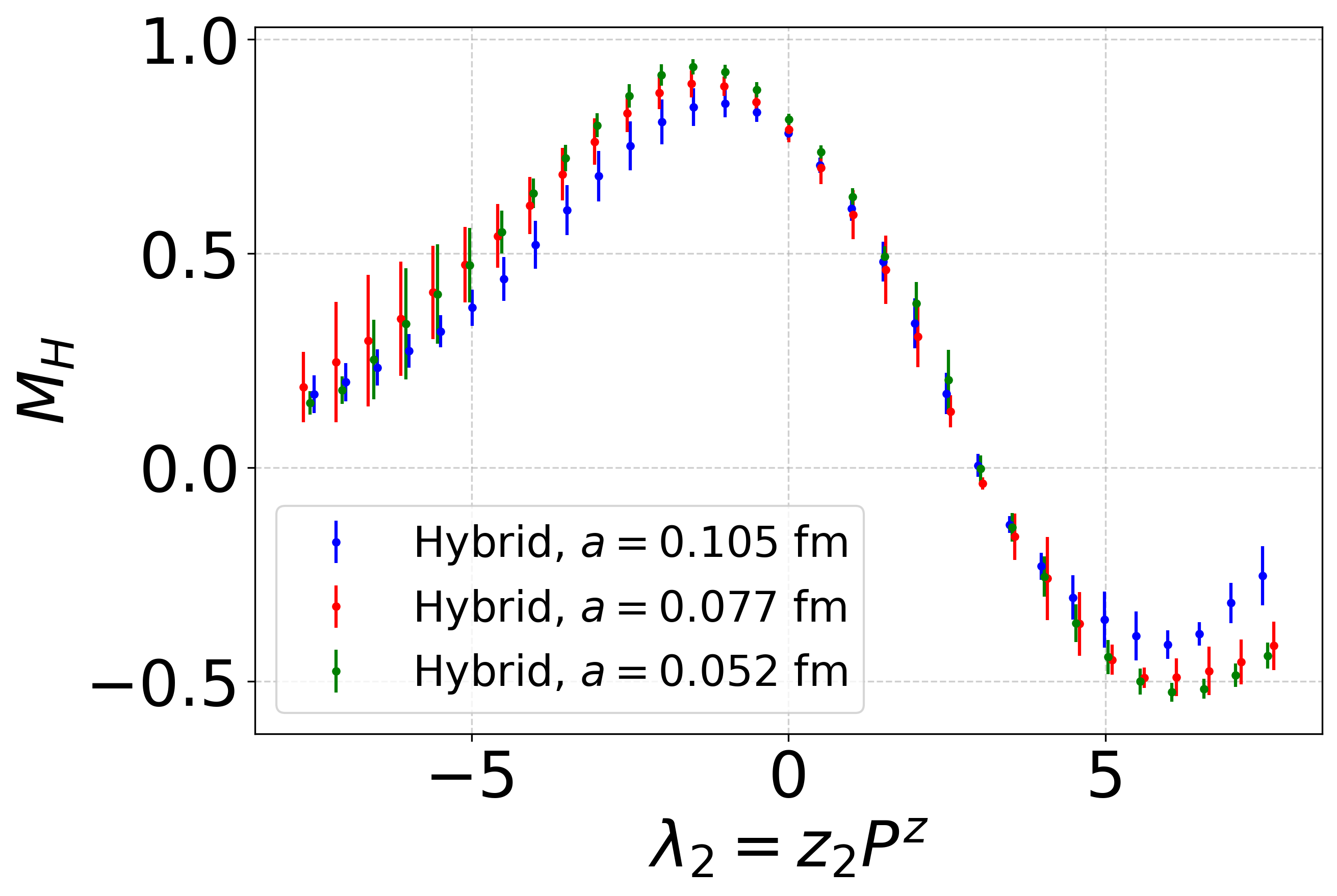}
    }
\caption{Results of the $\Lambda$ quasi-DA matrix elements in different schemes and with $P^z=2.0$ GeV, $z_1=0.200$ fm}
\label{fig:lambda_p4_z4}
\end{figure}

\begin{figure}[htbp]
\centering
\subfigure[\ Bare result of $\Lambda$ at $P=2.0$ GeV]{
    \centering
    \includegraphics[scale=0.185]{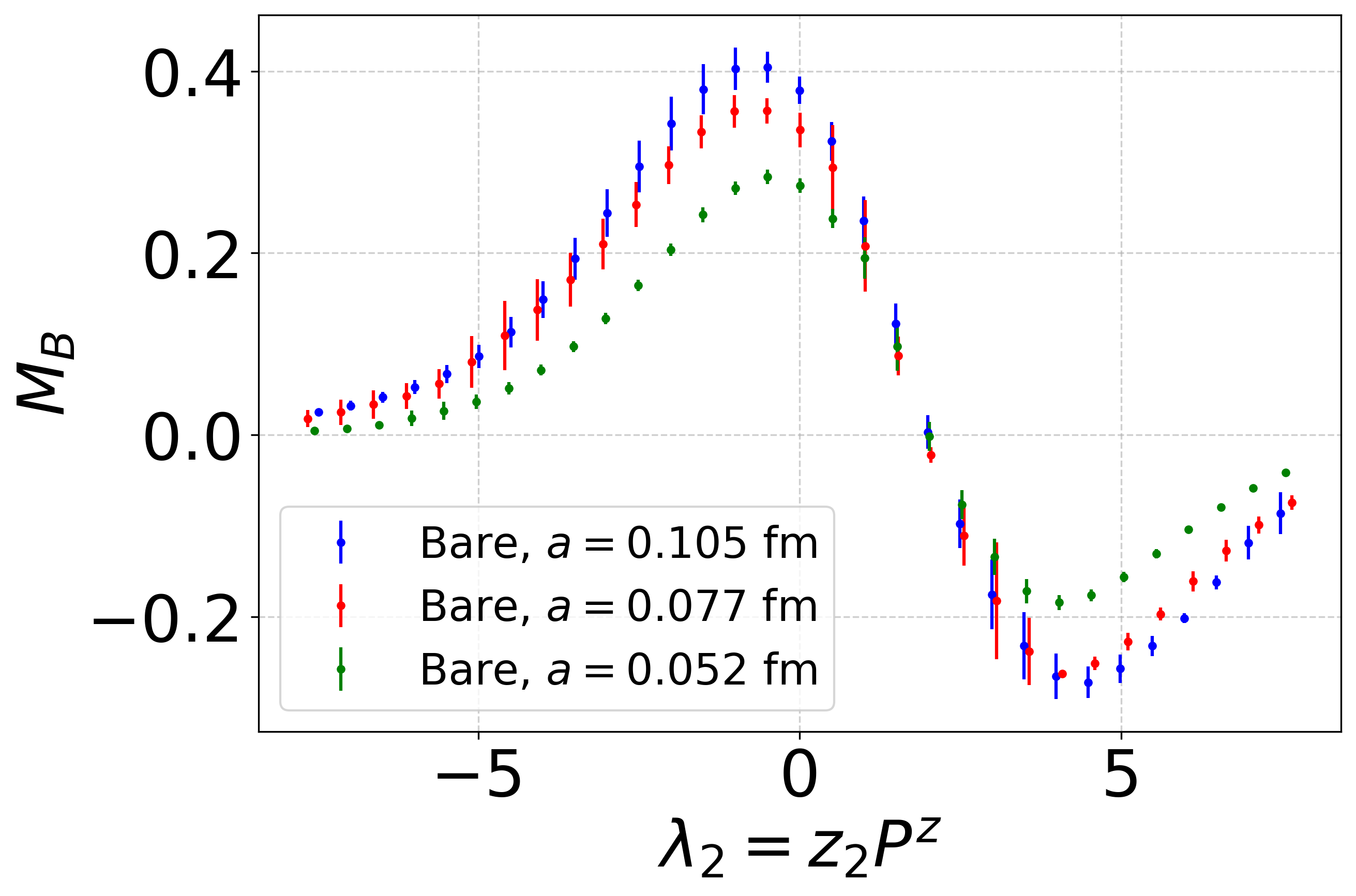}
    }
\vspace{0.0cm} 
\subfigure[\ Ratio scheme result of $\Lambda$ at $P=2.0$ GeV]{
    \centering
    \includegraphics[scale=0.185]{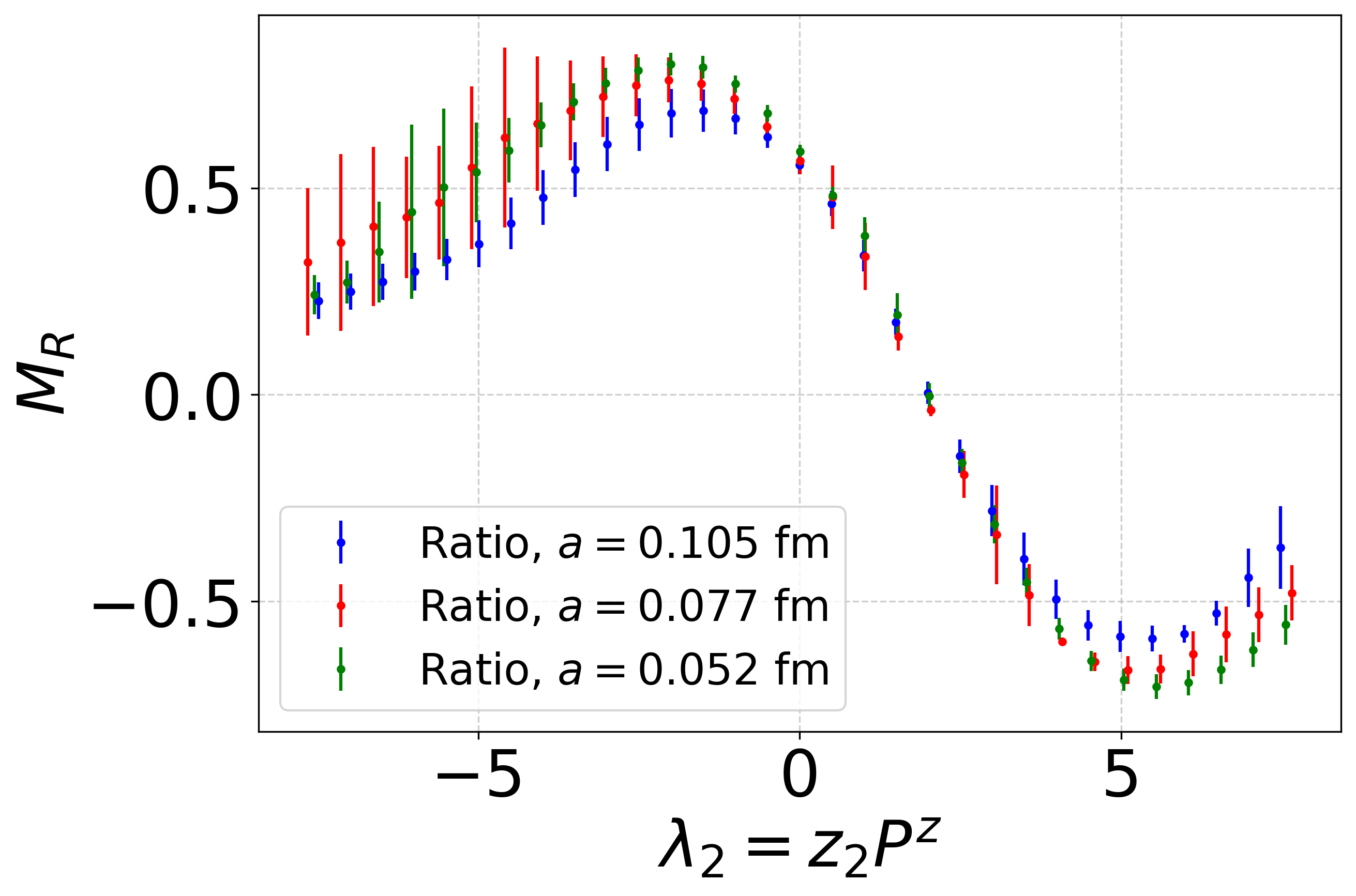}
    }
\vspace{0.0cm} 
\subfigure[\ Self scheme result of $\Lambda$ at $P=2.0$ GeV]{
    \centering
    \includegraphics[scale=0.185]{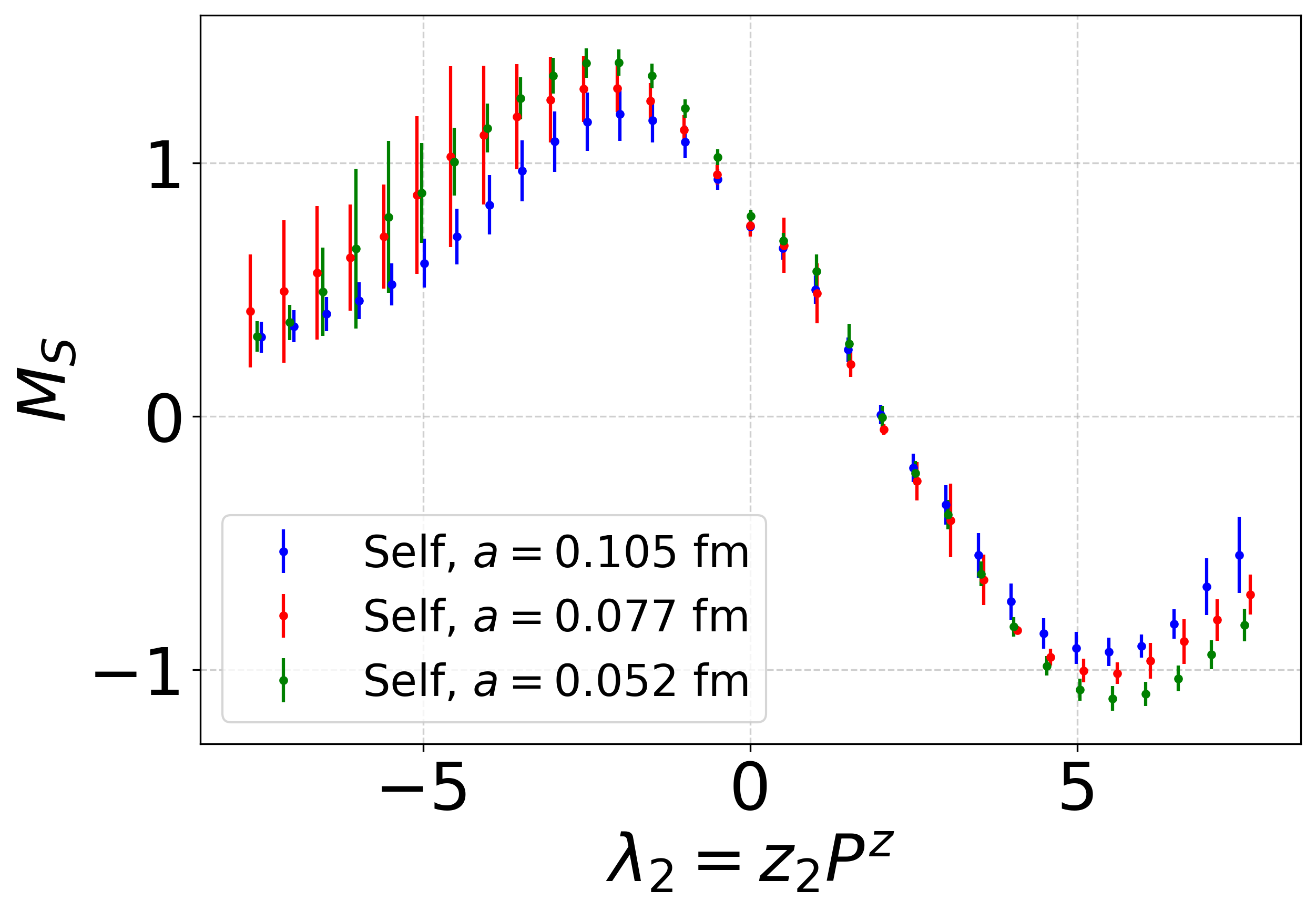}
    }
\vspace{0.0cm} 
\subfigure[\ Hybrid scheme result of $\Lambda$ at $P=2.0$ GeV]{
    \centering
    \includegraphics[scale=0.185]{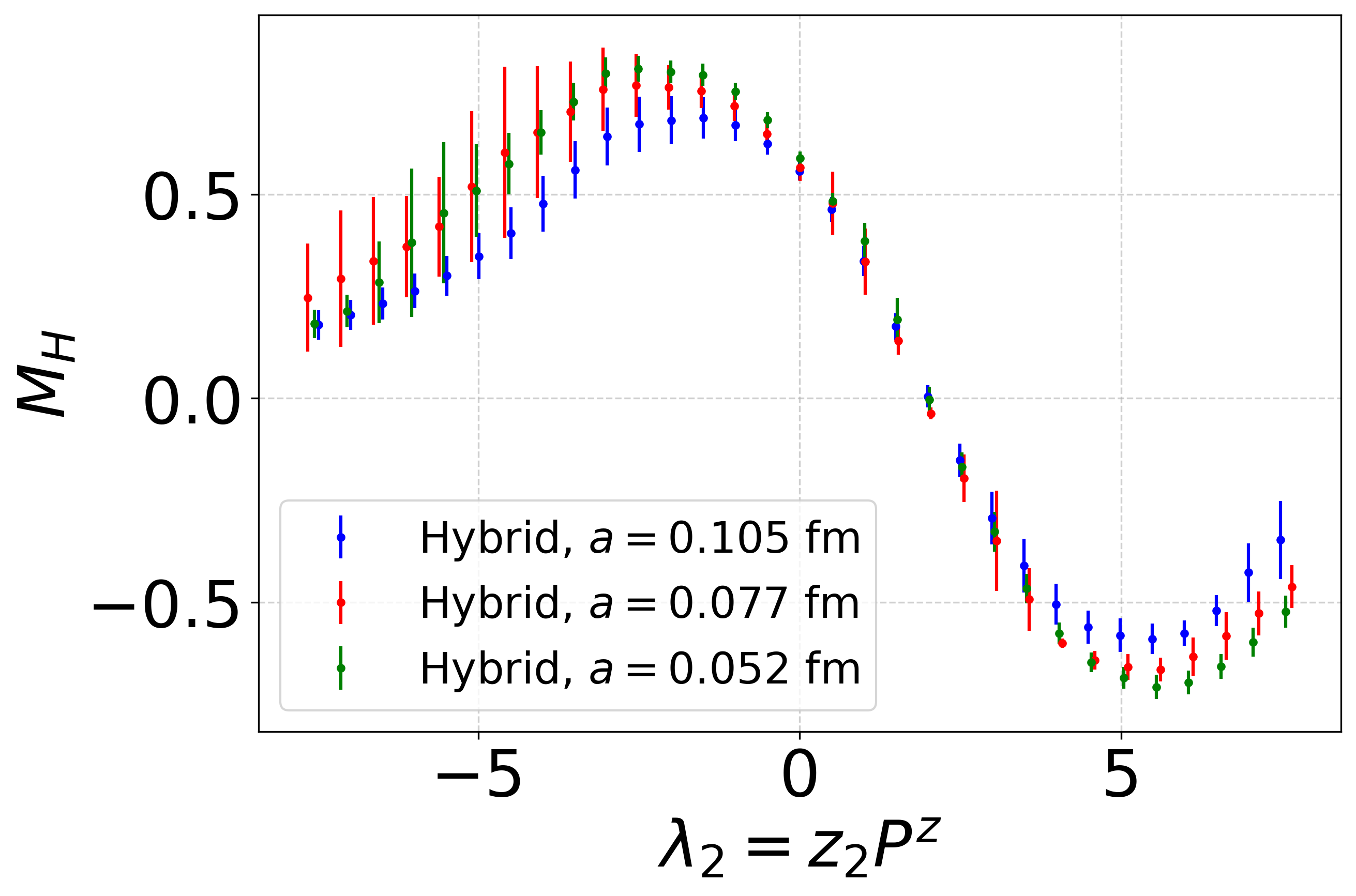}
    }
\caption{Results of the $\Lambda$ quasi-DA matrix elements in different schemes and with $P^z=2.0$ GeV, $z_1=0.300$ fm}
\label{fig:lambda_p4_z6}
\end{figure}

\begin{figure}[htbp]
\centering
\subfigure[\ Bare result of $\Lambda$ at $P=2.0$ GeV]{
    \centering
    \includegraphics[scale=0.185]{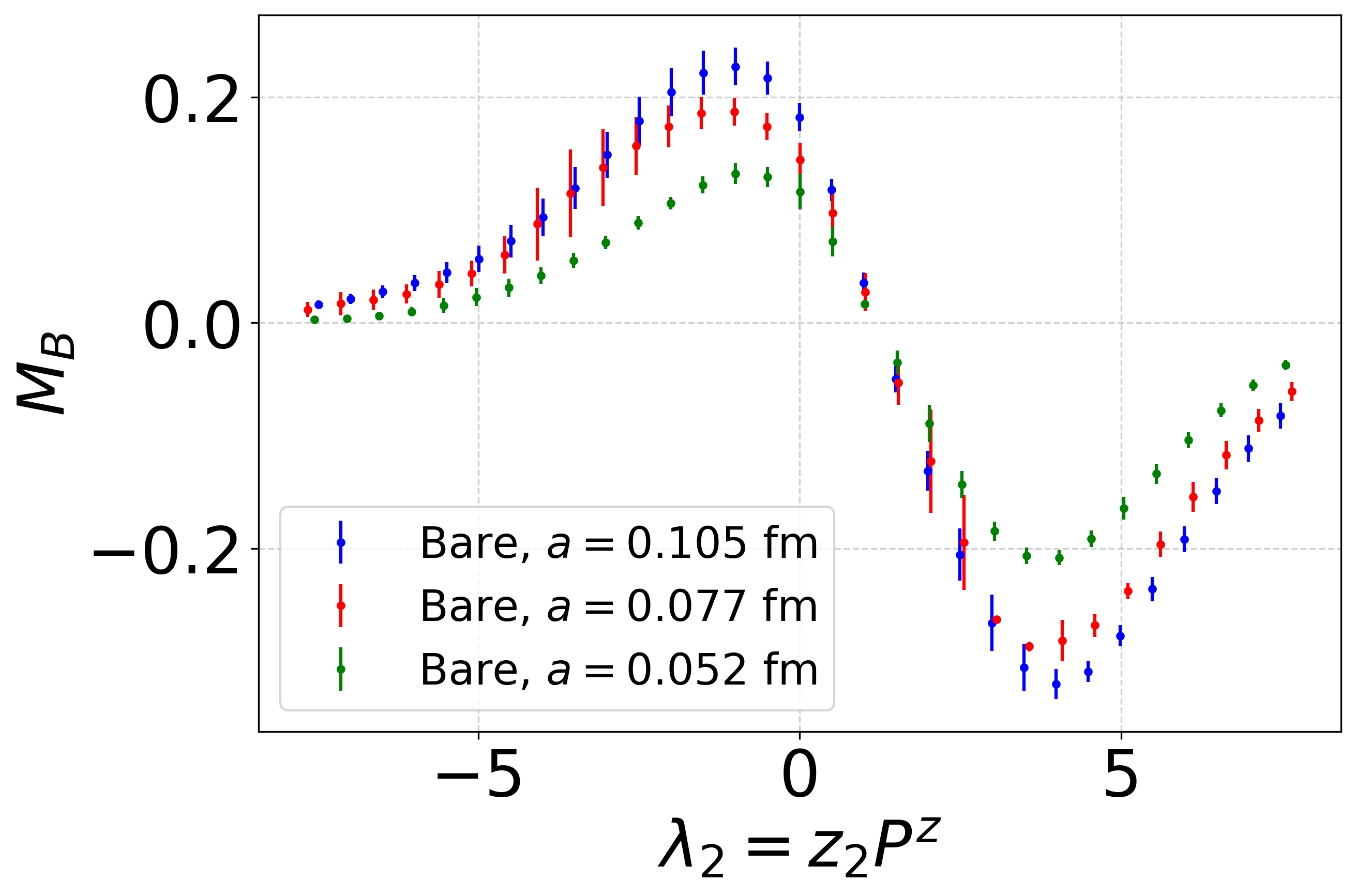}
    }
\vspace{0.0cm} 
\subfigure[\ Ratio scheme result of $\Lambda$ at $P=2.0$ GeV]{
    \centering
    \includegraphics[scale=0.185]{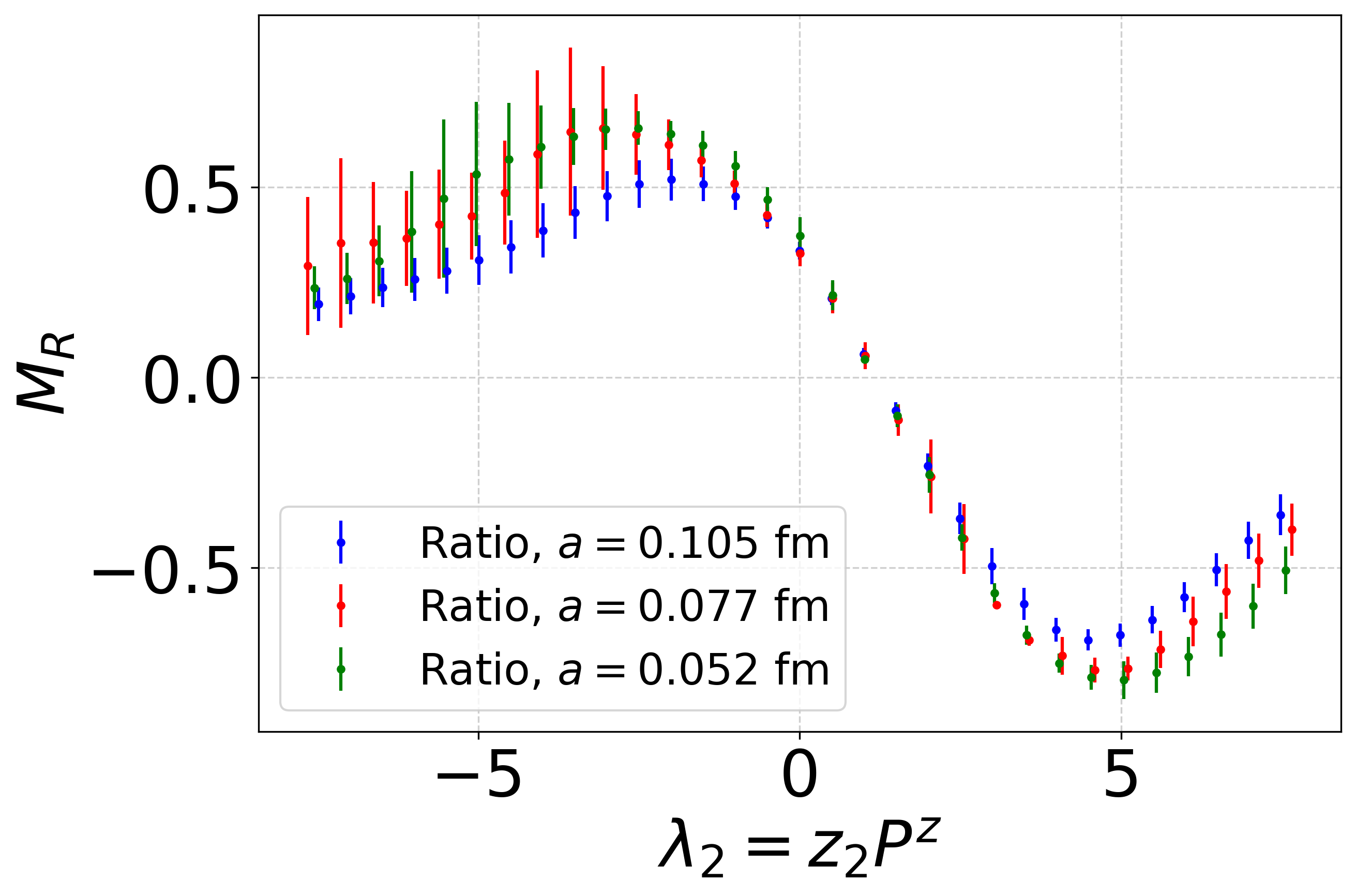}
    }
\vspace{0.0cm} 
\subfigure[\ Self scheme result of $\Lambda$ at $P=2.0$ GeV]{
    \centering
    \includegraphics[scale=0.185]{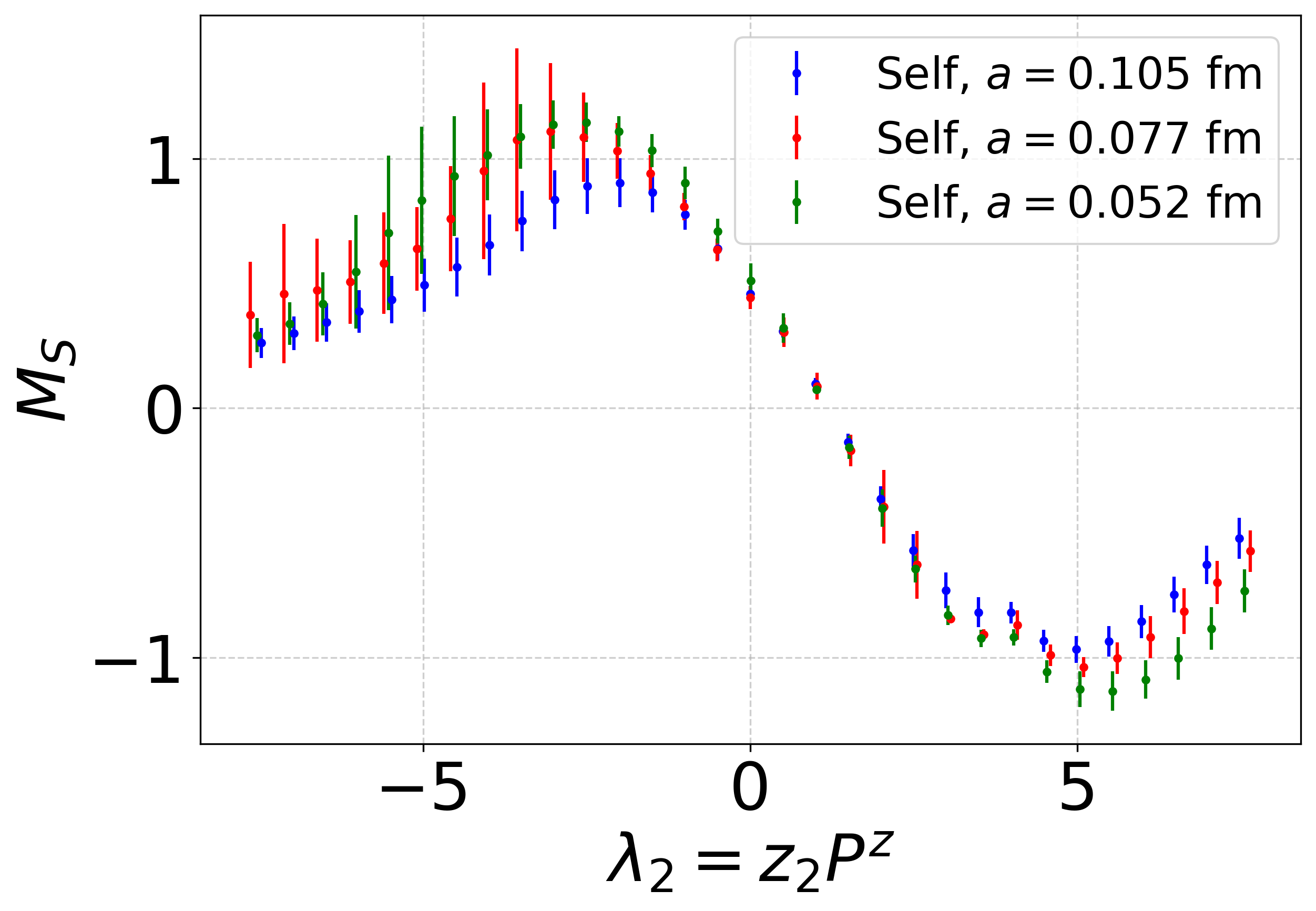}
    }
\vspace{0.0cm} 
\subfigure[\ Hybrid scheme result of $\Lambda$ at $P=2.0$ GeV]{
    \centering
    \includegraphics[scale=0.185]{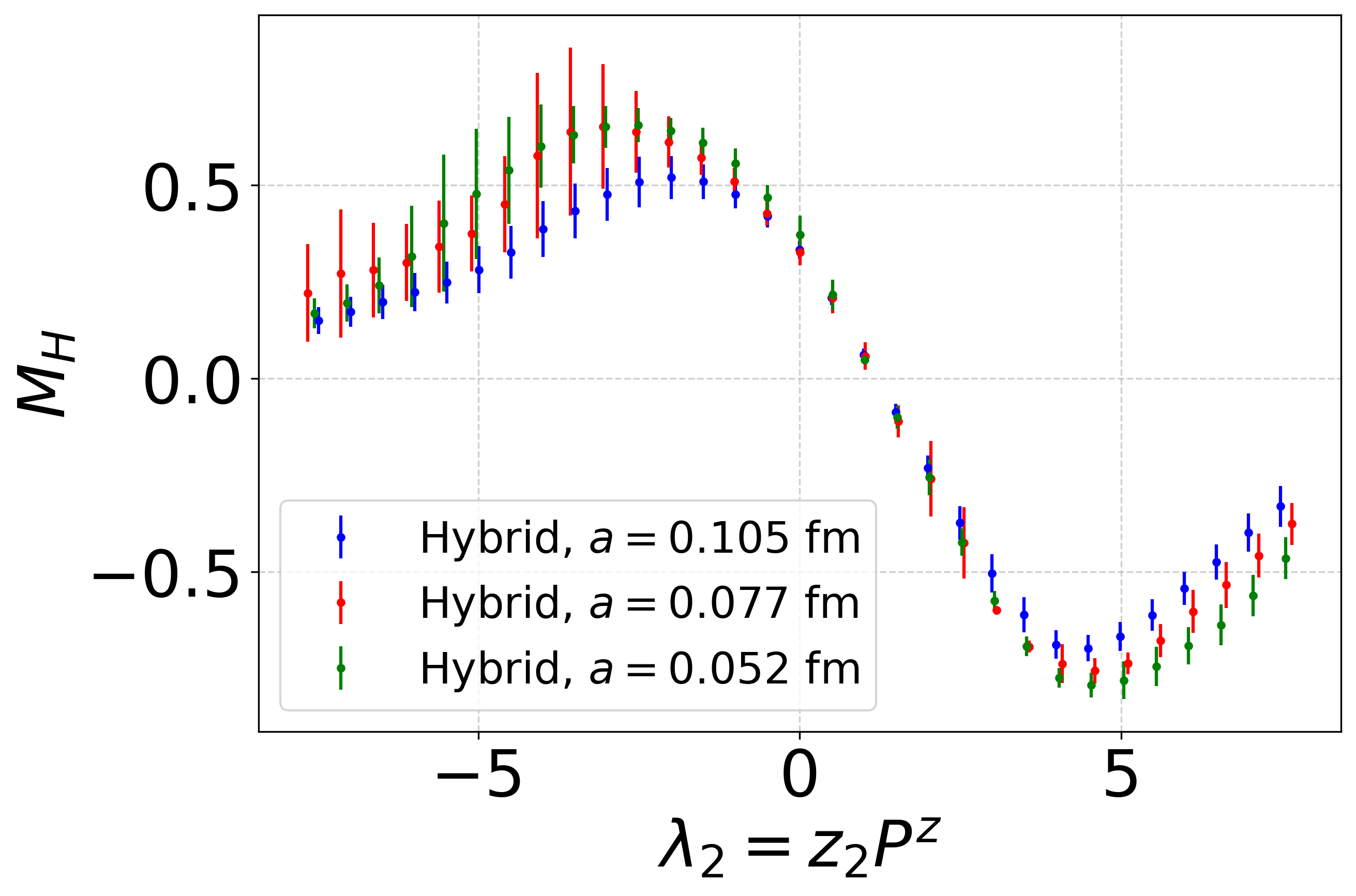}
    }
\caption{Results of the $\Lambda$ quasi-DA matrix elements in different schemes and with $P^z=2.0$ GeV, $z_1=0.400$ fm}
\label{fig:lambda_p4_z8}
\end{figure}

\begin{figure}[htbp]
\centering
\subfigure[\ Bare result of $\Lambda$ at $P=2.0$ GeV]{
    \centering
    \includegraphics[scale=0.185]{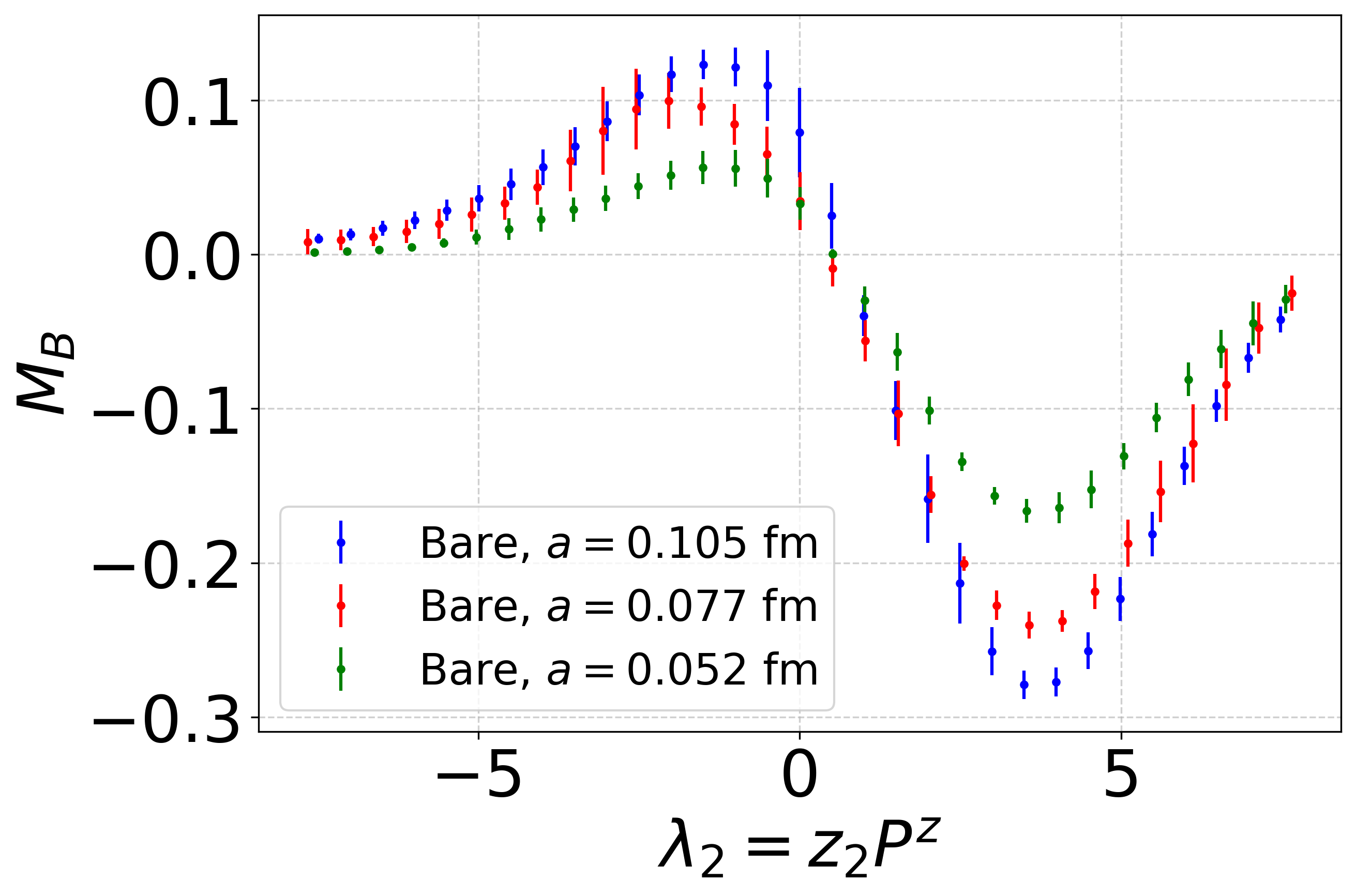}
    }
\vspace{0.0cm} 
\subfigure[\ Ratio scheme result of $\Lambda$ at $P=2.0$ GeV]{
    \centering
    \includegraphics[scale=0.185]{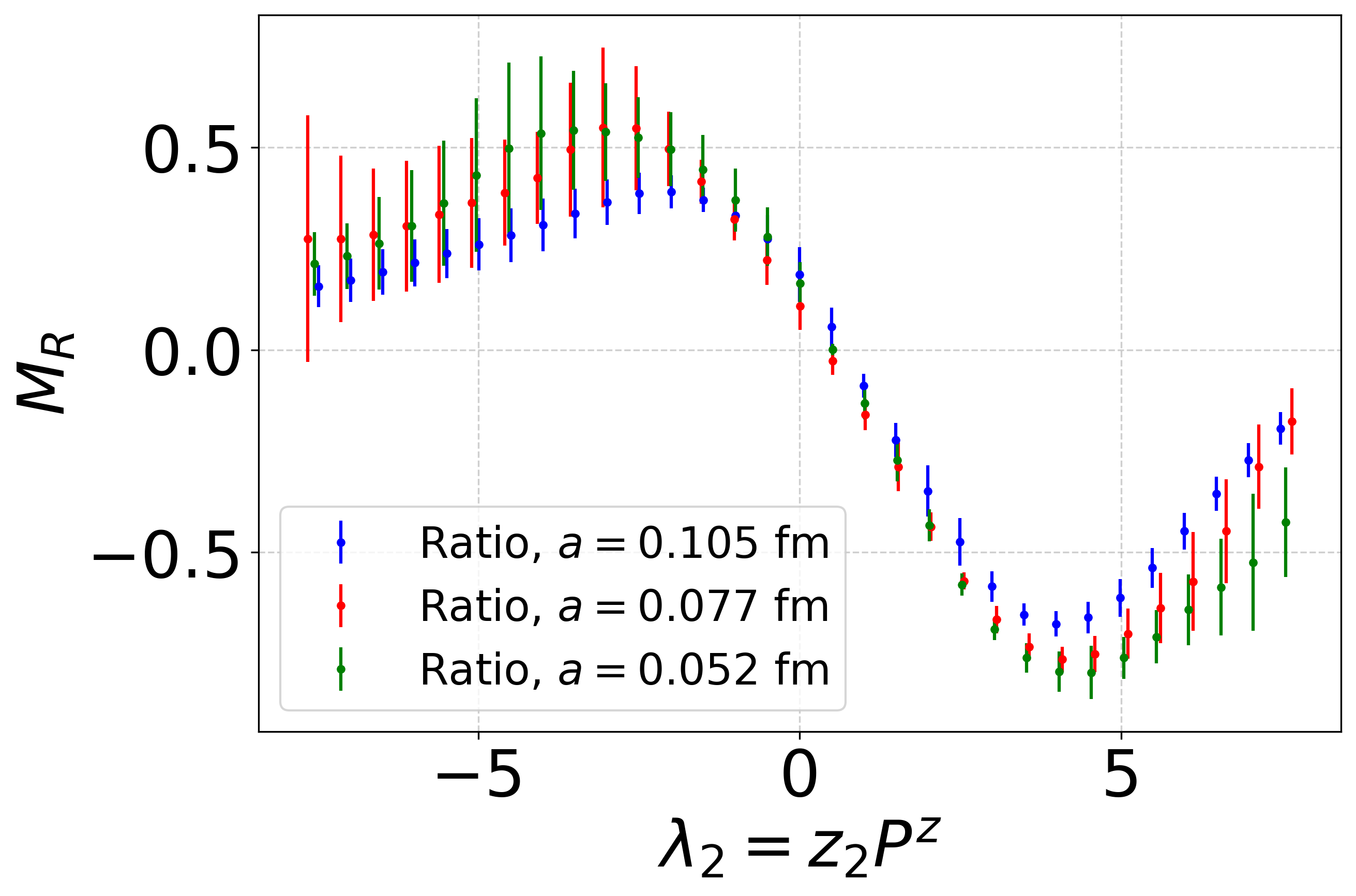}
    }
\vspace{0.0cm} 
\subfigure[\ Self scheme result of $\Lambda$ at $P=2.0$ GeV]{
    \centering
    \includegraphics[scale=0.185]{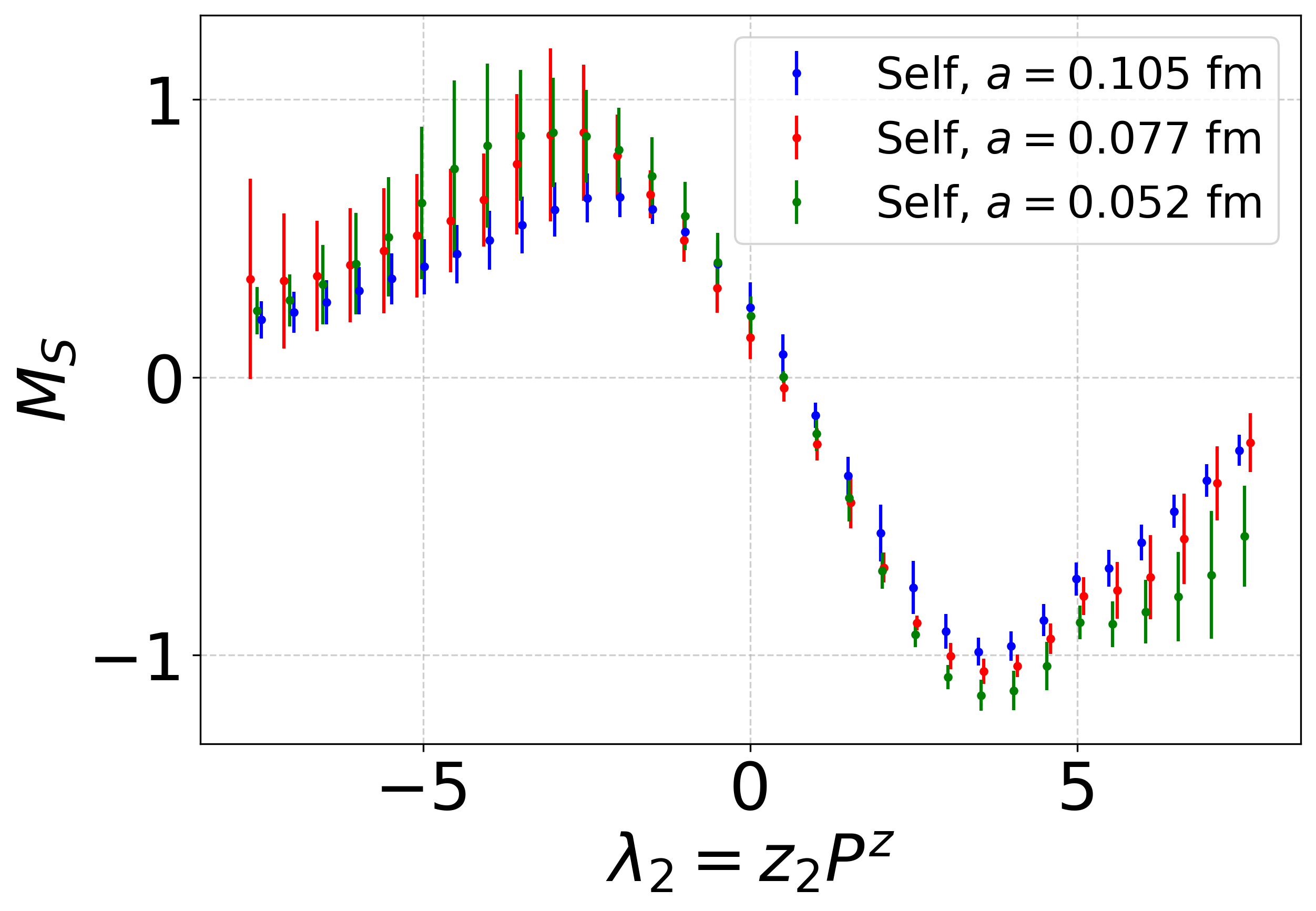}
    }
\vspace{0.0cm} 
\subfigure[\ Hybrid scheme result of $\Lambda$ at $P=2.0$ GeV]{
    \centering
    \includegraphics[scale=0.185]{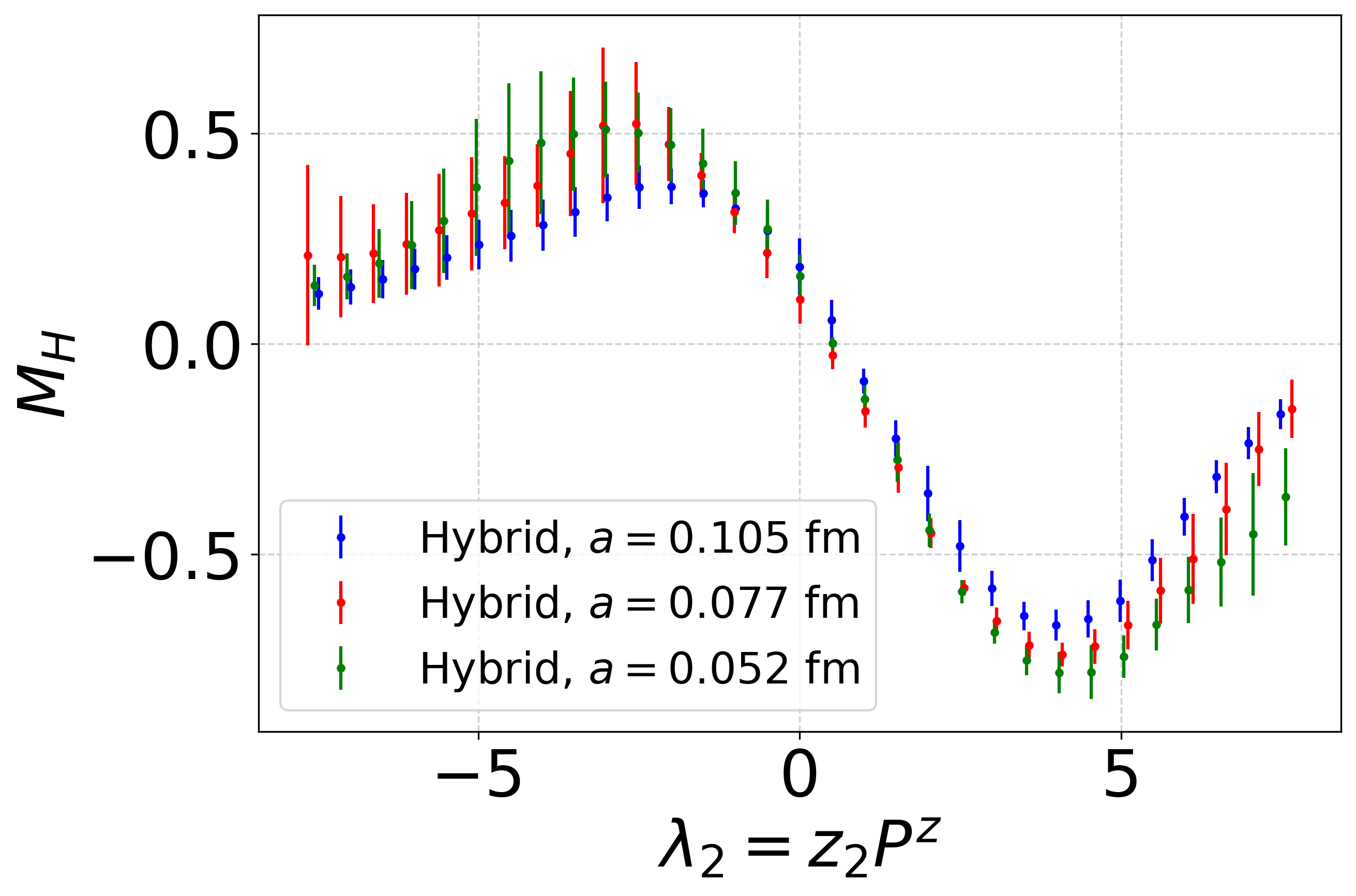}
    }
\caption{Results of the $\Lambda$ quasi-DA matrix elements in different schemes and with $P^z=2.0$ GeV, $z_1=0.500$ fm}
\label{fig:lambda_p4_z10}
\end{figure}

\clearpage

\subsection{More results of the proton (V-term) Quasi-DA at $P^z = 0.5$ GeV in different schemes}
\begin{figure}[htbp]
\centering
\subfigure[\ Bare result of proton at $P=0.5$ GeV]{
    \centering
    \includegraphics[scale=0.185]{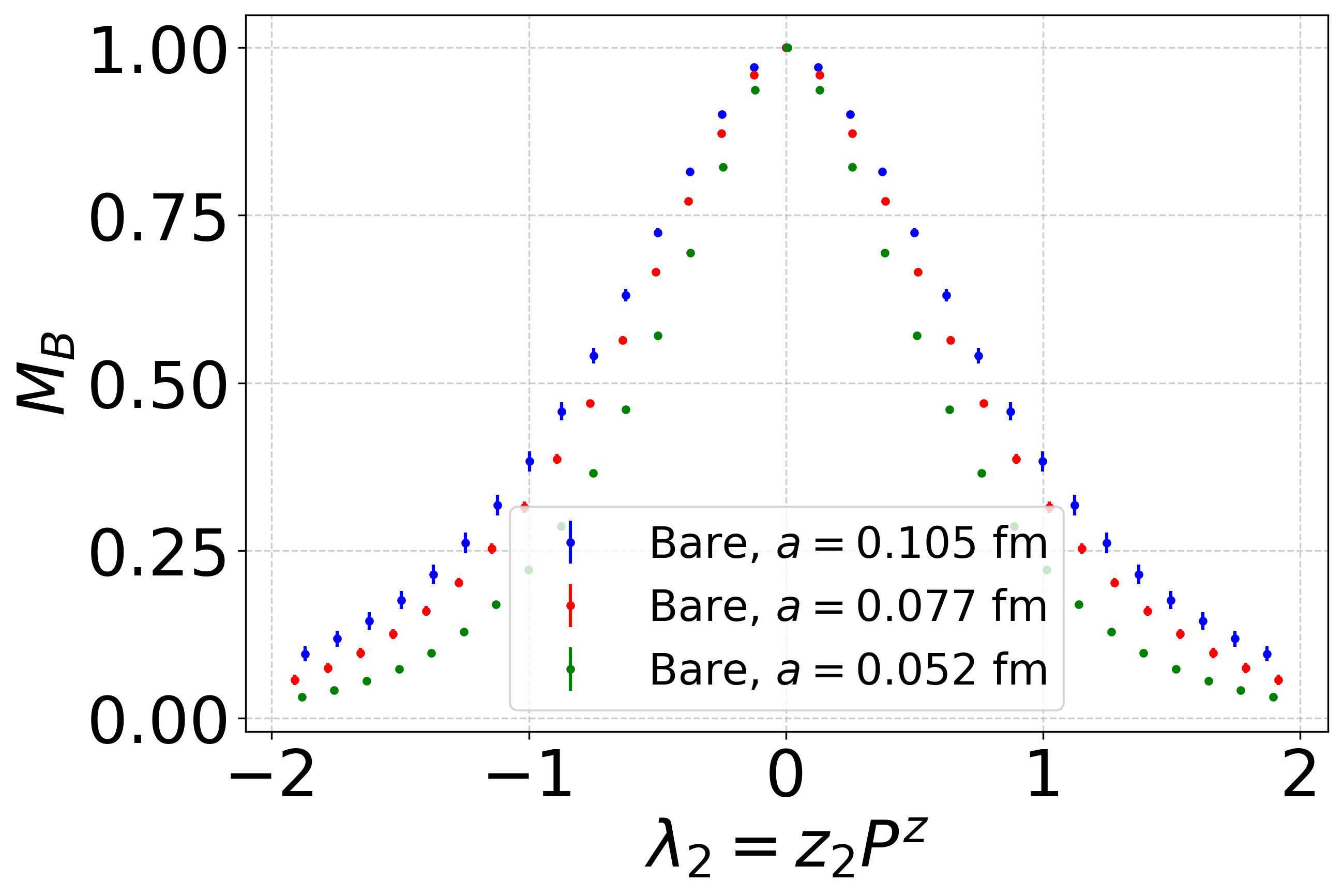}
    }
\vspace{0.0cm} 
\subfigure[\ Ratio scheme result of proton at $P=0.5$ GeV]{
    \centering
    \includegraphics[scale=0.185]{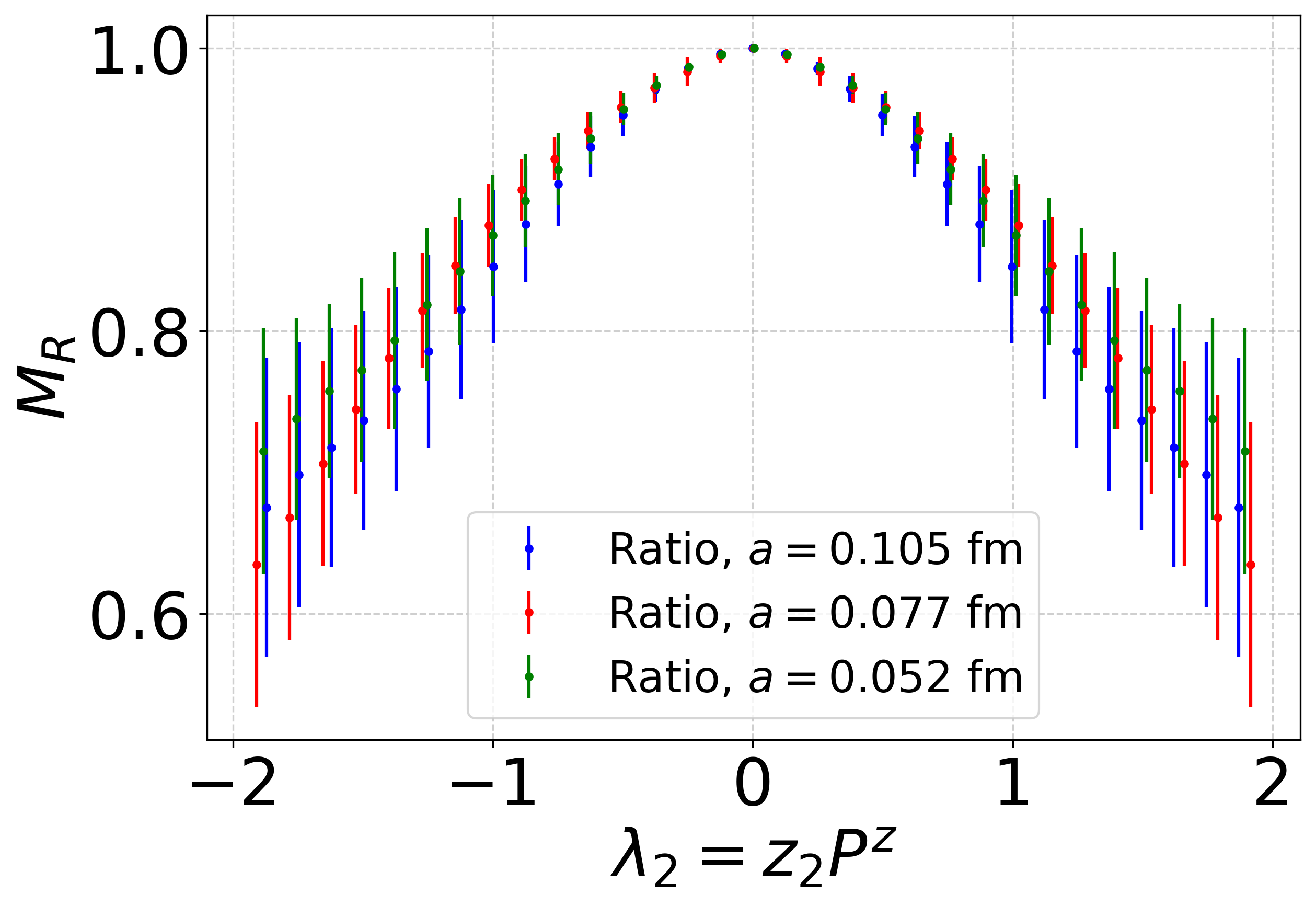}
    }
\vspace{0.0cm} 
\subfigure[\ Self scheme result of proton at $P=0.5$ GeV]{
    \centering
    \includegraphics[scale=0.185]{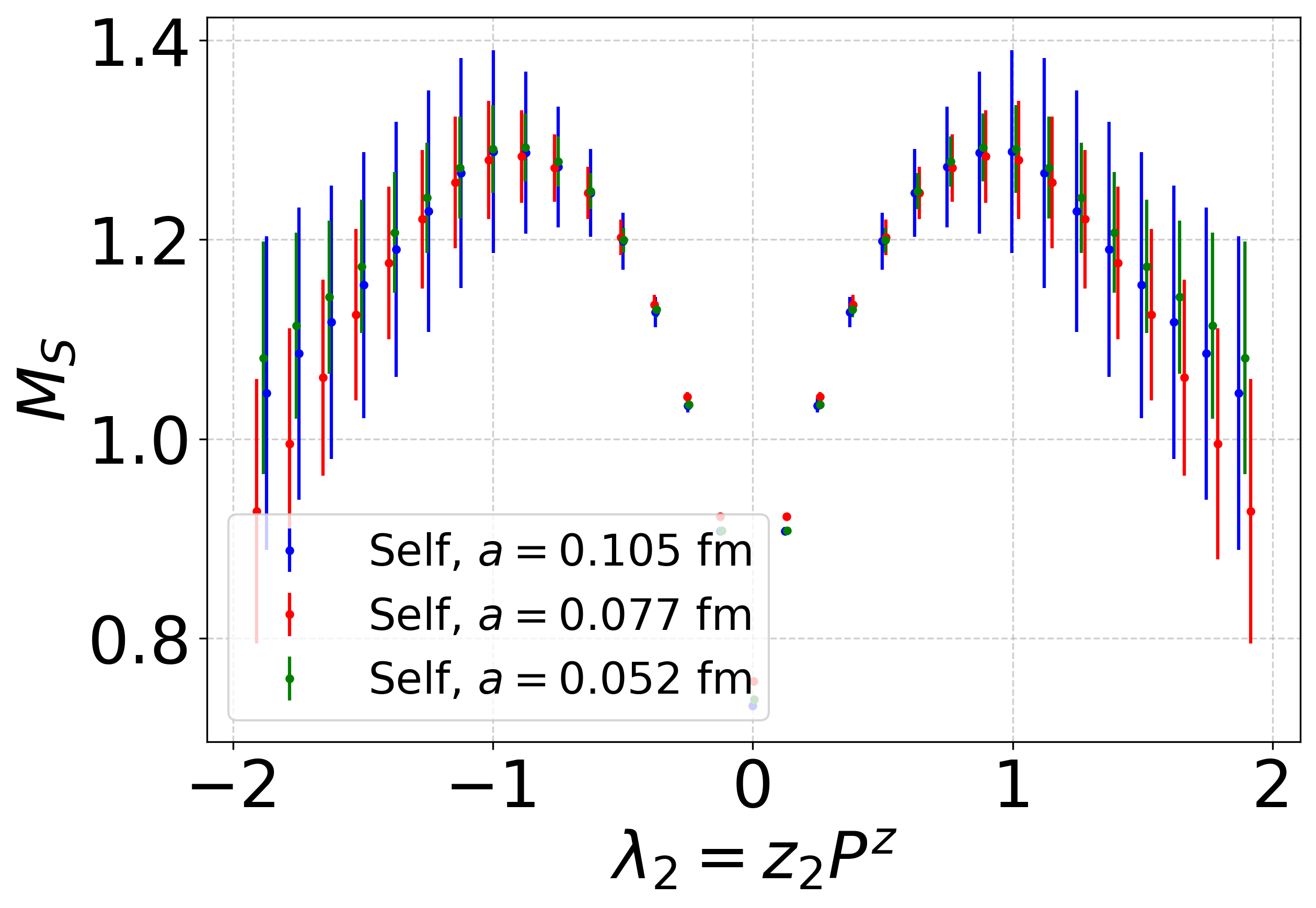}
    }
\vspace{0.0cm} 
\subfigure[\ Hybrid scheme result of proton at $P=0.5$ GeV]{
    \centering
    \includegraphics[scale=0.185]{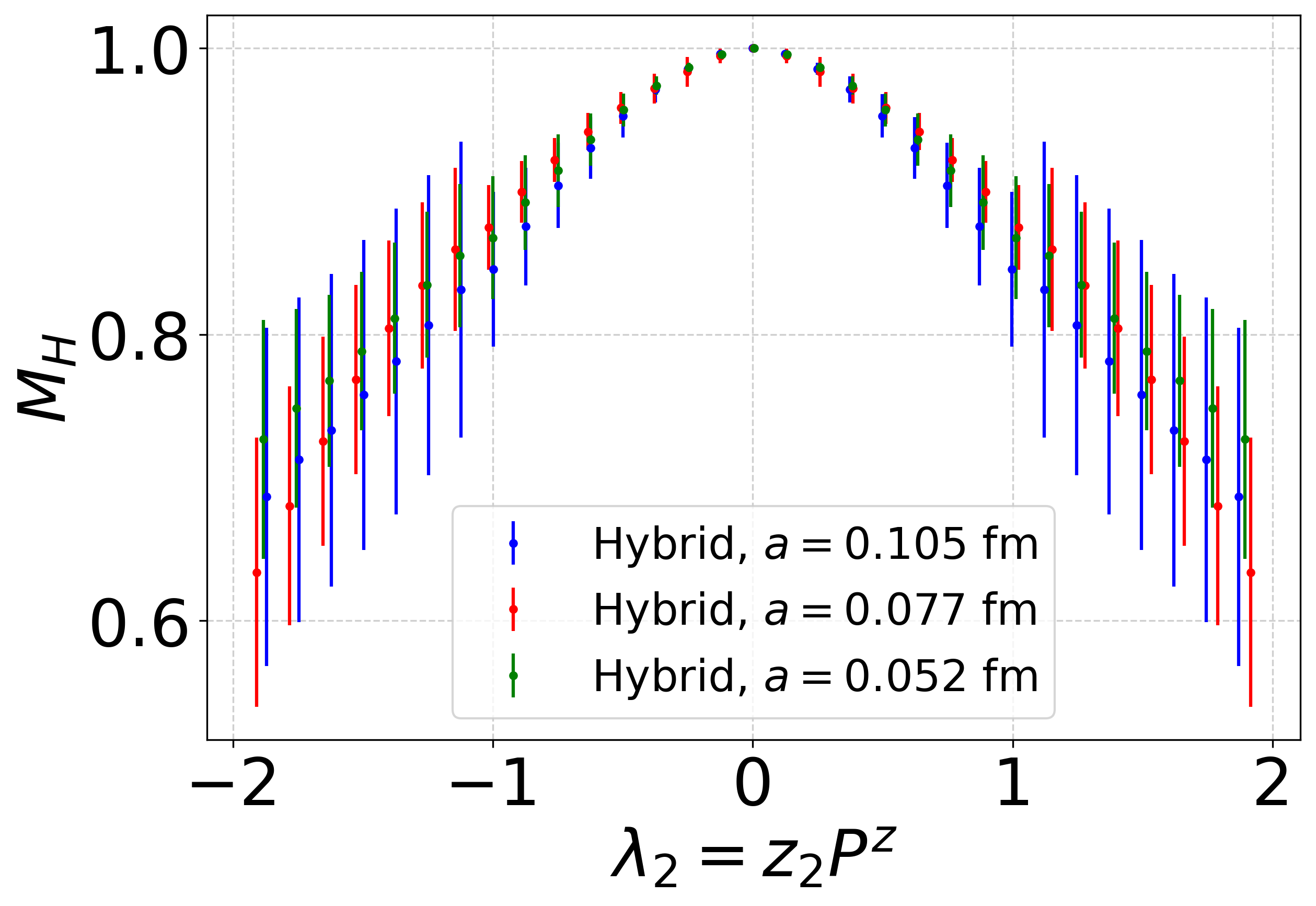}
    }
\caption{Results of the proton quasi-DA matrix elements in different schemes and with $P^z=0.5$ GeV, $z_1=0.000$ fm}
\label{fig:proton_p1_z0}
\end{figure}

\begin{figure}[htbp]
\centering
\subfigure[\ Bare result of proton at $P=0.5$ GeV]{
    \centering
    \includegraphics[scale=0.185]{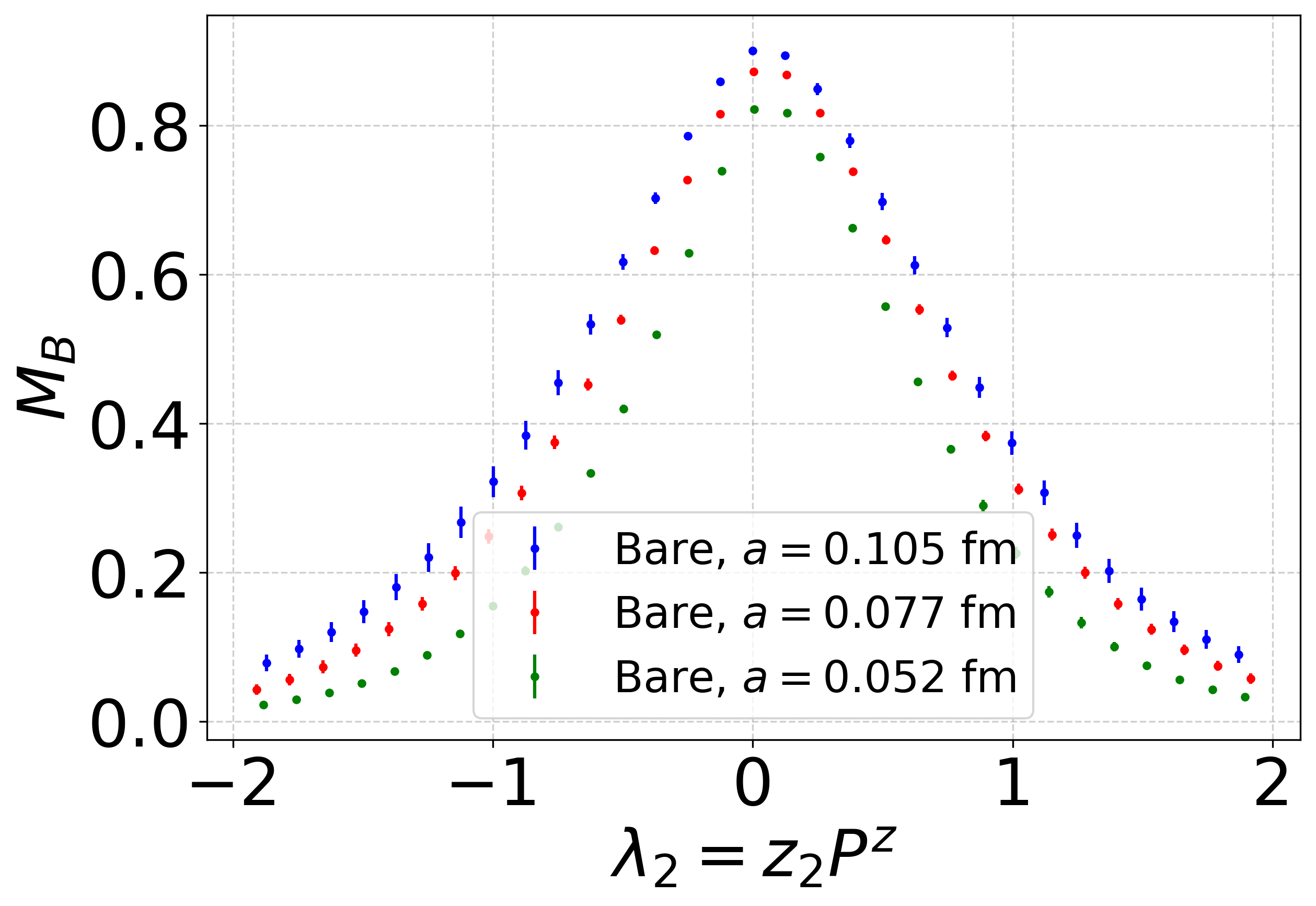}
    }
\vspace{0.0cm} 
\subfigure[\ Ratio scheme result of proton at $P=0.5$ GeV]{
    \centering
    \includegraphics[scale=0.185]{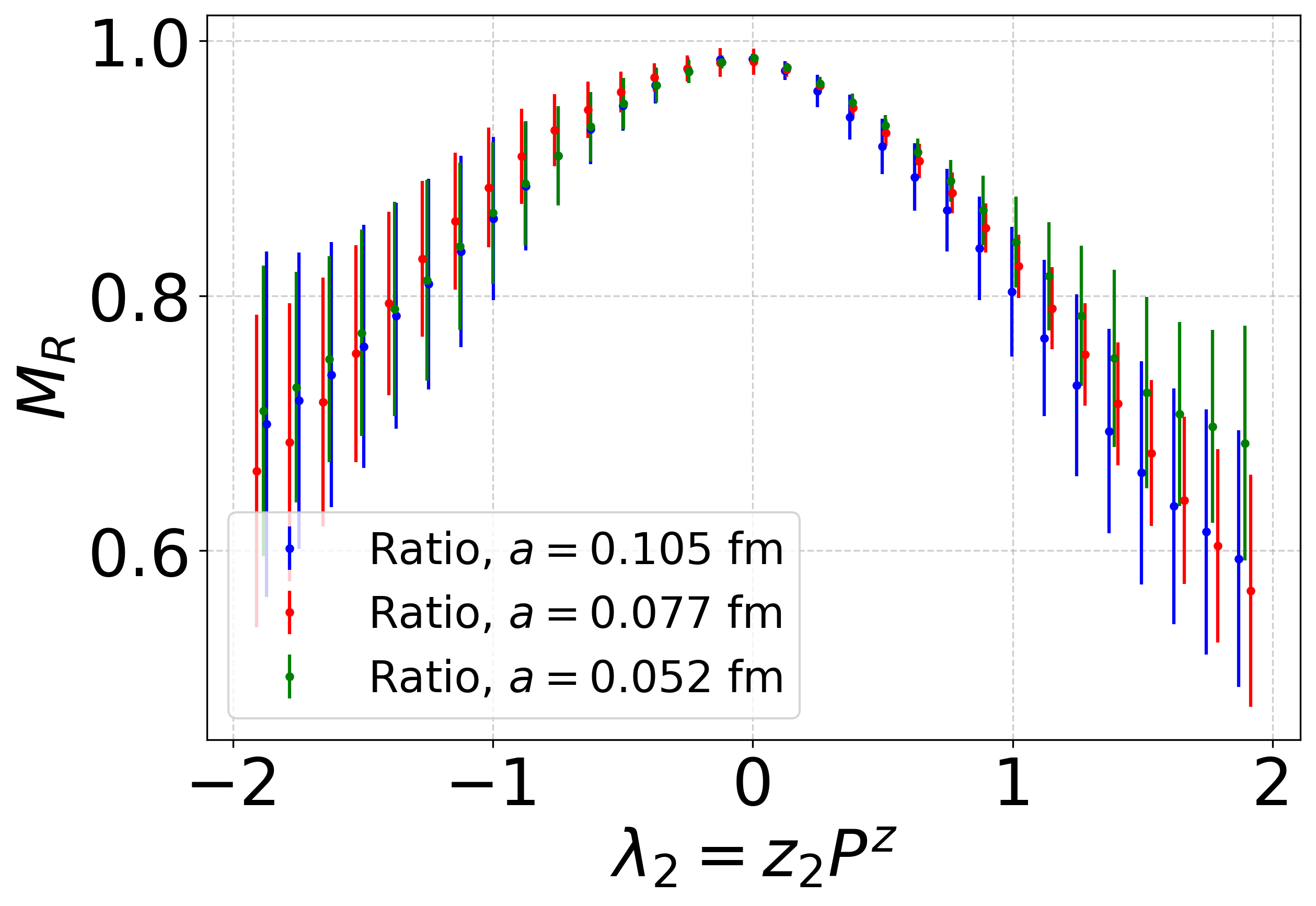}
    }
\vspace{0.0cm} 
\subfigure[\ Self scheme result of proton at $P=0.5$ GeV]{
    \centering
    \includegraphics[scale=0.185]{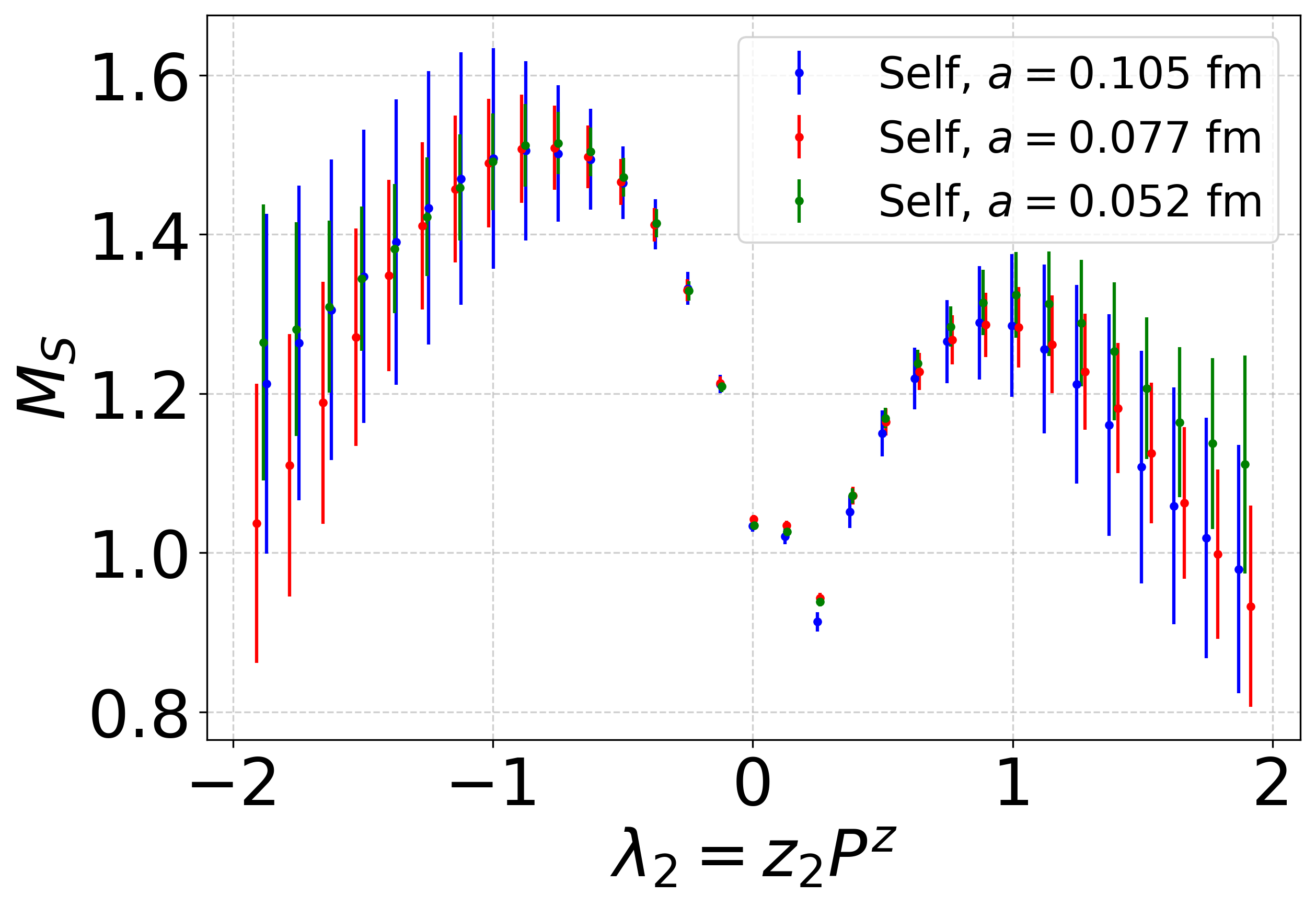}
    }
\vspace{0.0cm} 
\subfigure[\ Hybrid scheme result of proton at $P=0.5$ GeV]{
    \centering
    \includegraphics[scale=0.185]{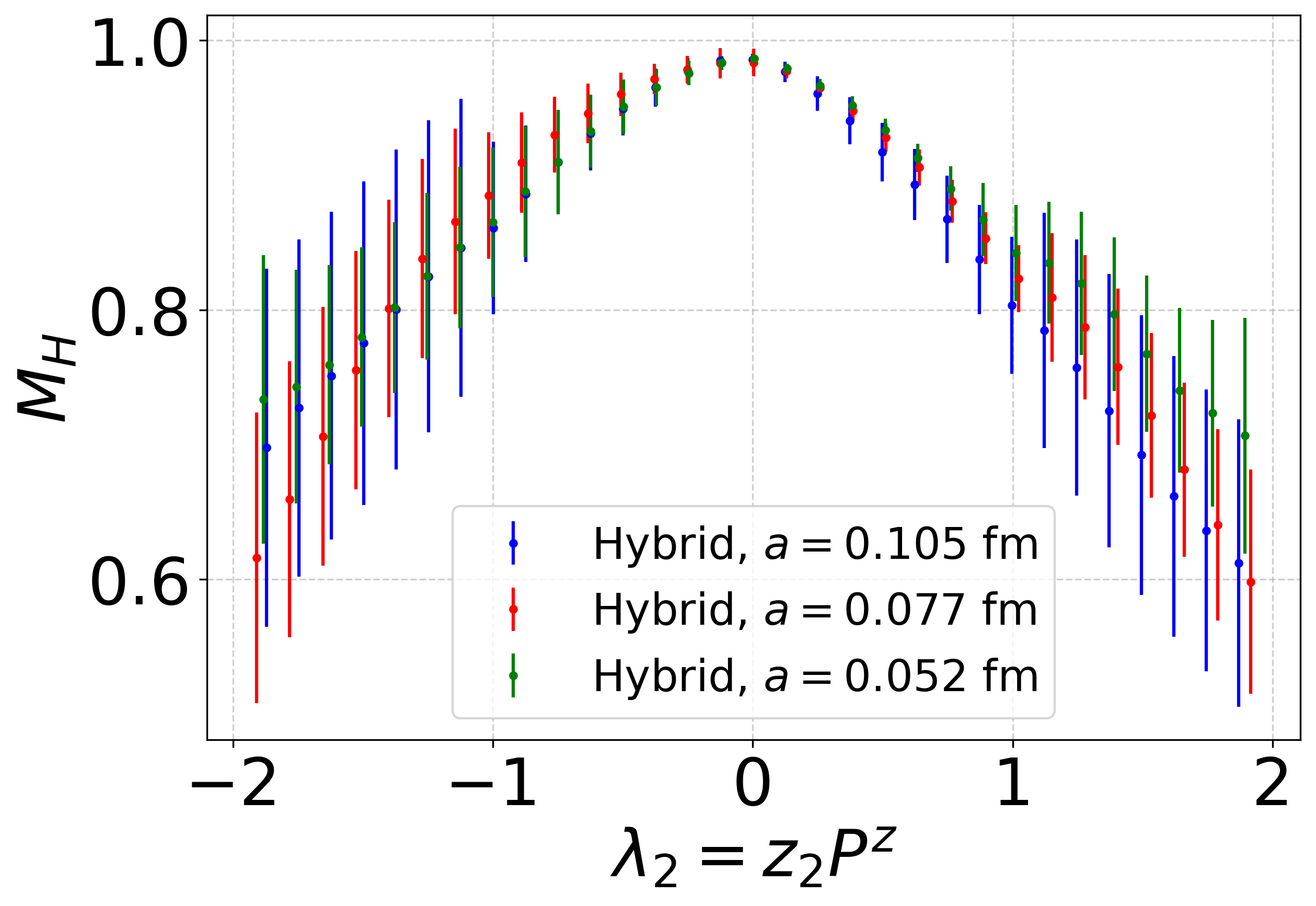}
    }
\caption{Results of the proton quasi-DA matrix elements in different schemes and with $P^z=0.5$ GeV, $z_1=0.100$ fm}
\label{fig:proton_p1_z2}
\end{figure}

\begin{figure}[htbp]
\centering
\subfigure[\ Bare result of proton at $P=0.5$ GeV]{
    \centering
    \includegraphics[scale=0.185]{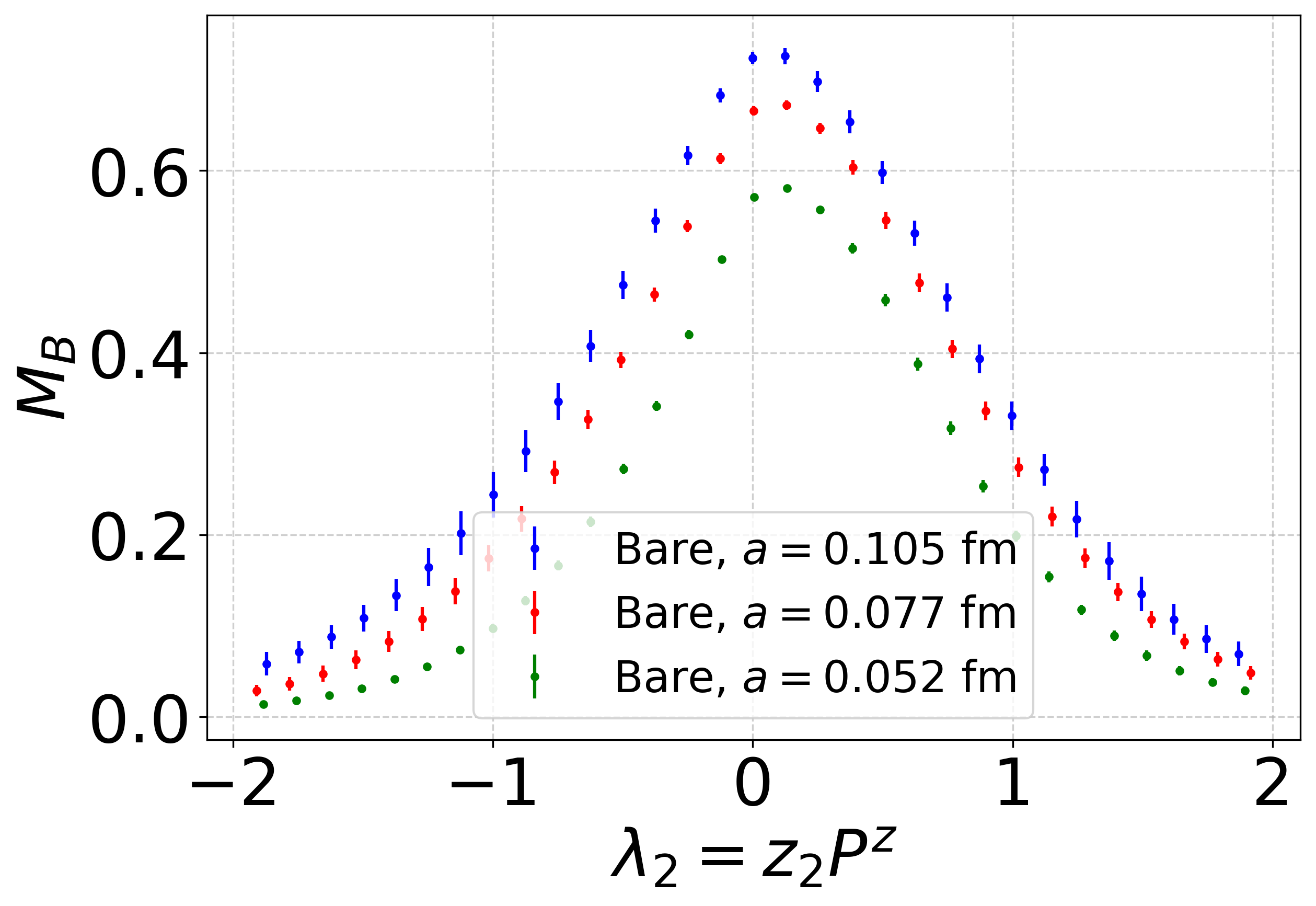}
    }
\vspace{0.0cm} 
\subfigure[\ Ratio scheme result of proton at $P=0.5$ GeV]{
    \centering
    \includegraphics[scale=0.185]{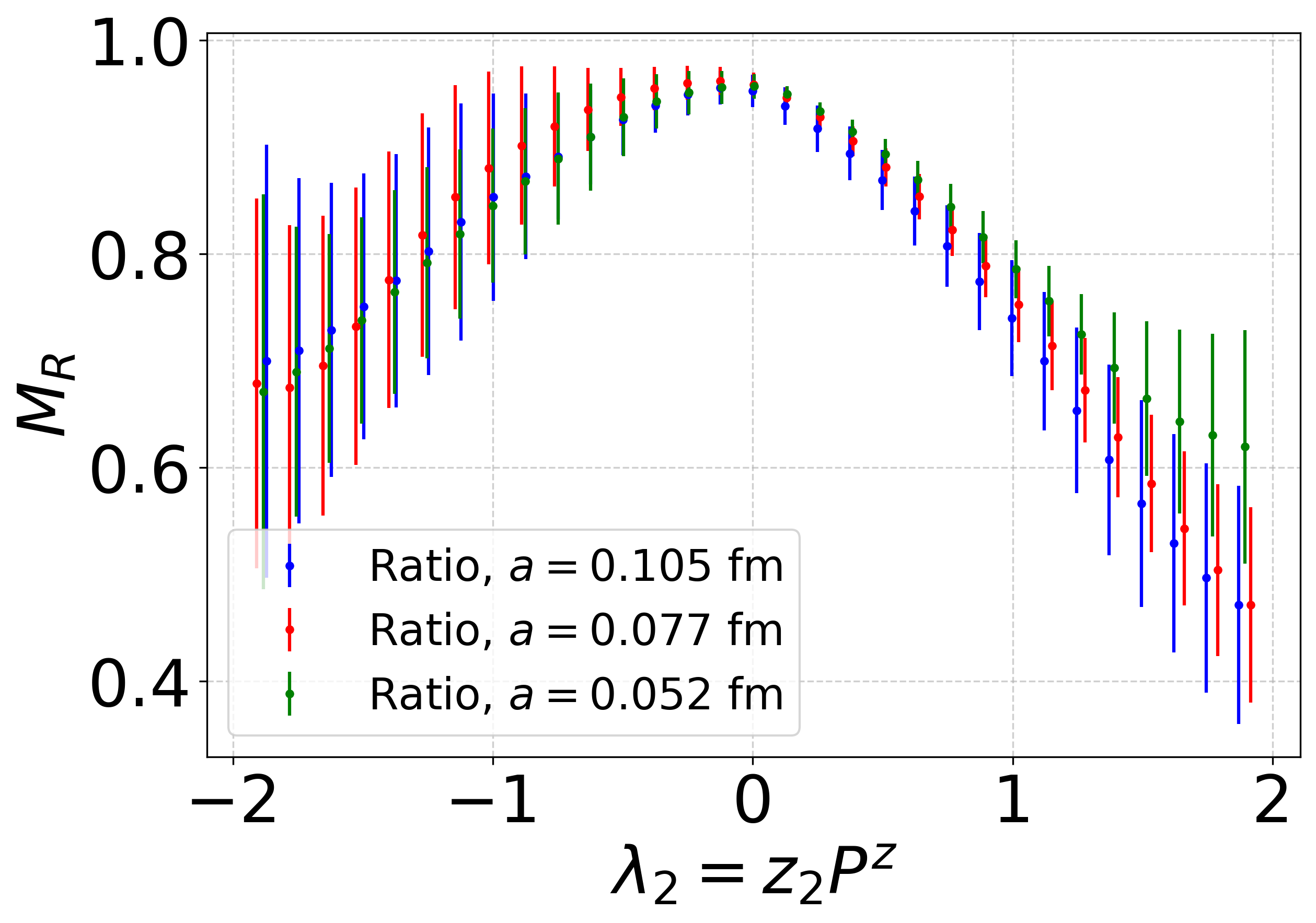}
    }
\vspace{0.0cm} 
\subfigure[\ Self scheme result of proton at $P=0.5$ GeV]{
    \centering
    \includegraphics[scale=0.185]{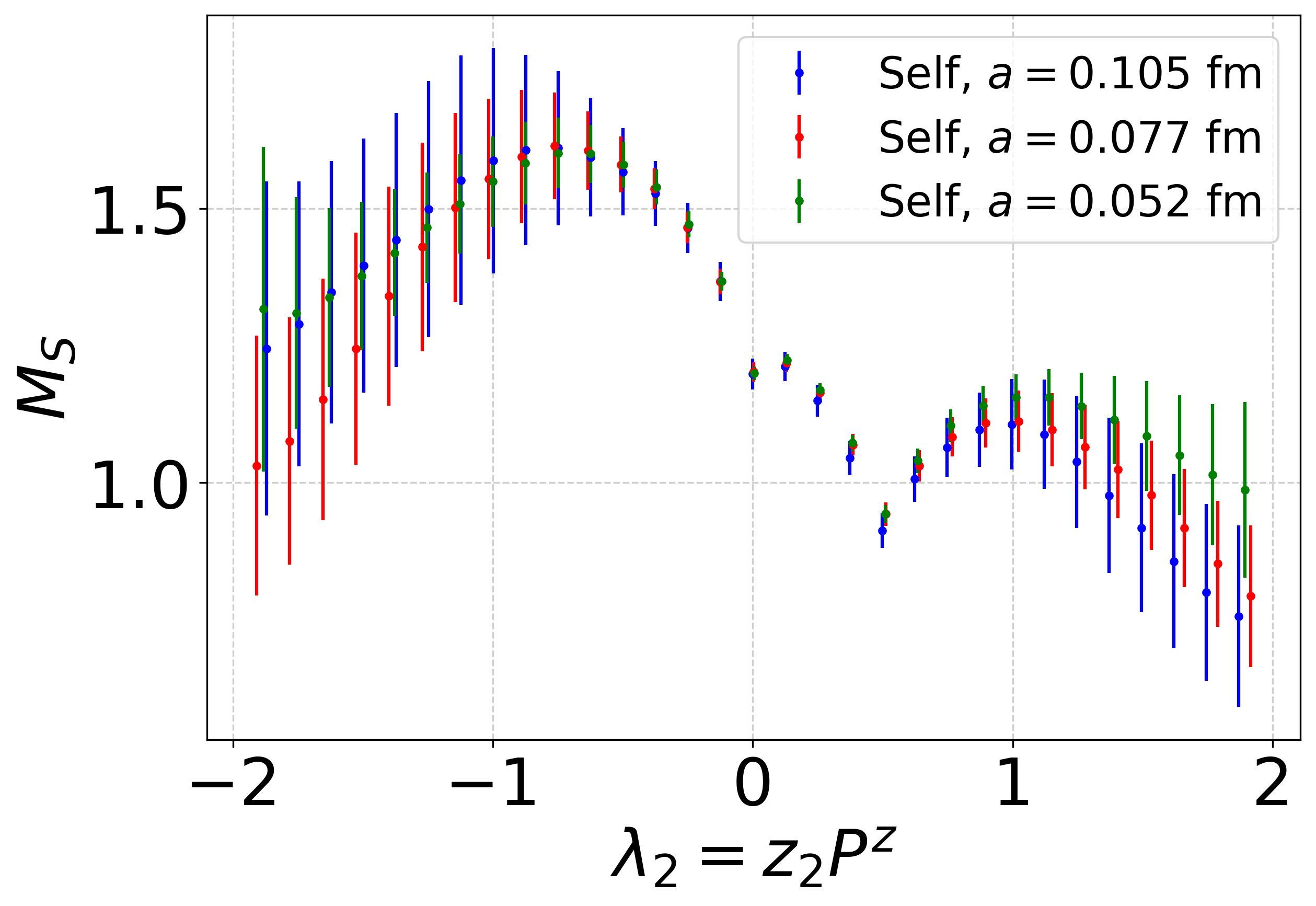}
    }
\vspace{0.0cm} 
\subfigure[\ Hybrid scheme result of proton at $P=0.5$ GeV]{
    \centering
    \includegraphics[scale=0.185]{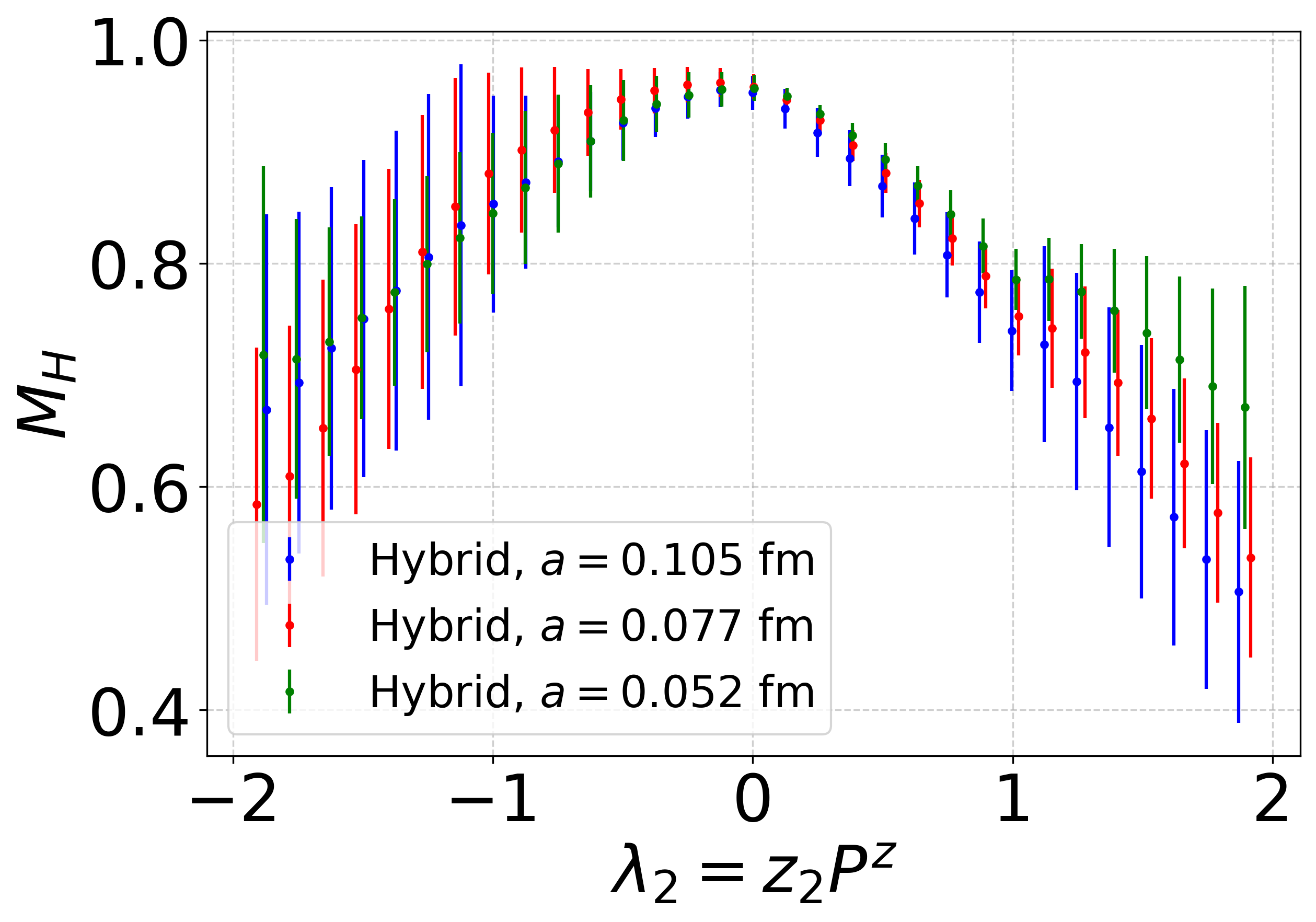}
    }
\caption{Results of the proton quasi-DA matrix elements in different schemes and with $P^z=0.5$ GeV, $z_1=0.200$ fm}
\label{fig:proton_p1_z4}
\end{figure}

\begin{figure}[htbp]
\centering
\subfigure[\ Bare result of proton at $P=0.5$ GeV]{
    \centering
    \includegraphics[scale=0.185]{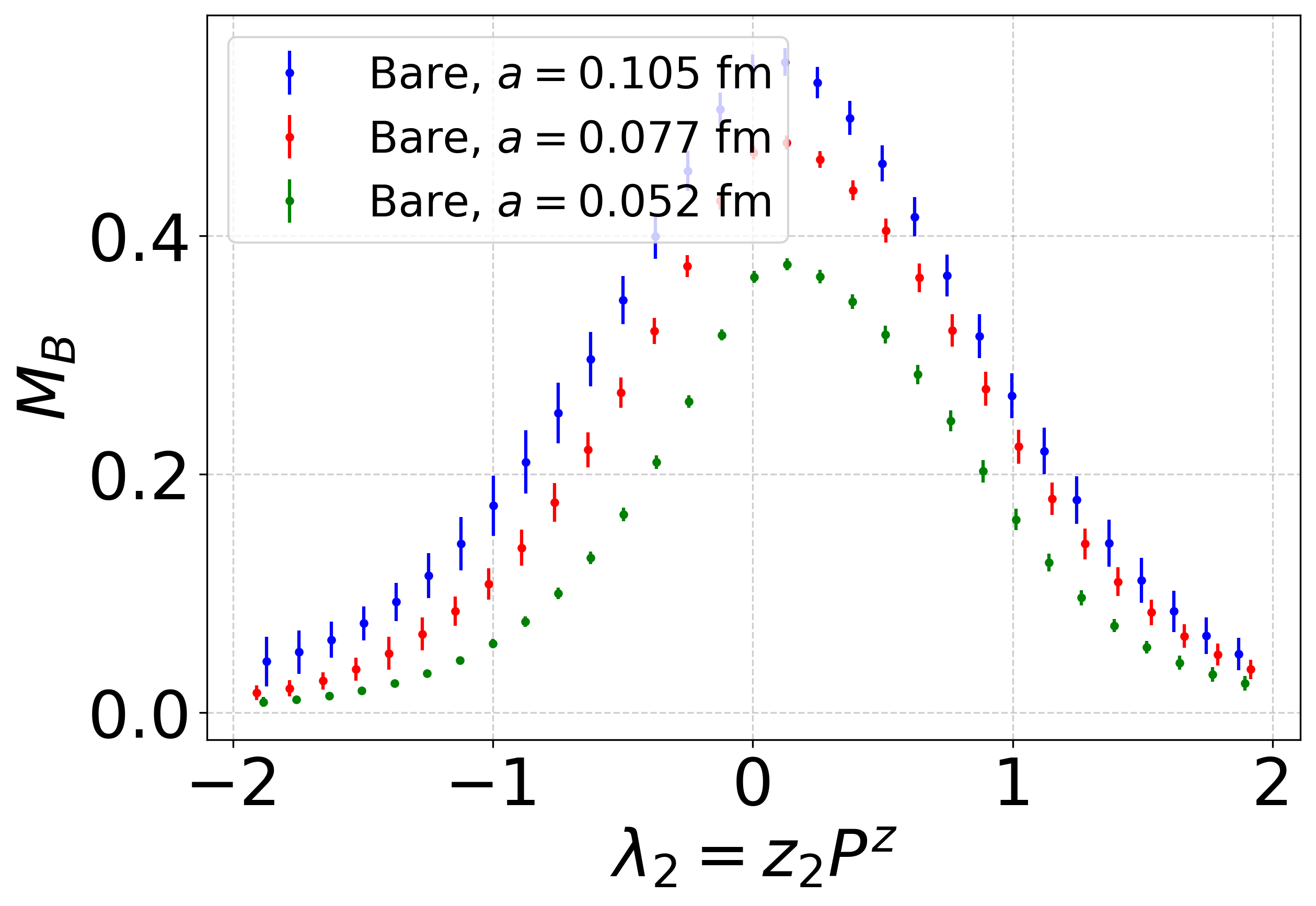}
    }
\vspace{0.0cm} 
\subfigure[\ Ratio scheme result of proton at $P=0.5$ GeV]{
    \centering
    \includegraphics[scale=0.185]{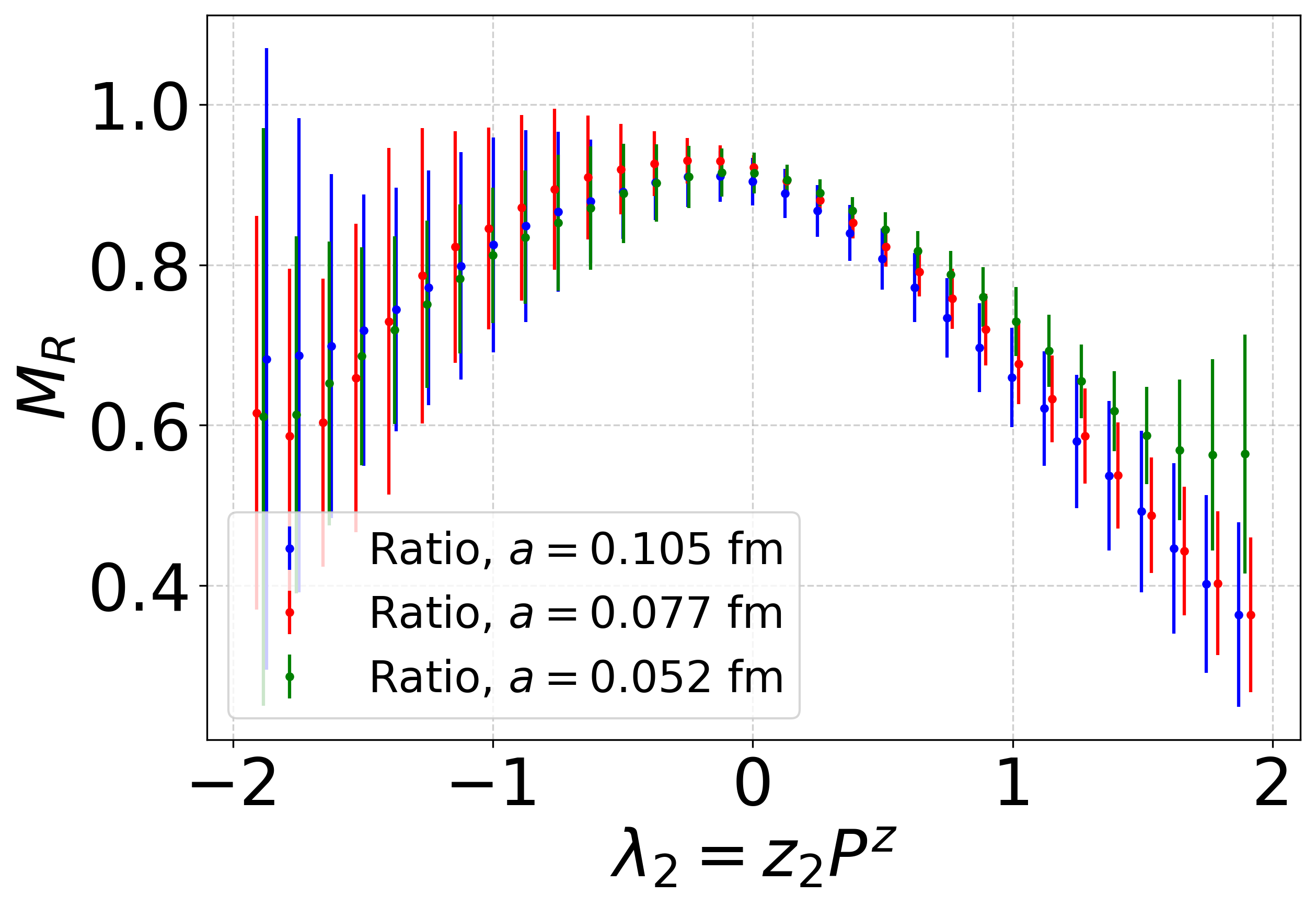}
    }
\vspace{0.0cm} 
\subfigure[\ Self scheme result of proton at $P=0.5$ GeV]{
    \centering
    \includegraphics[scale=0.185]{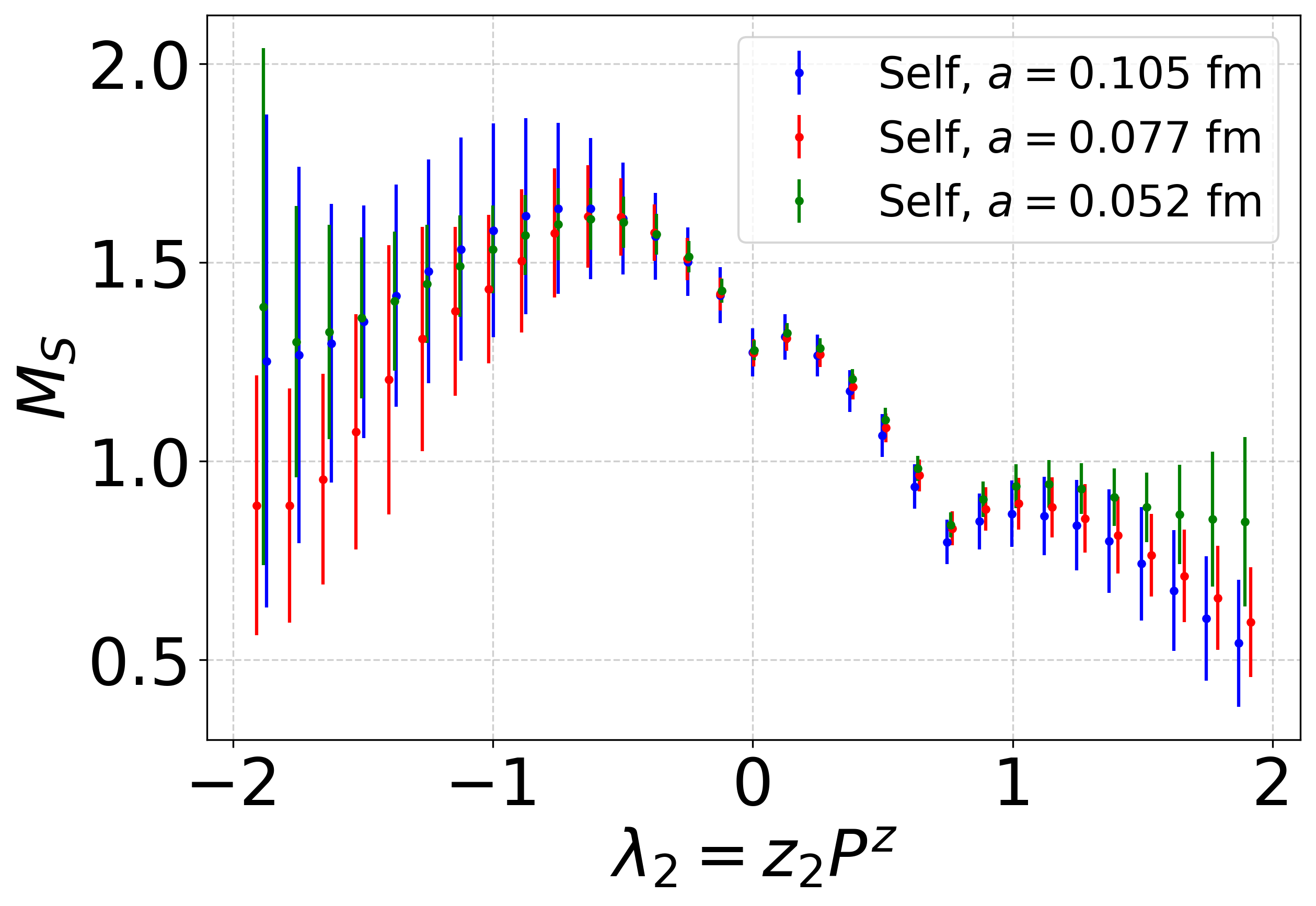}
    }
\vspace{0.0cm} 
\subfigure[\ Hybrid scheme result of proton at $P=0.5$ GeV]{
    \centering
    \includegraphics[scale=0.185]{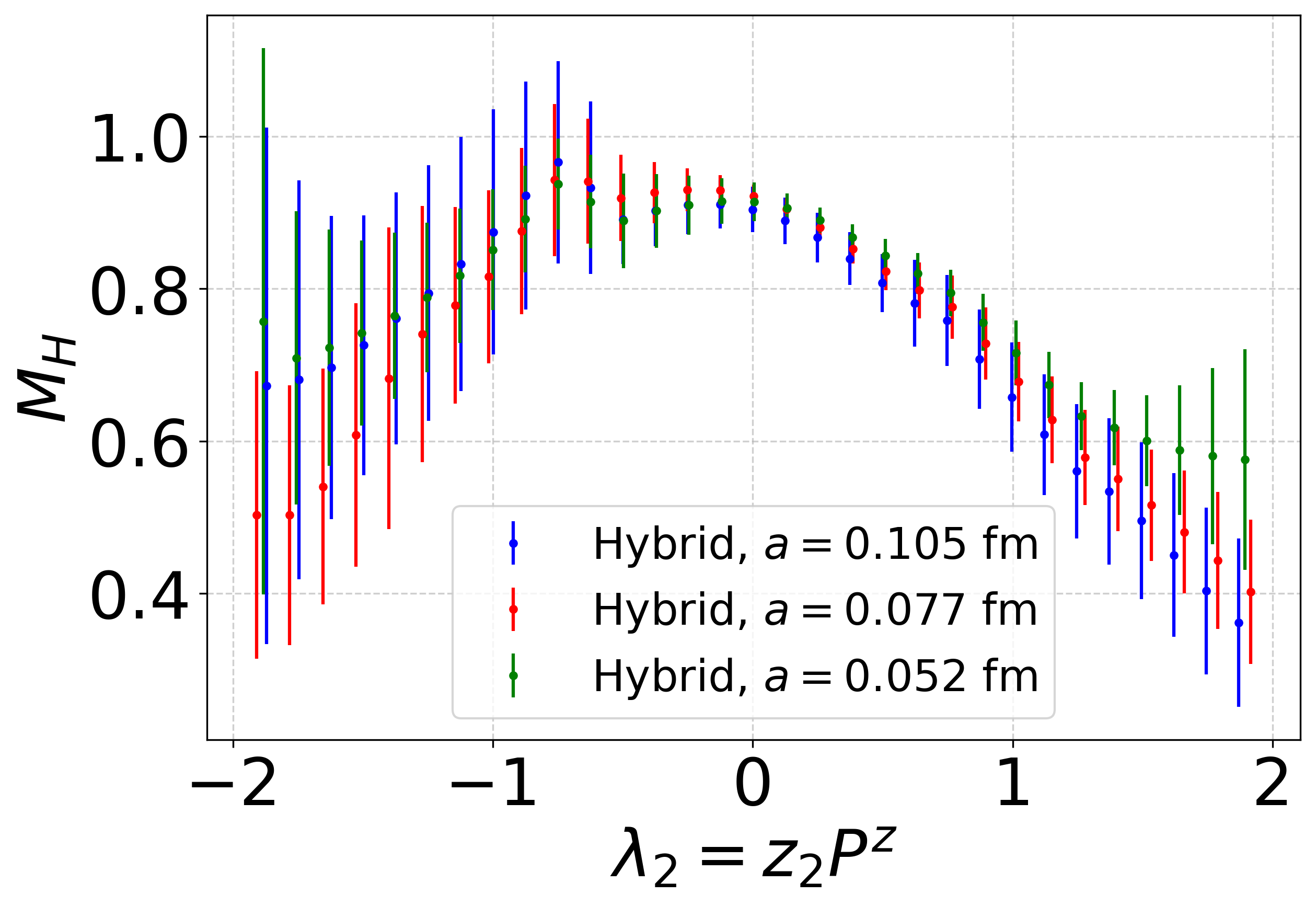}
    }
\caption{Results of the proton quasi-DA matrix elements in different schemes and with $P^z=0.5$ GeV, $z_1=0.300$ fm}
\label{fig:proton_p1_z6}
\end{figure}

\begin{figure}[htbp]
\centering
\subfigure[\ Bare result of proton at $P=0.5$ GeV]{
    \centering
    \includegraphics[scale=0.185]{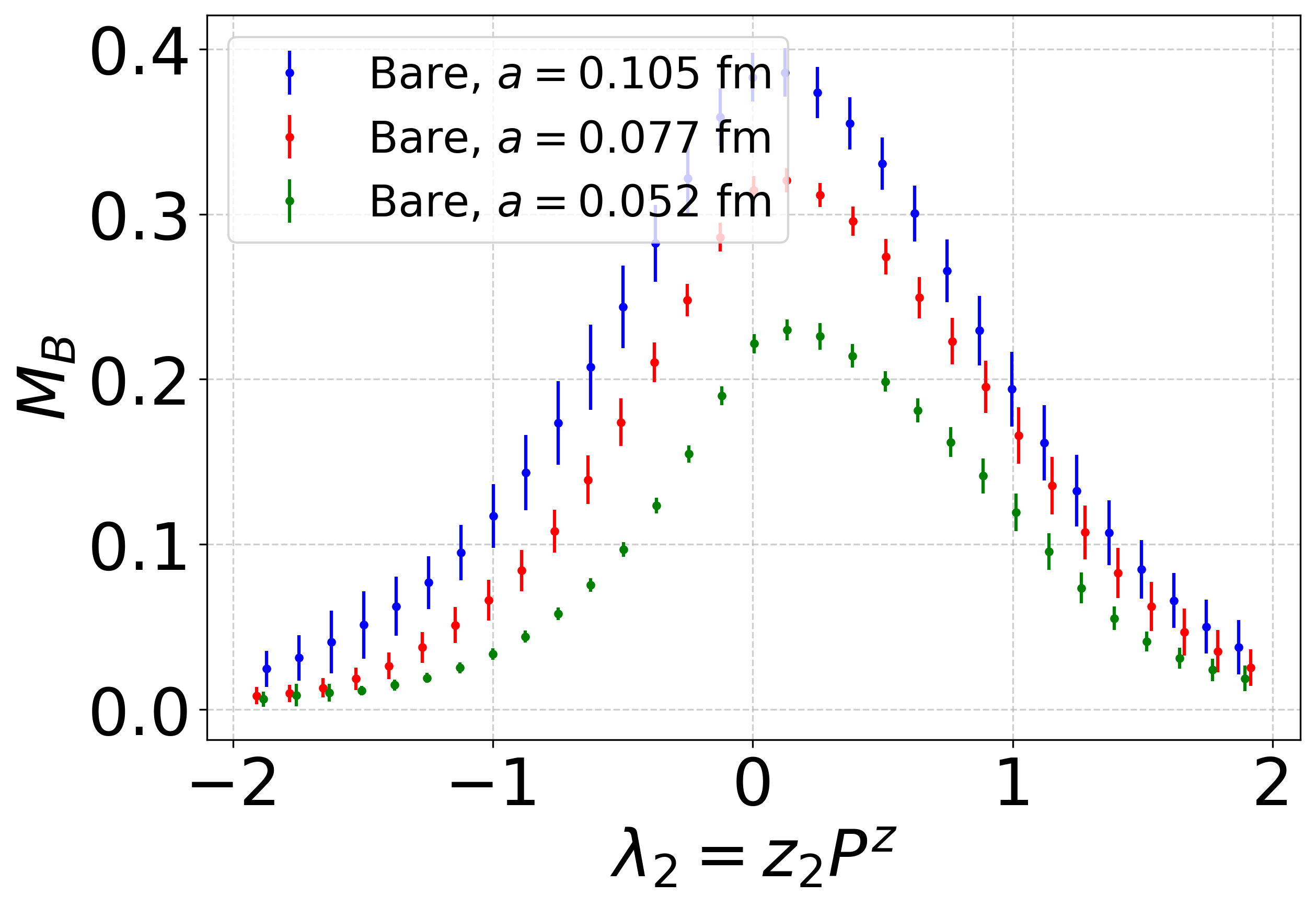}
    }
\vspace{0.0cm} 
\subfigure[\ Ratio scheme result of proton at $P=0.5$ GeV]{
    \centering
    \includegraphics[scale=0.185]{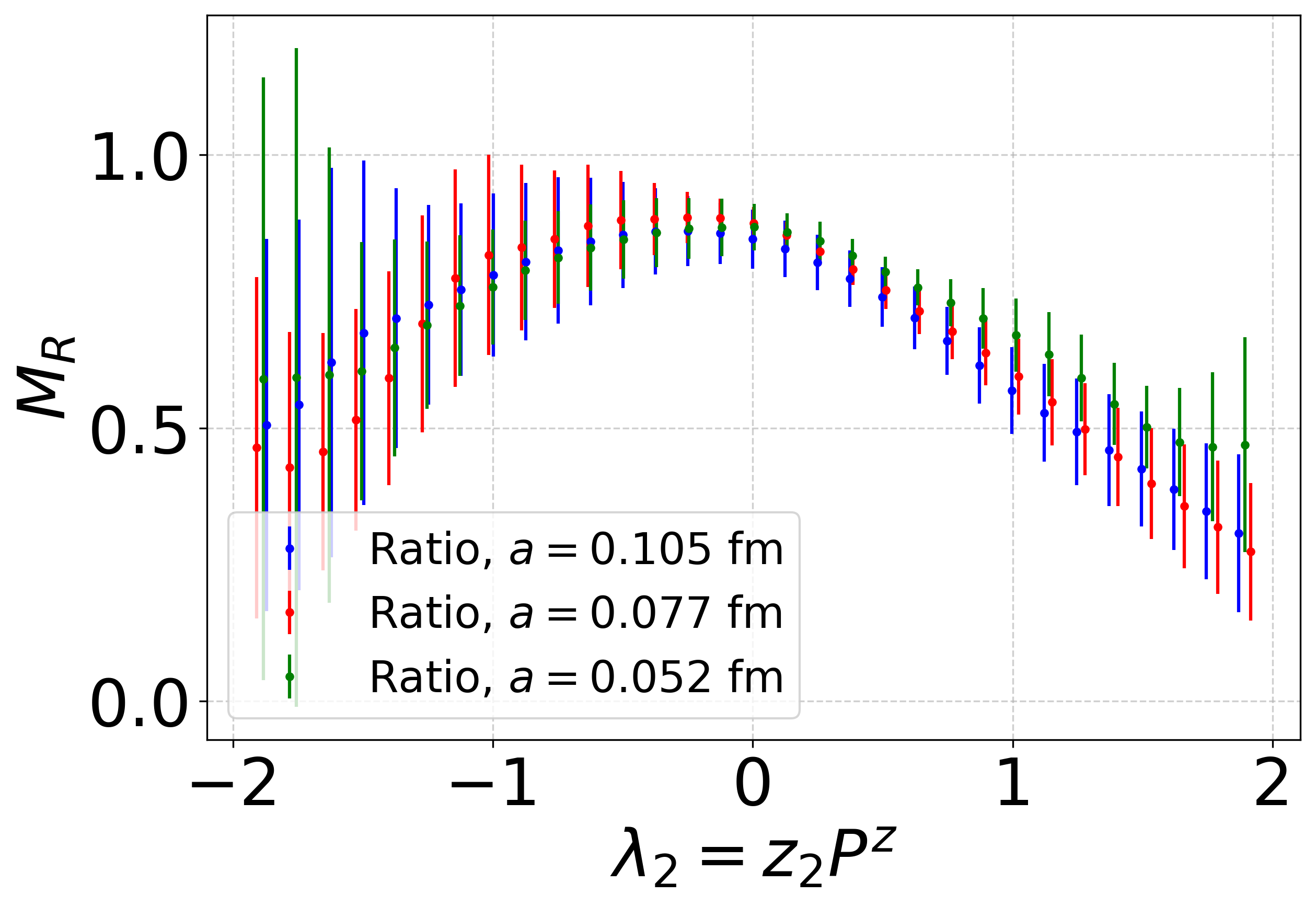}
    }
\vspace{0.0cm} 
\subfigure[\ Self scheme result of proton at $P=0.5$ GeV]{
    \centering
    \includegraphics[scale=0.185]{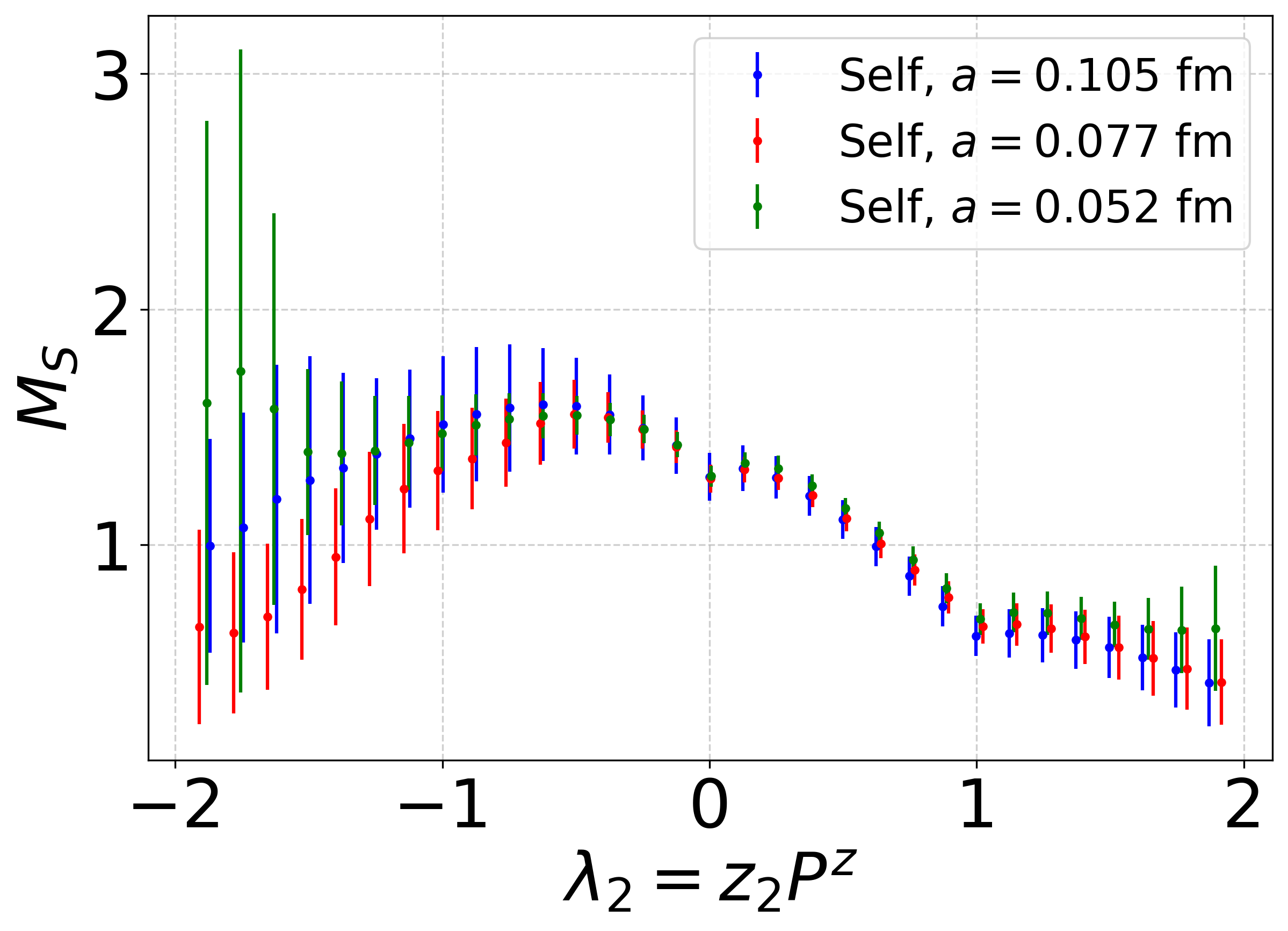}
    }
\vspace{0.0cm} 
\subfigure[\ Hybrid scheme result of proton at $P=0.5$ GeV]{
    \centering
    \includegraphics[scale=0.185]{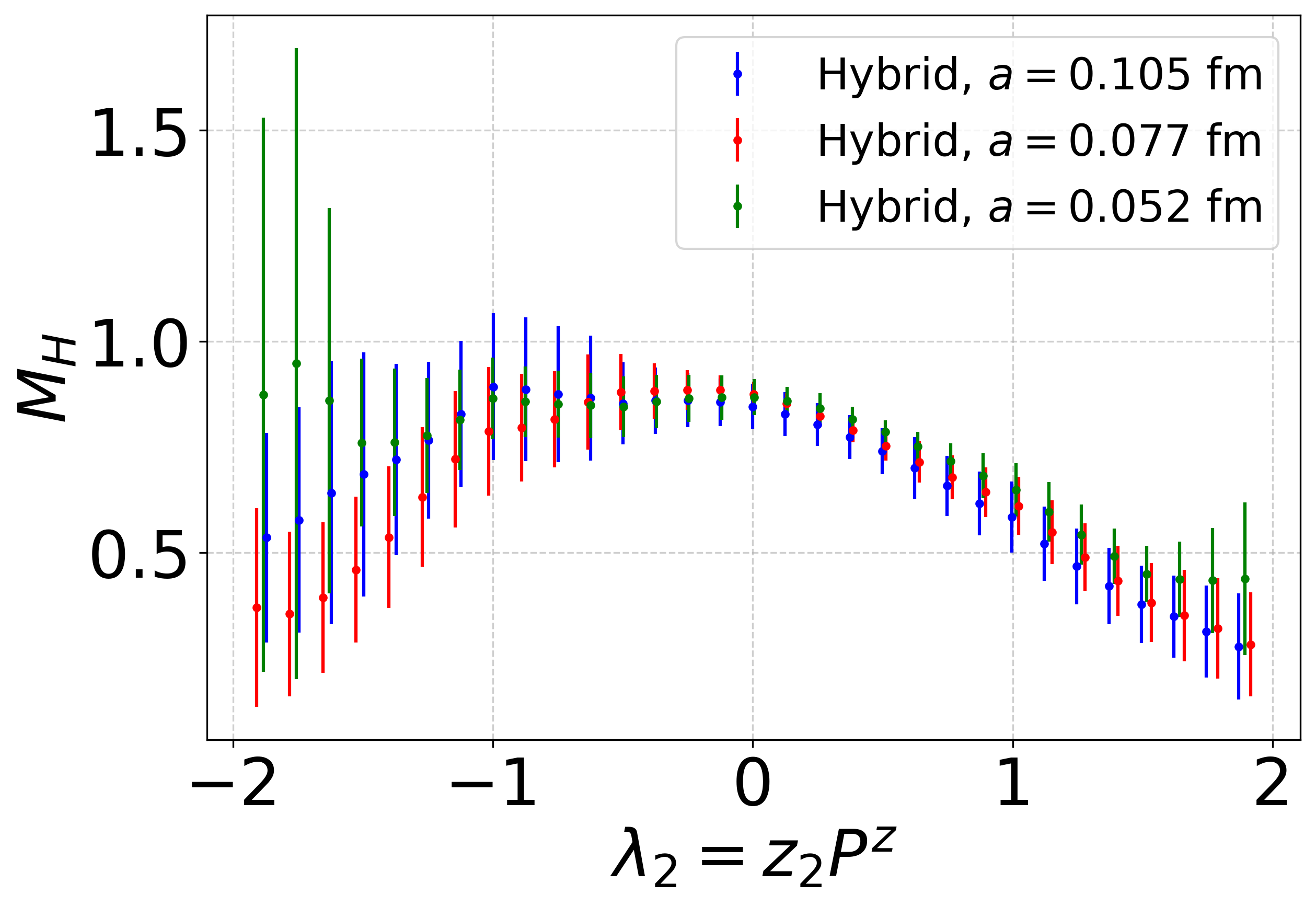}
    }
\caption{Results of the proton quasi-DA matrix elements in different schemes and with $P^z=0.5$ GeV, $z_1=0.400$ fm}
\label{fig:proton_p1_z8}
\end{figure}

\begin{figure}[htbp]
\centering
\subfigure[\ Bare result of proton at $P=0.5$ GeV]{
    \centering
    \includegraphics[scale=0.185]{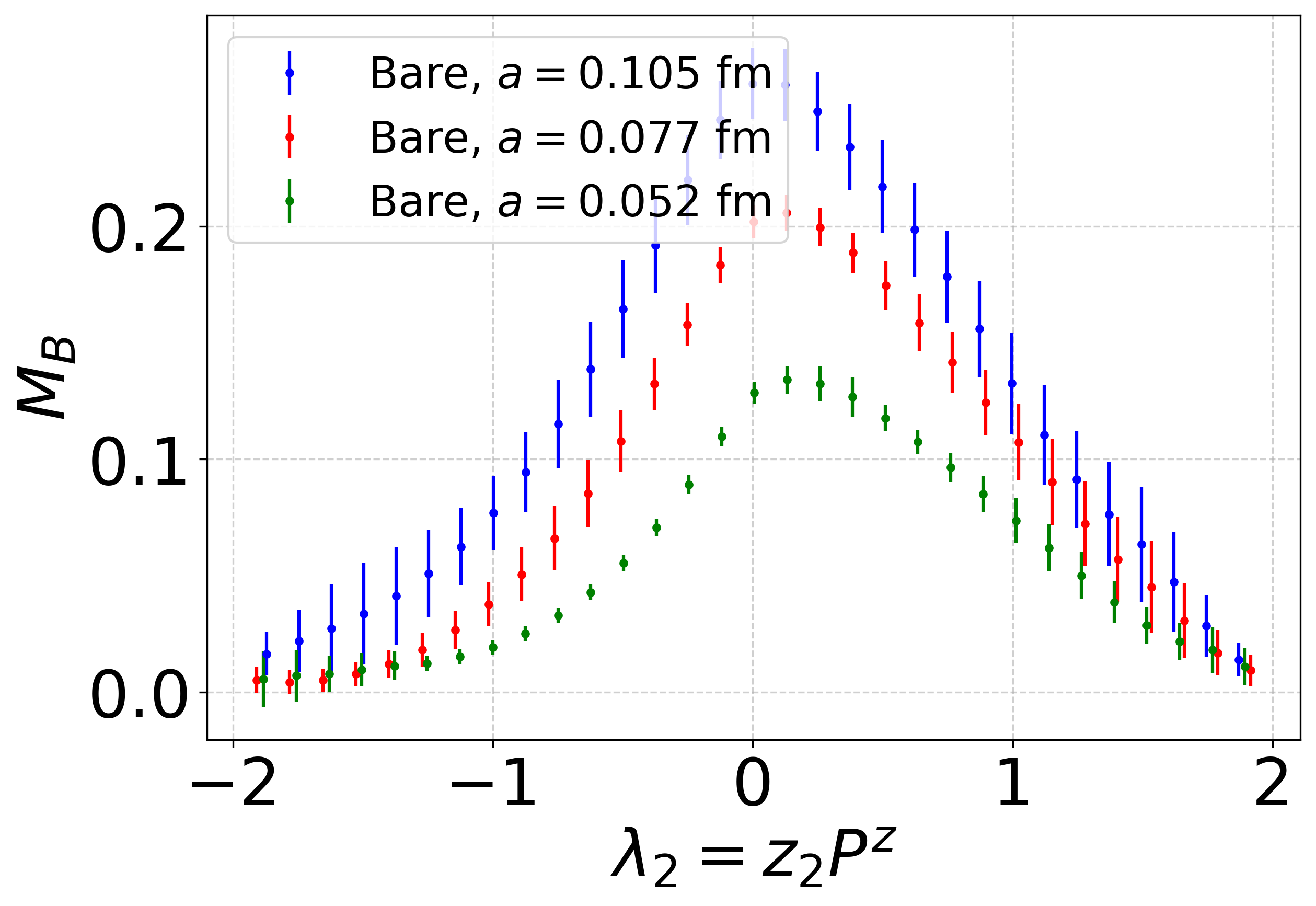}
    }
\vspace{0.0cm} 
\subfigure[\ Ratio scheme result of proton at $P=0.5$ GeV]{
    \centering
    \includegraphics[scale=0.185]{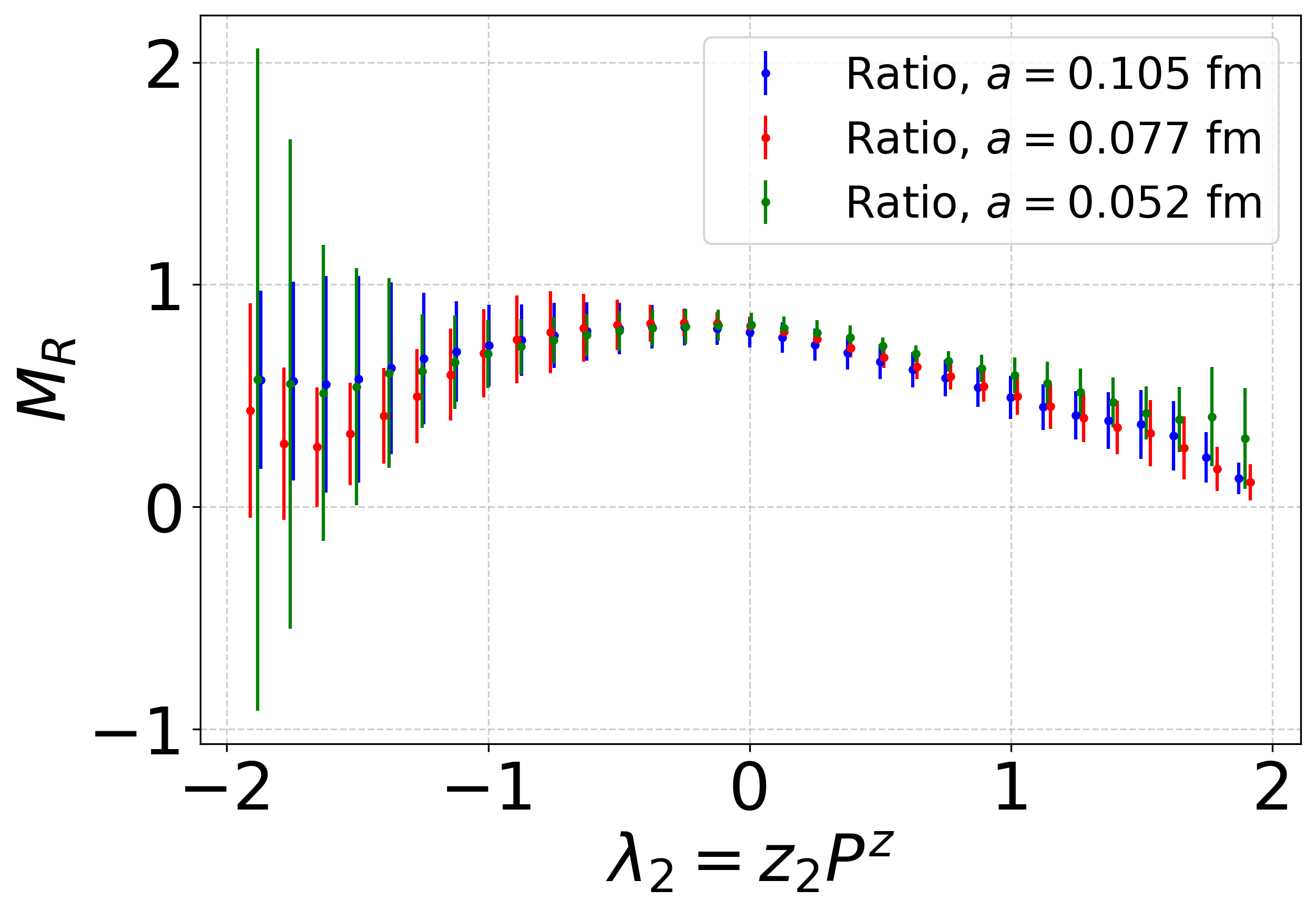}
    }
\vspace{0.0cm} 
\subfigure[\ Self scheme result of proton at $P=0.5$ GeV]{
    \centering
    \includegraphics[scale=0.185]{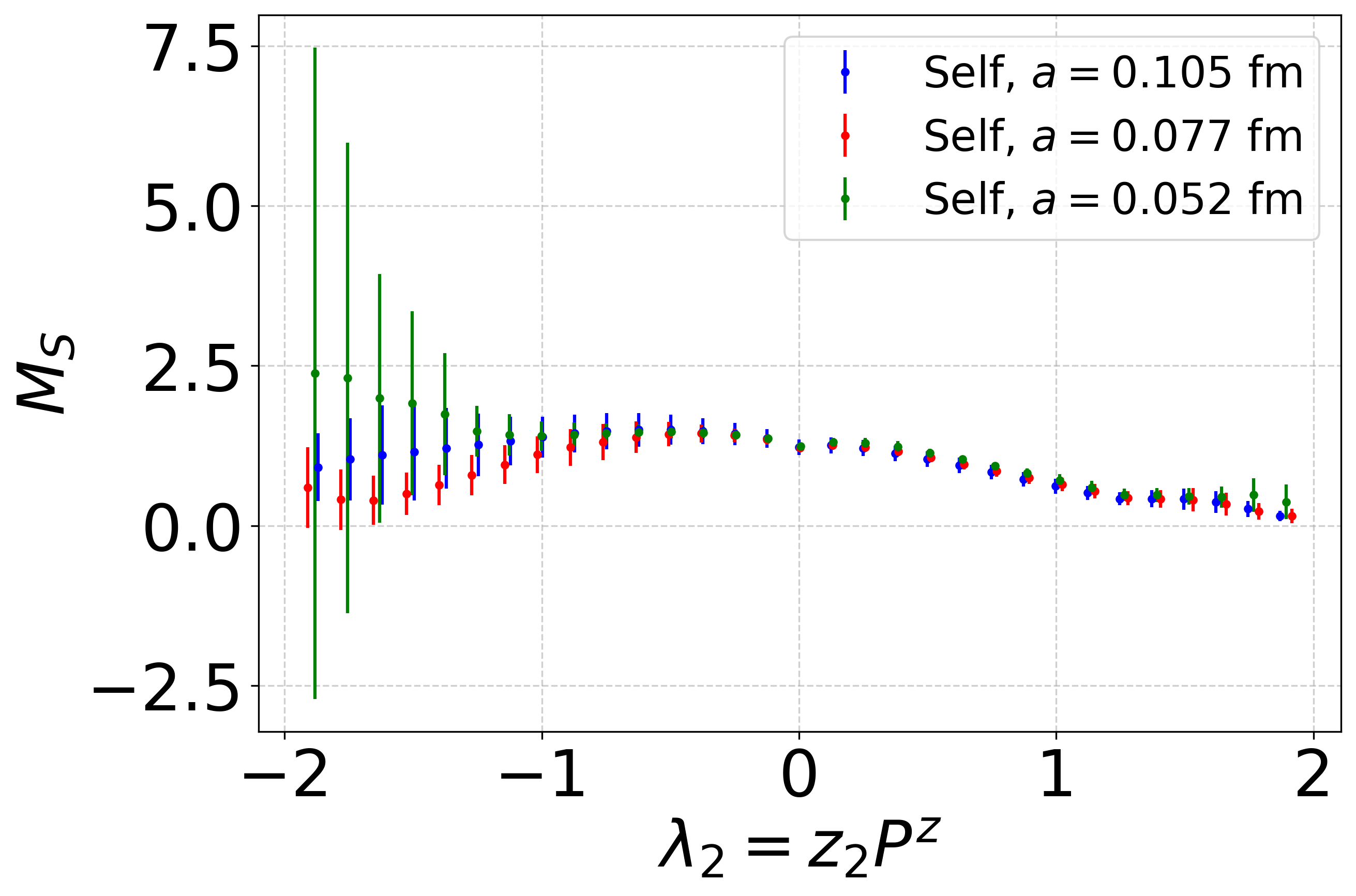}
    }
\vspace{0.0cm} 
\subfigure[\ Hybrid scheme result of proton at $P=0.5$ GeV]{
    \centering
    \includegraphics[scale=0.185]{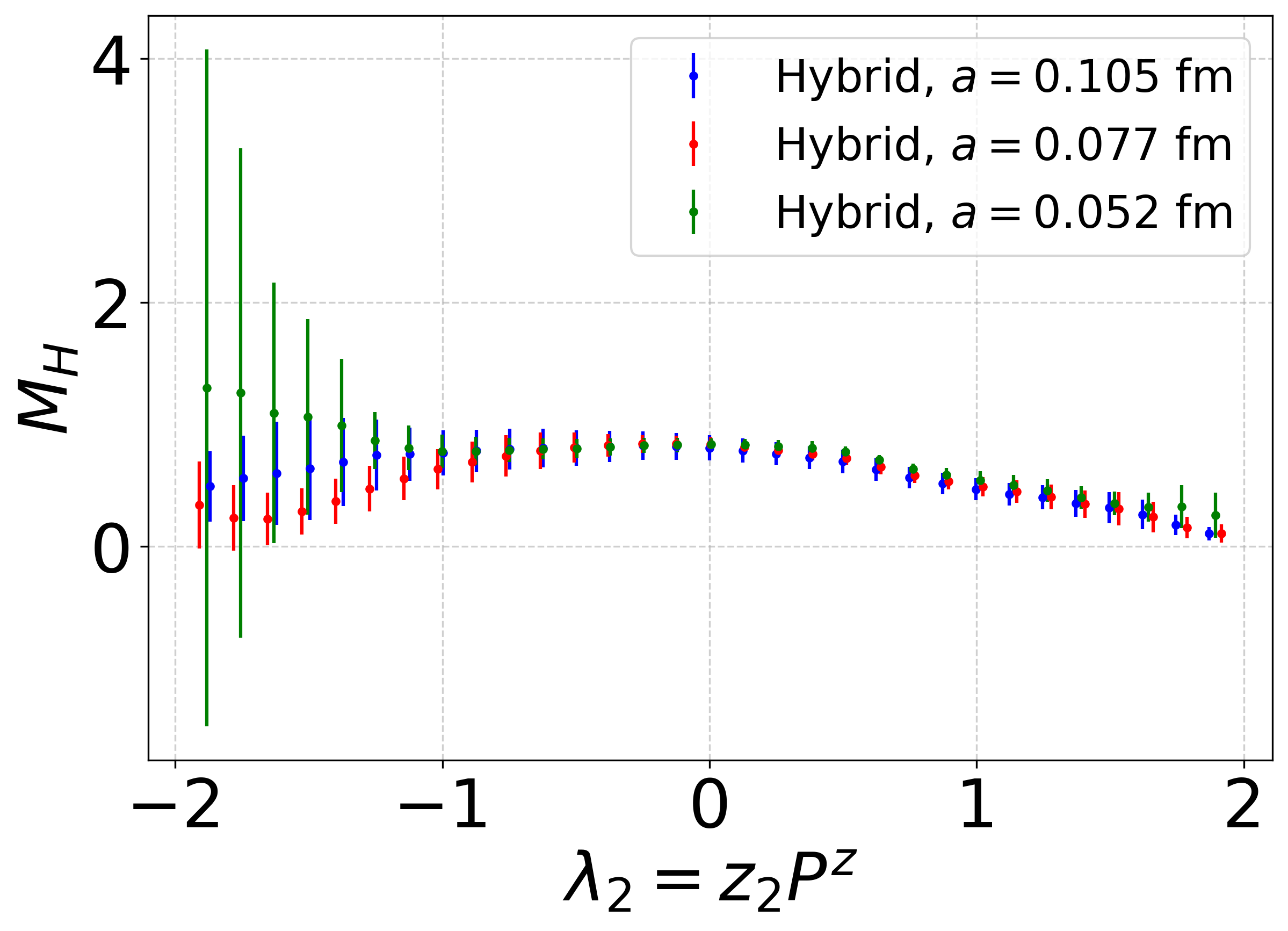}
    }
\caption{Results of the proton quasi-DA matrix elements in different schemes and with $P^z=0.5$ GeV, $z_1=0.500$ fm}
\label{fig:proton_p1_z10}
\end{figure}

\clearpage

\subsection{More results of the proton (V-term) Quasi-DA at $P^z = 2.0$ GeV in different schemes}
\begin{figure}[htbp]
\centering
\subfigure[\ Bare result of proton at $P=2.0$ GeV]{
    \centering
    \includegraphics[scale=0.185]{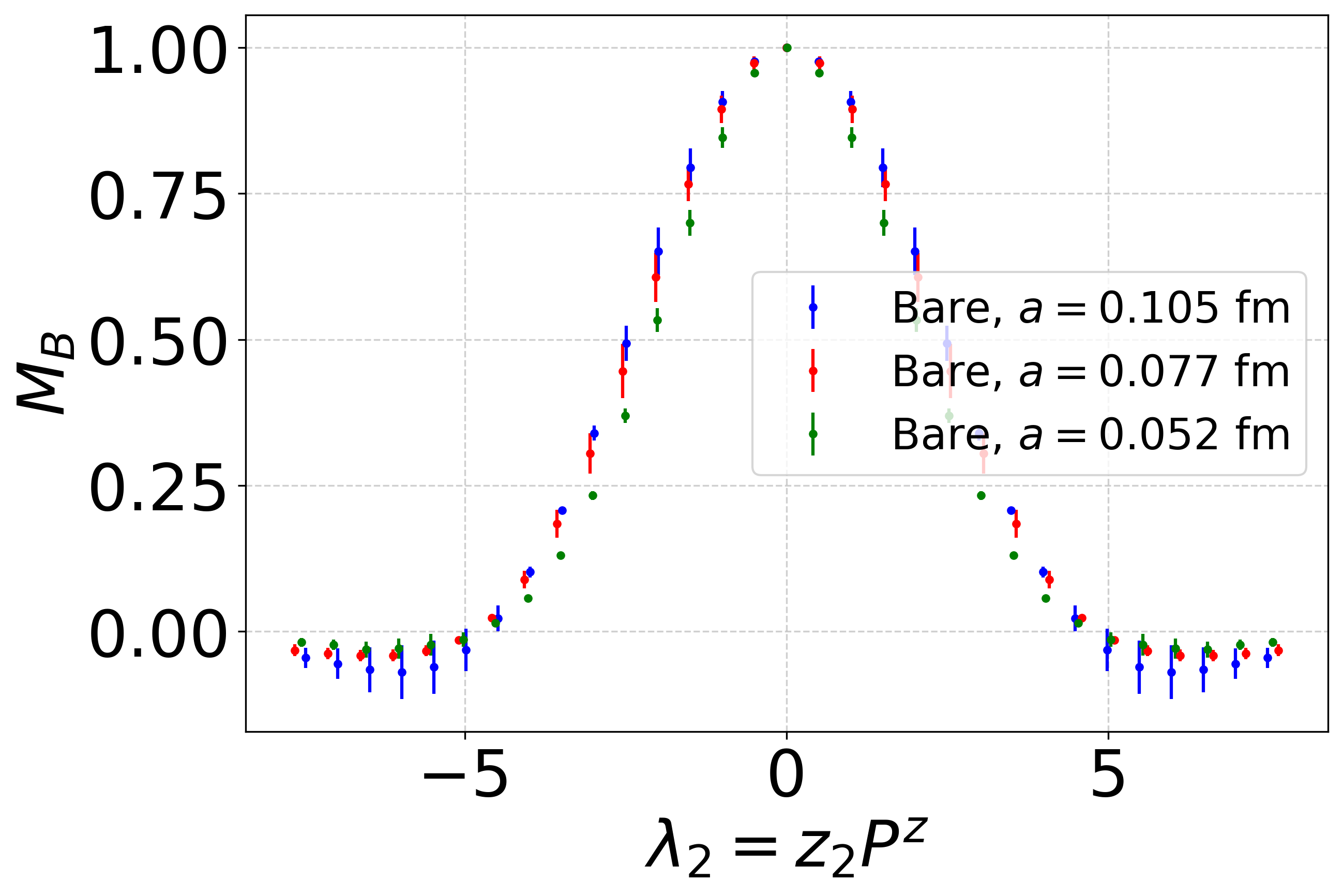}
    }
\vspace{0.0cm} 
\subfigure[\ Ratio scheme result of proton at $P=2.0$ GeV]{
    \centering
    \includegraphics[scale=0.185]{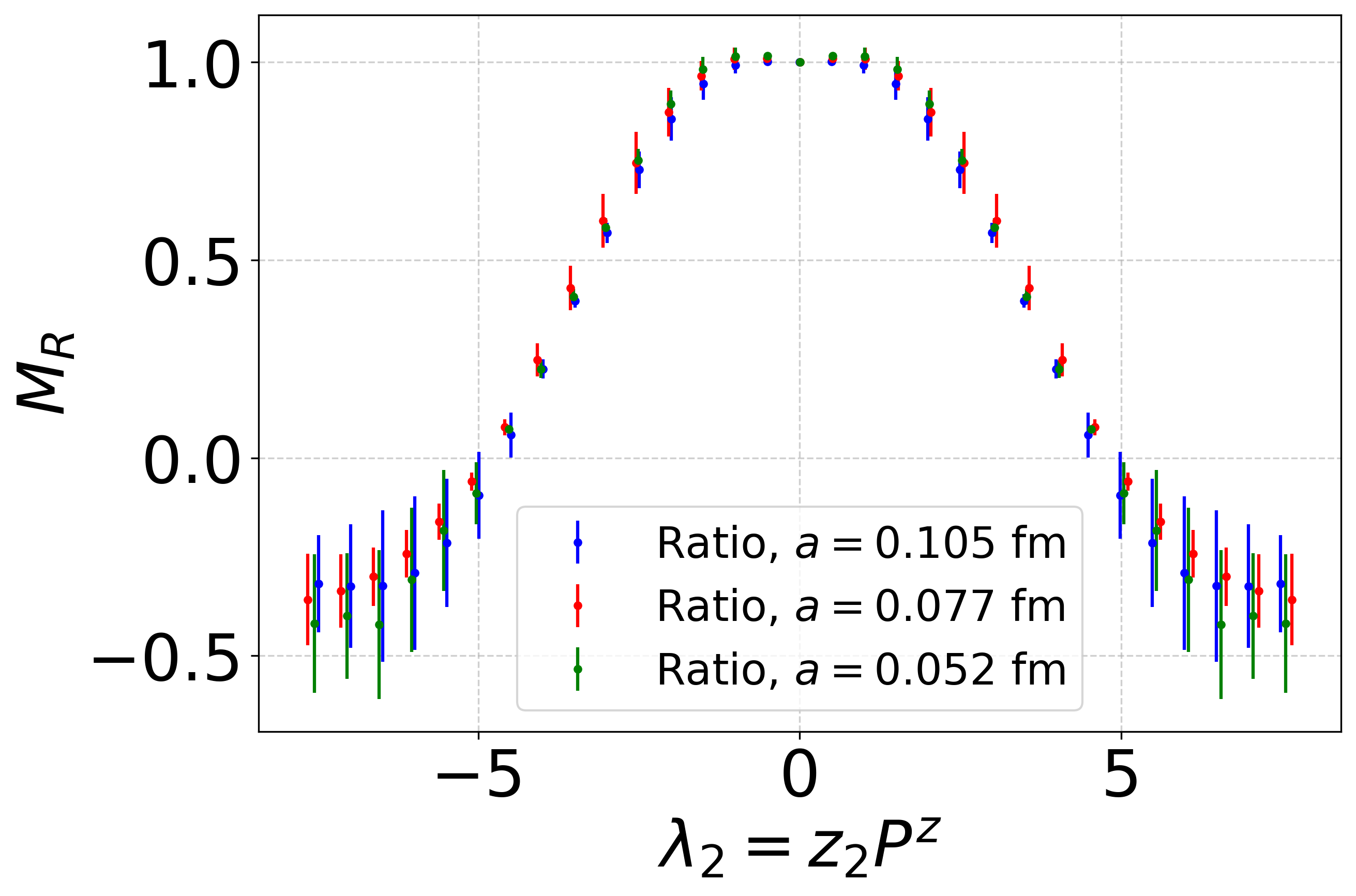}
    }
\vspace{0.0cm} 
\subfigure[\ Self scheme result of proton at $P=2.0$ GeV]{
    \centering
    \includegraphics[scale=0.185]{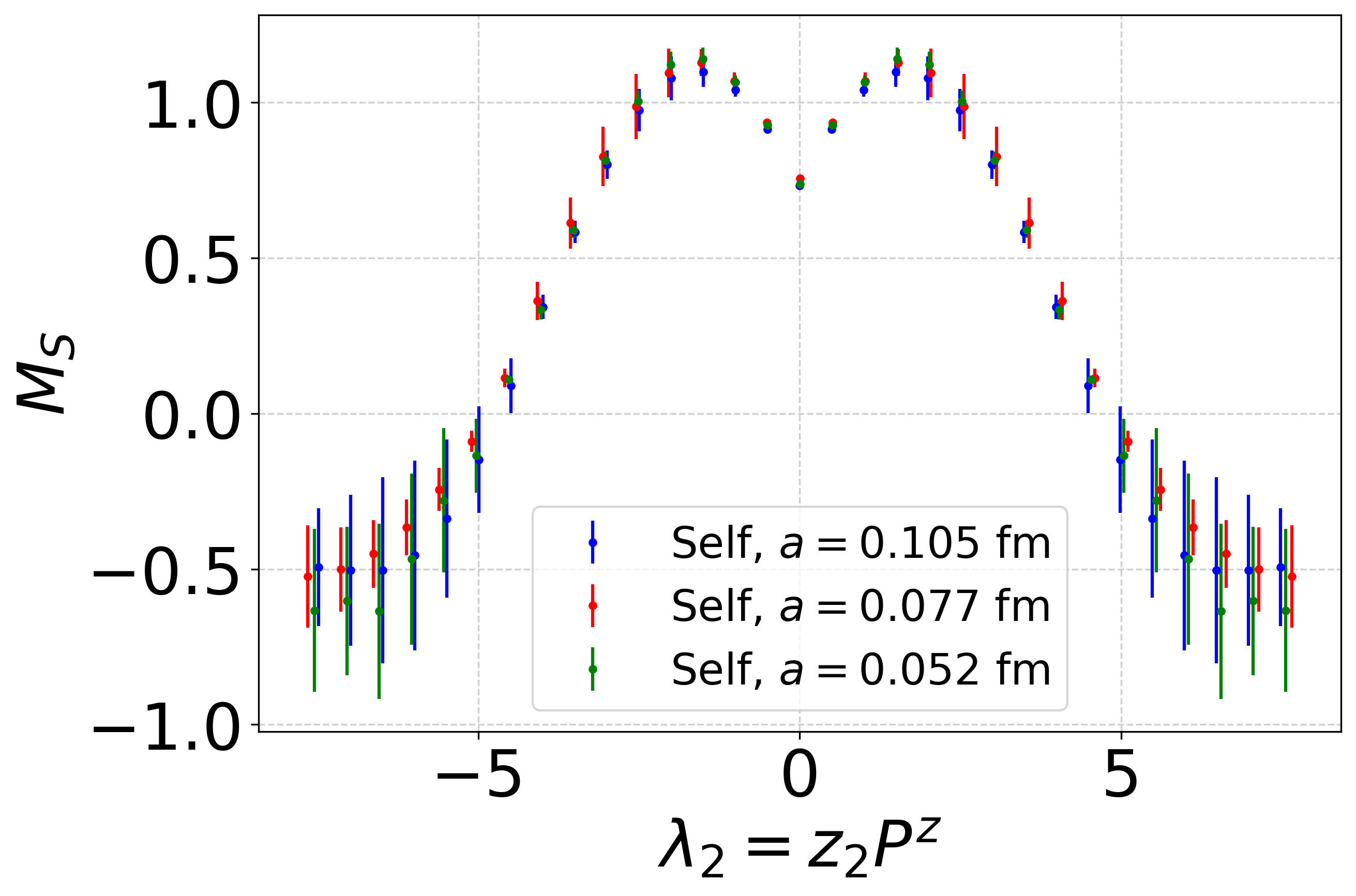}
    }
\vspace{0.0cm} 
\subfigure[\ Hybrid scheme result of proton at $P=2.0$ GeV]{
    \centering
    \includegraphics[scale=0.185]{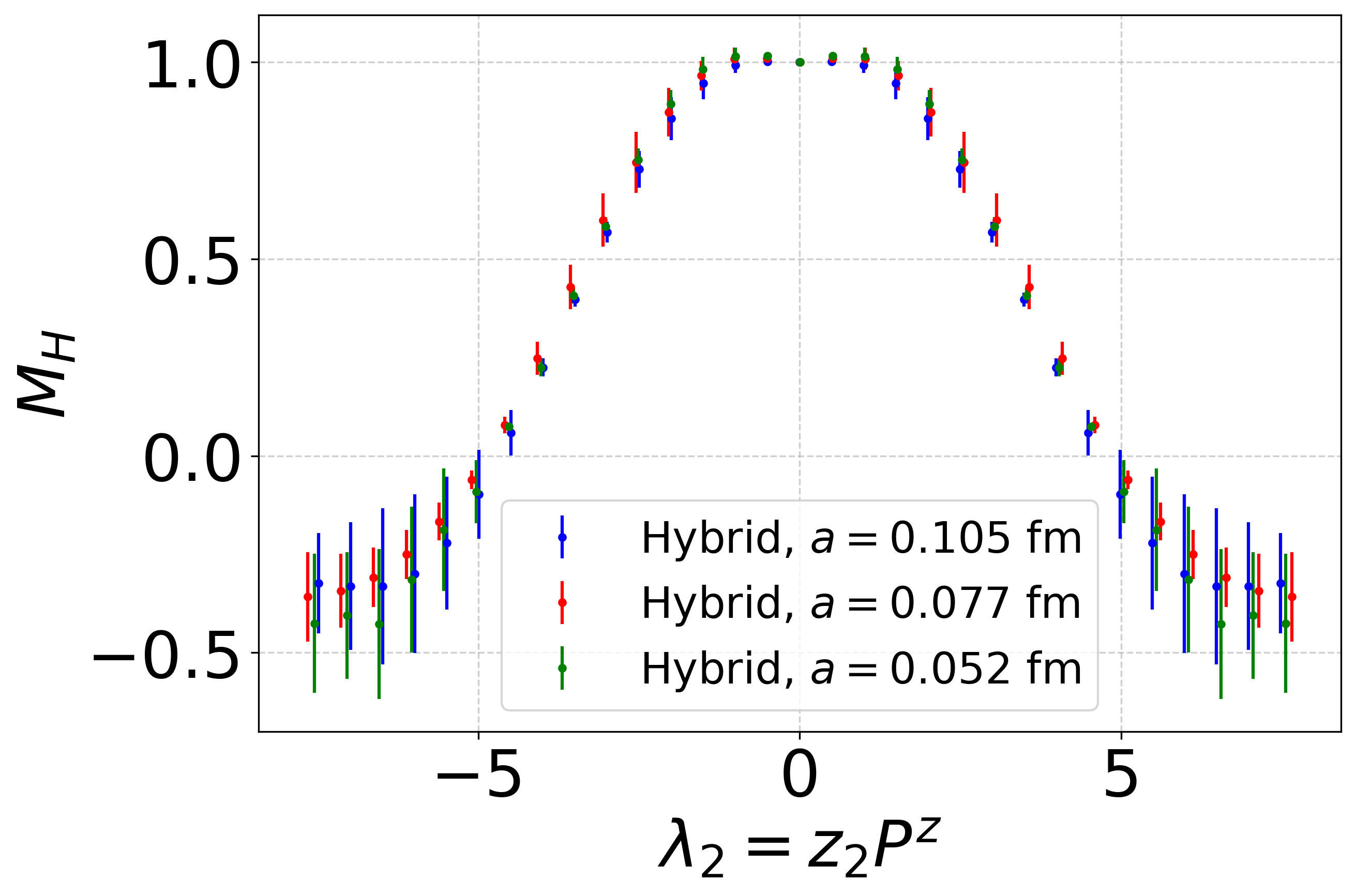}
    }
\caption{Results of the proton quasi-DA matrix elements in different schemes and with $P^z=2.0$ GeV, $z_1=0.000$ fm}
\label{fig:proton_p4_z0}
\end{figure}

\begin{figure}[htbp]
\centering
\subfigure[\ Bare result of proton at $P=2.0$ GeV]{
    \centering
    \includegraphics[scale=0.185]{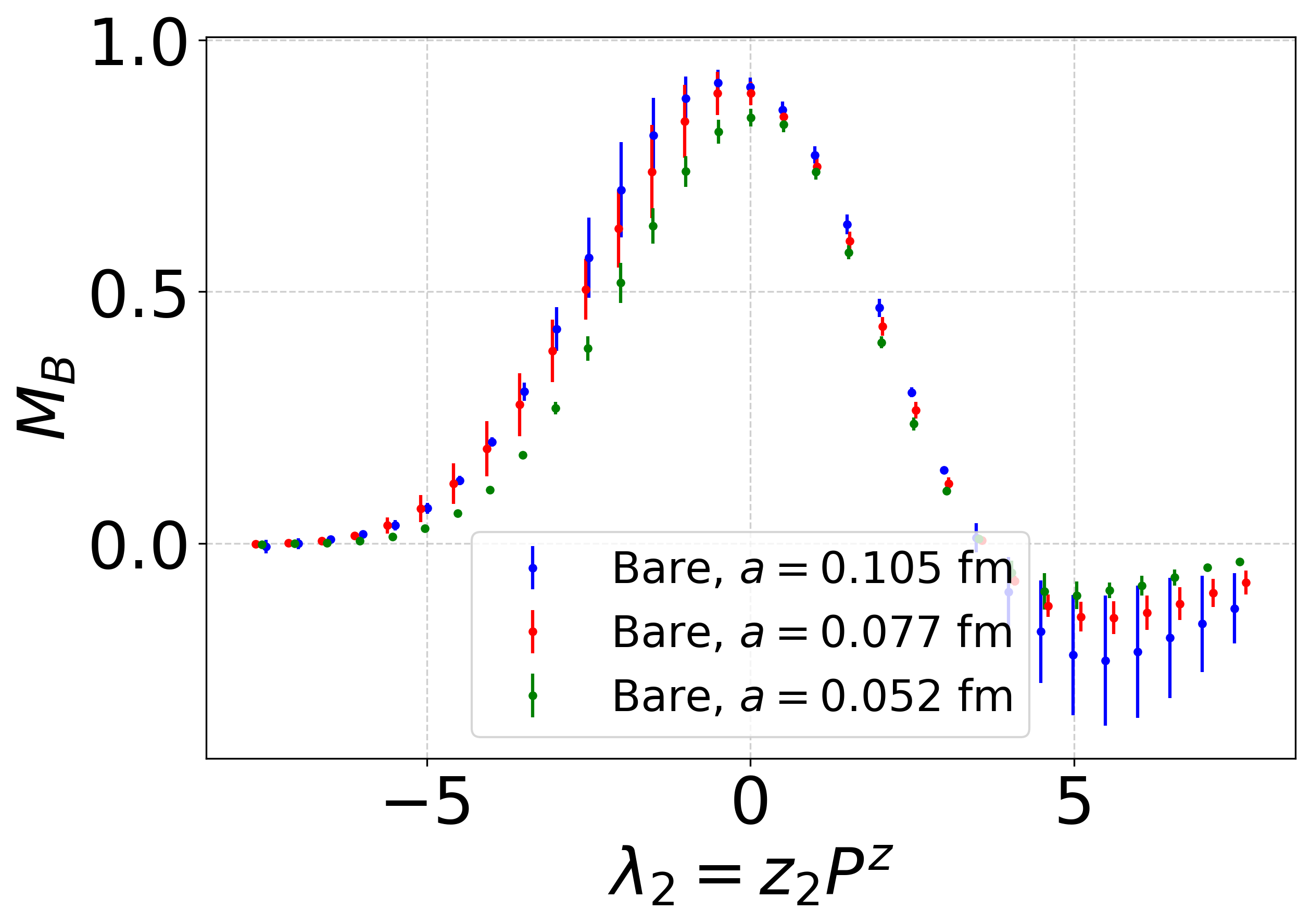}
    }
\vspace{0.0cm} 
\subfigure[\ Ratio scheme result of proton at $P=2.0$ GeV]{
    \centering
    \includegraphics[scale=0.185]{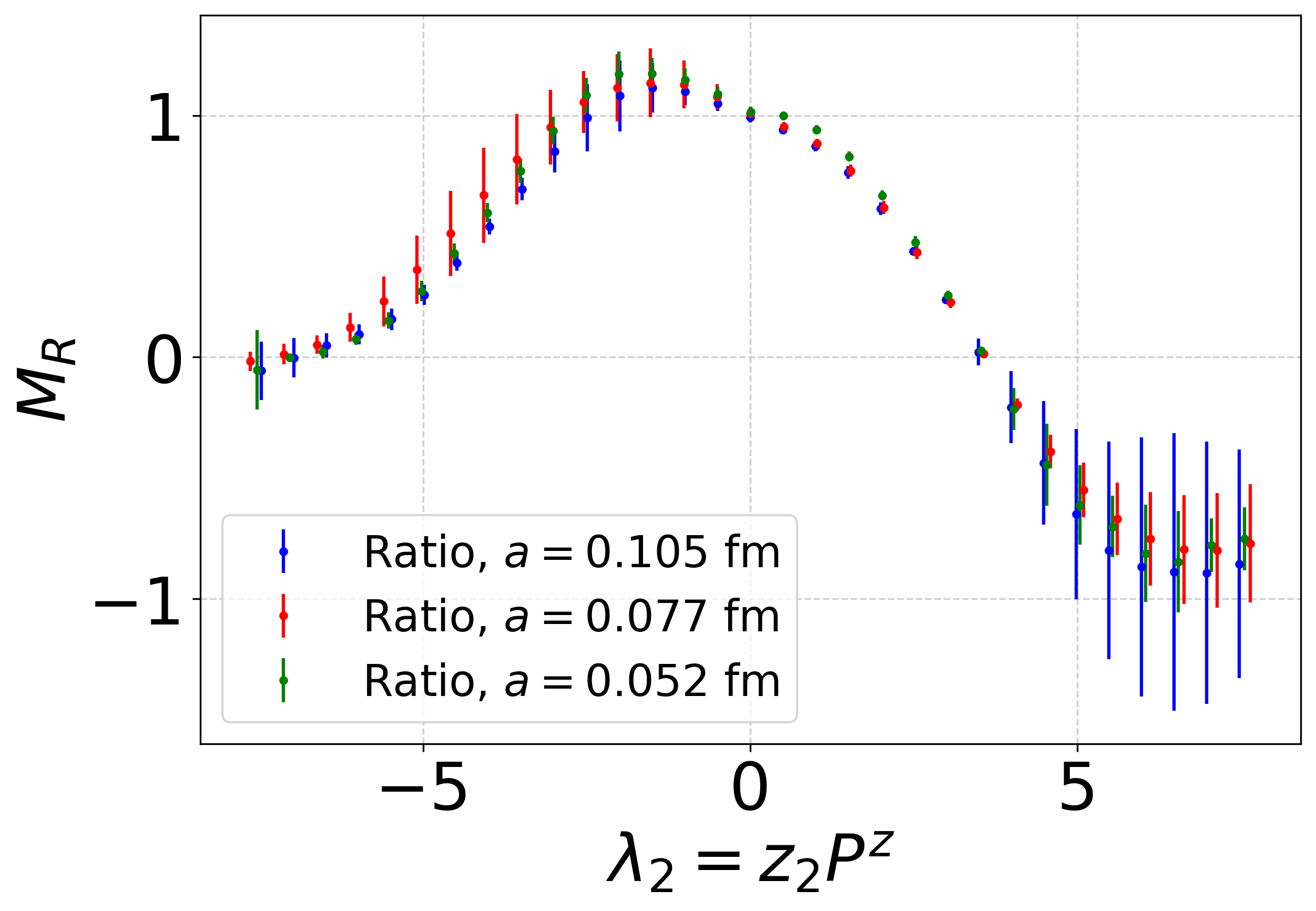}
    }
\vspace{0.0cm} 
\subfigure[\ Self scheme result of proton at $P=2.0$ GeV]{
    \centering
    \includegraphics[scale=0.185]{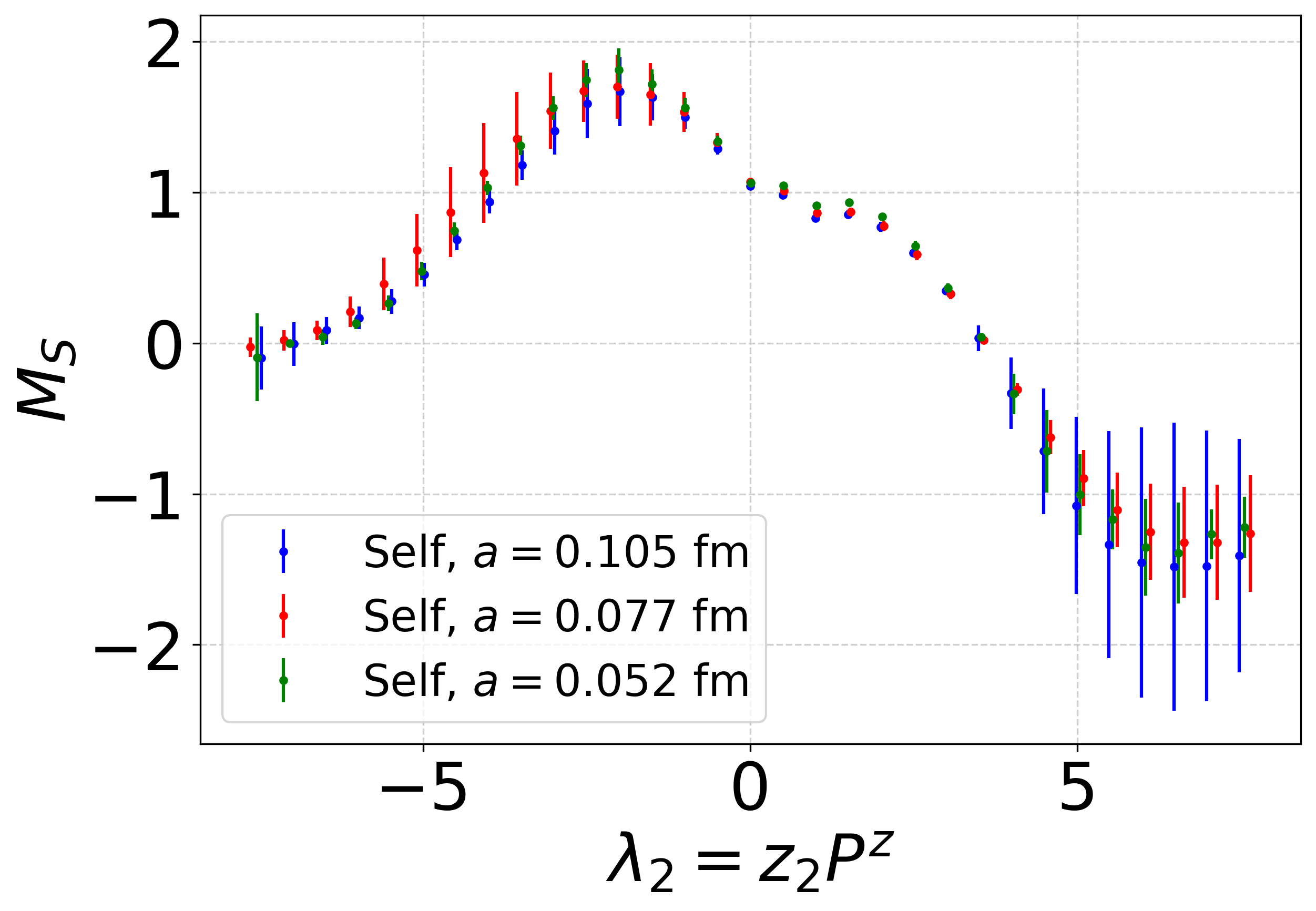}
    }
\vspace{0.0cm} 
\subfigure[\ Hybrid scheme result of proton at $P=2.0$ GeV]{
    \centering
    \includegraphics[scale=0.185]{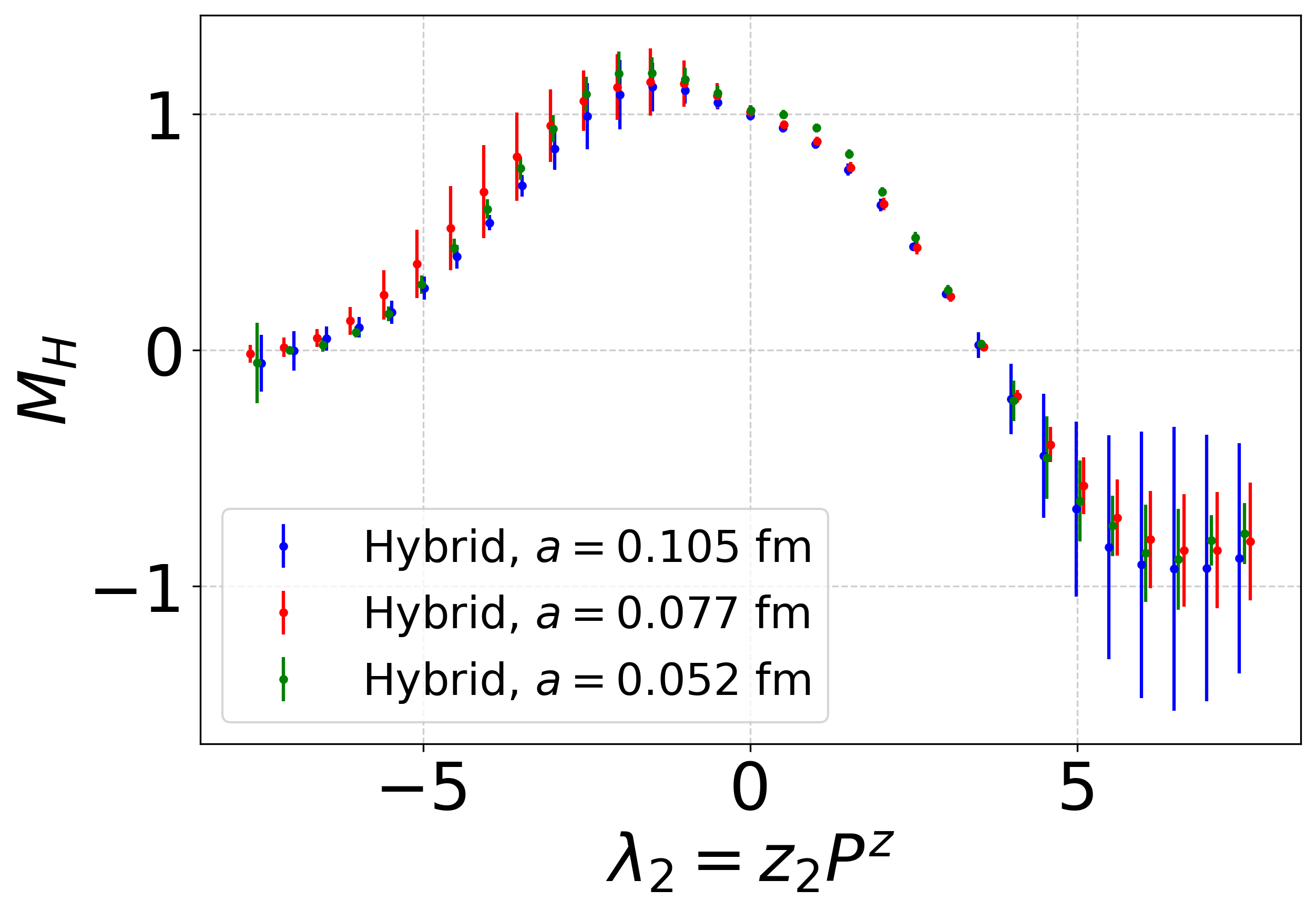}
    }
\caption{Results of the proton quasi-DA matrix elements in different schemes and with $P^z=2.0$ GeV, $z_1=0.100$ fm}
\label{fig:proton_p4_z2}
\end{figure}

\begin{figure}[htbp]
\centering
\subfigure[\ Bare result of proton at $P=2.0$ GeV]{
    \centering
    \includegraphics[scale=0.185]{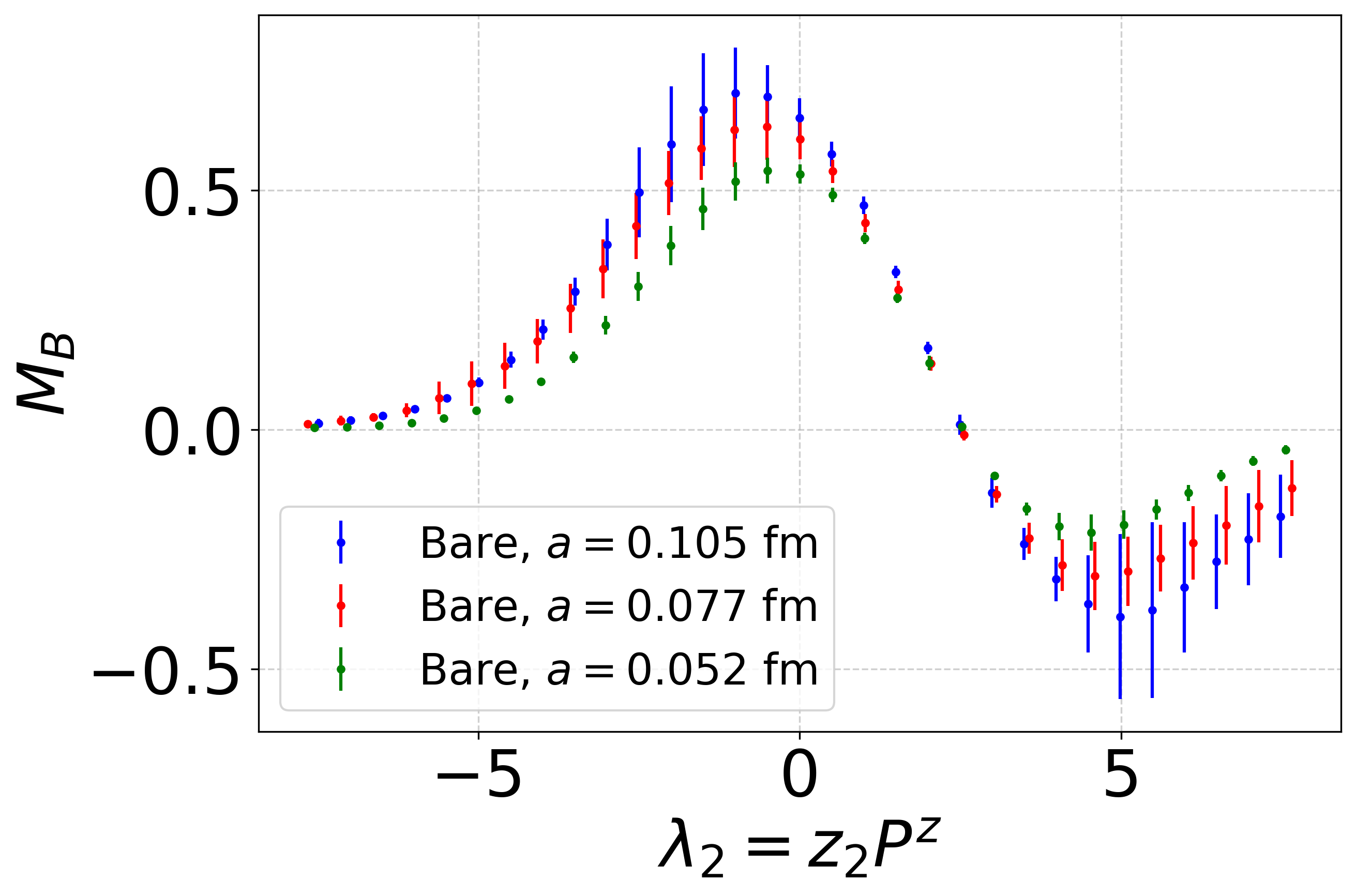}
    }
\vspace{0.0cm} 
\subfigure[\ Ratio scheme result of proton at $P=2.0$ GeV]{
    \centering
    \includegraphics[scale=0.185]{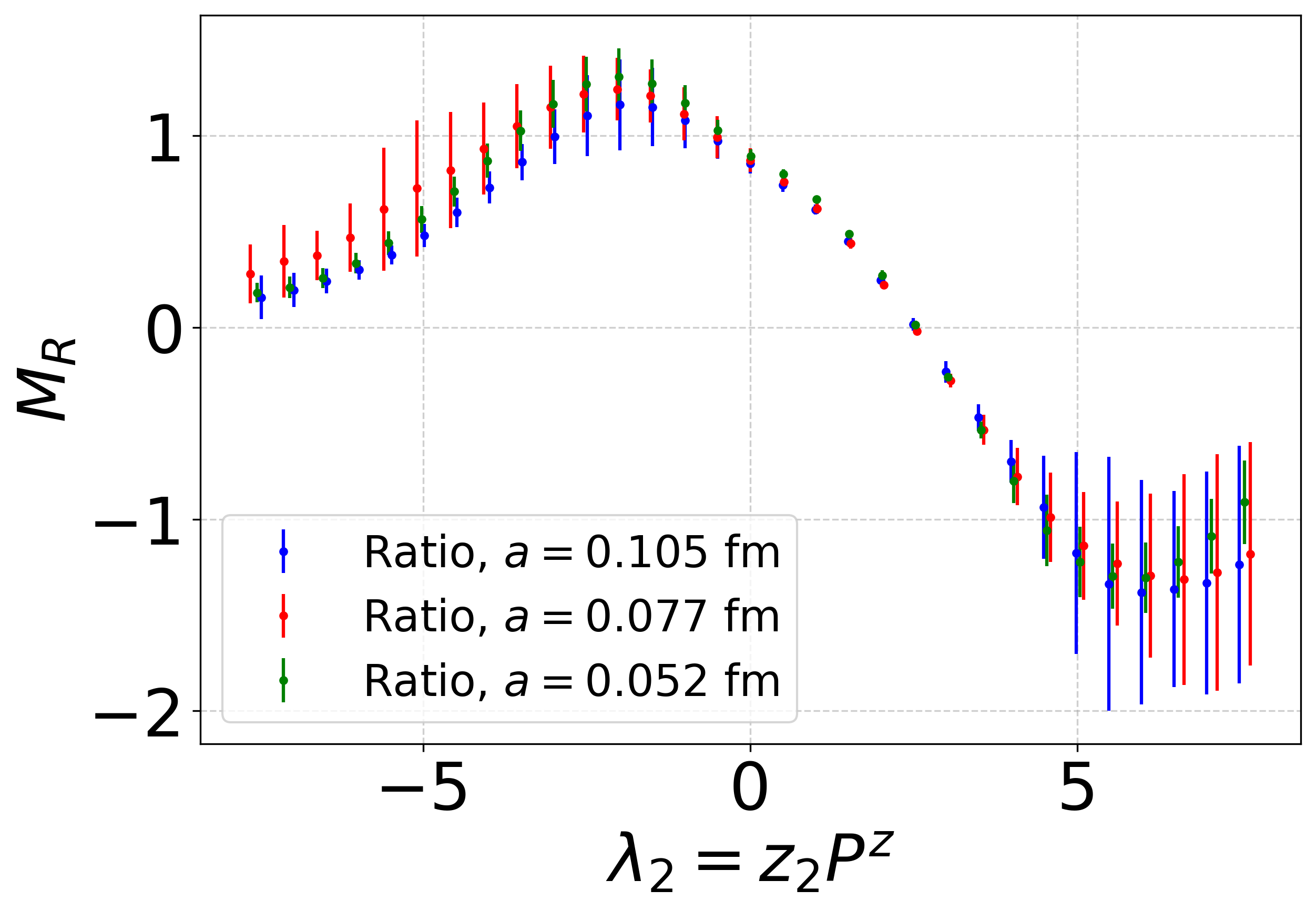}
    }
\vspace{0.0cm} 
\subfigure[\ Self scheme result of proton at $P=2.0$ GeV]{
    \centering
    \includegraphics[scale=0.185]{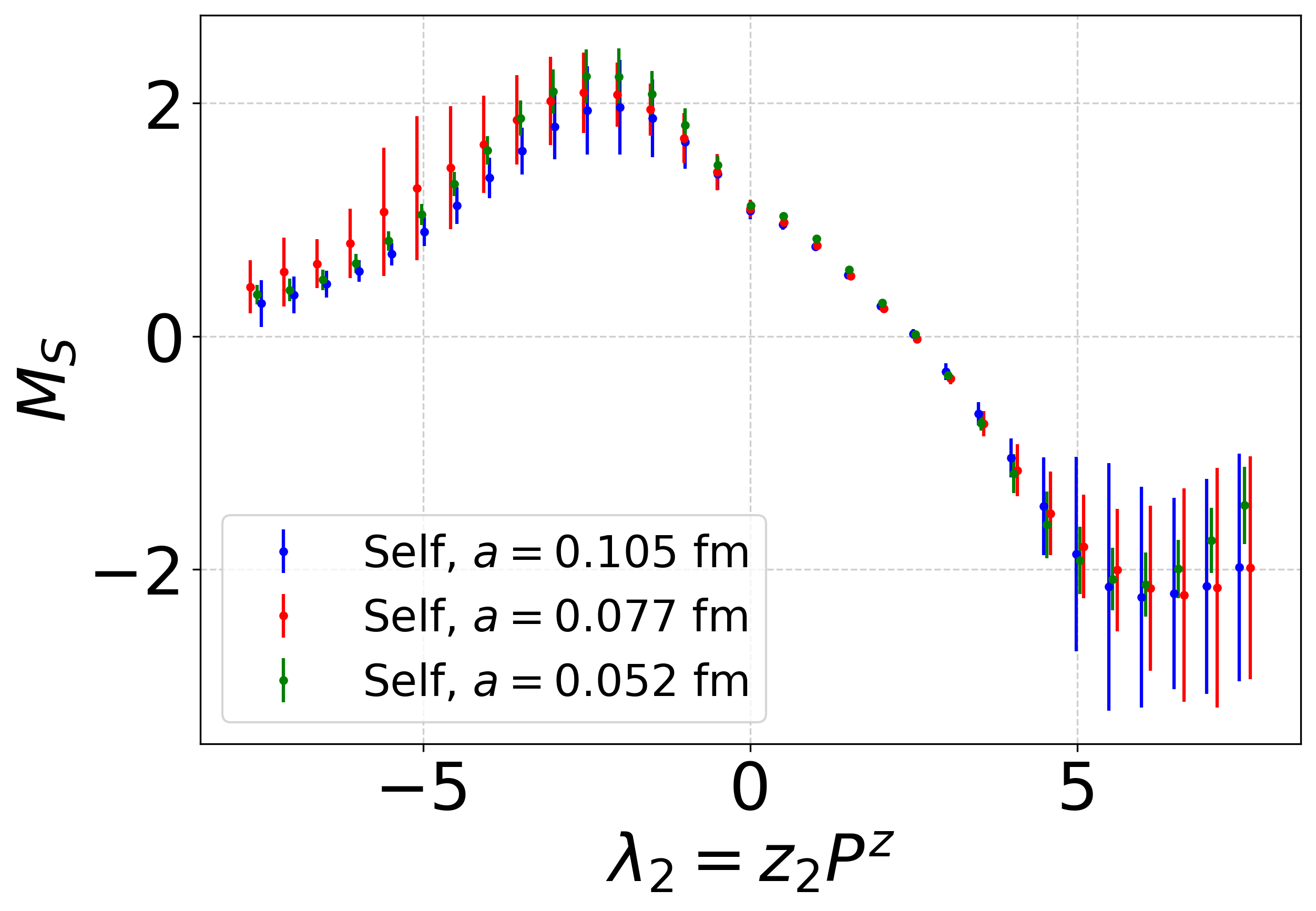}
    }
\vspace{0.0cm} 
\subfigure[\ Hybrid scheme result of proton at $P=2.0$ GeV]{
    \centering
    \includegraphics[scale=0.185]{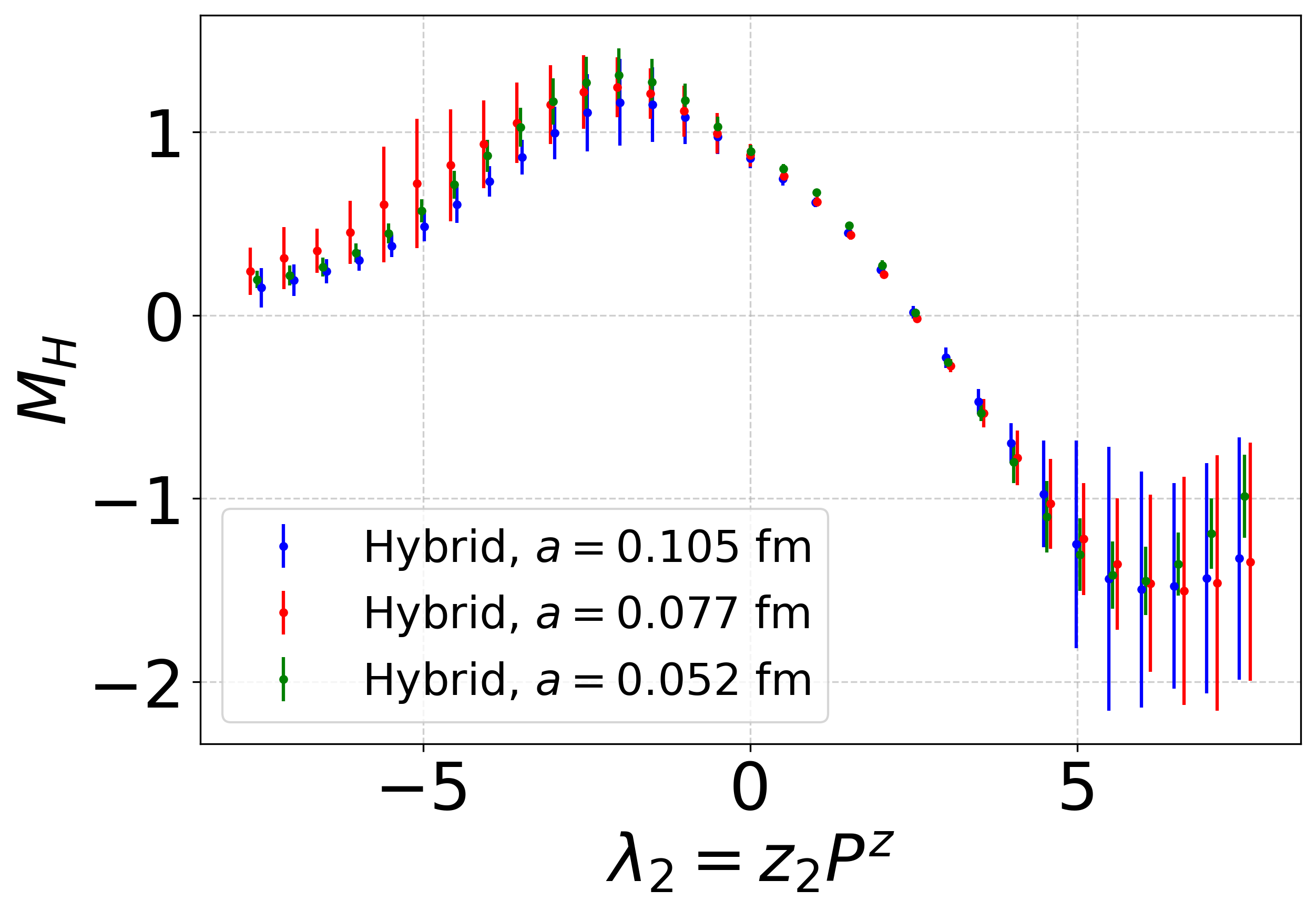}
    }
\caption{Results of the proton quasi-DA matrix elements in different schemes and with $P^z=2.0$ GeV, $z_1=0.200$ fm}
\label{fig:proton_p4_z4}
\end{figure}

\begin{figure}[htbp]
\centering
\subfigure[\ Bare result of proton at $P=2.0$ GeV]{
    \centering
    \includegraphics[scale=0.185]{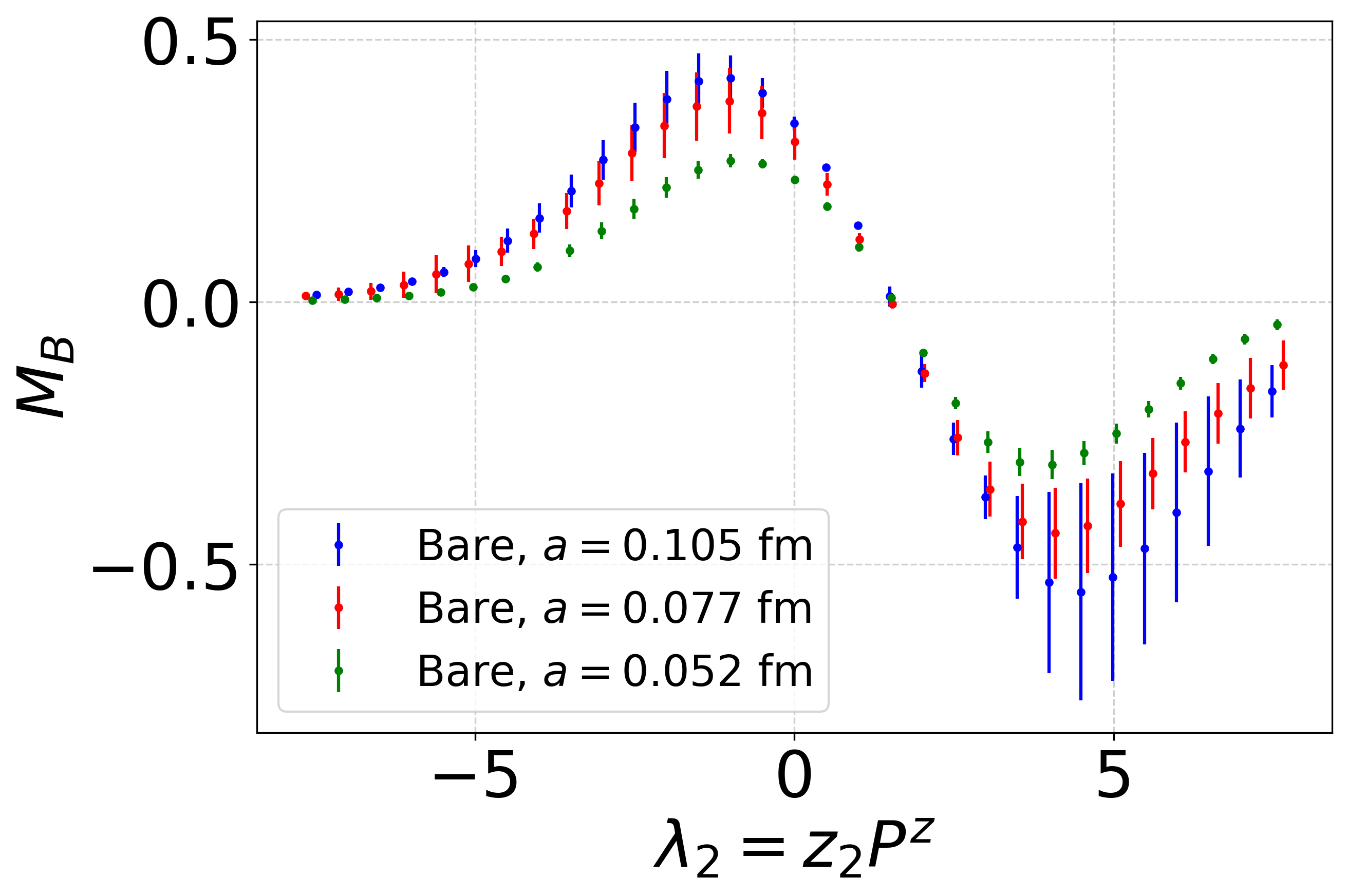}
    }
\vspace{0.0cm} 
\subfigure[\ Ratio scheme result of proton at $P=2.0$ GeV]{
    \centering
    \includegraphics[scale=0.185]{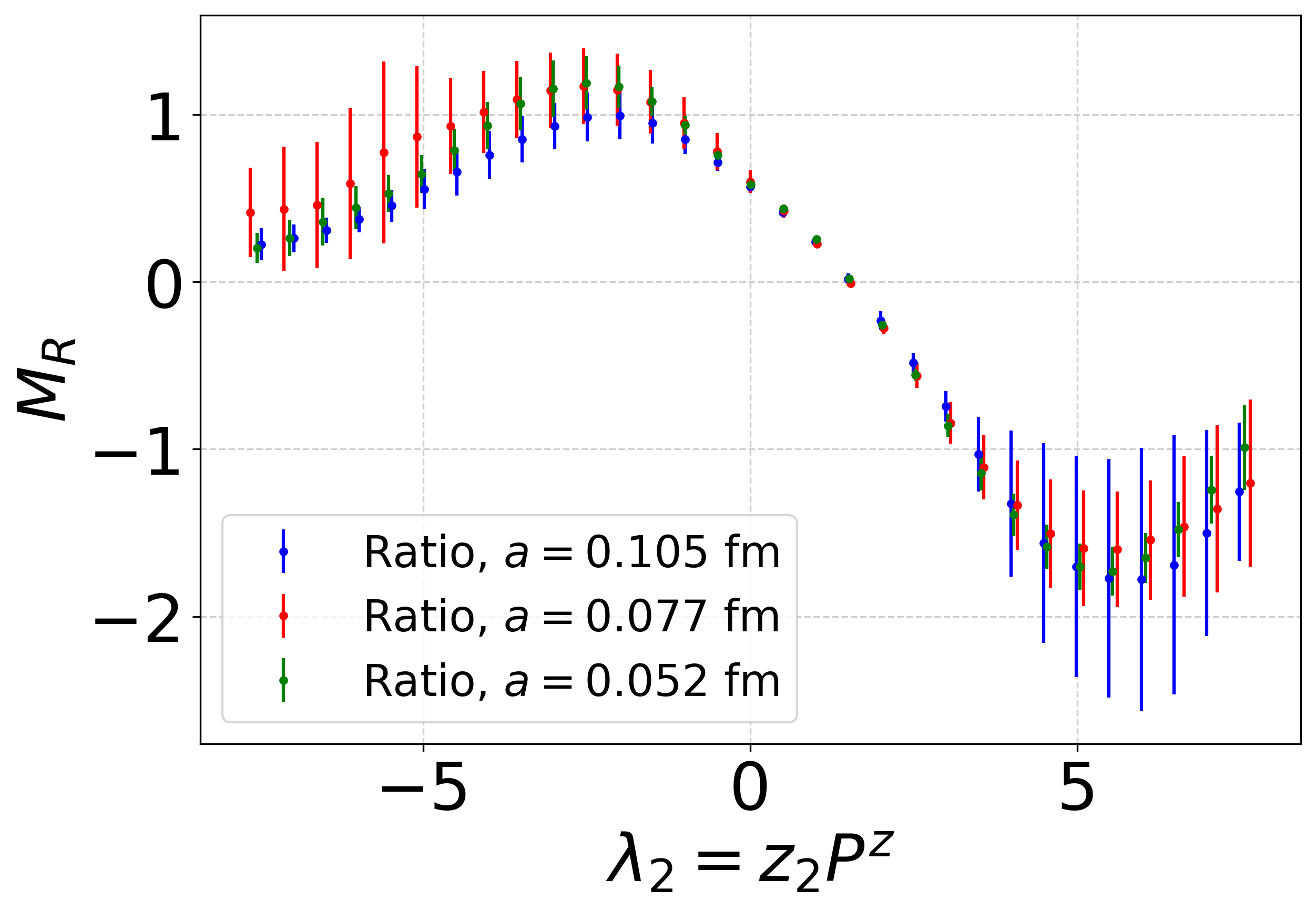}
    }
\vspace{0.0cm} 
\subfigure[\ Self scheme result of proton at $P=2.0$ GeV]{
    \centering
    \includegraphics[scale=0.185]{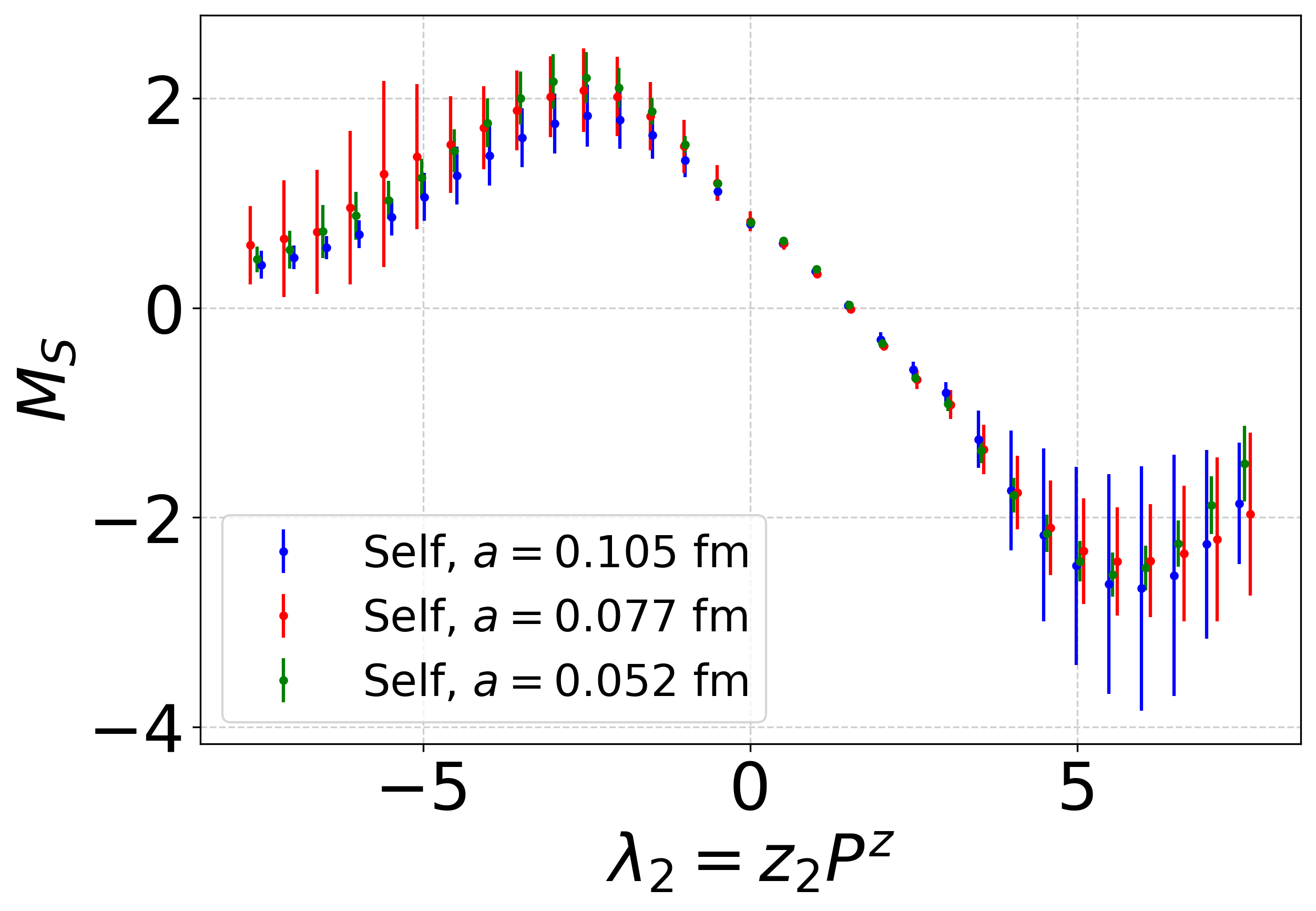}
    }
\vspace{0.0cm} 
\subfigure[\ Hybrid scheme result of proton at $P=2.0$ GeV]{
    \centering
    \includegraphics[scale=0.185]{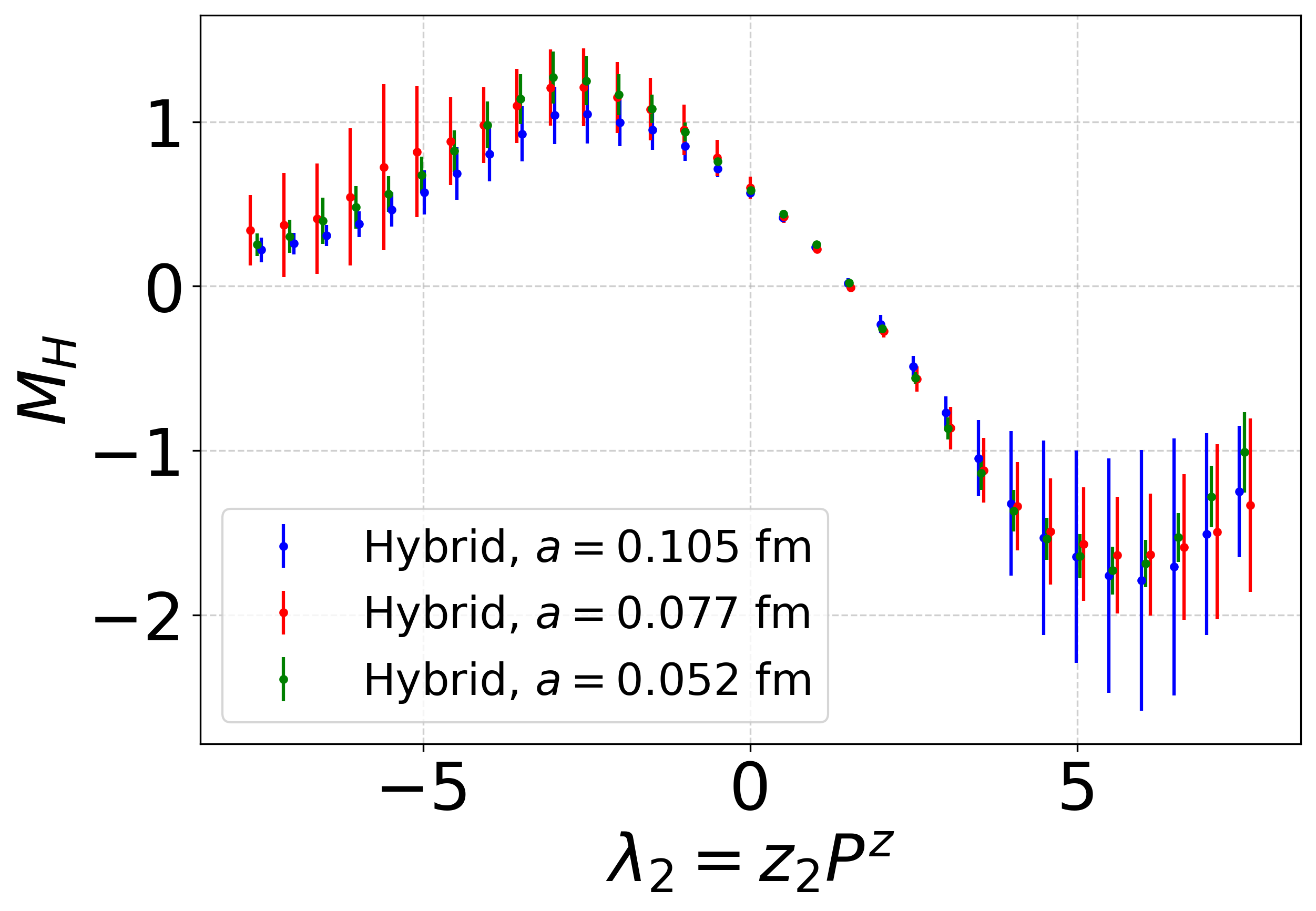}
    }
\caption{Results of the proton quasi-DA matrix elements in different schemes and with $P^z=2.0$ GeV, $z_1=0.300$ fm}
\label{fig:proton_p4_z6}
\end{figure}

\begin{figure}[htbp]
\centering
\subfigure[\ Bare result of proton at $P=2.0$ GeV]{
    \centering
    \includegraphics[scale=0.185]{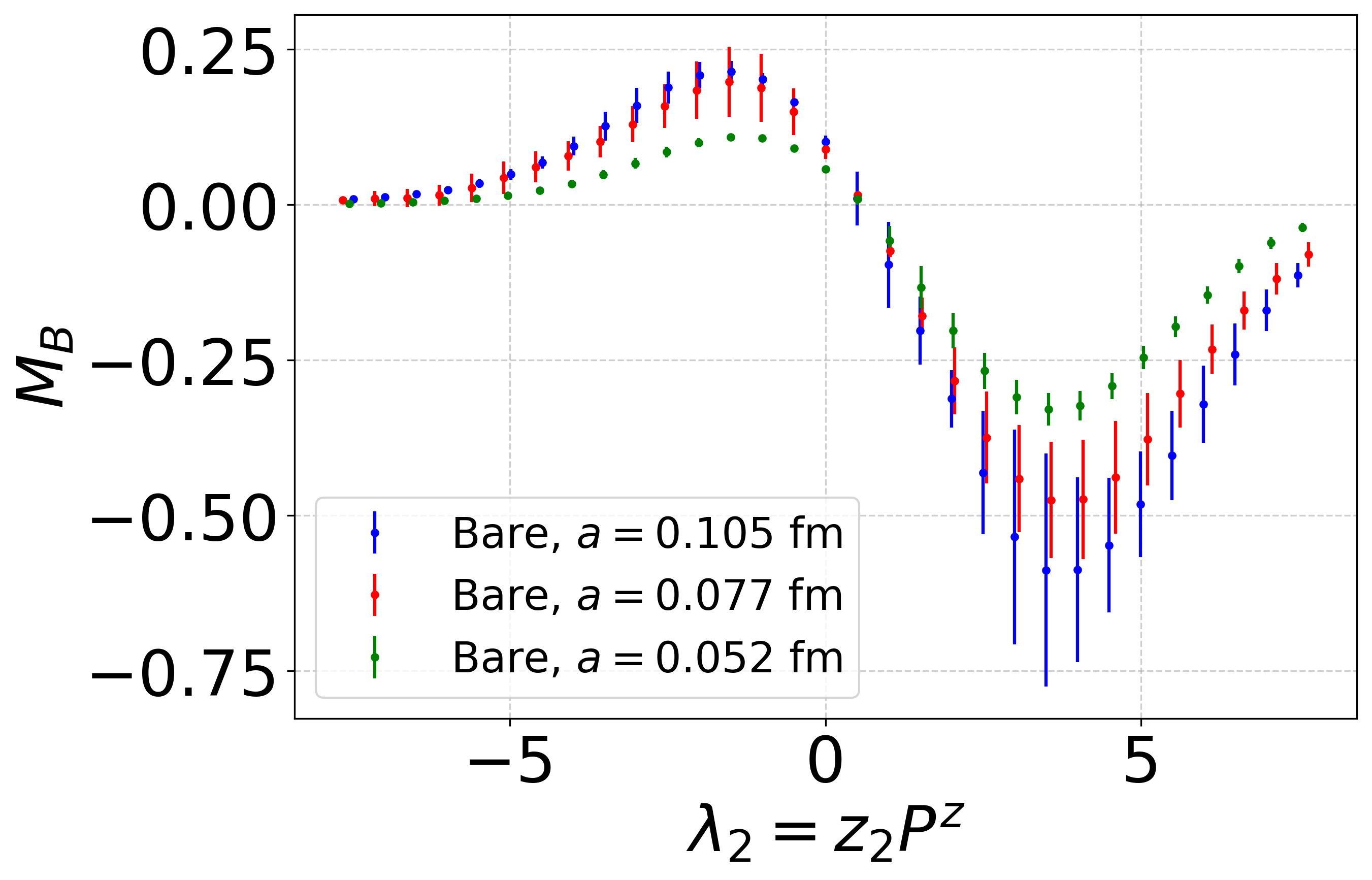}
    }
\vspace{0.0cm} 
\subfigure[\ Ratio scheme result of proton at $P=2.0$ GeV]{
    \centering
    \includegraphics[scale=0.185]{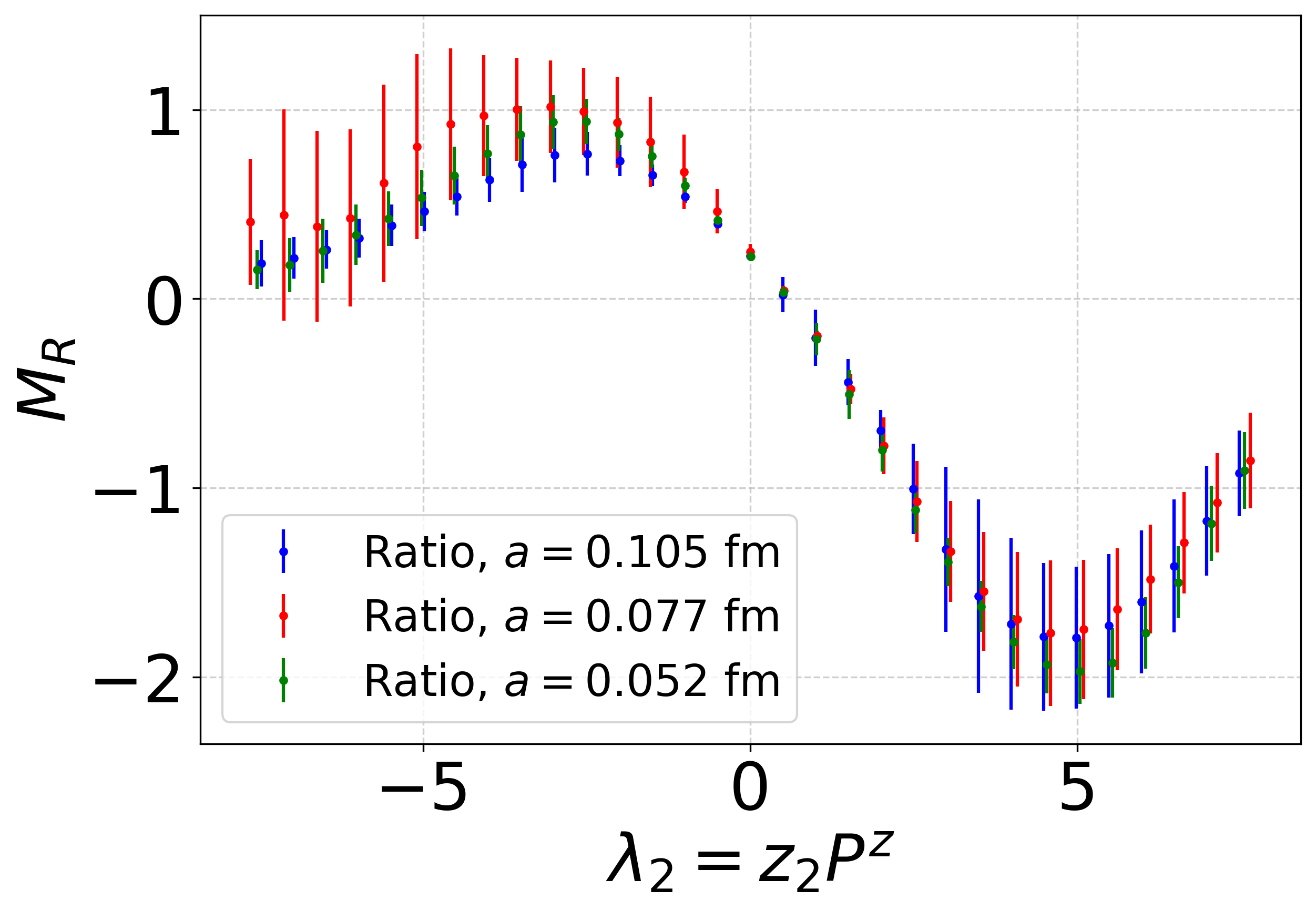}
    }
\vspace{0.0cm} 
\subfigure[\ Self scheme result of proton at $P=2.0$ GeV]{
    \centering
    \includegraphics[scale=0.185]{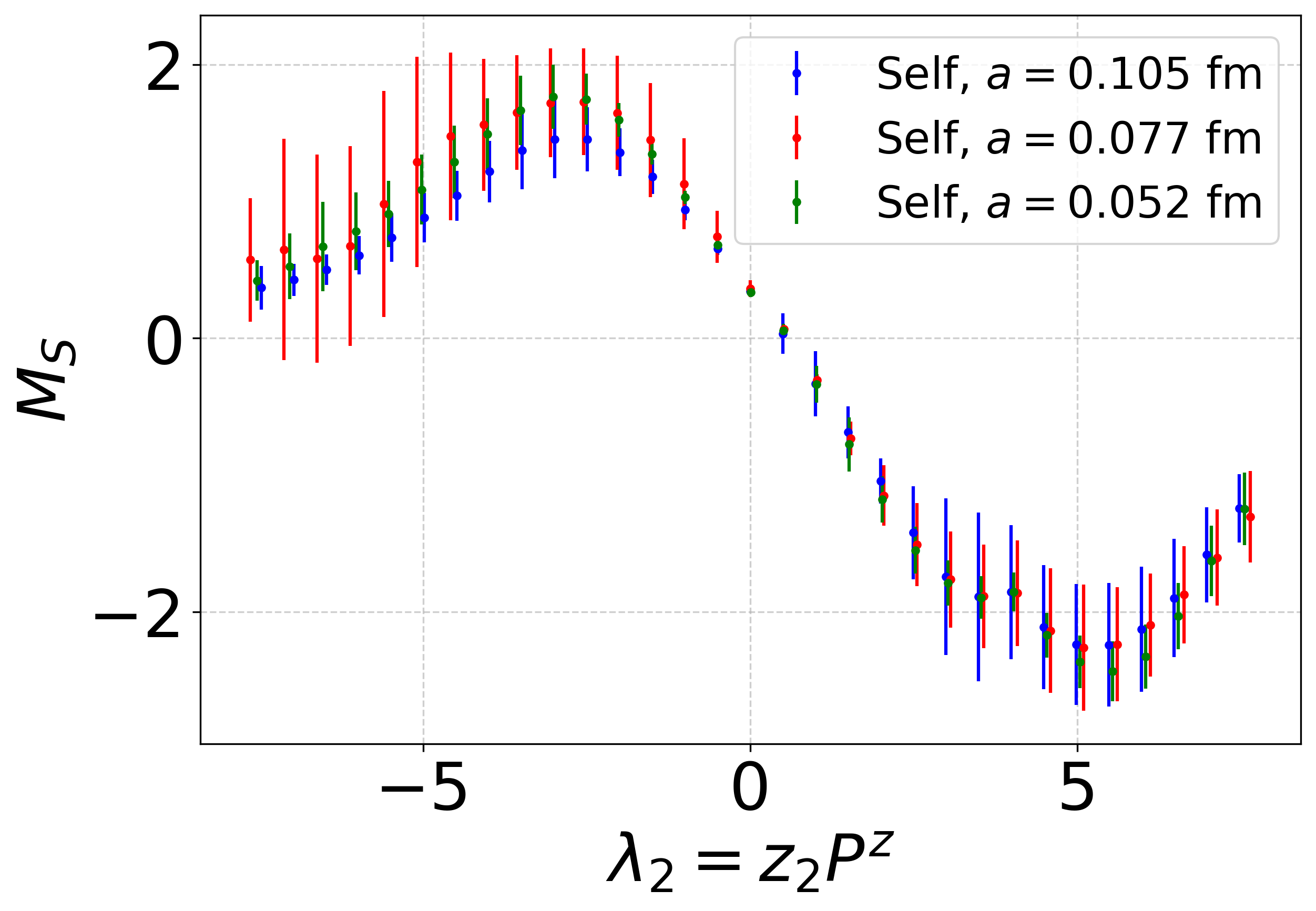}
    }
\vspace{0.0cm} 
\subfigure[\ Hybrid scheme result of proton at $P=2.0$ GeV]{
    \centering
    \includegraphics[scale=0.185]{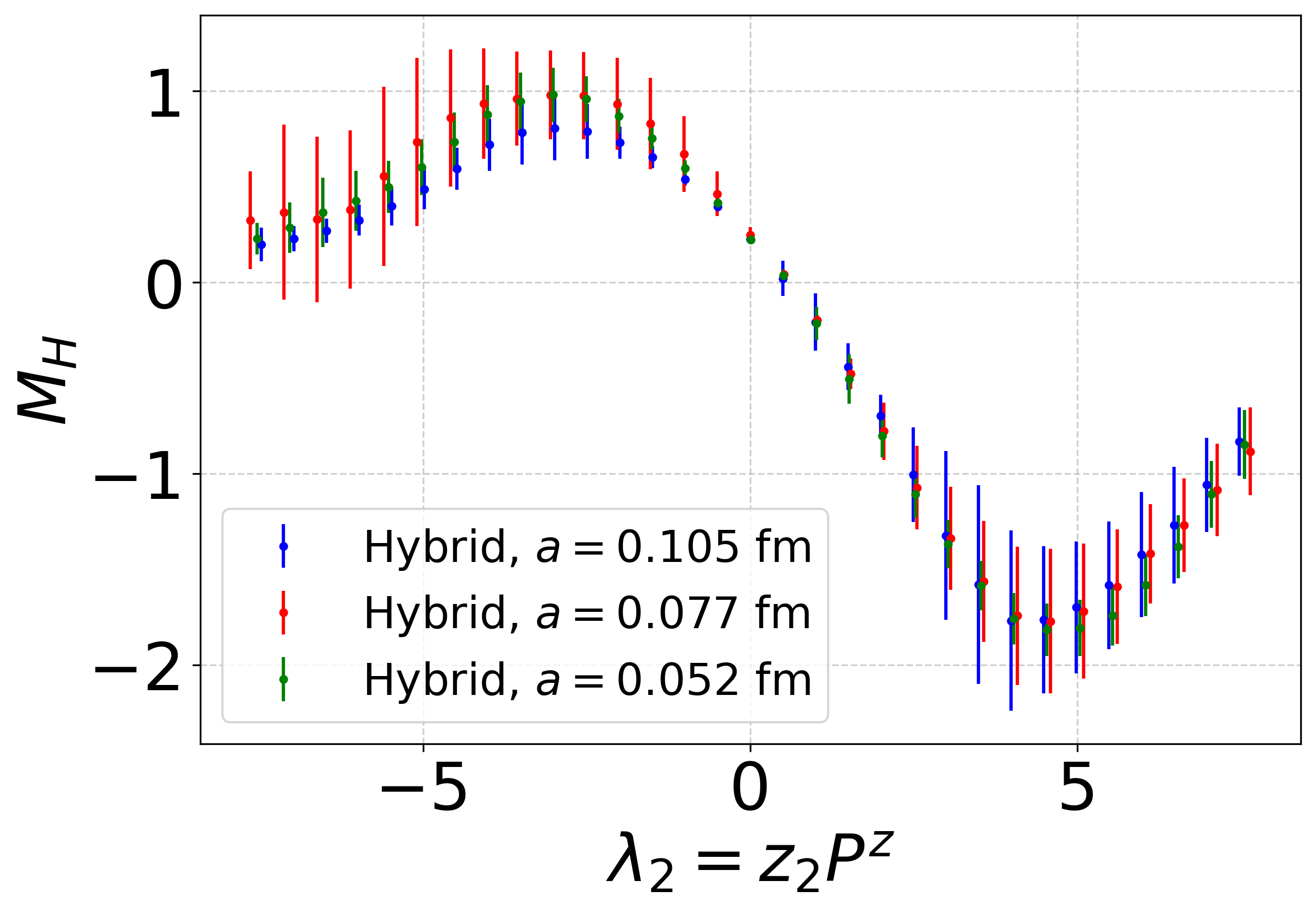}
    }
\caption{Results of the proton quasi-DA matrix elements in different schemes and with $P^z=2.0$ GeV, $z_1=0.400$ fm}
\label{fig:proton_p4_z8}
\end{figure}

\begin{figure}[htbp]
\centering
\subfigure[\ Bare result of proton at $P=2.0$ GeV]{
    \centering
    \includegraphics[scale=0.185]{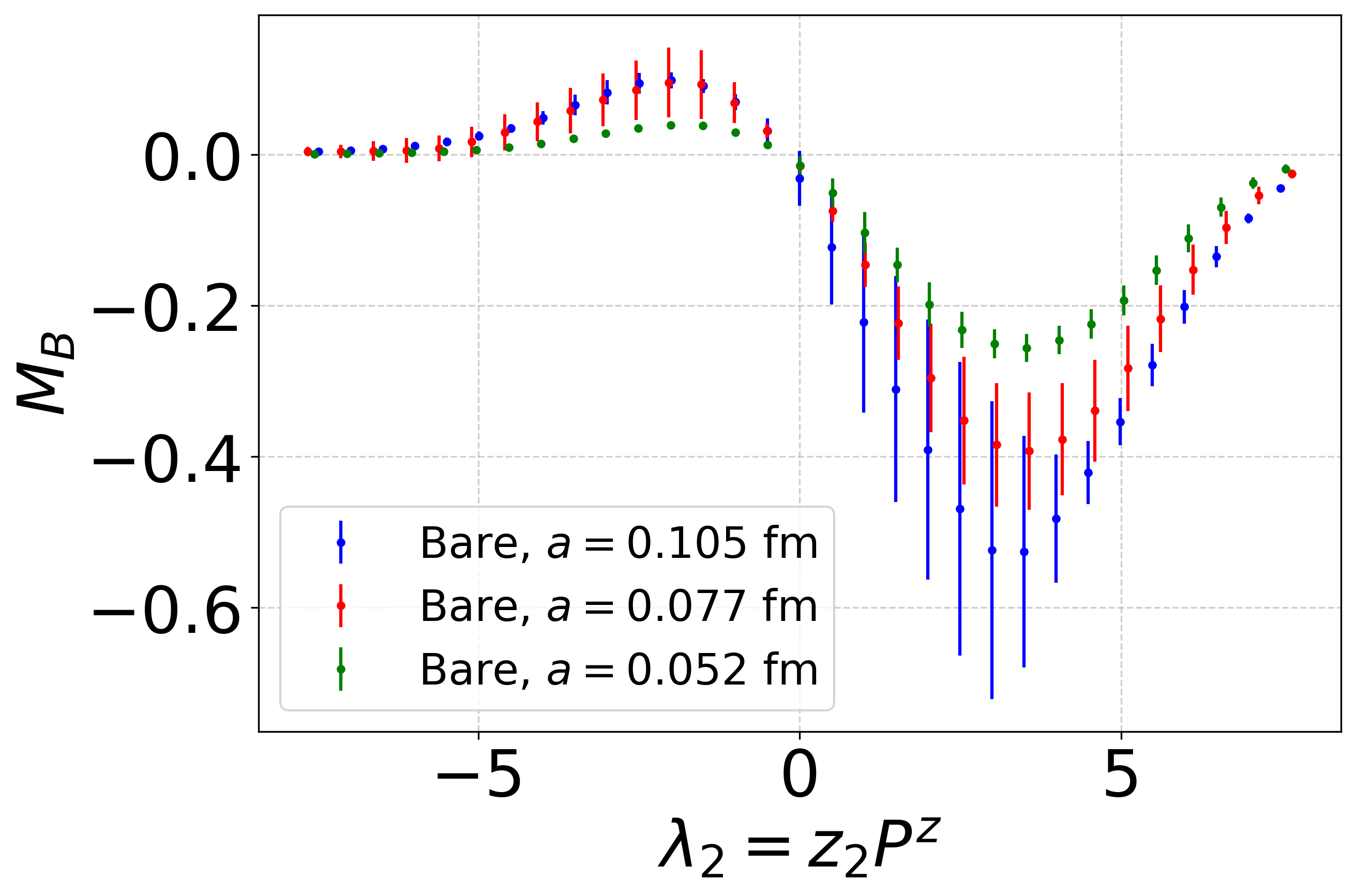}
    }
\vspace{0.0cm} 
\subfigure[\ Ratio scheme result of proton at $P=2.0$ GeV]{
    \centering
    \includegraphics[scale=0.185]{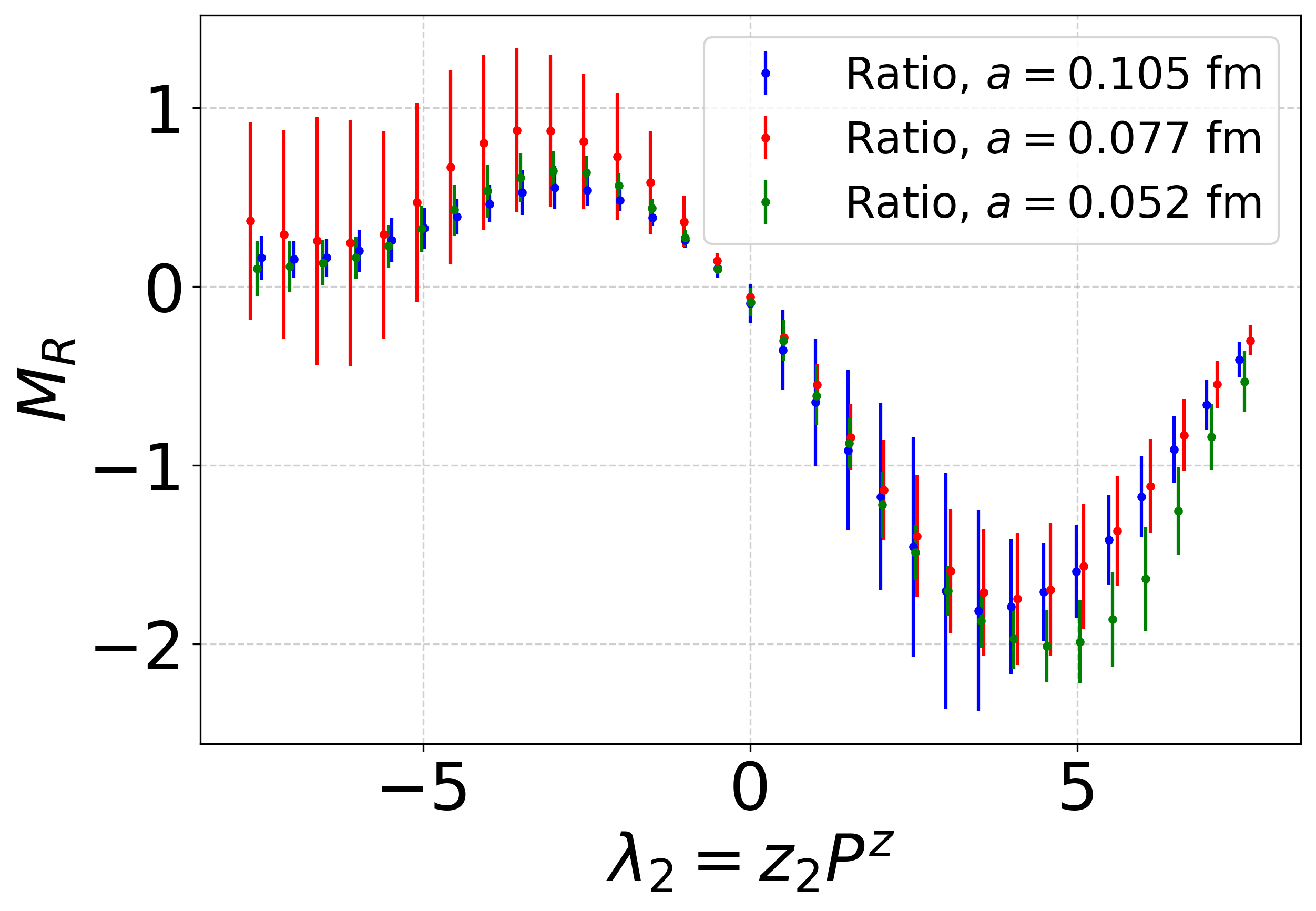}
    }
\vspace{0.0cm} 
\subfigure[\ Self scheme result of proton at $P=2.0$ GeV]{
    \centering
    \includegraphics[scale=0.185]{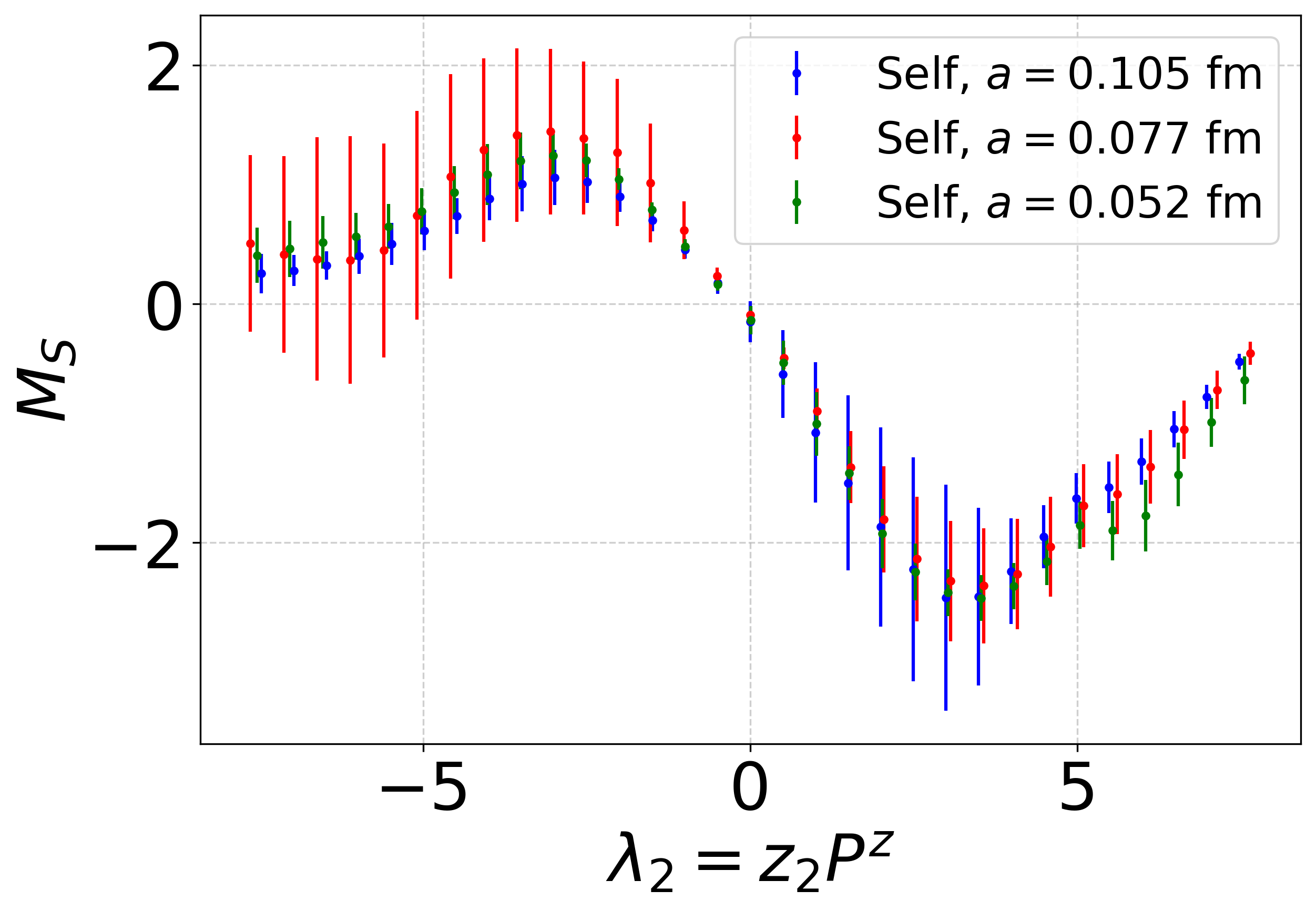}
    }
\vspace{0.0cm} 
\subfigure[\ Hybrid scheme result of proton at $P=2.0$ GeV]{
    \centering
    \includegraphics[scale=0.185]{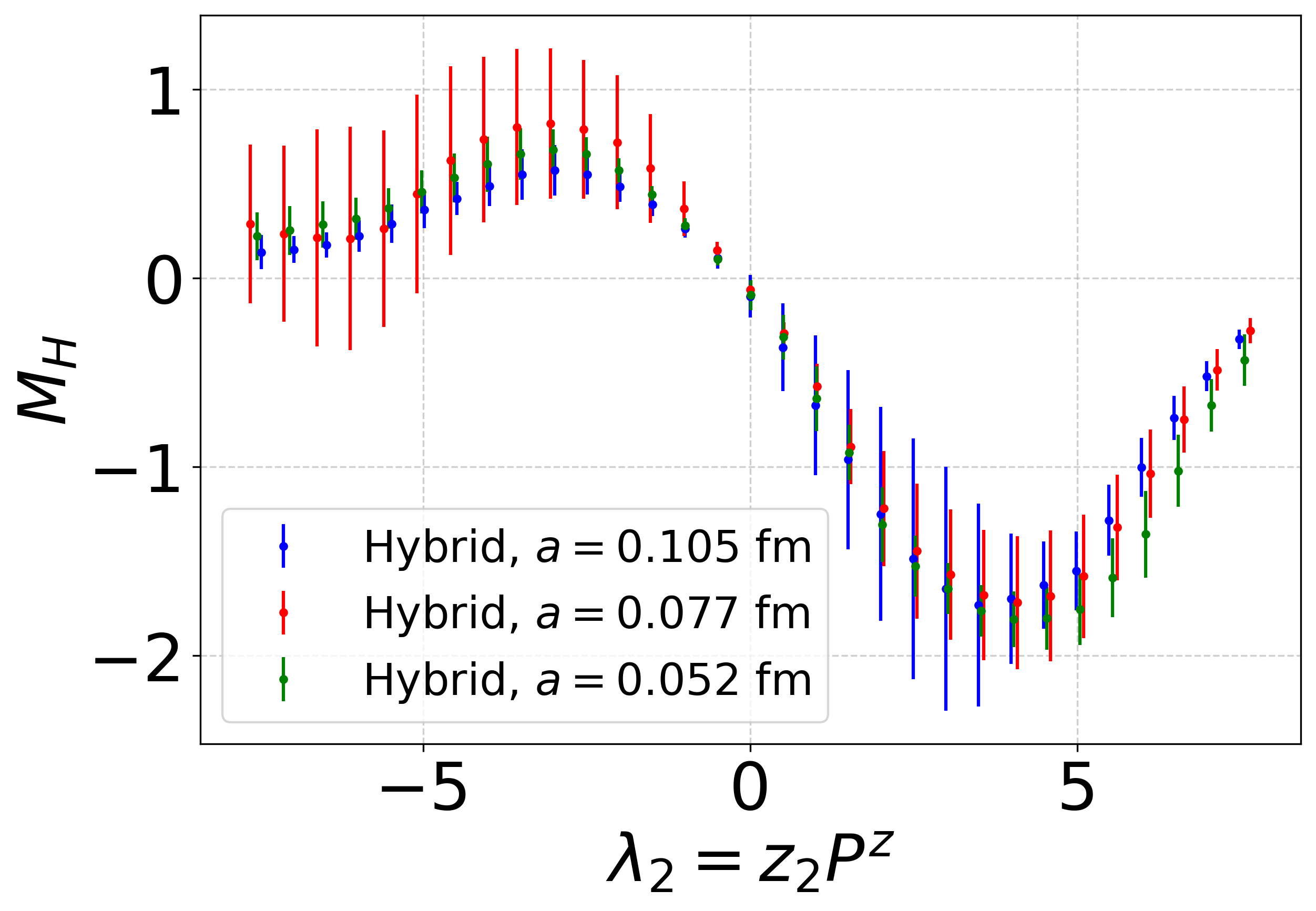}
    }
\caption{Results of the proton quasi-DA matrix elements in different schemes and with $P^z=2.0$ GeV, $z_1=0.500$ fm}
\label{fig:proton_p4_z10}
\end{figure}

\clearpage

\end{widetext}

\bibliography{Refs}

@article{LHCb:2025ray,
    author = "Aaij, Roel and others",
    collaboration = "LHCb",
    title = "{Observation of charge{\textendash}parity symmetry breaking in baryon decays}",
    eprint = "2503.16954",
    archivePrefix = "arXiv",
    primaryClass = "hep-ex",
    reportNumber = "LHCb-PAPER-2024-054, CERN-EP-2025-031",
    doi = "10.1038/s41586-025-09119-3",
    journal = "Nature",
    volume = "643",
    number = "8074",
    pages = "1223--1228",
    year = "2025"
}

@article{LHCb:2024yzj,
    author = "Aaij, R. and others",
    collaboration = "LHCb",
    title = "{Study of {\ensuremath{\Lambda}}b0 and {\ensuremath{\Xi}}b0 Decays to {\ensuremath{\Lambda}}h+h'- and Evidence for CP Violation in {\ensuremath{\Lambda}}b0{\textrightarrow}{\ensuremath{\Lambda}}K+K- Decays}",
    eprint = "2411.15441",
    archivePrefix = "arXiv",
    primaryClass = "hep-ex",
    reportNumber = "LHCb-PAPER-2024-043, CERN-EP-2024-281",
    doi = "10.1103/PhysRevLett.134.101802",
    journal = "Phys. Rev. Lett.",
    volume = "134",
    number = "10",
    pages = "101802",
    year = "2025"
}

@article{Wang:2015ndk,
    author = "Wang, Yu-Ming and Shen, Yue-Long",
    title = "{Perturbative Corrections to $\Lambda_b \to \Lambda$ Form Factors from QCD Light-Cone Sum Rules}",
    eprint = "1511.09036",
    archivePrefix = "arXiv",
    primaryClass = "hep-ph",
    reportNumber = "UWTHPH-2015-30",
    doi = "10.1007/JHEP02(2016)179",
    journal = "JHEP",
    volume = "02",
    pages = "179",
    year = "2016"
}

@article{Han:2022srw,
    author = "Han, Jia-Jie and Li, Ya and Li, Hsiang-nan and Shen, Yue-Long and Xiao, Zhen-Jun and Yu, Fu-Sheng",
    title = "{$\Lambda _b\rightarrow p$ transition form factors in perturbative QCD}",
    eprint = "2202.04804",
    archivePrefix = "arXiv",
    primaryClass = "hep-ph",
    doi = "10.1140/epjc/s10052-022-10642-0",
    journal = "Eur. Phys. J. C",
    volume = "82",
    number = "8",
    pages = "686",
    year = "2022"
}

@article{Huang:2022lfr,
    author = "Huang, Ke-Sheng and Liu, Wei and Shen, Yue-Long and Yu, Fu-Sheng",
    title = "{$\Lambda _b \rightarrow p, N^*(1535)$ form factors from QCD light-cone sum rules}",
    eprint = "2205.06095",
    archivePrefix = "arXiv",
    primaryClass = "hep-ph",
    doi = "10.1140/epjc/s10052-023-11349-6",
    journal = "Eur. Phys. J. C",
    volume = "83",
    number = "4",
    pages = "272",
    year = "2023"
}

@article{Han:2024kgz,
    author = "Han, Jia-Jie and Yu, Ji-Xin and Li, Ya and Li, Hsiang-nan and Wang, Jian-Peng and Xiao, Zhen-Jun and Yu, Fu-Sheng",
    title = "{Establishing CP Violation in b-Baryon Decays}",
    eprint = "2409.02821",
    archivePrefix = "arXiv",
    primaryClass = "hep-ph",
    doi = "10.1103/ynnx-f63h",
    journal = "Phys. Rev. Lett.",
    volume = "134",
    number = "22",
    pages = "221801",
    year = "2025"
}

@article{Lu:2025gjt,
    author = {Lu, Long-Shun and L{\"u}, Cai-Dian and Shen, Yue-Long and Wei, Yan-Bing},
    title = "{SCET sum rules for {\ensuremath{\Lambda}}$_{b}${\textrightarrow} {\ensuremath{\Lambda}}{\ensuremath{\ell}}$^{+}${\ensuremath{\ell}}$^{−}$, {\ensuremath{\Lambda}}{\ensuremath{\gamma}} decays}",
    eprint = "2506.21419",
    archivePrefix = "arXiv",
    primaryClass = "hep-ph",
    doi = "10.1007/JHEP09(2025)172",
    journal = "JHEP",
    volume = "09",
    pages = "172",
    year = "2025"
}

@article{Han:2025tvc,
    author = "Han, Jia-Jie and Yu, Ji-Xin and Li, Ya and Li, Hsiang-nan and Wang, Jian-Peng and Xiao, Zhen-Jun and Yu, Fu-Sheng",
    title = "{CP violation in two-body hadronic {\ensuremath{\Lambda}}b decays in the PQCD approach}",
    eprint = "2506.07197",
    archivePrefix = "arXiv",
    primaryClass = "hep-ph",
    doi = "10.1103/lvsn-v3xj",
    journal = "Phys. Rev. D",
    volume = "112",
    number = "5",
    pages = "053007",
    year = "2025"
}

@article{Wang:2011uv,
    author = "Wang, Wei",
    title = "{Factorization of Heavy-to-Light Baryonic Transitions in SCET}",
    eprint = "1112.0237",
    archivePrefix = "arXiv",
    primaryClass = "hep-ph",
    reportNumber = "DESY-11-228",
    doi = "10.1016/j.physletb.2012.01.036",
    journal = "Phys. Lett. B",
    volume = "708",
    pages = "119--126",
    year = "2012"
}

@article{Chen:2024fhj,
    author = "Chen, Long-Bin and Chen, Wen and Feng, Feng and Hu, Siwei and Jia, Yu",
    title = "{Next-to-Leading-Order QCD Corrections to Nucleon Dirac Form Factors}",
    eprint = "2406.19994",
    archivePrefix = "arXiv",
    primaryClass = "hep-ph",
    doi = "10.1103/r25y-rsz5",
    journal = "Phys. Rev. Lett.",
    volume = "135",
    number = "13",
    pages = "131903",
    year = "2025"
}

@article{Bali:2017ude,
    author = {Bali, G. S. and Braun, V. M. and G{\"o}ckeler, M. and Gruber, M. and Hutzler, F. and Korcyl, P. and Lang, B. and Sch{\"a}fer, A.},
    collaboration = "RQCD",
    title = "{Second moment of the pion distribution amplitude with the momentum smearing technique}",
    eprint = "1705.10236",
    archivePrefix = "arXiv",
    primaryClass = "hep-lat",
    doi = "10.1016/j.physletb.2017.08.077",
    journal = "Phys. Lett. B",
    volume = "774",
    pages = "91--97",
    year = "2017"
}

@article{RQCD:2019osh,
    author = {Bali, Gunnar S. and Braun, Vladimir M. and B{\"u}rger, Simon and G{\"o}ckeler, Meinulf and Gruber, Michael and Hutzler, Fabian and Korcyl, Piotr and Sch{\"a}fer, Andreas and Sternbeck, Andr{\'e} and Wein, Philipp},
    collaboration = "RQCD",
    title = "{Light-cone distribution amplitudes of pseudoscalar mesons from lattice QCD}",
    eprint = "1903.08038",
    archivePrefix = "arXiv",
    primaryClass = "hep-lat",
    doi = "10.1007/JHEP08(2019)065",
    journal = "JHEP",
    volume = "08",
    pages = "065",
    year = "2019",
    note = "[Addendum: JHEP 11, 037 (2020)]"
}

@article{Zhang:2017bzy,
    author = "Zhang, Jian-Hui and Chen, Jiunn-Wei and Ji, Xiangdong and Jin, Luchang and Lin, Huey-Wen",
    title = "{Pion Distribution Amplitude from Lattice QCD}",
    eprint = "1702.00008",
    archivePrefix = "arXiv",
    primaryClass = "hep-lat",
    doi = "10.1103/PhysRevD.95.094514",
    journal = "Phys. Rev. D",
    volume = "95",
    number = "9",
    pages = "094514",
    year = "2017"
}

@article{Chen:2017gck,
    author = {Zhang, Jian-Hui and Jin, Luchang and Lin, Huey-Wen and Sch{\"a}fer, Andreas and Sun, Peng and Yang, Yi-Bo and Zhang, Rui and Zhao, Yong and Chen, Jiunn-Wei},
    collaboration = "LP3",
    title = "{Kaon Distribution Amplitude from Lattice QCD and the Flavor SU(3) Symmetry}",
    eprint = "1712.10025",
    archivePrefix = "arXiv",
    primaryClass = "hep-ph",
    reportNumber = "MSUHEP-17-023",
    doi = "10.1016/j.nuclphysb.2018.12.020",
    journal = "Nucl. Phys. B",
    volume = "939",
    pages = "429--446",
    year = "2019"
}

@article{Zhang:2020gaj,
    author = "Zhang, Rui and Honkala, Carson and Lin, Huey-Wen and Chen, Jiunn-Wei",
    title = "{Pion and kaon distribution amplitudes in the continuum limit}",
    eprint = "2005.13955",
    archivePrefix = "arXiv",
    primaryClass = "hep-lat",
    reportNumber = "MSUHEP-20-010",
    doi = "10.1103/PhysRevD.102.094519",
    journal = "Phys. Rev. D",
    volume = "102",
    number = "9",
    pages = "094519",
    year = "2020"
}

@article{Gao:2022vyh,
    author = "Gao, Xiang and Hanlon, Andrew D. and Karthik, Nikhil and Mukherjee, Swagato and Petreczky, Peter and Scior, Philipp and Syritsyn, Sergey and Zhao, Yong",
    title = "{Pion distribution amplitude at the physical point using the leading-twist expansion of the quasi-distribution-amplitude matrix element}",
    eprint = "2206.04084",
    archivePrefix = "arXiv",
    primaryClass = "hep-lat",
    reportNumber = "JLAB-THY-22-3626",
    doi = "10.1103/PhysRevD.106.074505",
    journal = "Phys. Rev. D",
    volume = "106",
    number = "7",
    pages = "074505",
    year = "2022"
}

@article{Holligan:2023rex,
    author = "Holligan, Jack and Ji, Xiangdong and Lin, Huey-Wen and Su, Yushan and Zhang, Rui",
    title = "{Precision control in lattice calculation of x-dependent pion distribution amplitude}",
    eprint = "2301.10372",
    archivePrefix = "arXiv",
    primaryClass = "hep-lat",
    doi = "10.1016/j.nuclphysb.2023.116282",
    journal = "Nucl. Phys. B",
    volume = "993",
    pages = "116282",
    year = "2023"
}

@article{Hua:2020gnw,
    author = "Hua, Jun and Chu, Min-Huan and Sun, Peng and Wang, Wei and Xu, Ji and Yang, Yi-Bo and Zhang, Jian-Hui and Zhang, Qi-An",
    collaboration = "Lattice Parton",
    title = "{Distribution Amplitudes of K* and {\ensuremath{\phi}} at the Physical Pion Mass from Lattice QCD}",
    eprint = "2011.09788",
    archivePrefix = "arXiv",
    primaryClass = "hep-lat",
    doi = "10.1103/PhysRevLett.127.062002",
    journal = "Phys. Rev. Lett.",
    volume = "127",
    number = "6",
    pages = "062002",
    year = "2021"
}

@article{LatticeParton:2022zqc,
    author = "Hua, Jun and others",
    collaboration = "Lattice Parton",
    title = "{Pion and Kaon Distribution Amplitudes from Lattice QCD}",
    eprint = "2201.09173",
    archivePrefix = "arXiv",
    primaryClass = "hep-lat",
    doi = "10.1103/PhysRevLett.129.132001",
    journal = "Phys. Rev. Lett.",
    volume = "129",
    number = "13",
    pages = "132001",
    year = "2022"
}

@article{Baker:2024zcd,
    author = {Baker, Ethan and Bollweg, Dennis and Boyle, Peter and Clo{\"e}t, Ian and Gao, Xiang and Mukherjee, Swagato and Petreczky, Peter and Zhang, Rui and Zhao, Yong},
    title = "{Lattice QCD calculation of the pion distribution amplitude with domain wall fermions at physical pion mass}",
    eprint = "2405.20120",
    archivePrefix = "arXiv",
    primaryClass = "hep-lat",
    doi = "10.1007/JHEP07(2024)211",
    journal = "JHEP",
    volume = "07",
    pages = "211",
    year = "2024"
}

@article{Cloet:2024vbv,
    author = "Cloet, Ian and Gao, Xiang and Mukherjee, Swagato and Syritsyn, Sergey and Karthik, Nikhil and Petreczky, Peter and Zhang, Rui and Zhao, Yong",
    title = "{Lattice QCD calculation of x-dependent meson distribution amplitudes at physical pion mass with threshold logarithm resummation}",
    eprint = "2407.00206",
    archivePrefix = "arXiv",
    primaryClass = "hep-lat",
    doi = "10.1103/PhysRevD.110.114502",
    journal = "Phys. Rev. D",
    volume = "110",
    number = "11",
    pages = "114502",
    year = "2024"
}

@article{Chernyak:1981zz,
    author = "Chernyak, V. L. and Zhitnitsky, A. R.",
    title = "{Exclusive Decays of Heavy Mesons}",
    reportNumber = "IYF-81-75",
    doi = "10.1016/0550-3213(83)90251-1",
    journal = "Nucl. Phys. B",
    volume = "201",
    pages = "492",
    year = "1982",
    note = "[Erratum: Nucl.Phys.B 214, 547 (1983)]"
}

@article{CLEO:1997fho,
    author = "Gronberg, J. and others",
    collaboration = "CLEO",
    title = "{Measurements of the meson - photon transition form-factors of light pseudoscalar mesons at large momentum transfer}",
    eprint = "hep-ex/9707031",
    archivePrefix = "arXiv",
    reportNumber = "SLAC-PUB-9838, CLNS-97-1477, CLEO-97-7",
    doi = "10.1103/PhysRevD.57.33",
    journal = "Phys. Rev. D",
    volume = "57",
    pages = "33--54",
    year = "1998"
}

@article{BaBar:2009rrj,
    author = "Aubert, Bernard and others",
    collaboration = "BaBar",
    title = "{Measurement of the gamma gamma* ---{\ensuremath{>}} pi0 transition form factor}",
    eprint = "0905.4778",
    archivePrefix = "arXiv",
    primaryClass = "hep-ex",
    reportNumber = "SLAC-PUB-13641, BABAR-PUB-09-006",
    doi = "10.1103/PhysRevD.80.052002",
    journal = "Phys. Rev. D",
    volume = "80",
    pages = "052002",
    year = "2009"
}

@article{Belle:2012wwz,
    author = "Uehara, S. and others",
    collaboration = "Belle",
    title = "{Measurement of $\gamma \gamma^* \to \pi^0$ transition form factor at Belle}",
    eprint = "1205.3249",
    archivePrefix = "arXiv",
    primaryClass = "hep-ex",
    reportNumber = "BELLE-PREPRINT-2012-16, KEK-PREPRINT-2012-8",
    doi = "10.1103/PhysRevD.86.092007",
    journal = "Phys. Rev. D",
    volume = "86",
    pages = "092007",
    year = "2012"
}

@article{Chang:2013pq,
    author = "Chang, Lei and Cloet, I. C. and Cobos-Martinez, J. J. and Roberts, C. D. and Schmidt, S. M. and Tandy, P. C.",
    title = "{Imaging dynamical chiral symmetry breaking: pion wave function on the light front}",
    eprint = "1301.0324",
    archivePrefix = "arXiv",
    primaryClass = "nucl-th",
    doi = "10.1103/PhysRevLett.110.132001",
    journal = "Phys. Rev. Lett.",
    volume = "110",
    number = "13",
    pages = "132001",
    year = "2013"
}

@article{Cui:2020tdf,
    author = "Cui, Zhu-Fang and Ding, Minghui and Gao, Fei and Raya, Kh{\'e}pani and Binosi, Daniele and Chang, Lei and Roberts, Craig D and Rodr{\'\i}guez-Quintero, Jose and Schmidt, Sebastian M",
    title = "{Kaon and pion parton distributions}",
    doi = "10.1140/epjc/s10052-020-08578-4",
    journal = "Eur. Phys. J. C",
    volume = "80",
    number = "11",
    pages = "1064",
    year = "2020"
}

@article{Roberts:2021nhw,
    author = "Roberts, Craig D. and Richards, David G. and Horn, Tanja and Chang, Lei",
    title = "{Insights into the emergence of mass from studies of pion and kaon structure}",
    eprint = "2102.01765",
    archivePrefix = "arXiv",
    primaryClass = "hep-ph",
    reportNumber = "NJU-INP 034/21",
    doi = "10.1016/j.ppnp.2021.103883",
    journal = "Prog. Part. Nucl. Phys.",
    volume = "120",
    pages = "103883",
    year = "2021"
}

@article{Chernyak:1987nu,
    author = "Chernyak, V. L. and Ogloblin, A. A. and Zhitnitsky, I. R.",
    title = "{Wave Functions of Octet Baryons}",
    reportNumber = "IYF-87-136",
    doi = "10.1007/BF01557663",
    journal = "Yad. Fiz.",
    volume = "48",
    pages = "1410--1422",
    year = "1988"
}

@article{Bali:2015ykx,
    author = "Bali, Gunnar S. and others",
    title = "{Light-cone distribution amplitudes of the baryon octet}",
    eprint = "1512.02050",
    archivePrefix = "arXiv",
    primaryClass = "hep-lat",
    doi = "10.1007/JHEP02(2016)070",
    journal = "JHEP",
    volume = "02",
    pages = "070",
    year = "2016"
}

@article{RQCD:2019hps,
    author = "Bali, Gunnar S. and others",
    collaboration = "RQCD",
    title = "{Light-cone distribution amplitudes of octet baryons from lattice QCD}",
    eprint = "1903.12590",
    archivePrefix = "arXiv",
    primaryClass = "hep-lat",
    doi = "10.1140/epja/i2019-12803-6",
    journal = "Eur. Phys. J. A",
    volume = "55",
    number = "7",
    pages = "116",
    year = "2019"
}

@article{Bali:2024oxg,
    author = {Bali, G. S. and Braun, V. M. and B{\"u}rger, S. and G{\"o}ckeler, M. and Gruber, M. and Kaiser, F. and Kniehl, B. A. and Veretin, O. L. and Wein, P.},
    collaboration = "RQCD",
    title = "{Updated determination of light-cone distribution amplitudes of octet baryons in lattice QCD}",
    eprint = "2411.19091",
    archivePrefix = "arXiv",
    primaryClass = "hep-lat",
    doi = "10.1103/PhysRevD.111.094517",
    journal = "Phys. Rev. D",
    volume = "111",
    number = "9",
    pages = "094517",
    year = "2025"
}

@article{Ji:2013dva,
    author = "Ji, Xiangdong",
    title = "{Parton Physics on a Euclidean Lattice}",
    eprint = "1305.1539",
    archivePrefix = "arXiv",
    primaryClass = "hep-ph",
    doi = "10.1103/PhysRevLett.110.262002",
    journal = "Phys. Rev. Lett.",
    volume = "110",
    pages = "262002",
    year = "2013"
}

@article{Ji:2014gla,
    author = "Ji, Xiangdong",
    title = "{Parton Physics from Large-Momentum Effective Field Theory}",
    eprint = "1404.6680",
    archivePrefix = "arXiv",
    primaryClass = "hep-ph",
    doi = "10.1007/s11433-014-5492-3",
    journal = "Sci. China Phys. Mech. Astron.",
    volume = "57",
    pages = "1407--1412",
    year = "2014"
}

@article{Orginos:2017kos,
    author = "Orginos, Kostas and Radyushkin, Anatoly and Karpie, Joseph and Zafeiropoulos, Savvas",
    title = "{Lattice QCD exploration of parton pseudo-distribution functions}",
    eprint = "1706.05373",
    archivePrefix = "arXiv",
    primaryClass = "hep-ph",
    reportNumber = "JLAB-THY-17-2494",
    doi = "10.1103/PhysRevD.96.094503",
    journal = "Phys. Rev. D",
    volume = "96",
    number = "9",
    pages = "094503",
    year = "2017"
}

@article{Radyushkin:2017cyf,
    author = "Radyushkin, A. V.",
    title = "{Quasi-parton distribution functions, momentum distributions, and pseudo-parton distribution functions}",
    eprint = "1705.01488",
    archivePrefix = "arXiv",
    primaryClass = "hep-ph",
    reportNumber = "JLAB-THY-17-2455",
    doi = "10.1103/PhysRevD.96.034025",
    journal = "Phys. Rev. D",
    volume = "96",
    number = "3",
    pages = "034025",
    year = "2017"
}

@article{Ma:2014jla,
    author = "Ma, Yan-Qing and Qiu, Jian-Wei",
    title = "{Extracting Parton Distribution Functions from Lattice QCD Calculations}",
    eprint = "1404.6860",
    archivePrefix = "arXiv",
    primaryClass = "hep-ph",
    doi = "10.1103/PhysRevD.98.074021",
    journal = "Phys. Rev. D",
    volume = "98",
    number = "7",
    pages = "074021",
    year = "2018"
}

@article{Xiong:2013bka,
    author = "Xiong, Xiaonu and Ji, Xiangdong and Zhang, Jian-Hui and Zhao, Yong",
    title = "{One-loop matching for parton distributions: Nonsinglet case}",
    eprint = "1310.7471",
    archivePrefix = "arXiv",
    primaryClass = "hep-ph",
    doi = "10.1103/PhysRevD.90.014051",
    journal = "Phys. Rev. D",
    volume = "90",
    number = "1",
    pages = "014051",
    year = "2014"
}

@article{Alexandrou:2016eyt,
    author = "Alexandrou, Constantia and Cichy, Krzysztof and Constantinou, Martha and Hadjiyiannakou, Kyriakos and Jansen, Karl and Steffens, Fernanda and Wiese, Christian",
    title = "{Parton Distributions from Lattice QCD with Momentum Smearing}",
    eprint = "1612.08728",
    archivePrefix = "arXiv",
    primaryClass = "hep-lat",
    reportNumber = "DESY-16-225",
    doi = "10.22323/1.256.0151",
    journal = "PoS",
    volume = "LATTICE2016",
    pages = "151",
    year = "2016"
}

@article{Chen:2017mie,
    author = "Chen, Jiunn-Wei and Ishikawa, Tomomi and Jin, Luchang and Lin, Huey-Wen and Zhang, Jian-Hui and Zhao, Yong",
    collaboration = "LP3",
    title = "{Symmetry properties of nonlocal quark bilinear operators on a Lattice}",
    eprint = "1710.01089",
    archivePrefix = "arXiv",
    primaryClass = "hep-lat",
    reportNumber = "MSUHEP-17-015, MIT-CTP/4942, MIT-CTP-4942",
    doi = "10.1088/1674-1137/43/10/103101",
    journal = "Chin. Phys. C",
    volume = "43",
    number = "10",
    pages = "103101",
    year = "2019"
}

@article{Zhang:2017zfe,
    author = {Zhang, Jian-Hui and Jin, Luchang and Lin, Huey-Wen and Sch{\"a}fer, Andreas and Sun, Peng and Yang, Yi-Bo and Zhang, Rui and Zhao, Yong and Chen, Jiunn-Wei},
    collaboration = "LP3",
    title = "{Kaon Distribution Amplitude from Lattice QCD and the Flavor SU(3) Symmetry}",
    eprint = "1712.10025",
    archivePrefix = "arXiv",
    primaryClass = "hep-ph",
    reportNumber = "MSUHEP-17-023",
    doi = "10.1016/j.nuclphysb.2018.12.020",
    journal = "Nucl. Phys. B",
    volume = "939",
    pages = "429--446",
    year = "2019"
}

@article{Xu:2018mpf,
    author = "Xu, Ji and Zhang, Qi-An and Zhao, Shuai",
    title = "{Light-cone distribution amplitudes of vector meson in a large momentum effective theory}",
    eprint = "1804.01042",
    archivePrefix = "arXiv",
    primaryClass = "hep-ph",
    doi = "10.1103/PhysRevD.97.114026",
    journal = "Phys. Rev. D",
    volume = "97",
    number = "11",
    pages = "114026",
    year = "2018"
}

@article{Liu:2018hxv,
    author = "Liu, Yu-Sheng and Chen, Jiunn-Wei and Jin, Luchang and Li, Ruizi and Lin, Huey-Wen and Yang, Yi-Bo and Zhang, Jian-Hui and Zhao, Yong",
    title = "{Nucleon Transversity Distribution at the Physical Pion Mass from Lattice QCD}",
    eprint = "1810.05043",
    archivePrefix = "arXiv",
    primaryClass = "hep-lat",
    reportNumber = "MSUHEP-18-019, MIT-CTP/5033",
    month = "10",
    year = "2018"
}

@article{Wang:2019msf,
    author = "Wang, Wei and Wang, Yu-Ming and Xu, Ji and Zhao, Shuai",
    title = "{$B$-meson light-cone distribution amplitude from Euclidean quantities}",
    eprint = "1908.09933",
    archivePrefix = "arXiv",
    primaryClass = "hep-ph",
    reportNumber = "JLAB-THY-19-3025",
    doi = "10.1103/PhysRevD.102.011502",
    journal = "Phys. Rev. D",
    volume = "102",
    number = "1",
    pages = "011502",
    year = "2020"
}

@article{Zhang:2019qiq,
    author = "Zhang, Rui and Fan, Zhouyou and Li, Ruizi and Lin, Huey-Wen and Yoon, Boram",
    title = "{Machine-learning prediction for quasiparton distribution function matrix elements}",
    eprint = "1909.10990",
    archivePrefix = "arXiv",
    primaryClass = "hep-lat",
    reportNumber = "MSUHEP-19-021",
    doi = "10.1103/PhysRevD.101.034516",
    journal = "Phys. Rev. D",
    volume = "101",
    number = "3",
    pages = "034516",
    year = "2020"
}

@article{Chen:2020ody,
    author = "Chen, Long-Bin and Wang, Wei and Zhu, Ruilin",
    title = "{Next-to-Next-to-Leading Order Calculation of Quasiparton Distribution Functions}",
    eprint = "2006.14825",
    archivePrefix = "arXiv",
    primaryClass = "hep-ph",
    doi = "10.1103/PhysRevLett.126.072002",
    journal = "Phys. Rev. Lett.",
    volume = "126",
    number = "7",
    pages = "072002",
    year = "2021"
}

@article{Ji:2020brr,
    author = {Ji, Xiangdong and Liu, Yizhuang and Sch{\"a}fer, Andreas and Wang, Wei and Yang, Yi-Bo and Zhang, Jian-Hui and Zhao, Yong},
    title = "{A Hybrid Renormalization Scheme for Quasi Light-Front Correlations in Large-Momentum Effective Theory}",
    eprint = "2008.03886",
    archivePrefix = "arXiv",
    primaryClass = "hep-ph",
    doi = "10.1016/j.nuclphysb.2021.115311",
    journal = "Nucl. Phys. B",
    volume = "964",
    pages = "115311",
    year = "2021"
}

@article{Ji:2020ect,
    author = "Ji, Xiangdong and Liu, Yu-Sheng and Liu, Yizhuang and Zhang, Jian-Hui and Zhao, Yong",
    title = "{Large-momentum effective theory}",
    eprint = "2004.03543",
    archivePrefix = "arXiv",
    primaryClass = "hep-ph",
    doi = "10.1103/RevModPhys.93.035005",
    journal = "Rev. Mod. Phys.",
    volume = "93",
    number = "3",
    pages = "035005",
    year = "2021"
}

@article{LatticeParton:2020uhz,
    author = "Zhang, Qi-An and others",
    collaboration = "Lattice Parton",
    title = "{Lattice-QCD Calculations of TMD Soft Function Through Large-Momentum Effective Theory}",
    eprint = "2005.14572",
    archivePrefix = "arXiv",
    primaryClass = "hep-lat",
    doi = "10.22323/1.396.0477",
    journal = "Phys. Rev. Lett.",
    volume = "125",
    number = "19",
    pages = "192001",
    year = "2020"
}

@article{Lin:2020rxa,
    author = "Lin, Huey-Wen",
    title = "{Nucleon Tomography and Generalized Parton Distribution at Physical Pion Mass from Lattice QCD}",
    eprint = "2008.12474",
    archivePrefix = "arXiv",
    primaryClass = "hep-ph",
    reportNumber = "MSUHEP-20-014, MSUHEP-20-014",
    doi = "10.1103/PhysRevLett.127.182001",
    journal = "Phys. Rev. Lett.",
    volume = "127",
    number = "18",
    pages = "182001",
    year = "2021"
}

@article{Bhattacharya:2021moj,
    author = "Bhattacharya, Shohini and Cichy, Krzysztof and Constantinou, Martha and Metz, Andreas and Scapellato, Aurora and Steffens, Fernanda",
    title = "{Parton distribution functions beyond leading twist from lattice QCD: The hL(x) case}",
    eprint = "2107.02574",
    archivePrefix = "arXiv",
    primaryClass = "hep-lat",
    doi = "10.1103/PhysRevD.104.114510",
    journal = "Phys. Rev. D",
    volume = "104",
    number = "11",
    pages = "114510",
    year = "2021"
}

@article{Gao:2021hxl,
    author = "Gao, Xiang and Lee, Kyle and Mukherjee, Swagato and Shugert, Charles and Zhao, Yong",
    title = "{Origin and resummation of threshold logarithms in the lattice QCD calculations of PDFs}",
    eprint = "2102.01101",
    archivePrefix = "arXiv",
    primaryClass = "hep-ph",
    doi = "10.1103/PhysRevD.103.094504",
    journal = "Phys. Rev. D",
    volume = "103",
    number = "9",
    pages = "094504",
    year = "2021"
}

@article{LatticePartonLPC:2021gpi,
    author = "Huo, Yi-Kai and others",
    collaboration = "Lattice Parton (LPC)",
    title = "{Self-renormalization of quasi-light-front correlators on the lattice}",
    eprint = "2103.02965",
    archivePrefix = "arXiv",
    primaryClass = "hep-lat",
    doi = "10.1016/j.nuclphysb.2021.115443",
    journal = "Nucl. Phys. B",
    volume = "969",
    pages = "115443",
    year = "2021"
}

@article{Li:2021wvl,
    author = "Li, Yuan and others",
    title = "{Lattice QCD Study of Transverse-Momentum Dependent Soft Function}",
    eprint = "2106.13027",
    archivePrefix = "arXiv",
    primaryClass = "hep-lat",
    doi = "10.1103/PhysRevLett.128.062002",
    journal = "Phys. Rev. Lett.",
    volume = "128",
    number = "6",
    pages = "062002",
    year = "2022"
}

@article{Deng:2022gzi,
    author = "Deng, Zhi-Fu and Wang, Wei and Zeng, Jun",
    title = "{Transverse-momentum-dependent wave functions and soft functions at one-loop in large momentum effective theory}",
    eprint = "2207.07280",
    archivePrefix = "arXiv",
    primaryClass = "hep-th",
    doi = "10.1007/JHEP09(2022)046",
    journal = "JHEP",
    volume = "09",
    pages = "046",
    year = "2022"
}

@article{Gao:2022iex,
    author = "Gao, Xiang and Hanlon, Andrew D. and Karthik, Nikhil and Mukherjee, Swagato and Petreczky, Peter and Scior, Philipp and Shi, Shuzhe and Syritsyn, Sergey and Zhao, Yong and Zhou, Kai",
    title = "{Continuum-extrapolated NNLO valence PDF of the pion at the physical point}",
    eprint = "2208.02297",
    archivePrefix = "arXiv",
    primaryClass = "hep-lat",
    doi = "10.1103/PhysRevD.106.114510",
    journal = "Phys. Rev. D",
    volume = "106",
    number = "11",
    pages = "114510",
    year = "2022"
}

@article{Gao:2022uhg,
    author = "Gao, Xiang and Hanlon, Andrew D. and Holligan, Jack and Karthik, Nikhil and Mukherjee, Swagato and Petreczky, Peter and Syritsyn, Sergey and Zhao, Yong",
    title = "{Unpolarized proton PDF at NNLO from lattice QCD with physical quark masses}",
    eprint = "2212.12569",
    archivePrefix = "arXiv",
    primaryClass = "hep-lat",
    doi = "10.1103/PhysRevD.107.074509",
    journal = "Phys. Rev. D",
    volume = "107",
    number = "7",
    pages = "074509",
    year = "2023"
}

@article{LatticeParton:2022xsd,
    author = "Yao, Fei and others",
    collaboration = "Lattice Parton",
    title = "{Nucleon Transversity Distribution in the Continuum and Physical Mass Limit from Lattice QCD}",
    eprint = "2208.08008",
    archivePrefix = "arXiv",
    primaryClass = "hep-lat",
    doi = "10.1103/PhysRevLett.131.261901",
    journal = "Phys. Rev. Lett.",
    volume = "131",
    number = "26",
    pages = "261901",
    year = "2023"
}

@article{LatticePartonCollaborationLPC:2022myp,
    author = {He, Jin-Chen and Chu, Min-Huan and Hua, Jun and Ji, Xiangdong and Sch{\"a}fer, Andreas and Su, Yushan and Wang, Wei and Yang, Yi-Bo and Zhang, Jian-Hui and Zhang, Qi-An},
    collaboration = "Lattice Parton Collaboration (LPC)",
    title = "{Unpolarized transverse momentum dependent parton distributions of the nucleon from lattice QCD}",
    eprint = "2211.02340",
    archivePrefix = "arXiv",
    primaryClass = "hep-lat",
    doi = "10.1103/PhysRevD.109.114513",
    journal = "Phys. Rev. D",
    volume = "109",
    number = "11",
    pages = "114513",
    year = "2024"
}

@article{LatticePartonLPC:2022eev,
    author = "Chu, Min-Huan and others",
    collaboration = "Lattice Parton (LPC)",
    title = "{Nonperturbative determination of the Collins-Soper kernel from quasitransverse-momentum-dependent wave functions}",
    eprint = "2204.00200",
    archivePrefix = "arXiv",
    primaryClass = "hep-lat",
    doi = "10.1103/PhysRevD.106.034509",
    journal = "Phys. Rev. D",
    volume = "106",
    number = "3",
    pages = "034509",
    year = "2022"
}

@article{Zhang:2022xuw,
    author = "Zhang, Kuan and Ji, Xiangdong and Yang, Yi-Bo and Yao, Fei and Zhang, Jian-Hui",
    collaboration = "Lattice Parton (LPC)",
    title = "{Renormalization of Transverse-Momentum-Dependent Parton Distribution on the Lattice}",
    eprint = "2205.13402",
    archivePrefix = "arXiv",
    primaryClass = "hep-lat",
    doi = "10.1103/PhysRevLett.129.082002",
    journal = "Phys. Rev. Lett.",
    volume = "129",
    number = "8",
    pages = "082002",
    year = "2022"
}

@article{Zhu:2022bja,
    author = "Zhu, Ruilin and Ji, Yao and Zhang, Jian-Hui and Zhao, Shuai",
    title = "{Gluon transverse-momentum-dependent distributions from large-momentum effective theory}",
    eprint = "2209.05443",
    archivePrefix = "arXiv",
    primaryClass = "hep-ph",
    reportNumber = "TUM-HEP-1417/22, JLAB-THY-22-3720, JLAB-THY-22-3720",
    doi = "10.1007/JHEP02(2023)114",
    journal = "JHEP",
    volume = "02",
    pages = "114",
    year = "2023"
}

@article{Deng:2023csv,
    author = "Deng, Zhi-Fu and Han, Chao and Wang, Wei and Zeng, Jun and Zhang, Jia-Lu",
    title = "{Light-cone distribution amplitudes of a light baryon in large-momentum effective theory}",
    eprint = "2304.09004",
    archivePrefix = "arXiv",
    primaryClass = "hep-ph",
    doi = "10.1007/JHEP07(2023)191",
    journal = "JHEP",
    volume = "07",
    pages = "191",
    year = "2023"
}

@article{Ji:2023pba,
    author = "Ji, Xiangdong and Liu, Yizhuang and Su, Yushan",
    title = "{Threshold resummation for computing large-x parton distribution through large-momentum effective theory}",
    eprint = "2305.04416",
    archivePrefix = "arXiv",
    primaryClass = "hep-ph",
    doi = "10.1007/JHEP08(2023)037",
    journal = "JHEP",
    volume = "08",
    pages = "037",
    year = "2023"
}

@article{LatticeParton:2023xdl,
    author = "Chu, Min-Huan and others",
    collaboration = "Lattice Parton",
    title = "{Transverse-momentum-dependent wave functions of the pion from lattice QCD}",
    eprint = "2302.09961",
    archivePrefix = "arXiv",
    primaryClass = "hep-lat",
    doi = "10.1103/PhysRevD.109.L091503",
    journal = "Phys. Rev. D",
    volume = "109",
    number = "9",
    pages = "L091503",
    year = "2024"
}

@article{Zhao:2023ptv,
    author = "Zhao, Yong",
    title = "{Transverse Momentum Distributions from Lattice QCD without Wilson Lines}",
    eprint = "2311.01391",
    archivePrefix = "arXiv",
    primaryClass = "hep-ph",
    doi = "10.1103/PhysRevLett.133.241904",
    journal = "Phys. Rev. Lett.",
    volume = "133",
    number = "24",
    pages = "241904",
    year = "2024"
}

@article{Liu:2023onm,
    author = "Liu, Yizhuang and Su, Yushan",
    title = "{Renormalon cancellation and linear power correction to threshold-like asymptotics of space-like parton correlators}",
    eprint = "2311.06907",
    archivePrefix = "arXiv",
    primaryClass = "hep-ph",
    doi = "10.1007/JHEP02(2024)204",
    journal = "JHEP",
    volume = "2024",
    pages = "204",
    year = "2024"
}

@article{Avkhadiev:2024mgd,
    author = "Avkhadiev, Artur and Shanahan, Phiala E. and Wagman, Michael L. and Zhao, Yong",
    title = "{Determination of the Collins-Soper Kernel from Lattice QCD}",
    eprint = "2402.06725",
    archivePrefix = "arXiv",
    primaryClass = "hep-lat",
    reportNumber = "FERMILAB-PUB-24-0037-T, MIT-CTP/5677",
    doi = "10.1103/PhysRevLett.132.231901",
    journal = "Phys. Rev. Lett.",
    volume = "132",
    number = "23",
    pages = "231901",
    year = "2024"
}

@article{Good:2024iur,
    author = "Good, William and Hasan, Kinza and Lin, Huey-Wen",
    title = "{Toward the first gluon parton distribution from the LaMET}",
    eprint = "2409.02750",
    archivePrefix = "arXiv",
    primaryClass = "hep-lat",
    reportNumber = "MSUHEP-24-011",
    doi = "10.1088/1361-6471/ada815",
    journal = "J. Phys. G",
    volume = "52",
    number = "3",
    pages = "035105",
    year = "2025"
}

@article{Han:2024cht,
    author = "Han, Chao and Wang, Wei and Zhang, Jia-Lu and Zhang, Jian-Hui",
    title = "{Power corrections to quasidistribution amplitudes of a heavy meson}",
    eprint = "2408.13486",
    archivePrefix = "arXiv",
    primaryClass = "hep-ph",
    doi = "10.1103/PhysRevD.110.094038",
    journal = "Phys. Rev. D",
    volume = "110",
    number = "9",
    pages = "094038",
    year = "2024"
}

@article{Han:2024min,
    author = {Han, Xue-Ying and Hua, Jun and Ji, Xiangdong and L{\"u}, Cai-Dian and Wang, Wei and Xu, Ji and Zhang, Qi-An and Zhao, Shuai},
    title = "{Realistic method to access heavy meson light-cone distribution amplitudes from first-principle}",
    eprint = "2403.17492",
    archivePrefix = "arXiv",
    primaryClass = "hep-ph",
    doi = "10.1103/2t8s-w8t6",
    journal = "Phys. Rev. D",
    volume = "111",
    number = "11",
    pages = "L111503",
    year = "2025"
}

@article{Holligan:2024umc,
    author = "Holligan, Jack and Lin, Huey-Wen",
    title = "{Pion valence quark distribution at physical pion mass of N $_{f}$ = 2 + 1 + 1 lattice QCD}",
    eprint = "2404.14525",
    archivePrefix = "arXiv",
    primaryClass = "hep-lat",
    reportNumber = "MSUHEP-23-032",
    doi = "10.1088/1361-6471/ad3162",
    journal = "J. Phys. G",
    volume = "51",
    number = "6",
    pages = "065101",
    year = "2024"
}

@article{Holligan:2024wpv,
    author = "Holligan, Jack and Lin, Huey-Wen",
    title = "{Nucleon helicity parton distribution function in the continuum limit with self-renormalization}",
    eprint = "2405.18238",
    archivePrefix = "arXiv",
    primaryClass = "hep-lat",
    reportNumber = "MSUHEP-24-003",
    doi = "10.1016/j.physletb.2024.138731",
    journal = "Phys. Lett. B",
    volume = "854",
    pages = "138731",
    year = "2024"
}

@article{Ji:2024hit,
    author = "Ji, Xiangdong and Liu, Yizhuang and Su, Yushan and Zhang, Rui",
    title = "{Effects of threshold resummation for large-x PDF in large momentum effective theory}",
    eprint = "2410.12910",
    archivePrefix = "arXiv",
    primaryClass = "hep-ph",
    doi = "10.1007/JHEP03(2025)045",
    journal = "JHEP",
    volume = "03",
    pages = "045",
    year = "2025"
}

@article{Wang:2024wwa,
    author = "Wang, Wei and Xu, Ji and Zhang, Qi-An and Zhao, Shuai",
    title = "{Mass renormalization group of heavy meson light-cone distribution amplitude in QCD}",
    eprint = "2411.07101",
    archivePrefix = "arXiv",
    primaryClass = "hep-ph",
    month = "11",
    year = "2024"
}

@article{LatticeParton:2024zko,
    author = "Han, Xue-Ying and others",
    collaboration = "Lattice Parton",
    title = "{Calculation of heavy meson light-cone distribution amplitudes from lattice QCD}",
    eprint = "2410.18654",
    archivePrefix = "arXiv",
    primaryClass = "hep-lat",
    doi = "10.1103/PhysRevD.111.034503",
    journal = "Phys. Rev. D",
    volume = "111",
    number = "3",
    pages = "034503",
    year = "2025"
}

@article{Zhang:2024omt,
    author = "Zhang, Kuan and Huo, Yi-Kai and Ji, Xiangdong and Schaefer, Andreas and Shi, Chun-Jiang and Sun, Peng and Wang, Wei and Yang, Yi-Bo and Zhang, Jian-Hui",
    collaboration = "Lattice Parton",
    title = "{Impact of gauge fixing precision on the continuum limit of nonlocal quark-bilinear lattice operators}",
    eprint = "2405.14097",
    archivePrefix = "arXiv",
    primaryClass = "hep-lat",
    doi = "10.1103/PhysRevD.110.074505",
    journal = "Phys. Rev. D",
    volume = "110",
    number = "7",
    pages = "074505",
    year = "2024"
}

@article{Bollweg:2025iol,
    author = "Bollweg, Dennis and Gao, Xiang and He, Jinchen and Mukherjee, Swagato and Zhao, Yong",
    title = "{Transverse-momentum-dependent pion structures from lattice QCD: Collins-Soper kernel, soft factor, TMDWF, and TMDPDF}",
    eprint = "2504.04625",
    archivePrefix = "arXiv",
    primaryClass = "hep-lat",
    doi = "10.1103/j3n6-8kxy",
    journal = "Phys. Rev. D",
    volume = "112",
    number = "3",
    pages = "034501",
    year = "2025"
}

@article{Wang:2025uap,
    author = "Wang, Wei and Xu, Ji and Zhang, Qi-An and Zhao, Shuai",
    title = "{Factorization formula connecting the shape functions of heavy meson in QCD and heavy quark effective theory}",
    eprint = "2504.18018",
    archivePrefix = "arXiv",
    primaryClass = "hep-ph",
    doi = "10.1103/1547-t91t",
    journal = "Phys. Rev. D",
    volume = "112",
    number = "5",
    pages = "054044",
    year = "2025"
}

@article{Ji:2025mvk,
    author = "Ji, Yao and Yao, Fei and Zhang, Jian-Hui",
    title = "{Extracting Meson Distribution Amplitudes from Nonlocal Euclidean Correlations at Next-to-Next-to-Leading Order}",
    eprint = "2504.09367",
    archivePrefix = "arXiv",
    primaryClass = "hep-ph",
    month = "4",
    year = "2025"
}

@article{Chen:2025cxr,
    author = "Chen, Jiunn-Wei and others",
    title = "{LaMET's Asymptotic Extrapolation vs. Inverse Problem}",
    eprint = "2505.14619",
    archivePrefix = "arXiv",
    primaryClass = "hep-lat",
    month = "5",
    year = "2025"
}

@article{Radyushkin:2017lvu,
    author = "Radyushkin, A. V.",
    title = "{Quark pseudodistributions at short distances}",
    eprint = "1710.08813",
    archivePrefix = "arXiv",
    primaryClass = "hep-ph",
    reportNumber = "JLAB-THY-17-2579",
    doi = "10.1016/j.physletb.2018.04.023",
    journal = "Phys. Lett. B",
    volume = "781",
    pages = "433--442",
    year = "2018"
}

@article{Zhang:2018ggy,
    author = "Zhang, Jian-Hui and Chen, Jiunn-Wei and Monahan, Christopher",
    title = "{Parton distribution functions from reduced Ioffe-time distributions}",
    eprint = "1801.03023",
    archivePrefix = "arXiv",
    primaryClass = "hep-ph",
    doi = "10.1103/PhysRevD.97.074508",
    journal = "Phys. Rev. D",
    volume = "97",
    number = "7",
    pages = "074508",
    year = "2018"
}

@article{Karpie:2018zaz,
    author = "Karpie, Joseph and Orginos, Kostas and Zafeiropoulos, Savvas",
    title = "{Moments of Ioffe time parton distribution functions from non-local matrix elements}",
    eprint = "1807.10933",
    archivePrefix = "arXiv",
    primaryClass = "hep-lat",
    doi = "10.1007/JHEP11(2018)178",
    journal = "JHEP",
    volume = "11",
    pages = "178",
    year = "2018"
}

@article{Joo:2019jct,
    author = "Jo{\'o}, B{\'a}lint and Karpie, Joseph and Orginos, Kostas and Radyushkin, Anatoly and Richards, David and Zafeiropoulos, Savvas",
    title = "{Parton Distribution Functions from Ioffe time pseudo-distributions}",
    eprint = "1908.09771",
    archivePrefix = "arXiv",
    primaryClass = "hep-lat",
    doi = "10.1007/JHEP12(2019)081",
    journal = "JHEP",
    volume = "12",
    pages = "081",
    year = "2019"
}

@article{Joo:2019bzr,
    author = "Jo{\'o}, B{\'a}lint and Karpie, Joseph and Orginos, Kostas and Radyushkin, Anatoly V. and Richards, David G. and Sufian, Raza Sabbir and Zafeiropoulos, Savvas",
    title = "{Pion valence structure from Ioffe-time parton pseudodistribution functions}",
    eprint = "1909.08517",
    archivePrefix = "arXiv",
    primaryClass = "hep-lat",
    reportNumber = "JLAB-THY-19-3038",
    doi = "10.1103/PhysRevD.100.114512",
    journal = "Phys. Rev. D",
    volume = "100",
    number = "11",
    pages = "114512",
    year = "2019"
}

@article{HadStruc:2021qdf,
    author = "Egerer, Colin and others",
    collaboration = "HadStruc",
    title = "{Transversity parton distribution function of the nucleon using the pseudodistribution approach}",
    eprint = "2111.01808",
    archivePrefix = "arXiv",
    primaryClass = "hep-lat",
    reportNumber = "JLAB-THY-21-3521",
    doi = "10.1103/PhysRevD.105.034507",
    journal = "Phys. Rev. D",
    volume = "105",
    number = "3",
    pages = "034507",
    year = "2022"
}

@article{Bhat:2022zrw,
    author = "Bhat, Manjunath and Chomicki, Wojciech and Cichy, Krzysztof and Constantinou, Martha and Green, Jeremy R. and Scapellato, Aurora",
    title = "{Continuum limit of parton distribution functions from the pseudodistribution approach on the lattice}",
    eprint = "2205.07585",
    archivePrefix = "arXiv",
    primaryClass = "hep-lat",
    doi = "10.1103/PhysRevD.106.054504",
    journal = "Phys. Rev. D",
    volume = "106",
    number = "5",
    pages = "054504",
    year = "2022"
}

@article{Kovner:2024pwl,
    author = "Kovner, Daniel and Karpie, Joe and Orginos, Konstantinos and Radyushkin, Anatoly and Zafeiropoulos, Savvas",
    collaboration = "HadStruc",
    title = "{Extracting the Pion Distribution Amplitude from Lattice QCD through Pseudo-Distributions}",
    eprint = "2401.06858",
    archivePrefix = "arXiv",
    primaryClass = "hep-lat",
    reportNumber = "JLAB-THY-24-3982",
    doi = "10.22323/1.453.0300",
    journal = "PoS",
    volume = "LATTICE2023",
    pages = "300",
    year = "2024"
}

@article{Bhattacharya:2024qpp,
    author = "Bhattacharya, Shohini and Cichy, Krzysztof and Constantinou, Martha and Metz, Andreas and Nurminen, Niilo and Steffens, Fernanda",
    title = "{Generalized parton distributions from the pseudodistribution approach on the lattice}",
    eprint = "2405.04414",
    archivePrefix = "arXiv",
    primaryClass = "hep-lat",
    reportNumber = "LA-UR-24-23903",
    doi = "10.1103/PhysRevD.110.054502",
    journal = "Phys. Rev. D",
    volume = "110",
    number = "5",
    pages = "054502",
    year = "2024"
}

@article{HadStruc:2024rix,
    author = "Dutrieux, Herv{\'e} and Edwards, Robert G. and Egerer, Colin and Karpie, Joseph and Monahan, Christopher and Orginos, Kostas and Radyushkin, Anatoly and Richards, David and Romero, Eloy and Zafeiropoulos, Savvas",
    collaboration = "HadStruc",
    title = "{Towards unpolarized GPDs from pseudo-distributions}",
    eprint = "2405.10304",
    archivePrefix = "arXiv",
    primaryClass = "hep-lat",
    reportNumber = "JLAB-THY-24-4059, JLAB-THY-24-4059",
    doi = "10.1007/JHEP08(2024)162",
    journal = "JHEP",
    volume = "08",
    pages = "162",
    year = "2024"
}

@article{NieMiera:2025inn,
    author = "NieMiera, Alex and Good, William and Lin, Huey-Wen",
    title = "{Kaon gluon parton distribution and momentum fraction from 2+1+1 lattice QCD with high statistics}",
    eprint = "2506.03002",
    archivePrefix = "arXiv",
    primaryClass = "hep-lat",
    reportNumber = "MSUHEP-25-025",
    doi = "10.1103/89mq-gx33",
    journal = "Phys. Rev. D",
    volume = "112",
    number = "7",
    pages = "074504",
    year = "2025"
}

@article{Bali:2017gfr,
    author = "Bali, Gunnar S. and others",
    editor = "Della Morte, M. and Fritzsch, P. and G{\'a}miz S{\'a}nchez, E. and Pena Ruano, C.",
    title = "{Pion distribution amplitude from Euclidean correlation functions}",
    eprint = "1709.04325",
    archivePrefix = "arXiv",
    primaryClass = "hep-lat",
    doi = "10.1140/epjc/s10052-018-5700-9",
    journal = "Eur. Phys. J. C",
    volume = "78",
    number = "3",
    pages = "217",
    year = "2018"
}

@article{Sufian:2019bol,
    author = "Sufian, Raza Sabbir and Karpie, Joseph and Egerer, Colin and Orginos, Kostas and Qiu, Jian-Wei and Richards, David G.",
    title = "{Pion Valence Quark Distribution from Matrix Element Calculated in Lattice QCD}",
    eprint = "1901.03921",
    archivePrefix = "arXiv",
    primaryClass = "hep-lat",
    reportNumber = "JLAB-THY-19-2847",
    doi = "10.1103/PhysRevD.99.074507",
    journal = "Phys. Rev. D",
    volume = "99",
    number = "7",
    pages = "074507",
    year = "2019"
}

@article{Bali:2018spj,
    author = {Bali, Gunnar S. and Braun, Vladimir M. and Gl{\"a}{\ss}le, Benjamin and G{\"o}ckeler, Meinulf and Gruber, Michael and Hutzler, Fabian and Korcyl, Piotr and Sch{\"a}fer, Andreas and Wein, Philipp and Zhang, Jian-Hui},
    title = "{Pion distribution amplitude from Euclidean correlation functions: Exploring universality and higher-twist effects}",
    eprint = "1807.06671",
    archivePrefix = "arXiv",
    primaryClass = "hep-lat",
    doi = "10.1103/PhysRevD.98.094507",
    journal = "Phys. Rev. D",
    volume = "98",
    number = "9",
    pages = "094507",
    year = "2018"
}

@article{Sufian:2020vzb,
    author = "Sufian, Raza Sabbir and Egerer, Colin and Karpie, Joseph and Edwards, Robert G. and Jo{\'o}, B{\'a}lint and Ma, Yan-Qing and Orginos, Kostas and Qiu, Jian-Wei and Richards, David G.",
    title = "{Pion Valence Quark Distribution from Current-Current Correlation in Lattice QCD}",
    eprint = "2001.04960",
    archivePrefix = "arXiv",
    primaryClass = "hep-lat",
    reportNumber = "JLAB-THY-20-3131",
    doi = "10.1103/PhysRevD.102.054508",
    journal = "Phys. Rev. D",
    volume = "102",
    number = "5",
    pages = "054508",
    year = "2020"
}

@article{Zimmermann:2024zde,
    author = {Zimmermann, Christian and Sch{\"a}fer, Andreas},
    title = "{Valence quark PDFs of the proton from two-current correlations in lattice QCD}",
    eprint = "2405.07712",
    archivePrefix = "arXiv",
    primaryClass = "hep-lat",
    doi = "10.1103/PhysRevD.110.074503",
    journal = "Phys. Rev. D",
    volume = "110",
    number = "7",
    pages = "074503",
    year = "2024"
}

@article{LatticeParton:2024vck,
    author = "Chu, Min-Huan and others",
    collaboration = "Lattice Parton",
    title = "{Light cone distribution amplitude for the {\ensuremath{\Lambda}} baryon from lattice QCD}",
    eprint = "2411.12554",
    archivePrefix = "arXiv",
    primaryClass = "hep-lat",
    doi = "10.1103/PhysRevD.111.034510",
    journal = "Phys. Rev. D",
    volume = "111",
    number = "3",
    pages = "034510",
    year = "2025"
}

@article{Braun:1999te,
    author = "Braun, Vladimir M. and Derkachov, Sergey E. and Korchemsky, G. P. and Manashov, A. N.",
    title = "{Baryon distribution amplitudes in QCD}",
    eprint = "hep-ph/9902375",
    archivePrefix = "arXiv",
    reportNumber = "LPT-ORSAY-98-83, NORDITA-99-11-HE, SPBU-IP-99-04",
    doi = "10.1016/S0550-3213(99)00265-5",
    journal = "Nucl. Phys. B",
    volume = "553",
    pages = "355--426",
    year = "1999"
}

@article{Han:2024ucv,
    author = "Han, Chao and Wang, Wei and Zeng, Jun and Zhang, Jia-Lu",
    title = "{Lightcone and quasi distribution amplitudes for light octet and decuplet baryons}",
    eprint = "2404.04855",
    archivePrefix = "arXiv",
    primaryClass = "hep-ph",
    doi = "10.1007/JHEP07(2024)019",
    journal = "JHEP",
    volume = "07",
    pages = "019",
    year = "2024"
}

@article{Gao:2021dbh,
    author = "Gao, Xiang and Hanlon, Andrew D. and Mukherjee, Swagato and Petreczky, Peter and Scior, Philipp and Syritsyn, Sergey and Zhao, Yong",
    title = "{Lattice QCD Determination of the Bjorken-x Dependence of Parton Distribution Functions at Next-to-Next-to-Leading Order}",
    eprint = "2112.02208",
    archivePrefix = "arXiv",
    primaryClass = "hep-lat",
    doi = "10.1103/PhysRevLett.128.142003",
    journal = "Phys. Rev. Lett.",
    volume = "128",
    number = "14",
    pages = "142003",
    year = "2022"
}

@article{Chou:2022drv,
    author = "Chou, Chien-Yu and Chen, Jiunn-Wei",
    title = "{One-loop hybrid renormalization matching kernels for quasiparton distributions}",
    eprint = "2204.08343",
    archivePrefix = "arXiv",
    primaryClass = "hep-lat",
    doi = "10.1103/PhysRevD.106.014507",
    journal = "Phys. Rev. D",
    volume = "106",
    number = "1",
    pages = "014507",
    year = "2022"
}

@article{Hua:2022wop,
    author = "Hua, Jun and Chu, Min-Huan and Sun, Peng and Wang, Wei and Xu, Ji and Yang, Yi-Bo and Zhang, Jian-Hui and Zhang, Qi-An",
    collaboration = "Lattice Parton (LPC)",
    title = "{Distribution Amplitudes of K{\ensuremath{*}} and {\ensuremath{\varphi}} from Lattice QCD}",
    doi = "10.22323/1.396.0322",
    journal = "PoS",
    volume = "LATTICE2021",
    pages = "322",
    year = "2022"
}

@article{Su:2022fiu,
    author = "Su, Yushan and Holligan, Jack and Ji, Xiangdong and Yao, Fei and Zhang, Jian-Hui and Zhang, Rui",
    title = "{Resumming quark's longitudinal momentum logarithms in LaMET expansion of lattice PDFs}",
    eprint = "2209.01236",
    archivePrefix = "arXiv",
    primaryClass = "hep-ph",
    doi = "10.1016/j.nuclphysb.2023.116201",
    journal = "Nucl. Phys. B",
    volume = "991",
    pages = "116201",
    year = "2023"
}

@article{Ji:2022ezo,
    author = "Ji, Xiangdong",
    title = "{Large-Momentum Effective Theory vs. Short-Distance Operator Expansion: Contrast and Complementarity}",
    eprint = "2209.09332",
    archivePrefix = "arXiv",
    primaryClass = "hep-lat",
    doi = "10.34133/research.0695",
    journal = "Research",
    volume = "8",
    pages = "0695",
    year = "2025"
}

@article{Gao:2022ytj,
    author = "Gao, Xiang and Hanlon, Andrew D. and Mukherjee, Swagato and Petreczky, Peter and Scior, Philipp and Syritsyn, Sergey and Zhao, Yong",
    title = "{Lattice QCD Determination of the Bjorken-{\ensuremath{\boldsymbol{\mathit{x}}}} Dependence of PDFs at NNLO}",
    doi = "10.22323/1.430.0104",
    journal = "PoS",
    volume = "LATTICE2022",
    pages = "104",
    year = "2023"
}

@article{Ji:2022thb,
    author = "Yao, Fei and Ji, Yao and Zhang, Jian-Hui",
    title = "{Connecting Euclidean to light-cone correlations: from flavor nonsinglet in forward kinematics to flavor singlet in non-forward kinematics}",
    eprint = "2212.14415",
    archivePrefix = "arXiv",
    primaryClass = "hep-ph",
    reportNumber = "TUM-HEP-1446/22",
    doi = "10.1007/JHEP11(2023)021",
    journal = "JHEP",
    volume = "11",
    pages = "021",
    year = "2023"
}

@phdthesis{Zhang:2023tnc,
    author = "Zhang, Rui",
    title = "{Lattice Quantum Chromodynamics (QCD) Calculations of Parton Physics with Leading Power Accuracy in Large Momentum Expansion}",
    school = "Maryland U., College Park, Maryland U.",
    year = "2023"
}

@article{Zhang:2023bxs,
    author = "Zhang, Rui and Holligan, Jack and Ji, Xiangdong and Su, Yushan",
    title = "{Leading power accuracy in lattice calculations of parton distributions}",
    eprint = "2305.05212",
    archivePrefix = "arXiv",
    primaryClass = "hep-lat",
    doi = "10.1016/j.physletb.2023.138081",
    journal = "Phys. Lett. B",
    volume = "844",
    pages = "138081",
    year = "2023"
}

@article{Gao:2023lny,
    author = "Gao, Xiang and Liu, Wei-Yang and Zhao, Yong",
    title = "{Parton distributions from boosted fields in the Coulomb gauge}",
    eprint = "2306.14960",
    archivePrefix = "arXiv",
    primaryClass = "hep-ph",
    doi = "10.1103/PhysRevD.109.094506",
    journal = "Phys. Rev. D",
    volume = "109",
    number = "9",
    pages = "094506",
    year = "2024"
}

@article{Chen:2024rgi,
    author = "Chen, Chen and Geng, Yiqi and Liu, Liuming and Sun, Peng and Yang, Yi-Bo and Yao, Fei and Zhang, Jian-Hui and Zhang, Kuan",
    collaboration = "CLQCD, Lattice Parton",
    title = "{Parton distribution function of a deuteronlike dibaryon system from lattice QCD}",
    eprint = "2408.12819",
    archivePrefix = "arXiv",
    primaryClass = "hep-lat",
    doi = "10.1103/PhysRevD.111.074506",
    journal = "Phys. Rev. D",
    volume = "111",
    number = "7",
    pages = "074506",
    year = "2025"
}

@article{Holligan:2025ydm,
    author = "Holligan, Jack and Lin, Huey-Wen and Zhang, Rui and Zhao, Yong",
    title = "{Resummation for lattice QCD calculation of generalized parton distributions at nonzero skewness}",
    eprint = "2501.19225",
    archivePrefix = "arXiv",
    primaryClass = "hep-ph",
    reportNumber = "MSUHEP-24-023",
    doi = "10.1007/JHEP07(2025)241",
    journal = "JHEP",
    volume = "207",
    pages = "241",
    year = "2025"
}

@article{Good:2025daz,
    author = "Good, William and Yao, Fei and Lin, Huey-Wen",
    title = "{First Nucleon Gluon PDF from Large Momentum Effective Theory}",
    eprint = "2505.13321",
    archivePrefix = "arXiv",
    primaryClass = "hep-lat",
    month = "5",
    year = "2025"
}

@article{Han:2023xbl,
    author = "Han, Chao and Su, Yushan and Wang, Wei and Zhang, Jia-Lu",
    title = "{Hybrid renormalization for quasi distribution amplitudes of a light baryon}",
    eprint = "2308.16793",
    archivePrefix = "arXiv",
    primaryClass = "hep-ph",
    doi = "10.1007/JHEP12(2023)044",
    journal = "JHEP",
    volume = "12",
    pages = "044",
    year = "2023"
}

@article{Han:2023hgy,
    author = "Han, Chao and Zhang, Jialu",
    title = "{Light baryon spatial correlators at short distances}",
    eprint = "2311.02669",
    archivePrefix = "arXiv",
    primaryClass = "hep-ph",
    doi = "10.1103/PhysRevD.109.014034",
    journal = "Phys. Rev. D",
    volume = "109",
    number = "1",
    pages = "014034",
    year = "2024"
}

@article{Chen:2016fxx,
    author = "Chen, Jiunn-Wei and Ji, Xiangdong and Zhang, Jian-Hui",
    title = "{Improved quasi parton distribution through Wilson line renormalization}",
    eprint = "1609.08102",
    archivePrefix = "arXiv",
    primaryClass = "hep-ph",
    doi = "10.1016/j.nuclphysb.2016.12.004",
    journal = "Nucl. Phys. B",
    volume = "915",
    pages = "1--9",
    year = "2017"
}

@article{Ji:2017oey,
    author = "Ji, Xiangdong and Zhang, Jian-Hui and Zhao, Yong",
    title = "{Renormalization in Large Momentum Effective Theory of Parton Physics}",
    eprint = "1706.08962",
    archivePrefix = "arXiv",
    primaryClass = "hep-ph",
    doi = "10.1103/PhysRevLett.120.112001",
    journal = "Phys. Rev. Lett.",
    volume = "120",
    number = "11",
    pages = "112001",
    year = "2018"
}

@article{Ishikawa:2017faj,
    author = "Ishikawa, Tomomi and Ma, Yan-Qing and Qiu, Jian-Wei and Yoshida, Shinsuke",
    title = "{Renormalizability of quasiparton distribution functions}",
    eprint = "1707.03107",
    archivePrefix = "arXiv",
    primaryClass = "hep-ph",
    doi = "10.1103/PhysRevD.96.094019",
    journal = "Phys. Rev. D",
    volume = "96",
    number = "9",
    pages = "094019",
    year = "2017"
}

@article{Green:2017xeu,
    author = "Green, Jeremy and Jansen, Karl and Steffens, Fernanda",
    title = "{Nonperturbative Renormalization of Nonlocal Quark Bilinears for Parton Quasidistribution Functions on the Lattice Using an Auxiliary Field}",
    eprint = "1707.07152",
    archivePrefix = "arXiv",
    primaryClass = "hep-lat",
    reportNumber = "DESY-17-109, DESY 17-109",
    doi = "10.1103/PhysRevLett.121.022004",
    journal = "Phys. Rev. Lett.",
    volume = "121",
    number = "2",
    pages = "022004",
    year = "2018"
}

@article{Ji:1995tm,
    author = "Ji, Xiang-Dong",
    title = "{Matching perturbative and nonperturbative physics with power accuracy in heavy quark effective theory}",
    eprint = "hep-ph/9507322",
    archivePrefix = "arXiv",
    reportNumber = "MIT-CTP-2453",
    month = "7",
    year = "1995"
}

@article{Beneke:1998ui,
    author = "Beneke, M.",
    title = "{Renormalons}",
    eprint = "hep-ph/9807443",
    archivePrefix = "arXiv",
    reportNumber = "CERN-TH-98-233",
    doi = "10.1016/S0370-1573(98)00130-6",
    journal = "Phys. Rept.",
    volume = "317",
    pages = "1--142",
    year = "1999"
}

@article{Bauer:2011ws,
    author = "Bauer, Clemens and Bali, Gunnar S. and Pineda, Antonio",
    title = "{Compelling Evidence of Renormalons in QCD from High Order Perturbative Expansions}",
    eprint = "1111.3946",
    archivePrefix = "arXiv",
    primaryClass = "hep-ph",
    doi = "10.1103/PhysRevLett.108.242002",
    journal = "Phys. Rev. Lett.",
    volume = "108",
    pages = "242002",
    year = "2012"
}

@article{Bali:2013pla,
    author = "Bali, Gunnar S. and Bauer, Clemens and Pineda, Antonio and Torrero, Christian",
    title = "{Perturbative expansion of the energy of static sources at large orders in four-dimensional SU(3) gauge theory}",
    eprint = "1303.3279",
    archivePrefix = "arXiv",
    primaryClass = "hep-lat",
    doi = "10.1103/PhysRevD.87.094517",
    journal = "Phys. Rev. D",
    volume = "87",
    pages = "094517",
    year = "2013"
}

@article{LatticePartonCollaborationLPC:2021xdx,
    author = "Huo, Yi-Kai and others",
    collaboration = "Lattice Parton (LPC)",
    title = "{Self-renormalization of quasi-light-front correlators on the lattice}",
    eprint = "2103.02965",
    archivePrefix = "arXiv",
    primaryClass = "hep-lat",
    doi = "10.1016/j.nuclphysb.2021.115443",
    journal = "Nucl. Phys. B",
    volume = "969",
    pages = "115443",
    year = "2021"
}

@article{CLQCD:2023sdb,
    author = "Hu, Zhi-Cheng and others",
    collaboration = "CLQCD",
    title = "{Quark masses and low-energy constants in the continuum from the tadpole-improved clover ensembles}",
    eprint = "2310.00814",
    archivePrefix = "arXiv",
    primaryClass = "hep-lat",
    doi = "10.1103/PhysRevD.109.054507",
    journal = "Phys. Rev. D",
    volume = "109",
    number = "5",
    pages = "054507",
    year = "2024"
}

@article{Liu:2022gxf,
    author = "Liu, Hang and He, Jinchen and Liu, Liuming and Sun, Peng and Wang, Wei and Yang, Yi-Bo and Zhang, Qi-An",
    title = "{Exploring hidden-charm and hidden-strange hexaquark states from lattice QCD}",
    eprint = "2207.00183",
    archivePrefix = "arXiv",
    primaryClass = "hep-lat",
    doi = "10.1007/s11433-023-2205-0",
    journal = "Sci. China Phys. Mech. Astron.",
    volume = "67",
    number = "1",
    pages = "211011",
    year = "2024"
}

@article{Xing:2022ijm,
    author = "Xing, Hanyang and Liang, Jian and Liu, Liuming and Sun, Peng and Yang, Yi-Bo",
    title = "{First observation of the hidden-charm pentaquarks on lattice}",
    eprint = "2210.08555",
    archivePrefix = "arXiv",
    primaryClass = "hep-lat",
    month = "10",
    year = "2022"
}

@article{Yan:2024yuq,
    author = "Yan, Haobo and Liu, Chuan and Liu, Liuming and Meng, Yu and Xing, Hanyang",
    title = "{Pion mass dependence in D{\ensuremath{\pi}} scattering and the D0*(2300) resonance from lattice QCD}",
    eprint = "2404.13479",
    archivePrefix = "arXiv",
    primaryClass = "hep-lat",
    doi = "10.1103/PhysRevD.111.014503",
    journal = "Phys. Rev. D",
    volume = "111",
    number = "1",
    pages = "014503",
    year = "2025"
}

@article{Zhang:2021oja,
    author = "Zhang, Qi-An and others",
    title = "{First lattice QCD calculation of semileptonic decays of charmed-strange baryons {\ensuremath{\Xi}}$_{c}$ *}",
    eprint = "2103.07064",
    archivePrefix = "arXiv",
    primaryClass = "hep-lat",
    doi = "10.1088/1674-1137/ac2b12",
    journal = "Chin. Phys. C",
    volume = "46",
    number = "1",
    pages = "011002",
    year = "2022"
}

@article{Liu:2023pwr,
    author = "Liu, Hang and Wang, Wei and Zhang, Qi-An",
    title = "{Improved method to determine the {\ensuremath{\Xi}}c-{\ensuremath{\Xi}}c' mixing}",
    eprint = "2309.05432",
    archivePrefix = "arXiv",
    primaryClass = "hep-ph",
    doi = "10.1103/PhysRevD.109.036037",
    journal = "Phys. Rev. D",
    volume = "109",
    number = "3",
    pages = "036037",
    year = "2024"
}

@article{Bali:2016lva,
    author = {Bali, Gunnar S. and Lang, Bernhard and Musch, Bernhard U. and Sch{\"a}fer, Andreas},
    title = "{Novel quark smearing for hadrons with high momenta in lattice QCD}",
    eprint = "1602.05525",
    archivePrefix = "arXiv",
    primaryClass = "hep-lat",
    doi = "10.1103/PhysRevD.93.094515",
    journal = "Phys. Rev. D",
    volume = "93",
    number = "9",
    pages = "094515",
    year = "2016"
}

@article{Hasenfratz:2001hp,
    author = "Hasenfratz, Anna and Knechtli, Francesco",
    title = "{Flavor symmetry and the static potential with hypercubic blocking}",
    eprint = "hep-lat/0103029",
    archivePrefix = "arXiv",
    reportNumber = "COLO-HEP-462",
    doi = "10.1103/PhysRevD.64.034504",
    journal = "Phys. Rev. D",
    volume = "64",
    pages = "034504",
    year = "2001"
}

@article{DeGrand:2002vu,
    author = "DeGrand, Thomas A. and Hasenfratz, Anna and Kovacs, Tamas G.",
    title = "{Improving the chiral properties of lattice fermions}",
    eprint = "hep-lat/0211006",
    archivePrefix = "arXiv",
    doi = "10.1103/PhysRevD.67.054501",
    journal = "Phys. Rev. D",
    volume = "67",
    pages = "054501",
    year = "2003"
}

@article{Zhang:2025hyo,
    author = "Zhang, Rui and Grebe, Anthony V. and Hackett, Daniel C. and Wagman, Michael L. and Zhao, Yong",
    title = "{Kinematically enhanced interpolating operators for boosted hadrons}",
    eprint = "2501.00729",
    archivePrefix = "arXiv",
    primaryClass = "hep-lat",
    reportNumber = "FERMILAB-PUB-24-0968-T",
    doi = "10.1103/6dh4-6k4t",
    journal = "Phys. Rev. D",
    volume = "112",
    number = "5",
    pages = "L051502",
    year = "2025"
}

@article{Jay:2020jkz,
    author = "Jay, William I. and Neil, Ethan T.",
    title = "{Bayesian model averaging for analysis of lattice field theory results}",
    eprint = "2008.01069",
    archivePrefix = "arXiv",
    primaryClass = "stat.ME",
    reportNumber = "FERMILAB-PUB-20-374-T",
    doi = "10.1103/PhysRevD.103.114502",
    journal = "Phys. Rev. D",
    volume = "103",
    pages = "114502",
    year = "2021"
}

@article{Xiong:2025obq,
    author = "Xiong, Ao-Sheng and Hua, Jun and Wei, Ting and Yu, Fu-Sheng and Zhang, Qi-An and Zheng, Yong",
    title = "{Ill-Posedness in Limited Discrete Fourier Inversion and Regularization for Quasi Distributions in LaMET}",
    eprint = "2506.16689",
    archivePrefix = "arXiv",
    primaryClass = "hep-lat",
    month = "6",
    year = "2025"
}

@article{Jiang:2024lto,
    author = "Jiang, Xiangyu and Shi, Chunjiang and Chen, Ying and Gong, Ming and Yang, Yi-Bo",
    title = "{Use QUDA for lattice QCD calculation with Python}",
    eprint = "2411.08461",
    archivePrefix = "arXiv",
    primaryClass = "hep-lat",
    month = "11",
    year = "2024"
}

@article{Clark:2009wm,
    author = "Clark, M. A. and Babich, R. and Barros, K. and Brower, R. C. and Rebbi, C.",
    collaboration = "QUDA",
    title = "{Solving Lattice QCD systems of equations using mixed precision solvers on GPUs}",
    eprint = "0911.3191",
    archivePrefix = "arXiv",
    primaryClass = "hep-lat",
    doi = "10.1016/j.cpc.2010.05.002",
    journal = "Comput. Phys. Commun.",
    volume = "181",
    pages = "1517--1528",
    year = "2010"
}

@inproceedings{Babich:2011np,
    author = "Babich, R. and Clark, M. A. and Joo, B. and Shi, G. and Brower, R. C. and Gottlieb, S.",
    collaboration = "QUDA",
    title = "{Scaling lattice QCD beyond 100 GPUs}",
    booktitle = "{International Conference for High Performance Computing, Networking, Storage and Analysis}",
    eprint = "1109.2935",
    archivePrefix = "arXiv",
    primaryClass = "hep-lat",
    doi = "10.1145/2063384.2063478",
    month = "9",
    year = "2011"
}

@inproceedings{Clark:2016rdz,
    author = "Clark, M. A. and Jo{\'o}, B{\'a}lint and Strelchenko, Alexei and Cheng, Michael and Gambhir, Arjun and Brower, Richard. C.",
    collaboration = "QUDA",
    title = "{Accelerating lattice QCD multigrid on GPUs using fine-grained parallelization}",
    booktitle = "{International Conference for High Performance Computing, Networking, Storage and Analysis}",
    eprint = "1612.07873",
    archivePrefix = "arXiv",
    primaryClass = "hep-lat",
    reportNumber = "FERMILAB-CONF-16-638-CD",
    doi = "10.5555/3014904.3014995",
    month = "12",
    year = "2016"
}

@article{Bi:2020wpt,
    author = "Bi, Yu-Jiang and Xiao, Yi and Guo, Wei-Yi and Gong, Ming and Sun, Peng and Xu, Shun and Yang, Yi-Bo",
    title = "{Lattice QCD package GWU-code and QUDA with HIP}",
    eprint = "2001.05706",
    archivePrefix = "arXiv",
    primaryClass = "hep-lat",
    doi = "10.22323/1.363.0286",
    journal = "PoS",
    volume = "LATTICE2019",
    pages = "286",
    year = "2020"
}

@article{Braun:2000kw,
    author = "Braun, V. and Fries, R. J. and Mahnke, N. and Stein, E.",
    title = "{Higher twist distribution amplitudes of the nucleon in QCD}",
    eprint = "hep-ph/0007279",
    archivePrefix = "arXiv",
    reportNumber = "TPR-00-10",
    doi = "10.1016/S0550-3213(00)00516-2",
    journal = "Nucl. Phys. B",
    volume = "589",
    pages = "381--409",
    year = "2000",
    note = "[Erratum: Nucl.Phys.B 607, 433--433 (2001)]"
}

@article{FlavourLatticeAveragingGroupFLAG:2024oxs,
    author = "Aoki, Y. and others",
    collaboration = "Flavour Lattice Averaging Group (FLAG)",
    title = "{FLAG Review 2024}",
    eprint = "2411.04268",
    archivePrefix = "arXiv",
    primaryClass = "hep-lat",
    reportNumber = "CERN-TH-2024-192, FERMILAB-PUB-24-0785-T",
    month = "11",
    year = "2024"
}
\bibliographystyle{apsrev4-1}

\end{document}